\documentclass[12pt,twoside]{book}
\usepackage{fleqn,espcrc1}
\usepackage{axodraw}
\usepackage{epsf}
\usepackage{epsfig}
\usepackage{graphicx}
\newcommand{\eqnzero}{\setcounter{equation}{0}} 
\def\text{\textstyle}
\newcommand{\litwo}{\mbox{Li}_2}
\newcommand{\lithree}{\mbox{Li}_3}
\newcommand{\SUBR}[1]{\bigskip\fbox{\tt #1}\bigskip}

\newcommand{\BHANG}{{\tt BHANG}}

\newcommand{\nn}{\noindent}

\newcommand{\bq}{\begin{equation}}
\newcommand{\eq}{\end{equation}}
\newcommand{\ba}{\begin{eqnarray}}
\newcommand{\ea}{\end{eqnarray}}
\newcommand{\nl}{\nonumber\\}
\newcommand{\nll}{\nonumber\\}

\newcommand{\LEPI}{LEP1}
\newcommand{\LEPII}{LEP2}
\newcommand {\zf}{{\tt ZFITTER}}
\newcommand{\ee}{$e^+e^-$}
\newcommand{\oalf}{\mbox{${\cal O}(\alpha) \:$}}

\newcommand{\BS}{\bigskip}
\newcommand{\Z}{$Z$}

\newcommand{\EG}{{ e.g. }}

\newcommand{\RS}{$\sqrt{s}$}

\newcommand{\FF}{$f\bar{f}$}
\newcommand{\zetaz}{\mbox{$\zeta(2)$}}
\newcommand{\zetad}{\mbox{$\zeta(3)$}}
\newcommand{\rt} {\mbox{$r_{t}   $}}

\newcommand{\rw} {\mbox{$r_{\sss{W}}  $}}
\newcommand{\rz} {\mbox{$r_{\sss{Z}}  $}}

\newcommand{\Rw} {\mbox{$R_{\sss{W}}  $}}

\newcommand{\Rz} {\mbox{$R_{\sss{Z}}  $}}

\newcommand{\Litwo}{\mbox{${\rm{Li}}_{2}$}}
\newcommand{\nf }{\mbox{$n_f  $}}
\newcommand{\nfS}{\mbox{$n^2_f$}}
\newcommand{\mq }{\mbox{$m_q  $}}
\newcommand{\mqS}{\mbox{$m^2_q$}}
\newcommand{\mqQ}{\mbox{$m^4_q$}}
\newcommand{\mqX}{\mbox{$m^6_q$}}
\newcommand{\mqp}{\mbox{$m'_q $}}

\newcommand{\mqpQ}{\mbox{$m'^4_q$}}
\newcommand{\ms }{\mbox{$m_s  $}}

\newcommand{\mc }{\mbox{$m_c  $}}
\newcommand{\mcS}{\mbox{$m^2_c$}}
\newcommand{\mcQ}{\mbox{$m^4_c$}}
\newcommand{\mb }{\mbox{$m_b  $}}
\newcommand{\mbS}{\mbox{$m^2_b$}}
\newcommand{\mbQ}{\mbox{$m^4_b$}}

\newcommand{\Mc }{\mbox{$M_c  $}}

\newcommand{\MqS}{\mbox{$M^2_q$}}

\newcommand{\McS}{\mbox{$M^2_c$}}
\newcommand{\MbS}{\mbox{$M^2_b$}}

\newcommand{\MSB}{\overline{MS}}

\newcommand{\LMSBn }{\mbox{$\Lambda^{(\nf)}_{\overline{\mathrm{MS}}}$}}
\newcommand{\LMSBnS}{\mbox{$\left(\Lambda^{(\nf)}_{\overline{\mathrm{MS}}}
\right)^2$}}
\newcommand{\LMSBnml }{\mbox{$\Lambda^{(\nf-1)}_{\overline{\mathrm{MS}}}$}}

\newcommand{\LMSBt }{\mbox{$\Lambda^{(3)}_{\overline{\mathrm{MS}}}$}}

\newcommand{\LMSBf }{\mbox{$\Lambda^{(4)}_{\overline{\mathrm{MS}}}$}}

\newcommand{\LMSBv }{\mbox{$\Lambda^{(5)}_{\overline{\mathrm{MS}}}$}}

\newcommand{\als }{\alpha_{_S}}

\newcommand{\alsS}{\alpha^2_{_S}}
\newcommand{\ztwo}{\zeta(2)}
\newcommand{\ztri}{\zeta(3)}
\newcommand{\zfor}{\zeta(4)}
\newcommand{\zfiv}{\zeta(5)}
\def\gn{\Gamma_{\nu}}

\def\gmu{\Gamma_{\mu}}
\def\gt{\Gamma_{\tau}}
\def\gl{\Gamma_{l}}
\def\gu{\Gamma_{u}}
\def\gd{\Gamma_{d}}
\def\gc{\Gamma_{c}}

\def\gz{\Gamma_{\sss{Z}}}
\def\gh{\Gamma_{h}}
\def\gi{\Gamma_{\rm{inv}}}        
\def\afb{\mbox{$A_{_{FB}}$}}
\def\barf{\overline f}

\def\barb{\overline b}
\def\bart{\overline t}
\def\barc{\overline c}
\def\dr{\Delta r}

\def\Szz{\Sigma_{_{ZZ}}}
\def\Sww{\Sigma_{_{WW}}}

\def\Pgg{\Pi_{\gamma\gamma}}
\def\gfd{\gamma_5}

\def\srt{\sqrt{2}}

\def\s0h{\sigma^0_h}
\def\nl{\nonumber \\}

\newcommand{\bqa}{\begin{eqnarray}}
\newcommand{\eqa}{\end{eqnarray}}
\newcommand{\ds }{\displaystyle}

\newcommand{\gQ}{\ph{\scriptscriptstyle{Q}}}
\newcommand{\gL}{\ph{\scriptscriptstyle{L}}}
\newcommand{\gZQ}{\ph{\scriptscriptstyle{(Z)Q}}}
\newcommand{\gZL}{\ph{\scriptscriptstyle{(Z)L}}}
\newcommand{\ZQ}{\scriptscriptstyle{ZQ}}
\newcommand{\ZL}{\scriptscriptstyle{ZL}}
\newcommand{\QL}{\scriptscriptstyle{QL}}
\newcommand{\LQ}{\scriptscriptstyle{LQ}}
\newcommand{\zg}{{\scriptscriptstyle{Z}}\ph}
\newcommand{\sss}[1]{\scriptscriptstyle{#1}}

\newcommand{\gf}{G_{\mu}}
\newcommand{\gfs}{G^2_{\mu}}
\newcommand{\ph}{\gamma}
\newcommand{\ab}{A}
\newcommand{\zb}{Z}
\newcommand{\wb}{W}
\newcommand{\hb}{H}
\newcommand{\fe}{e}
\newcommand{\fbe}{{\bar{e}}}
\newcommand{\ff}{f}
\newcommand{\ffp}{f'}
\newcommand{\fep}{e^{+}}
\newcommand{\fem}{e^{-}}
\newcommand{\fnue}{\nu_e}
\newcommand{\barl}{\overline{l}}
\newcommand{\fu}{u}
\newcommand{\fd}{d}
\newcommand{\fc}{c}
\newcommand{\fs}{s}
\newcommand{\ft}{t}
\newcommand{\fb}{b}
\newcommand{\ffb}{b}
\newcommand{\fl}{l}
\newcommand{\fq}{q}
\newcommand{\flm}{\mu}
\newcommand{\flmp}{\mu^{+}}
\newcommand{\flmm}{\mu^{-}}
\newcommand{\flt}{\tau}
\newcommand{\gap}{\lpar 1+\gamma_5\rpar}
\newcommand{\gadi}[1]{\gamma_{#1}}
\newcommand{\cff}[5]{C_{#1}\lpar #2;#3,#4,#5\rpar}    
\newcommand{\mws}{M^2_{\sss{W}}}
\newcommand{\mzs}{M^2_{\sss{Z}}}
\newcommand{\mzc}{M^3_{\sss{Z}}}
\newcommand{\mhs}{M^2_{\sss{H}}}
\newcommand{\mts}{m^2_{t}}
\newcommand{\mvs}{M^2_{_V}}

\newcommand{\mf }{m^2_f}

\newcommand{\mt }{m^2_t}
\newcommand{\mes}{m^2_e}

\newcommand{\mwq }{M^4_{\sss{W}}}
\newcommand{\mzq }{M^4_{\sss{Z}}}

\newcommand{\mtq }{m^4_{t}}
\newcommand{\mhl }{M_{\sss{H}}}

\newcommand{\mwl }{M_{\sss{W}}}
\newcommand{\mzl }{M_{\sss{Z}}}
\newcommand{\mfl }{m_f}
\newcommand{\mtl }{m_t}

\newcommand{\mfpl}{m_{f'}}

\newcommand{\mfs }{m^2_f}
\newcommand{\mls }{m^2_l}
\newcommand{\uml}{ m   _{t}}
\newcommand{\um }{ m^2_{t} }
\newcommand{\umf}{ m^4_{t} }
\newcommand{\wml}{ M  _{\sss{W}}}
\newcommand{\zml}{ M  _{\sss{Z}}}
\newcommand{\dml}{ m   {_f}}
\newcommand{\dms}{ m^2_{_f}}
\newcommand{\hml}{ M  _{\sss{H}}}
\newcommand{\rtc}{ r^3_  t }
\newcommand{\sla}[1]{/\!\!\!#1}
\newcommand{\spd}{\partial}
\newcommand{\rws}{r^2_{\sss{W}}}
\newcommand{\rzs}{r^2_{\sss{Z}}}
\newcommand{\asums}[1]{\sum_{#1}}
\newcommand{\cf}{c_f}
\newcommand{\Nf}{N_f}     
\newcommand{\fbf}{{\bar{f}}}
\newcommand{\Vvert}{V}
\newcommand{\tpfi}{\lpar 2\pi\rpar^4\ib}
\newcommand{\zcont}[2]{z_{#1}^{#2}}
\newcommand{\Vverti}[3]{V_{#1}^{#2}\lpar{#3}\rpar}
\newcommand{\lpar}{\left(}                            
\newcommand{\rpar}{\right)}
\newcommand{\lrbr}{\left[}
\newcommand{\rrbr}{\right]}
\newcommand{\lcbr}{\left\{}
\newcommand{\rcbr}{\right\}} 
\newcommand{\ib  }{i}
\newcommand{\qf  }{Q_f  }
\newcommand{\qfs }{Q^2_f}
\newcommand{\qfc }{Q^3_f}
\newcommand{\qe  }{Q_e  }
\newcommand{\qes }{Q^2_e}
\newcommand{\qep }{Q_{e'}}
\newcommand{\qfp }{Q_{f'}}

\newcommand{\gbs }{g^2}
\newcommand{\gbc }{g^3}
\newcommand{\Gverti}[3]{G_{#1}^{{#2}}\lpar{#3}\rpar}
\newcommand{\Zverti}[3]{Z_{#1}^{{#2}}\lpar{#3}\rpar}
\newcommand{\vverti}[3]{F^{#1}_{_{#2}}\lpar{#3}\rpar}
\newcommand{\vvertil}[3]{F^{#1}_{#2}\lpar{#3}\rpar}
\newcommand{\cvetri}[3]{{\cal{F}}^{#1}_{_{#2}}\lpar{#3}\rpar}
\newcommand{\cvetril}[3]{{\cal{F}}^{#1}_{#2}\lpar{#3}\rpar}
\newcommand{\hvetri}[3]{{\hat{\cal{F}}}^{#1}_{_{#2}}\lpar{#3}\rpar}
\newcommand{\averti}[3]{{\bar{F}}^{#1}_{_{#2}}\lpar{#3}\rpar}
\newcommand{\avetri}[3]{{\overline{\cal{F}}}^{#1}_{_{#2}}\lpar{#3}\rpar}

\newcommand{\fverti}[2]{F^{#1}_{#2}}
\newcommand{\bDz}[2]{{\cal{D}}^{{#1},F}_{_{\zb}}\lpar{#2}\rpar}
\newcommand{\gadu}[1]{\gamma_{#1}}
\newcommand{\siw }{s_{\sss{W}}}           
\newcommand{\cow }{c_{\sss{W}}}
\newcommand{\siws}{s^2_{\sss{W}}}
\newcommand{\cows}{c^2_{\sss{W}}}
\newcommand{\siwc}{s^3_{\sss{W}}}

\newcommand{\siwf}{s^4_{\sss{W}}}
\newcommand{\cowf}{c^4_{\sss{W}}}
\newcommand{\vpa}[2]{\sigma_{#1}^{#2}}
\newcommand{\vma}[2]{\delta_{#1}^{#2}}
\newcommand{\vc }[1]{v_{#1}}
\newcommand{\ac }[1]{a_{#1}}
\newcommand{\vcs}[1]{v^2_{#1}}
\newcommand{\acs}[1]{a^2_{#1}}
\newcommand{\Rvaz}[1]{g^{#1}_{\sss{\zb}}}
\newcommand{\rvab}[1]{{\bar{g}}_{#1}}
\newcommand{\rvabs}[1]{{\bar{g}}^2_{#1}}
\newcommand{\rab }[1]{{\bar{a}}_{#1}}
\newcommand{\rvb }[1]{{\bar{v}}_{#1}}
\newcommand{\rabs}[1]{{\bar{a}}^2_{#1}}
\newcommand{\rva }[1]{g_{#1}}

\newcommand{\Rva }[1]{G_{#1}}
\newcommand{\Rvac}[1]{G^{*}_{#1}}
\newcommand{\Rvah }[1]{{\hat{G}}_{#1}}
\newcommand{\Rvahc}[1]{{\hat{G}}^{*}_{#1}}

\newcommand{\vaeII}{{\bar{g}}^2_{\fe}+{\bar{a}}^2_{\fe}}
\newcommand{\vafII}{{\bar{g}}^2_{\ff}+{\bar{a}}^2_{\ff}}
\newcommand{\four}{{\bar{g}}_{\fe}{\bar{a}}_{\fe}{\bar{g}}_{\ff}{\bar{a}}_{\ff}}
\newcommand{\tcie}{I^{(3)}_e}
\newcommand{\tcif}{I^{(3)}_f}
\newcommand{\rhoe}{\rho_{\fe}}
\newcommand{\rhoi}[1]{\rho_{#1}}
\newcommand{\rhois}[1]{\rho^2_{#1}}
\newcommand{\rhopi}[1]{\rho'_{#1}}
\newcommand{\rhobi} [1]{{\bar{\rho}}_{#1}}
\newcommand{\rhohi} [1]{{\hat{\rho}}_{#1}}

\newcommand{\saff}[1]{A_{#1}}             
\newcommand{\aff}[2]{A_{#1}\lpar #2\rpar}                   
\newcommand{\sbff}[1]{B_{#1}}             
\newcommand{\sfbff}[1]{B^{F}_{#1}}
\newcommand{\bff}[4]{B_{#1}\lpar #2;#3,#4\rpar}             
\newcommand{\fbff}[4]{B^{F}_{#1}\lpar #2;#3,#4\rpar}        
\newcommand{\scff}[1]{C_{#1}}             
\newcommand{\sdff}[1]{D_{#1}}                 
\newcommand{\dff}[7]{D_{#1}\lpar #2,#3;#4,#5,#6,#7\rpar}       
\newcommand{\delrho}[1]{{\Delta \rho}^{#1}}
\newcommand{\fbu}{{\overline{u}}}
\newcommand{\fbd}{{\overline{d}}}
\newcommand{\wbm}{W^{-}}
\newcommand{\wbp}{W^{+}}
\newcommand{\rts}{r^2_{\ft}}
\newcommand{\Lnrt}{\ln{\rt}}
\newcommand{\Rws}{R^2_{_{\wb}}}
\newcommand{\Rwc}{R^3_{_{\wb}}}
\newcommand{\Rzs}{R^2_{\sss{Z}} }

\newcommand{\rhz}{r_{_{\zb}}}
\newcommand{\rhzs}{r^2_{_{\zb}}}
\newcommand{\boxc}[2]{{\cal{B}}_{#1}^{#2}}
\newcommand{\hboxc}[3]{\hat{{\cal{B}}}_{#1}^{#2}\lpar{#3}\rpar}

\newcommand{\Trmoms}{-s}
\newcommand{\sewti}[2]{w_{#1}^{#2}}
\newcommand{\Dz}[2]{{\cal{D}}_{_{\zb}}^{#1}\lpar{#2}\rpar}

\newcommand{\Pzg}{\Pi_{\zg}}
\newcommand{\hDz}[2]{{\hat{\cal{D}}}^{{#1},F}_{_{\zb}}\lpar{#2}\rpar}
    \newcommand{\tHs}{\mu}
    \newcommand{\tHss}{\mu^2}
    \newcommand{\stwl}{s_{\sss{W}}  }
    \newcommand{\ctwl}{c_{\sss{W}}  }
    \newcommand{\stws}{s^2_{\sss{W}}}
    \newcommand{\stwf}{s^4_{\sss{W}}}
    \newcommand{\ctws}{c^2_{\sss{W}}}
    \newcommand{\ctwf}{c^4_{\sss{W}}}
 \newcommand{\qd }{Q_f  }
 \newcommand{\qds}{Q^2_f}
 \newcommand{\qdc}{Q^3_f}
  \newcommand{\vmad }{\delta_f  }
  \newcommand{\vmads}{\delta^2_f}
  \newcommand{\vpau }{\sigma_{f'}}
  \newcommand{\vmaes}{\delta^2_e }
  \newcommand{\zaOfer}{\Pi^{\fer,F}_{\gamma\gamma}(0)}
\newcommand{\bos}{\rm{bos}}
\newcommand{\fer}{\rm{fer}}
\newcommand{\pole}{\left( \dlt-\Lnw \right)}
\newcommand{\deltar}{ \Delta r}
\newcommand{\Trqf}{ \sum_f c_f Q^2_f }
\newcommand{\qum}{ \left| Q_{f'} \right| }

\newcommand{\zmz}{\Sigma^{'}_{_{\zb\zb}}(\mzs)}                          
\newcommand{\wwUf}{w^F_{_{W}}}
\newcommand{\Lnw }{\ln \frac{\mws}{\mu^2}}   
\newcommand{\Lnfw}{\ln \frac{\dms}{\mws} }
\newcommand{\Lnew}{\ln \frac{\mes}{\mws} }

\newcommand{\au }{a_{f'}}
\newcommand{\vd }{v_f}
\newcommand{\ad }{a_f}
\newcommand{\ve }{v_e}  
\newcommand{\tvpad}{(3v^2_f +a^2_f)}
\newcommand{\tvpae}{(3 v^2_e +a^2_e)}
\newcommand{\vpad }{\sigma_f}

\newcommand{\vpadpa}{\sigma^a_{f^{'}}}
\newcommand{\vpada }{\sigma^a_f}

\newcommand{\pd}[1]{\partial_{#1}}
\newcommand{\asum}[3]{\sum_{#1=#2}^{#3}}
\newcommand{\lkall}[3]{\lambda\lpar#1,#2,#3\rpar}       
\newcommand{\mind}[1]{m_{#1}}
\newcommand{\minds}[1]{m^2_{#1}}

\newcommand{\mindf}[1]{m^4_{#1}}
\newcommand{\Mind}[1]{M_{#1}}
\newcommand{\Minds}[1]{M^2_{#1}}

\newcommand{\imom}{q}

\newcommand{\imoms}{q^2}
\newcommand{\pmom}{p}

\newcommand{\pmoms}{p^2}
\newcommand{\pmomq}{p^4}

\newcommand{\pmomi}[1]{p_{#1}}

\newcommand{\dlt}{\displaystyle{\frac{1}{\epsb}}}
\newcommand{\epsh}{\hat\varepsilon}
\newcommand{\epsb}{\bar\varepsilon}

\newcommand{\ep}{\epsilon}
\newcommand{\chapt}[1]{Chapter~\ref{#1}}
\newcommand{\chaptsc}[2]{Chapter~\ref{#1} and \ref{#2}}
\newcommand{\eqn}[1]{Eq.~(\ref{#1})}
\newcommand{\eqns}[2]{Eqs.~(\ref{#1})--(\ref{#2})}
\newcommand{\eqnss}[1]{Eqs.~(\ref{#1})}
\newcommand{\eqnsc}[2]{Eqs.~(\ref{#1}) and (\ref{#2})}
\newcommand{\eqnst}[3]{Eqs.~(\ref{#1}), (\ref{#2}) and (\ref{#3})}

\newcommand{\tbn}[1]{Tab.~\ref{#1}}
\newcommand{\tabn}[1]{Tab.~\ref{#1}}

\newcommand{\fig}[1]{Fig.~\ref{#1}}
\newcommand{\figs}[2]{Figs.~\ref{#1}--\ref{#2}}

\newcommand{\sect}[1]{Section~\ref{#1}}
\newcommand{\sects}[2]{Sections~\ref{#1} and \ref{#2}}
\newcommand{\subsect}[1]{Subsection~\ref{#1}}
\newcommand{\appendx}[1]{Appendix~\ref{#1}}
\newcommand{\sman}{s}
\newcommand{\tman}{t}
\newcommand{\uman}{u}

\newcommand{\smanp}{s'}

\newcommand{\smans}{s^2}

\newcommand{\gspi}{\frac{g^2}{16\pi^2}}

\newcommand{\drrem}{\deltar_{\rm rem}}
\newcommand{\deltarremho}{\deltar^{ho}_{\rm rem}}
\newcommand{\dalpha}{\Delta\alpha} 
\newcommand{\dalphav}{\Delta\alpha^{(5)}(\mzs)} 
\newcommand{\dalphat}{\Delta\alpha^{\ft}(\mzs)} 
\newcommand{\Reb}{{\rm{Re}}}
\newcommand{\Imb}{{\rm{Im}}}
\newcommand{\pir}[1]{\Pi^{\rm{\sss{R}}}\lpar #1\rpar}
\newcommand{\mqs}{m^2_{q}}
\newcommand{\ord}[1]{{\cal O}\lpar#1\rpar}
\newcommand{\prot}{p}
\newcommand{\aprot}{{\bar{p}}}

\newcommand{\dalhv}{\Delta\alpha^{(5)}_{\had}(\mzs)}
\newcommand{\dall}{\Delta\alpha_{\lep}}
\newcommand{\lep}{{l}}
\newcommand{\had}{{h}}
\newcommand{\drho}{\Delta\rho}
\newcommand{\drhov}{\delta\rho}
\newcommand{\dkapv}{\delta\kappa}

\newcommand{\drhovb}{\delta{\hat{\rho}}}
\newcommand{\EW}{\sss{\rm{EW}}}
\newcommand{\seffsf}[1]{\sin^2\theta^{#1}_{\rm{eff}}}
\newcommand{\drh}{\Delta{\hat{r}}}
\newcommand{\rZf}{\rho^{\ff}_{\sss{Z}}}
\newcommand{\rZl}{\rho^{\fl}_{\sss{Z}}}

\newcommand{\kZf}{\kappa^{\ff}_{\sss{Z}}}
\newcommand{\rZdf}[1]{\rho^{#1}_{\sss{Z}}}
\newcommand{\kZdf}[1]{\kappa^{#1}_{\sss{Z}}}
\newcommand{\wt}{ w_t}
\newcommand{\zt}{ z_t}
\newcommand{\Ht}{ h_t}
\newcommand{\xts}{x^2_t}
\newcommand{\gel}{\Gamma_{\fe}}
\newcommand{\gff}{\Gamma_{\ff}}
\newcommand{\gll}{\Gamma_{\fl}}

\newcommand{\gbq}{\Gamma_{\fb}}
\newcommand{\gzs}{\Gamma^2_{\sss{Z}}}
\newcommand{\drhigs }{\deltar^{H}_{\rm{res}}}
\newcommand{\drhigsa}{\deltar^{H,\alpha}_{\rm{res}}}
\newcommand{\drhigsg}{\deltar^{H,G}_{\rm{res}}}
\newcommand{\Ksc}{K_{\rm{scale}}}
\newcommand{\afba}[1]{A^{#1}_{_{\rm FB}}}

\newcommand{\Pzga}[2]{\Pi^{#1}_{_{\zb\ab}}\lpar#2\rpar}
\newcommand{\vfwi}[1]{\sigma^{a}_{#1}}
\newcommand{\vfwsi}[1]{\lpar\sigma^{a}_{#1}\rpar^2}
\newcommand{\qb}{Q_b}
\newcommand{\qt}{Q_t}
\newcommand{\qus}{Q^2_u}

\newcommand{\qts}{Q^2_t}
\newcommand{\xvar}{x}
\newcommand{\xvars}{x^2}
\newcommand{\rvar}{r}
\newcommand{\rvari}[1]{r_{#1}}

\newcommand{\lpoli}[1]{\lambda_{#1}}
\newcommand{\hpoli}[1]{h_{#1}}
\newcommand{\reni}[1]{R_{#1}}
\newcommand{\renis}[1]{R^2_{#1}}
\newcommand{\sreni}[1]{\sqrt{R_{#1}}}
\newcommand{\kappai}[1]{\kappa_{#1}}
\newcommand{\Imsi}[1]{I^2_{#1}}
\newcommand{\bgz}[1]{{\overline{\Gamma}}_{#1}}
\newcommand{\pgz}[1]{\Gamma_{#1}}
\newcommand{\intfx}[1]{\int_{\scriptstyle 0}^{\scriptstyle 1}\,d#1}
\newcommand{\intmomi}[2]{\int\,d^{#1}#2}

\begin{document}
\thispagestyle{empty}
\begin{flushleft}
DESY 99--070   
\\
August 1999
\\
hep-ph/9908433 
\\
v.2 (02/2000)
\end{flushleft}
\begin{center}
\bigskip
{\LARGE \zf \,\,\,  {\large  \bf v.6.21}
\\ \vspace*{.3cm}}
{\Large \bf
A Semi-Analytical Program for Fermion Pair \\
Production  in  \ee\  Annihilation
\vspace*{2.0cm}
}
 
 
D. Bardin$^{1,\dag}$, 
M. Bilenky$^{2}$,
P. Christova$^{1,3,*}$,
M. Jack$^{4}$,
\\
\medskip
L. Kalinovskaya$^{1,\dag}$,
A. Olchevski$^{1,5}$,
S. Riemann$^{4}$,  
T. Riemann$^{4,\ddag}$ 
\vspace*{0.5cm}
\\
{
$^{1}$
Laboratory of Nuclear Problems,
Joint Institute for Nuclear Research, 
Dubna, Russia  \\ 
\medskip
$^{2}$
Institute of Physics, Academy of Sciences, 
Prague, Czech Republik\\
\medskip
$^{3}$
Faculty of Physics, Bishop Preslavsky University, Shoumen,
Bulgaria \\
\medskip
$^{4}$
DESY Zeuthen, Germany \\
\medskip
$^{5}$
EP Division, CERN, Geneva, Switzerland  \\
} 
\medskip
\end{center}
\vfill
\begin{center}

{\bf\large Abstract}
\end{center}
{\small
We describe \zf, a Fortran program based on a semi-analytical
approach to fermion pair production in \ee\ annihilation at a wide
range of centre-of-mass energies, including the PETRA, TRISTAN,
LEP1/SLC, and LEP2 energies. 
A flexible treatment of complete ${\cal O}(\alpha)$
QED corrections and of some higher order contributions
is made possible with three calculational chains containing
different realistic sets of
restrictions in the photon phase space.
Numerical integrations are at most one-dimensional.
Complete ${\cal O}(\alpha)$ weak loop corrections supplemented by
selected higher-order terms may be included.  
The program calculates $\Delta r$, the $Z$ width, differential cross-sections,
total cross-sections, integrated forward-backward asymmetries,  
left-right asymmetries, and for $\tau$~pair production also final-state
polarization effects.  
Various interfaces allow fits to be performed
with different sets of free parameters.
}  

\bigskip

\centerline{\large \em  Submitted to Computer Physics Communications}

\bigskip

\footnoterule
\noindent
{\footnotesize 
$^{\dag}$ Supported by European Union with grant INTAS-93-744
and by German-JINR Heisenberg-Landau program
\\
$^{\ddag}$ Supported by European Union with grant CHRX-CT920004
\\
$^{*}$ Supported by European Union with grant CIPD-CT940016
and by Bulgarian Foundation for Scientific Researches with grant
$\Phi$--620/1996}
 
\setcounter{page}{0}

 
\thispagestyle{empty}
\title{
\centerline{\tt{ZFITTER} v.6.21}
A Semi-Analytical Program for Fermion Pair
Production  in  \ee\  Annihilation
}

\author{%
D. Bardin
\address{%
Laboratory of Nuclear Problems,
Joint Institute for Nuclear Research, 
Dubna, Russia}
\thanks{%
Supported by European Union with grant INTAS-93-744
and by German-JINR Heisenberg-Landau program
},
M. Bilenky
\address{%
Institute of Physics, Academy of Sciences, Prague, Czech Republik}, 
P. Christova~$^a$
\address{%
Faculty of Physics, Bishop Preslavsky University, Shoumen,
Bulgaria}
\thanks{%
Supported by European Union with grant CIPD-CT940016
and by Bulgarian Foundation for Scientific Researches with grant
$\Phi$--620/1996},
M. Jack
\address{%
DESY Zeuthen, Germany}, 
A. Olchevski~$^a$
\address{%
EP Division, CERN, Geneva, Switzerland},
L.~Kalinovskaya~$^a$
\thanks{%
Supported by European Union with grant INTAS-93-744
and by German-JINR Heisenberg-Landau program},
S. Riemann~$^d$
        and 
T. Riemann~$^d$
\thanks{%
Supported by European Union with grant CHRX-CT920004}
}
       
\maketitle

\vspace*{2.0cm}

\begin{abstract}
We describe \zf, a Fortran program based on a semi-analytical
approach to fermion pair production in \ee\ annihilation at a wide
range of centre-of-mass energies, including the PETRA, TRISTAN,
LEP1/SLC, and LEP2 energies. 
A flexible treatment of complete ${\cal O}(\alpha)$
QED corrections and of some higher order contributions
is made possible with three calculational chains containing
different realistic sets of
restrictions in the photon phase space.
Numerical integrations are at most one-dimensional.
Complete ${\cal O}(\alpha)$ weak loop corrections supplemented by
selected higher-order terms may be included.  
The program calculates $\Delta r$, the $Z$ width, differential cross-sections,
total cross-sections, integrated forward-backward asymmetries,  
left-right asymmetries, and for $\tau$~pair production also final-state
polarization effects.  
Various interfaces allow fits to be performed
with different sets of free parameters.
\end{abstract}
\thispagestyle{empty}

\newpage

\noindent{\bf PROGRAM SUMMARY}
\vspace{10pt}

\noindent{\sl Title of the program:} {\tt ZFITTER} version 6.21 (26 July 1999)

\noindent{\sl Computer:} any computer with {\tt FORTRAN 77} compiler

\noindent{\sl Operating system:}
{\tt UNIX}, program tested e.g. under {\tt HP-UX} and {\tt Linux}, but also
 under {\tt IBM}, {\tt IBM PC}, {\tt VMS}, {\tt APOLLO}, and {\tt SUN}

\noindent{\sl Programming language used:}
{\tt FORTRAN 77} 

\noindent{\sl High-speed storage required:}  $<$ 2 MB

\noindent{\sl No. of cards in combined program and test deck:}
about 24,000

\noindent{\sl Keywords:}
Quantum electrodynamics (QED), Standard Model, electroweak interactions,
heavy boson $Z$, $e^+e^-$-annihilation,
radiative corrections, initial-state radiation (ISR),
final-state radiation (FSR), QED interference,
LEP1, LEP2, Linear Collider, TESLA

\noindent{\sl Nature of the physical problem:}
Fermion pair production  is important for the study of the properties
of the $Z$-boson and for precision tests
of the Standard Model at LEP and future linear colliders at higher energies.
QED corrections and combined electroweak and QCD corrections have to
be calculated  for this purpose with high precision, including higher
order effects.  
For multi-parameter fits a program is needed with sufficient
flexibility and also high calculational speed.
{\tt ZFITTER} combines the two aspects by at most one-dimensional
numerical integrations and a variety of flags, defining the physics
contents used. 
The Standard Model predictions are typically at the per mille
precision level, sometimes better.

\noindent{\sl Method of solution:}
Numerical integration of analytical formulae.

\noindent{\sl Restrictions on the complexity of the problem:}
Fermion pair production is described below the top quark production
threshold. 
Photonic corrections are taken into account with relatively simple
cuts on photon energy, or the energies and 
acollinearity of the two fermions, and {\em one} fermion production angle.
Bhabha scattering is treated poorly.

\noindent{\sl Typical running time:}
On a Pentium II PC installation (400 MHz), Linux 2.0.34,
approximately 140 sec are needed to run the standard test with subroutine
{\tt ZFTEST}. 
This result is for a {\em default/recommended} 
setting of the input parameters, with {\em all} corrections in the
Standard Model switched {\em on}.
{\tt ZFTEST} computes 12 cross-sections and cross-section asymmetries
for 8 energies with 5 interfaces, i.e. about 360 cross-sections in 140
seconds. 

\newpage 
  \tableofcontents 
  \listoffigures  
  \listoftables    
\clearpage
\addcontentsline{toc}{section}{Introduction}
\def\theequation{\Roman{section}.\arabic{equation}}
\def\thefigure{\Roman{section}.\arabic{figure}}
\def\thetable{\Roman{section}.\arabic{table}}
\setcounter{section}{1}
\section*{Introduction\label{intro}}
\markboth{Introduction}{Introduction}
\setcounter{equation}{0}
The \zf\ project was started in 1983 by D. Bardin, O. Fedorenko and
T. Riemann.
A first program package, {\tt ZBIZON},
was shortly used in 1989 at LEP by the DELPHI and
L3 Collaborations but got immediately replaced by the \zf\ package.

The Fortran program \zf~\cite{Bardin:1992jc1}, with the packages {\tt
  DIZET}~\cite{Bardin:1989tq1} and {\tt BHANG}, 
was originally intended for the description
of fermion  pair production around the $Z$ resonance at the \ee\
colliders \LEPI\ and SLC
in the Standard Model using the on mass shell renormalization 
scheme~\cite{Bardin:1980fet,Bardin:1982svt}. 
The 1989 version of {\tt DIZET} was described in~\cite{Bardin:1989tqt}, and the
corresponding 1992 versions \zf\ v.4.5, {\tt DIZET} v.4.04, and 
{\tt BHANG} in~\cite{Bardin:1992jc}.
The actual version 6.21 of \zf\
  \cite{zfitter:v6.21,Bardin:1989dit,Bardin:1991fut,%
Bardin:1991det,Christova:1999cct}  
was improved in many respects and is now also intended for
considerably 
smaller  and higher energies thus covering 
two fermion production physics ranging from PETRA to \LEPII:  
\begin{equation}
e^+e^- \longrightarrow f {\bar f}({\rm{n}} \gamma), \,\,\, f =
\mu, \nu_{\mu}, \tau, \nu_{\tau}, u, d, c, s, b.
\label{firsteq}
\end{equation}
Applications at higher energies have not been carefully tested so far.

\zf, when it is used with the {\tt DIZET} package allows for 
the description of data in the Standard Model.
\zf\ calculates radiative corrections to
the muon decay 
constant~\cite{%
Bardin:1980fet,Bardin:1982svt,Akhundov:1986fct,Bardin:1989aa,Bardin:1995a2},
the decay 
width of the $Z$ boson~\cite{Akhundov:1986fct,Bardin:1989aa,Bardin:1995a2}, 
and  improved Born  
cross-sections~\cite{Bardin:1989dit,Bardin:1989aa,Bardin:1995a2} for
reaction~\eqn{firsteq} with virtual electromagnetic, electroweak, and QCD
corrections.   
Photonic corrections with different cuts and for different observables are 
described as convolutions of improved Born cross-sections with
radiator functions (flux functions)
~\cite{Bardin:1989cwt,Bardin:1991fut,Bardin:1991det,Christova:1999cct}.

The Fortran program package \zf\ consists of three parts:
\begin{itemize}
\item
The main package \zf\ itself.
\\
The authors of \zf\ are:
\\
D. Bardin,  
M. Bilenky (1987-1994), 
A. Chizhov (1987-1991),  
P. Christova (since 1997), 
O. Fedorenko (1990),
M. Jack (since 1997),
L. Kalinovskaya (since 1997),
A. Olshevsky,
S. Riemann, 
T. Riemann, 
M. Sachwitz (1987-1991), 
A. Sazonov (1987-1991),
Yu. Sedykh (1989-1991), 
I. Sheer (1991-1992),
L. Vertogradov (1990).
\item
The package \zf\ calculates the virtual corrections in the Standard
Model with the Fortran program {\tt DIZET}.
\\
The authors of {\tt DIZET} are:
\\
A. Akhundov (1985-1989),  
D. Bardin,  
M. Bilenky (1987-1994),
P. Christova, 
L. Kalinovskaya (since 1997),
S. Riemann, 
T. Riemann,
M. Sachwitz (1987-1991), 
H. Vogt (1989). 
\item
The package \zf\ has a branch for Bhabha scattering, the Fortran program 
{\tt BHANG}. 
{\tt BHANG} calculates QED corrections and determines the improved Born
cross-section with the aid of {\tt DIZET}. 
\\
The author of {\tt BHANG} is M. Bilenky.
\end{itemize} 
This article describes {\zf}, version 6.21 and
{\tt DIZET}, version 6.21, and the parts of {\tt BHANG}, version 4.67,
which determine the improved Born
cross-section~\cite{Bardin:1991xet,Bardin:1995a2} for the reaction 
\begin{equation}
e^+e^- \longrightarrow e^+ e^- .
\label{seceq}
\end{equation}

Alternatively to the Standard Model, several so-called 
``semi--model-independent''
approaches may be used in \zf, see \sect{ewcoup}.
In addition, it can be used to fit the experimental data 
with the $S$-matrix approach, package
{\tt SMATASY}~\cite{Kirsch:1995cf1,Kirsch:1995cf,Leike:1991pq,Riemann:1992gv}
and to theories that go beyond the Standard Model,
package {\tt ZEFIT}~\cite{SRiemann:1997aa,Leike:1992uf}.
These packages are to be run together with \zf.

\begin{figure}[thbp] 
\setlength{\unitlength}{1mm}
\begin{picture}(140,45)(-20,0)
\put(30,20){\vector(1,0){34}}
\put(100,20){\vector(-1,0){34}}
\put(65,20.0){\vector(-2,-1){34}}
\put(65,20.0){\vector(2,1){34}}
\put(78,23){$\vartheta$}
\put(25,20){e$^+$}
\put(25,2){$f$}
\put(103,20){e$^-$}
\put(103,38){$\bar{f}$}
\end{picture}
\caption
[Scattering angle $\vartheta$]
{\it Scattering angle $\vartheta$
\label{fig.theta}
}
\end{figure}
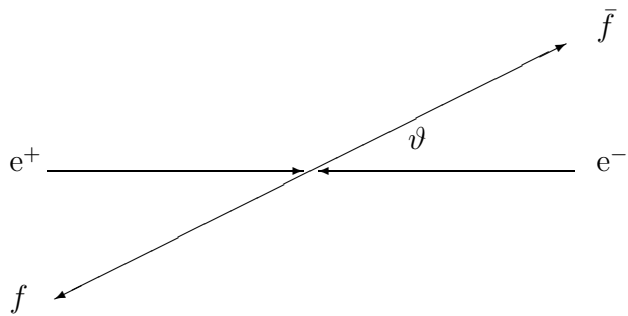

\zf\ is based on a semi-analytical approach to radiative corrections.
It relies on formulae which are either differential in the scattering
angle $\vartheta$ as defined in  
\fig{fig.theta} or, alternatively, are analytically integrated
over a finite angular region.  

\bigskip

\zf\ calculates:
\begin{itemize}
\item
$\Delta r$ -- Standard Model corrections to $G_{\mu}$, the muon decay
constant  
\item $M_W$ -- the $W$ boson mass from $M_Z, M_H$, other masses, and
  $\Delta r$ 
\item
$\Gamma_Z = \sum_f \Gamma_f$ --  total and partial $Z$ boson
decay widths 
\item
${d\sigma}/{d\cos\vartheta}$ -- differential cross-sections 
\item
$\sigma_T$ -- total cross-sections
\item
$A_{FB}$ -- forward-backward asymmetries
\item
$A_{LR}$ -- left-right asymmetries
\item
$A_{pol}, A_{FB}^{pol}$ -- final state polarization effects for $\tau$
leptons  
\end{itemize}
Total cross-sections and asymmetries
may be calculated in a non-symmetric angular interval: 
\ba
\nl
\sigma_T(c_1,c_2)
&=&  \int_{c_1}^{c_2} d\cos\vartheta
                     \frac{d\sigma}{d\cos\vartheta},
\label{sigmatn}
\\ \nl
A_{FB}(c_1,c_2)
&=&  \frac{\sigma_{FB}(c_1,c_2)}{\sigma_T(c_1,c_2)},
\label{afbafb}
\ea
where
\bq
\sigma_{FB}(c_1,c_2)
=
\left[  \int_{0}^{c_2} d\cos\vartheta
      - \int_{c_1}^{0} d\cos\vartheta \right]
    \frac{d\sigma}{d\cos\vartheta}.
\label{sigafbn}
\eq
Analogous definitions of other asymmetries are given in \sect{IBORNA}.
Two cases of special physical interest are:
\ba
\sigma_T(c)
&=&  \int_{-c}^{c} d\cos\vartheta
                     \frac{d\sigma}{d\cos\vartheta},
\label{sigmat}
\\ \nl
\sigma_{FB}(c)
&=&
\left[  \int_{0}^{c} d\cos\vartheta
      - \int_{-c}^{0} d\cos\vartheta \right]
    \frac{d\sigma}{d\cos\vartheta}.
\label{sigafb}
\ea
In \zf, the expressions~\eqn{sigmatn} and~\eqn{afbafb} are
constructed from the following integral:
\bq
\sigma(0,c) \equiv
\int_{0}^{c} d\cos\vartheta     \frac{d\sigma}{d\cos\vartheta}
= \frac{1}{2} \left[ \sigma_T(c) + \sigma_{FB}(c) \right].
\label{binsec}
\eq

\noindent
Corresponding to 
$c_1 = \cos\vartheta_1 < \cos\vartheta < c_2 = \cos\vartheta_2$, 
the maximal and minimal polar angles for the integrated cross-sections
are:
\ba
\label{ang1}
 {\tt ANG1} &=& \vartheta_2,
\\
 {\tt ANG0} &=& \vartheta_1.
\label{ang0}
\ea
For a symmetrical angular region, it is 
\ba
\vartheta_2 &=& 180^{\circ} - \vartheta_1.
\label{ang01}
\ea
If the angles related by \eqn{ang01}, an internal flag {\tt ISYM} is set equal
to 1 in subroutine {\tt ZCUT} and the integrated quantities 
$\sigma_{T,FB}(c) = [\sigma(0,c) \pm \sigma(0,-c)]$ 
are simply twice the contributions calculated by calls to subroutines
 {\tt SFAST} or {\tt SCUT}.
We mention this here since in  {\tt SFAST} and {\tt SCUT} the cross-section
contributions are by a factor two smaller than the (integrated) corresponding
quantities in subroutine {\tt COSCUT} from the branch for the calculation of
angular distributions.


With \zf\, cross-sections may be calculated in two ways: without and with 
photonic corrections.
The Born approximation and,  if electroweak corrections are applied,
the improved Born approximation, are based on analytic formulae. 
Photonic corrections are realized 
as {\em one-dimensional numeric integrations} over the invariant mass of the
fermion pair:
\ba
s' \equiv (p_f + p_{\bar f})^2 = \left( 1 -
\frac{2 E_{\gamma}}{\sqrt{s}}\right) s, 
\label{sprime}
\ea
where $p_f$ ($p_{\bar f}$) is the fermion (antifermion) momentum and
$E_{\gamma}$ the energy of the emitted photon.
Often we will also use a related variable $v$:
\ba
\Delta = v_{\max} \geq v &=& \frac{2 E_{\gamma}}{\sqrt{s}} = 1 - \frac{s'}{s}.
\label{defv}
\ea
The cut conditions for $\Delta$, $s'_{\min}$, and $E_{\gamma}^{\max}$ are
trivially related by~\eqn{defv}.  

The photonic corrections are implemented
by convoluting the corresponding (improved) Born cross-sections 
$\sigma^{0}_A$, $A = T, FB$, with photonic radiator functions $R^a_A$,
$a = ini, int, ini+fin$. 
For initial state radiation, e.g., the correction  to the differential
cross-section is:    
\bq
\frac{d \sigma^{ini}}{d\cos\vartheta} \sim \int_{4m_f^2\leq
  s'_{\min}}^{s} \frac{ds'}{s} 
 \left[   R_T^{ini}(s,s',\cos\vartheta) \,   \sigma_T^{0}(s') 
+
   R_{FB}^{ini}(s,s',\cos\vartheta) \,   \sigma_{FB}^{0}(s') \right].
\label{sigexa}
\eq
The QED radiators are defined in
\chapt{ch-phot}. 
The pseudo-observables are defined for the Standard Model in \chapt{ch-PO} 
and the improved Born cross-sections in \chapt{ch-IBA}. 
%
The latter two Chapters make intensive use of one-loop amplitudes 
collected in 
\appendx{app-ew}. 
Initialization and use of the program are described in 
\sect{dizetug}
and 
\sect{zfguide}.
\sect{interfaces} documents the various {\em interfaces}. 
\sect{zupars} contains a description of the contents of some of the
common blocks of \zf\ and 
\appendx{ch-zftest} an example of the use
of the program.

\zf\ is a self-contained package, but contains some routines of other
authors: 
\begin{itemize}
\item Subroutines {\tt SIMPS} \cite{Silin:19xy}, written by I. Silin, and {\tt
    FDSIMP} \cite{Sedykh:19xy}, written by Yu. Sedykh, perform self
  adapting one-dimensional numerical integrations without and with mapping. 
\item Subroutines {\tt TRILOG} and {\tt S12} \cite{Matsuura:1987},
written by T. Matsuura, calculate polylogarithms $\lithree$ and $S_{2,1}$. 
\item Package with main subroutine {\tt hadr5}, version 24/02/1995,
  written by F. Jegerlehner \cite{Jegerlehner:1995ZZ}, calculates the
contribution of light 
  hadrons to the hadronic vacuum polarization~\cite{Eidelman:1995ny}
  (default, flag {\tt VPOL}=1).
\item Another calculation of this contribution is subroutine {\tt
    HADRQQ}, written by H. Burk\-hardt
\cite{Burkhardt:1989ZZ,Burkhardt:1989ky} (flag 
  {\tt VPOL}=3). 
\item Package {\tt m2tcor}, version 2.0 (Oct 1996),
  written by 
  G. Degrassi \cite{Degrassi:1996ZZ}, calculates the electroweak
  corrections of the order  
  ${\cal{O}}(G_{\mu}^2 m_t^2 M_Z^2)$  and 
  ${\cal{O}}(G_{\mu}^2 m_t^4)$
  to $\Delta r$, $\sin^2\theta^{\rm{lept}}_{\rm{eff}}$
  and to the partial $Z$ boson widths
\cite{Degrassi:1994a0,Degrassi:1995ae,Degrassi:1995mc,Degrassi:1996mg,%
Degrassi:1997ps,Degrassi:1999jd}   (default, flag {\tt AMT4}=4).   
\item Package {\tt pairho.f}, version 16/07/99, written by 
A.~Arbuzov~\cite{Arbuzov:1999uq}, 
contains a rather complete 
  treatment of higher order initial state pair production corrections 
  to $e^{+}e^{-}$ annihilation (default, flag {\tt ISPP}=2). 
\item A package for the computation of functions $V(r),\;A(r),$ and $F(x)$,
   needed for the calculation of $\ord{\alpha\als}$  
   corrections to bosonic self-energies, written by B. Kniehl 
   \cite{Kniehl:1990yc} (default, flag {\tt IQCD}=3).
\end{itemize}

Finally we would like to draw the attention of the reader to the
comprehensive presentation of many subtleties of the underlying
formulae given in \cite{BardinPassarino:1999}, and to some quite
recent numerical comparisons with the results of other programmes
given in \cite{Jack:1999af,Christova:2000zu}.


\vfill
\vspace{-1cm}
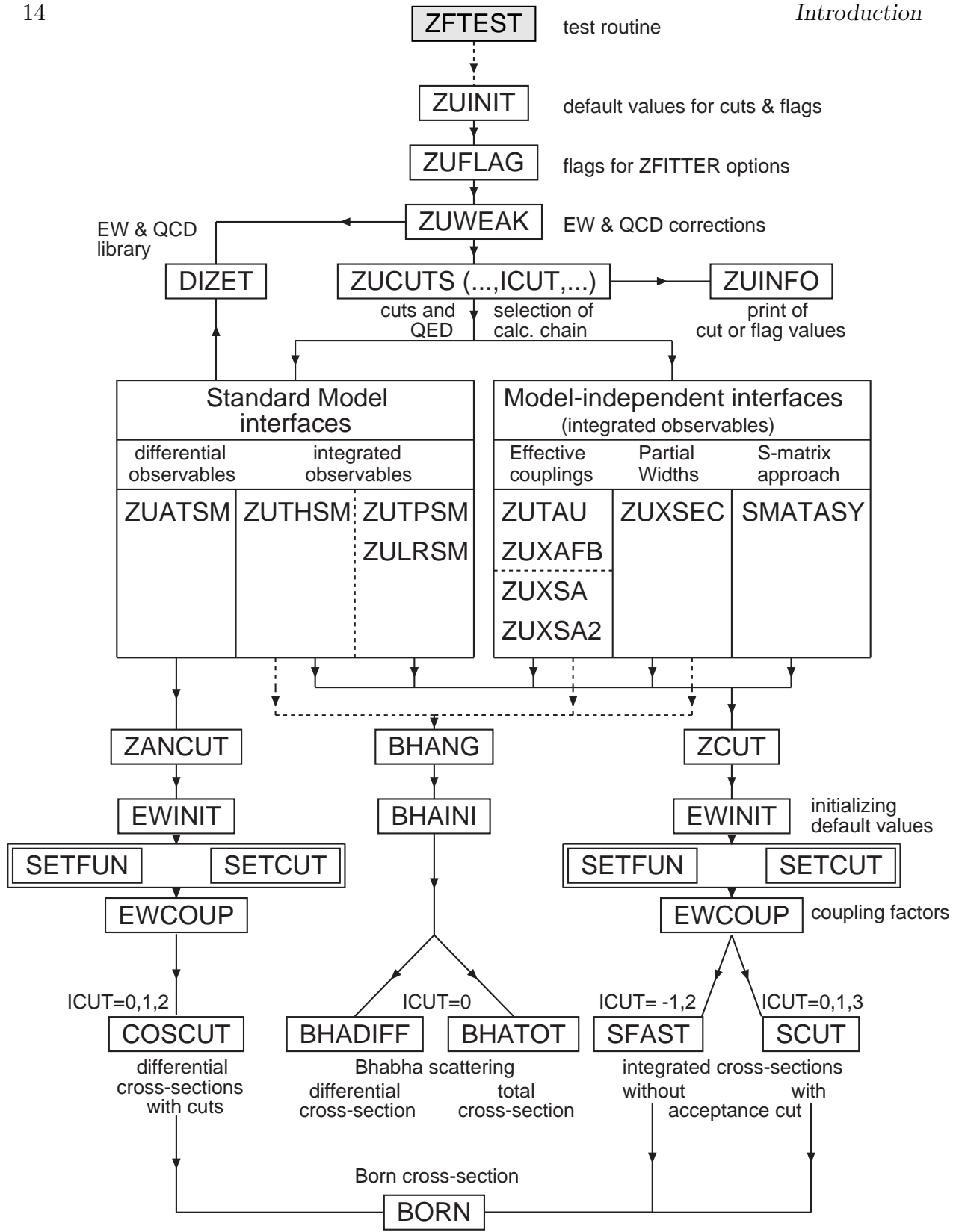
\begin{figure}[b]
\begin{center}
\setlength{\unitlength}{1pt}
\SetWidth{0.8}
\begin{picture}(400,540)(0,0)
\SetPFont{Helvetica}{13}
\SetScale{1.0}
\thicklines
%
\ArrowLine(200,520)(200,490)
\ArrowLine(200,490)(200,460)
\ArrowLine(200,460)(200,430)
\ArrowLine(200,460)(70,460)
\ArrowLine(240,430)(350,430)
\ArrowLine(70,380)(70,430)
\Line(70,430)(70,460)
\ArrowLine(200,425)(200,400)
\Line(110,400)(300,400)
\ArrowLine(110,400)(110,380)
\ArrowLine(300,400)(300,380)
\ArrowLine(50,240)(50,203)
\ArrowLine(50,186)(50,168) 
\ArrowLine(50,100.5)(50,59.5)
\Line(-35,124)(135,124)
\Line(-35,146)(135,146)
\Line(-35,124)(-35,146)
\Line(135,124)(135,146)
\ArrowLine(50,150)(50,146)
\ArrowLine(50,124)(50,118)
\DashLine(100,240)(100,225){2}
\DashArrowLine(100,225)(100,211){2}
\DashLine(100,211)(180,211){2}
\DashLine(250,240)(250,225){2}
\DashArrowLine(250,225)(250,211){2}
\DashLine(250,211)(180,211){2}
\DashLine(310,240)(310,225){2}
\DashArrowLine(310,225)(310,211){2}
\DashLine(310,211)(180,211){2}
\ArrowLine(180,211)(180,203)
\ArrowLine(180,186)(180,168)
\ArrowLine(180,152)(180,100)
\ArrowLine(180,100)(140,59)
\ArrowLine(180,100)(220,59)
\ArrowLine(120,240)(120,225)
\ArrowLine(170,240)(170,225)
\ArrowLine(230,240)(230,225)
\ArrowLine(290,240)(290,225)
\ArrowLine(360,240)(360,225)
\Line(120,225)(360,225)
\ArrowLine(330,100)(345,59)
\ArrowLine(330,100)(315,59)
\ArrowLine(330,150)(330,146)
\ArrowLine(330,124)(330,118)
\ArrowLine(50,9)(50,-40)
\Line(50,-40)(180,-40)
\ArrowLine(290,15)(290,-40)
\Line(290,-40)(180,-40)
\ArrowLine(370,15)(370,-40)
\Line(370,-40)(180,-40)
\BText(180,-40){BORN}
%
\DashArrowLine(200,555)(200,520){2}
\GText(200,560){0.9}{ZFTEST}
\BText(200,520){ZUINIT}
\BText(200,490){ZUFLAG}
\BText(200,460){ZUWEAK}
\BText(200,430){ZUCUTS (...,ICUT,...)}
\BText(350,430){ZUINFO}
\BText(70,430){DIZET}
%
\Line(20,240)(200,240)
\Line(20,240)(20,380)
\Line(200,240)(200,380)
\Line(20,380)(200,380)
\Line(20,350)(200,350)
\Line(20,325)(200,325)
\Line(80,240)(80,325)
\DashLine(140,240)(140,325){2}
%
%
%
\Line(210,240)(400,240)
\Line(210,240)(210,380)
\Line(400,240)(400,380)
\Line(210,380)(400,380)
\Line(210,350)(400,350)
\Line(210,325)(400,325)
\Line(270,240)(270,325)
\DashLine(210,284)(270,284){2}
\Line(330,240)(330,325)
\PText(24,309)(0)[lb]{ZUATSM}
\PText(84,309)(0)[lb]{ZUTHSM}
\PText(144,309)(0)[lb]{ZUTPSM}
\PText(144,289)(0)[lb]{ZULRSM}
\PText(214,309)(0)[lb]{ZUTAU}
\PText(214,289)(0)[lb]{ZUXAFB}
\PText(214,269)(0)[lb]{ZUXSA}
\PText(214,249)(0)[lb]{ZUXSA2}
\PText(274,309)(0)[lb]{ZUXSEC}
\PText(335,309)(0)[lb]{SMATASY}
\SetPFont{Helvetica}{13}
\PText(65,367)(0)[lb]{Standard  Model}
\PText(82,354)(0)[lb]{interfaces}
\SetPFont{Helvetica}{10}
\PText(29,340)(0)[lb]{differential}
\PText(26,329)(0)[lb]{observables}
\PText(119,340)(0)[lb]{integrated}
\PText(115,329)(0)[lb]{observables}
\SetPFont{Helvetica}{13}
\PText(215,367)(0)[lb]{Model-independent  interfaces}
\SetPFont{Helvetica}{10}
\PText(244,354)(0)[lb]{(integrated  observables)}
\PText(218,340)(0)[lb]{Effective}
\PText(218,329)(0)[lb]{couplings}
\PText(283,340)(0)[lb]{Partial} 
\PText(283,329)(0)[lb]{Widths}
\PText(343,340)(0)[lb]{S-matrix}
\PText(343,329)(0)[lb]{approach}
\SetPFont{Helvetica}{13}
%
\BText(50,195){ZANCUT}
\BText(50,160){EWINIT}
\BText(0,135){SETFUN}
\BText(100,135){SETCUT}
\BText(50,110){EWCOUP}
\BText(50,50){COSCUT}
\BText(180,195){BHANG}
\BText(180,160){BHAINI}
\BText(140,50){BHADIFF}
\BText(220,50){BHATOT}
\ArrowLine(330,225)(330,203)
\ArrowLine(330,186)(330,168) 
\Line(245,124)(415,124)
\Line(245,146)(415,146)
\Line(245,124)(245,146)
\Line(415,124)(415,146)
\BText(330,195){ZCUT}
\BText(330,160){EWINIT}
\BText(280,135){SETFUN}
\BText(380,135){SETCUT}
\BText(330,110){EWCOUP}
\BText(288,50){SFAST}
\BText(370,50){SCUT}
%
\SetPFont{Helvetica}{10}
\PText(245,555)(0)[lb]{test routine}
\PText(245,515)(0)[lb]{default values for cuts & flags}
\PText(245,485)(0)[lb]{flags for ZFITTER options}
\PText(10,453)(0)[lb]{EW & QCD}
\PText(10,443)(0)[lb]{library}
\PText(245,455)(0)[lb]{EW & QCD corrections}
\PText(153,412)(0)[lb]{cuts and} 
\PText(210,412)(0)[lb]{selection of}
\PText(338,412)(0)[lb]{print of} 
\PText(312,402)(0)[lb]{cut or flag values}
\PText(168,402)(0)[lb]{QED} 
\PText(210,402)(0)[lb]{calc. chain}
\PText(-5,63)(0)[lb]{ICUT=0,1,2}
\PText(164,63)(0)[lb]{ICUT=0}
\PText(263,63)(0)[lb]{ICUT= -1,2}
\PText(345,63)(0)[lb]{ICUT=0,1,3}
\PText(30,30)(0)[lb]{differential}
\PText(20,20)(0)[lb]{cross-sections}
\PText(35,10)(0)[lb]{with cuts}
\PText(140,30)(0)[lb]{Bhabha scattering}
\PText(117,18)(0)[lb]{differential}
\PText(112,8)(0)[lb]{cross-section}
\PText(212,18)(0)[lb]{total}
\PText(192,8)(0)[lb]{cross-section}
\PText(275,30)(0)[lb]{integrated cross-sections}
\PText(275,18)(0)[lb]{without}
\PText(298,8)(0)[lb]{acceptance cut}
\PText(360,18)(0)[lb]{with}
%
\PText(370,162)(0)[lb]{initializing} 
\PText(370,152)(0)[lb]{default values}
\PText(370,108)(0)[lb]{coupling factors}
\PText(140,-25)(0)[lb]{Born cross-section}
\end{picture}
\end{center}
\vspace*{2.0cm}
\caption
[Logical structure of the package \zf]
{\it
Logical structure of the package \zf \label{zf_flow_main}}
\end{figure} 
\input{flow_zcut} 
\begin{figure}
\vspace{1cm}
\hspace{2.5cm}
\unitlength=1.00mm
\special{em:linewidth 0.4pt}
\linethickness{0.4pt}
\begin{picture}(94.00,201.07)
\linethickness{1pt}
\put(39.34,104.99){\line(1,0){20.00}}
\put(59.34,104.99){\line(0,-1){10.00}}
\put(59.34,94.99){\line(-1,0){20.00}}
\put(49.34,99.99){\makebox(0,0)[cc]{SOFT}}
\put(39.34,94.99){\line(0,1){10.00}}
\put(39.34,104.99){\line(1,0){20.00}}
\put(59.34,104.99){\line(0,-1){10.00}}
\put(59.34,94.99){\line(-1,0){20.00}}
\put(39.34,94.99){\line(0,1){10.00}}
\put(39.34,142.33){\line(1,0){20.00}}
\put(59.34,142.33){\line(0,-1){10.00}}
\put(59.34,132.33){\line(-1,0){20.00}}
\put(39.34,132.33){\line(0,1){10.00}}
\put(49.34,137.33){\makebox(0,0)[cc]{ZETCOS}}
\put(39.35,188.66){\line(1,0){20.00}}
\put(59.35,188.66){\line(0,-1){10.00}}
\put(59.35,178.33){\line(-1,0){20.00}}
\put(39.35,178.66){\line(0,1){10.00}}
\put(49.35,183.66){\makebox(0,0)[cc]{EWINIT}}
\put(39.35,157.67){\line(1,0){20.00}}
\put(59.35,157.67){\line(0,-1){10.00}}
\put(59.35,147.67){\line(-1,0){20.00}}
\put(49.35,152.67){\makebox(0,0)[cc]{EWCOUP}}
\put(39.35,147.67){\line(0,1){10.00}}
\put(59.35,178.33){\line(-1,0){20.00}}
\put(39.35,157.67){\line(1,0){20.00}}
\put(59.35,157.67){\line(0,-1){10.00}}
\put(59.35,147.67){\line(-1,0){20.00}}
\put(39.35,147.67){\line(0,1){10.00}}
\linethickness{0.5pt}
\linethickness{1.0pt}
\put(81.34,94.99){\line(-1,0){20.00}}
\put(71.34,99.99){\makebox(0,0)[cc]{HARD}}
\put(61.34,94.99){\line(0,1){10.00}}
\put(61.34,104.99){\line(1,0){20.00}}
\put(81.34,104.99){\line(0,-1){10.00}}
\put(81.34,94.99){\line(-1,0){20.00}}
\put(61.34,94.99){\line(0,1){10.00}}
\linethickness{0.5pt}
\linethickness{1.0pt}
\put(15.01,121.33){\line(1,0){20.00}}
\put(35.01,121.33){\line(0,-1){10.00}}
\put(35.01,111.33){\line(-1,0){20.00}}
\put(15.01,111.33){\line(0,1){10.00}}
\put(25.01,116.33){\makebox(0,0)[cc]{BORN}}
\put(39.35,201.33){\line(1,0){20.00}}
\put(59.35,201.33){\line(0,-1){10.00}}
\put(59.35,191.33){\line(-1,0){20.00}}
\put(39.35,191.33){\line(0,1){10.00}}
\put(49.35,196.33){\makebox(0,0)[cc]{ZANCUT}}
\put(39.35,201.33){\line(1,0){20.00}}
\put(59.35,201.33){\line(0,-1){10.00}}
\put(59.35,191.33){\line(-1,0){20.00}}
\put(39.35,191.33){\line(0,1){10.00}}
\put(39.34,121.33){\line(1,0){20.00}}
\put(59.34,121.33){\line(0,-1){10.00}}
\put(59.34,111.33){\line(-1,0){20.00}}
\put(39.34,111.33){\line(0,1){10.00}}
\put(61.34,121.33){\line(1,0){20.00}}
\put(81.34,121.33){\line(0,-1){10.00}}
\put(81.34,111.33){\line(-1,0){20.00}}
\put(61.34,111.33){\line(0,1){10.00}}
\put(49.34,116.33){\makebox(0,0)[cc]{BOXIN}}
\put(71.34,116.33){\makebox(0,0)[cc]{SOFTIN }}
\put(64.01,148.00){\line(1,0){20.00}}
\put(84.01,148.00){\line(0,-1){10.00}}
\put(84.01,138.00){\line(-1,0){20.00}}
\put(64.01,138.00){\line(0,1){10.00}}
\put(74.01,143.00){\makebox(0,0)[cc]{COSCUT}}
\put(28.34,173.33){\line(1,0){20.00}}
\put(48.34,173.33){\line(0,-1){10.00}}
\put(48.34,163.33){\line(-1,0){20.00}}
\put(28.34,163.33){\line(0,1){10.00}}
\put(50.34,173.33){\line(1,0){20.00}}
\put(70.34,173.33){\line(0,-1){10.00}}
\put(70.34,163.33){\line(-1,0){20.00}}
\put(50.34,163.33){\line(0,1){10.00}}
\put(38.34,168.33){\makebox(0,0)[cc]{SETFUN}}
\put(60.34,168.33){\makebox(0,0)[cc]{SETCUT}}
\put(25.01,100.00){\makebox(0,0)[cc]{BORN}}
\put(15.01,95.00){\line(0,1){10.00}}
\put(15.01,105.00){\line(1,0){20.00}}
\put(35.01,105.00){\line(0,-1){10.00}}
\put(35.01,95.00){\line(-1,0){20.00}}
\linethickness{0.5pt}
\put(36.01,94.00){\line(0,1){12.00}}
\put(36.01,106.00){\line(-1,0){22.00}}
\put(14.01,106.00){\line(0,-1){12.00}}
\put(14.01,94.00){\line(1,0){22.00}}
\put(82.34,110.33){\line(-1,0){44.00}}
\put(38.34,110.33){\line(0,1){12.00}}
\put(38.34,122.33){\line(1,0){44.00}}
\put(82.34,122.33){\line(0,-1){12.00}}
\put(49.34,191.33){\line(0,-1){2.67}}
\put(14.01,110.33){\line(0,1){12.00}}
\put(14.01,122.33){\line(1,0){22.00}}
\put(36.01,122.33){\line(0,-1){12.00}}
\put(14.01,110.33){\line(1,0){22.00}}
\put(27.34,162.33){\line(1,0){44.00}}
\put(71.34,162.33){\line(0,1){12.00}}
\put(71.34,174.33){\line(-1,0){44.00}}
\put(27.34,174.33){\line(0,-1){12.00}}
\put(49.34,178.33){\line(0,-1){4.00}}
\put(49.34,162.33){\line(0,-1){4.67}}
\put(34.01,42.67){\line(0,1){10.00}}
\put(44.01,47.67){\makebox(0,0)[cc]{SETDEL}}
\put(54.01,52.67){\line(0,-1){10.00}}
\put(34.01,42.67){\line(0,1){10.00}}
\put(26.67,39.33){\line(1,0){20.00}}
\put(46.67,39.33){\line(0,-1){10.00}}
\put(26.67,29.33){\line(0,1){10.00}}
\put(36.67,34.33){\makebox(0,0)[cc]{HCUT}}
\put(26.67,39.33){\line(1,0){20.00}}
\put(46.67,39.33){\line(0,-1){10.00}}
\put(26.67,29.33){\line(0,1){10.00}}
\put(34.02,52.99){\line(1,0){20.00}}
\put(54.02,52.99){\line(0,-1){10.00}}
\put(54.02,42.66){\line(-1,0){20.00}}
\put(34.02,42.99){\line(0,1){10.00}}
\put(34.02,52.99){\line(1,0){20.00}}
\put(54.02,52.99){\line(0,-1){10.00}}
\put(34.02,42.99){\line(0,1){10.00}}
\put(34.01,64.66){\line(1,0){20.00}}
\put(54.01,64.66){\line(0,-1){10.00}}
\put(54.01,54.66){\line(-1,0){20.00}}
\put(34.01,54.66){\line(0,1){10.00}}
\put(44.01,59.66){\makebox(0,0)[cc]{SETR}}
\put(26.67,68.33){\line(0,1){10.00}}
\put(46.67,78.33){\line(0,-1){10.00}}
\put(26.67,68.33){\line(0,1){10.00}}
\put(26.68,78.66){\line(1,0){20.00}}
\put(46.68,78.66){\line(0,-1){10.00}}
\put(46.68,68.33){\line(-1,0){20.00}}
\put(26.68,68.66){\line(0,1){10.00}}
\put(36.68,73.66){\makebox(0,0)[cc]{acol.f}}
\put(26.68,78.66){\line(1,0){20.00}}
\put(46.68,78.66){\line(0,-1){10.00}}
\put(26.68,68.66){\line(0,1){10.00}}
\put(26.67,29.00){\line(1,0){20.00}}
\put(26.00,28.00){\dashbox{1.00}(38.67,38.00)[cc]{}}
\put(25.00,40.34){\dashbox{1.00}(41.00,39.67)[cc]{}}
\put(47.00,36.00){\makebox(0,0)[lc]{{\tt ICUT=0,1}}}
\put(49.01,75.67){\makebox(0,0)[lc]{{\tt ICUT=2}}}
\put(46.67,74.00){\line(1,0){32.33}}
\put(79.00,74.00){\line(0,1){21.00}}
\put(79.00,74.00){\line(0,-1){40.00}}
\put(79.00,34.00){\line(-1,0){32.33}}
\put(59.33,153.00){\line(1,0){14.67}}
\put(74.00,153.00){\line(0,-1){5.00}}
\put(59.33,140.00){\line(1,0){4.67}}
\put(69.00,138.00){\line(0,-1){11.00}}
\put(69.00,127.00){\line(-1,0){44.00}}
\put(25.00,127.00){\line(0,-1){4.67}}
\put(79.00,138.00){\line(0,-1){15.67}}
\put(84.00,141.00){\line(1,0){4.00}}
\put(88.00,141.00){\line(0,-1){33.00}}
\put(88.00,108.00){\line(-1,0){39.00}}
\put(49.00,108.00){\line(0,-1){3.00}}
\put(81.33,100.33){\line(1,0){6.67}}
\put(88.00,100.33){\line(0,1){7.67}}
\put(25.00,90.00){\line(1,0){24.00}}
\put(49.00,90.00){\line(0,1){5.00}}
\put(25.00,86.00){\line(1,0){46.33}}
\put(71.33,86.00){\line(0,1){9.00}}
\put(25.00,86.00){\line(0,1){8.00}}
\put(13.0,15.3){\makebox(0,0)[lc]{{\bf CONVOLUTION }}}
\put(9.00,10.00){\dashbox{2.00}(85.00,99.00)[lb]{    }}
\end{picture}
\caption[Logical structure of subroutine {\tt ZANCUT}]
{\centering{Logical structure of subroutine {\tt ZANCUT}{\hspace{1cm}}}}
\label{structZANCUT}
\end{figure}
\clearpage 
\input{flow_ewcoup} 
\input{flow_born} 
\begin{figure}
\hspace{2.5cm}
\unitlength=1.00mm
\special{em:linewidth 0.4pt}
\linethickness{0.4pt}
\begin{picture}(133.33,240.33)
\linethickness{1pt}
\put(38.35,224.66){\line(1,0){20.00}}
\put(58.35,224.66){\line(0,-1){10.00}}
\put(58.35,214.33){\line(-1,0){20.00}}
\put(38.35,214.66){\line(0,1){10.00}}
\put(58.35,214.33){\line(-1,0){20.00}}
\linethickness{0.5pt}
\linethickness{1.0pt}
\linethickness{0.5pt}
\linethickness{1.0pt}
\linethickness{0.5pt}
\put(38.35,208.33){\line(1,0){20.00}}
\put(58.35,208.33){\line(0,-1){10.00}}
\put(58.35,198.33){\line(-1,0){20.00}}
\put(38.35,198.33){\line(0,1){10.00}}
\put(38.35,208.33){\line(1,0){20.00}}
\put(58.35,208.33){\line(0,-1){10.00}}
\put(58.35,198.33){\line(-1,0){20.00}}
\put(38.35,198.33){\line(0,1){10.00}}
\put(77.67,126.66){\line(1,0){20.00}}
\put(97.67,126.66){\line(0,-1){10.00}}
\put(97.67,116.66){\line(-1,0){20.00}}
\put(77.67,116.66){\line(0,1){10.00}}
\put(99.67,126.66){\line(1,0){20.00}}
\put(119.67,126.66){\line(0,-1){10.00}}
\put(119.67,116.66){\line(-1,0){20.00}}
\put(99.67,116.66){\line(0,1){10.00}}
\put(76.67,115.66){\line(1,0){44.00}}
\put(120.67,115.66){\line(0,1){12.00}}
\put(120.67,127.66){\line(-1,0){44.00}}
\put(76.67,127.66){\line(0,-1){12.00}}
\put(38.35,224.66){\line(1,0){20.00}}
\put(58.35,224.66){\line(0,-1){10.00}}
\put(58.35,214.33){\line(-1,0){20.00}}
\put(38.35,214.66){\line(0,1){10.00}}
\put(58.35,214.33){\line(-1,0){20.00}}
\put(38.35,110.00){\line(1,0){20.00}}
\put(58.35,110.00){\line(0,-1){10.00}}
\put(58.35,99.67){\line(-1,0){20.00}}
\put(38.35,100.00){\line(0,1){10.00}}
\put(58.35,99.67){\line(-1,0){20.00}}
\put(38.35,110.00){\line(1,0){20.00}}
\put(58.35,110.00){\line(0,-1){10.00}}
\put(58.35,99.67){\line(-1,0){20.00}}
\put(38.35,100.00){\line(0,1){10.00}}
\put(58.35,99.67){\line(-1,0){20.00}}
\put(-11.65,65.00){\line(1,0){20.00}}
\put(8.35,65.00){\line(0,-1){10.00}}
\put(8.35,54.67){\line(-1,0){20.00}}
\put(-11.65,55.00){\line(0,1){10.00}}
\put(8.35,54.67){\line(-1,0){20.00}}
\put(-11.65,65.00){\line(1,0){20.00}}
\put(8.35,65.00){\line(0,-1){10.00}}
\put(8.35,54.67){\line(-1,0){20.00}}
\put(-11.65,55.00){\line(0,1){10.00}}
\put(8.35,54.67){\line(-1,0){20.00}}
\put(25.35,65.00){\line(1,0){20.00}}
\put(45.68,65.00){\line(0,-1){10.00}}
\put(45.35,54.67){\line(-1,0){20.00}}
\put(25.35,55.00){\line(0,1){10.00}}
\put(45.35,54.67){\line(-1,0){20.00}}
\put(25.35,65.00){\line(1,0){20.00}}
\put(45.35,54.67){\line(-1,0){20.00}}
\put(45.35,54.67){\line(-1,0){20.00}}
\put(52.02,65.00){\line(1,0){20.00}}
\put(72.35,65.00){\line(0,-1){10.00}}
\put(72.02,54.67){\line(-1,0){20.00}}
\put(52.02,55.00){\line(0,1){10.00}}
\put(72.02,54.67){\line(-1,0){20.00}}
\put(52.02,65.00){\line(1,0){20.00}}
\put(72.02,54.67){\line(-1,0){20.00}}
\put(72.02,54.67){\line(-1,0){20.00}}
\put(88.35,65.00){\line(1,0){20.00}}
\put(108.35,65.00){\line(0,-1){10.00}}
\put(108.35,54.67){\line(-1,0){20.00}}
\put(88.35,55.00){\line(0,1){10.00}}
\put(108.35,54.67){\line(-1,0){20.00}}
\put(88.35,65.00){\line(1,0){20.00}}
\put(108.35,65.00){\line(0,-1){10.00}}
\put(108.35,54.67){\line(-1,0){20.00}}
\put(88.35,55.00){\line(0,1){10.00}}
\put(108.35,54.67){\line(-1,0){20.00}}
\put(47.99,99.67){\line(0,-1){10.00}}
\put(48.33,219.33){\makebox(0,0)[cc]{CONST1}}
\put(48.33,203.33){\makebox(0,0)[cc]{DAL5H}}
\put(48.33,104.67){\makebox(0,0)[cc]{QCDCOF}}
\put(-1.67,60.00){\makebox(0,0)[cc]{ROKAPN}}
\put(35.33,60.00){\makebox(0,0)[cc]{ROKAPP}}
\put(62.00,60.00){\makebox(0,0)[cc]{FOKAPP}}
\put(98.33,60.00){\makebox(0,0)[cc]{ROKANC}}
\put(38.35,240.33){\line(1,0){20.00}}
\put(58.35,240.33){\line(0,-1){10.00}}
\put(58.35,230.33){\line(-1,0){20.00}}
\put(38.35,230.33){\line(0,1){10.00}}
\put(38.35,240.33){\line(1,0){20.00}}
\put(58.35,240.33){\line(0,-1){10.00}}
\put(58.35,230.33){\line(-1,0){20.00}}
\put(38.35,230.33){\line(0,1){10.00}}
\put(38.35,240.33){\line(1,0){20.00}}
\put(58.35,240.33){\line(0,-1){10.00}}
\put(58.35,230.33){\line(-1,0){20.00}}
\put(38.35,230.33){\line(0,1){10.00}}
\put(38.35,240.33){\line(1,0){20.00}}
\put(58.35,240.33){\line(0,-1){10.00}}
\put(58.35,230.33){\line(-1,0){20.00}}
\put(38.35,230.33){\line(0,1){10.00}}
\put(48.33,235.33){\makebox(0,0)[cc]{{\bf DIZET}}}
\put(68.35,208.33){\line(1,0){20.00}}
\put(88.35,208.33){\line(0,-1){10.00}}
\put(88.35,198.33){\line(-1,0){20.00}}
\put(68.35,198.33){\line(0,1){10.00}}
\put(68.35,208.33){\line(1,0){20.00}}
\put(88.35,208.33){\line(0,-1){10.00}}
\put(88.35,198.33){\line(-1,0){20.00}}
\put(68.35,198.33){\line(0,1){10.00}}
\put(78.33,203.33){\makebox(0,0)[cc]{hadr5.f}}
\put(38.35,181.33){\line(1,0){20.00}}
\put(58.35,181.33){\line(0,-1){10.00}}
\put(58.35,171.33){\line(-1,0){20.00}}
\put(38.35,171.33){\line(0,1){10.00}}
\put(38.35,181.33){\line(1,0){20.00}}
\put(58.35,181.33){\line(0,-1){10.00}}
\put(58.35,171.33){\line(-1,0){20.00}}
\put(38.35,171.33){\line(0,1){10.00}}
\put(68.35,181.33){\line(1,0){20.00}}
\put(88.35,181.33){\line(0,-1){10.00}}
\put(88.35,171.33){\line(-1,0){20.00}}
\put(68.35,171.33){\line(0,1){10.00}}
\put(68.35,181.33){\line(1,0){20.00}}
\put(88.35,181.33){\line(0,-1){10.00}}
\put(88.35,171.33){\line(-1,0){20.00}}
\put(68.35,171.33){\line(0,1){10.00}}
\put(109.01,181.33){\line(1,0){20.00}}
\put(129.01,181.33){\line(0,-1){10.00}}
\put(129.01,171.00){\line(-1,0){20.00}}
\put(109.01,171.33){\line(0,1){10.00}}
\put(129.01,171.00){\line(-1,0){20.00}}
\put(109.01,165.00){\line(1,0){20.00}}
\put(129.01,165.00){\line(0,-1){10.00}}
\put(129.01,155.00){\line(-1,0){20.00}}
\put(109.01,155.00){\line(0,1){10.00}}
\put(109.01,165.00){\line(1,0){20.00}}
\put(129.01,165.00){\line(0,-1){10.00}}
\put(129.01,155.00){\line(-1,0){20.00}}
\put(109.01,155.00){\line(0,1){10.00}}
\put(109.01,181.33){\line(1,0){20.00}}
\put(129.01,181.33){\line(0,-1){10.00}}
\put(129.01,171.00){\line(-1,0){20.00}}
\put(109.01,171.33){\line(0,1){10.00}}
\put(129.01,171.00){\line(-1,0){20.00}}
\put(109.01,197.00){\line(1,0){20.00}}
\put(129.01,197.00){\line(0,-1){10.00}}
\put(129.01,187.00){\line(-1,0){20.00}}
\put(109.01,187.00){\line(0,1){10.00}}
\put(109.01,197.00){\line(1,0){20.00}}
\put(129.01,197.00){\line(0,-1){10.00}}
\put(129.01,187.00){\line(-1,0){20.00}}
\put(109.01,187.00){\line(0,1){10.00}}
\put(109.01,197.00){\line(1,0){20.00}}
\put(129.01,197.00){\line(0,-1){10.00}}
\put(129.01,187.00){\line(-1,0){20.00}}
\put(109.01,187.00){\line(0,1){10.00}}
\put(109.01,197.00){\line(1,0){20.00}}
\put(129.01,197.00){\line(0,-1){10.00}}
\put(129.01,187.00){\line(-1,0){20.00}}
\put(109.01,187.00){\line(0,1){10.00}}
\put(48.33,176.33){\makebox(0,0)[cc]{XFOTF3}}
\put(38.35,143.33){\line(1,0){20.00}}
\put(58.35,143.33){\line(0,-1){10.00}}
\put(58.35,133.33){\line(-1,0){20.00}}
\put(38.35,133.33){\line(0,1){10.00}}
\put(38.35,143.33){\line(1,0){20.00}}
\put(58.35,143.33){\line(0,-1){10.00}}
\put(58.35,133.33){\line(-1,0){20.00}}
\put(38.35,133.33){\line(0,1){10.00}}
\put(48.33,138.33){\makebox(0,0)[cc]{QCDCOF}}
\put(38.35,126.67){\line(1,0){20.00}}
\put(58.35,126.67){\line(0,-1){10.00}}
\put(58.35,116.67){\line(-1,0){20.00}}
\put(38.35,116.67){\line(0,1){10.00}}
\put(38.35,126.67){\line(1,0){20.00}}
\put(58.35,126.67){\line(0,-1){10.00}}
\put(58.35,116.67){\line(-1,0){20.00}}
\put(38.35,116.67){\line(0,1){10.00}}
\put(48.33,121.67){\makebox(0,0)[cc]{SETCON}}
\put(35.00,84.67){\line(0,-1){12.00}}
\put(35.00,68.67){\line(0,-1){3.67}}
\put(61.66,84.67){\line(0,-1){11.67}}
\put(61.66,68.67){\line(0,-1){3.67}}
\put(-10.66,52.67){\makebox(0,0)[lc]{{\tiny for neutrino }}}
\put(-10.66,51.00){\makebox(0,0)[lc]{{\tiny reactions}}}
\put(89.33,52.67){\makebox(0,0)[lc]{{\tiny for $e^+e^-\to f \bar{f}$}}}
\put(42.00,52.67){\makebox(0,0)[lc]{{\tiny for $Z$-decay}}}
\put(61.66,71.00){\makebox(0,0)[cc]{{\tt AMT4=4}}}
\put(35.00,70.67){\makebox(0,0)[cc]{{\tt AMT4 $\neq 4$}}}
\put(87.66,121.67){\makebox(0,0)[cc]{CONST2}}
\put(109.66,121.67){\makebox(0,0)[cc]{SEARCH}}
\put(78.33,176.33){\makebox(0,0)[cc]{DALPHL}}
\put(119.00,192.00){\makebox(0,0)[cc]{ALQCDS}}
\put(118.66,176.33){\makebox(0,0)[cc]{ALQCD}}
\put(118.66,160.00){\makebox(0,0)[cc]{AQCDBK}}
\put(48.00,230.33){\line(0,-1){5.67}}
\put(48.00,214.33){\line(0,-1){6.00}}
\put(58.33,203.67){\line(1,0){10.00}}
\put(48.00,198.33){\line(0,-1){17.00}}
\put(58.33,176.33){\line(1,0){10.00}}
\put(88.33,176.67){\line(1,0){2.00}}
\put(90.66,192.33){\line(0,-1){32.00}}
\put(90.66,160.33){\line(1,0){2.67}}
\put(90.66,176.67){\line(1,0){2.67}}
\put(90.66,192.33){\line(1,0){2.67}}
\put(106.66,192.33){\line(1,0){2.33}}
\put(109.00,176.67){\line(-1,0){2.33}}
\put(106.66,160.33){\line(1,0){2.33}}
\put(100.00,192.33){\makebox(0,0)[cc]{{\tt QCDC=1}}}
\put(100.00,176.67){\makebox(0,0)[cc]{{\tt QCDC=2}}}
\put(100.00,160.33){\makebox(0,0)[cc]{{\tt QCDC=3}}}
\put(48.00,171.33){\line(0,-1){28.00}}
\put(48.00,133.33){\line(0,-1){6.67}}
\put(48.00,116.67){\line(0,-1){6.67}}
\put(58.33,121.67){\line(1,0){18.33}}
\put(-20.00,40.00){\dashbox{1.00}(136.67,35.00)[lb]{  }}
\put(48.00,45.00){\makebox(0,0)[cc]{EW form factors}}
\put(58.35,89.67){\line(0,-1){10.00}}
\put(38.35,79.67){\line(0,1){10.00}}
\put(58.35,79.34){\line(-1,0){20.00}}
\put(38.35,89.67){\line(1,0){20.00}}
\put(58.35,89.67){\line(0,-1){10.00}}
\put(58.35,79.34){\line(-1,0){20.00}}
\put(38.35,79.67){\line(0,1){10.00}}
\put(58.35,79.34){\line(-1,0){20.00}}
\put(48.33,84.34){\makebox(0,0)[cc]{ZWRATE}}
\put(35.00,84.67){\line(1,0){3.33}}
\put(58.33,84.67){\line(1,0){3.33}}
\put(15.00,152.00){\dashbox{1.00}(118.33,59.33)[lb]{  }}
\put(20.00,160.33){\makebox(0,0)[lc]{$\Delta\alpha$}}
\put(49.00,196.33){\makebox(0,0)[lc]{{\tiny $\Delta\alpha_h$}}}
\put(69.33,169.33){\makebox(0,0)[lc]{{\tiny $\Delta\alpha_l$}}}
\put(110.00,198.67){\makebox(0,0)[lc]{{\tiny 
 $\Delta\alpha^{\alpha\alpha_s}$}}}
\end{picture}
\vspace{-1.5cm}
\caption[Logical structure of the package {\tt DIZET}]{
\centering{\it Logical structure of the package {\tt DIZET}\hspace{1cm}}}
\label{structDIZET}
\end{figure}
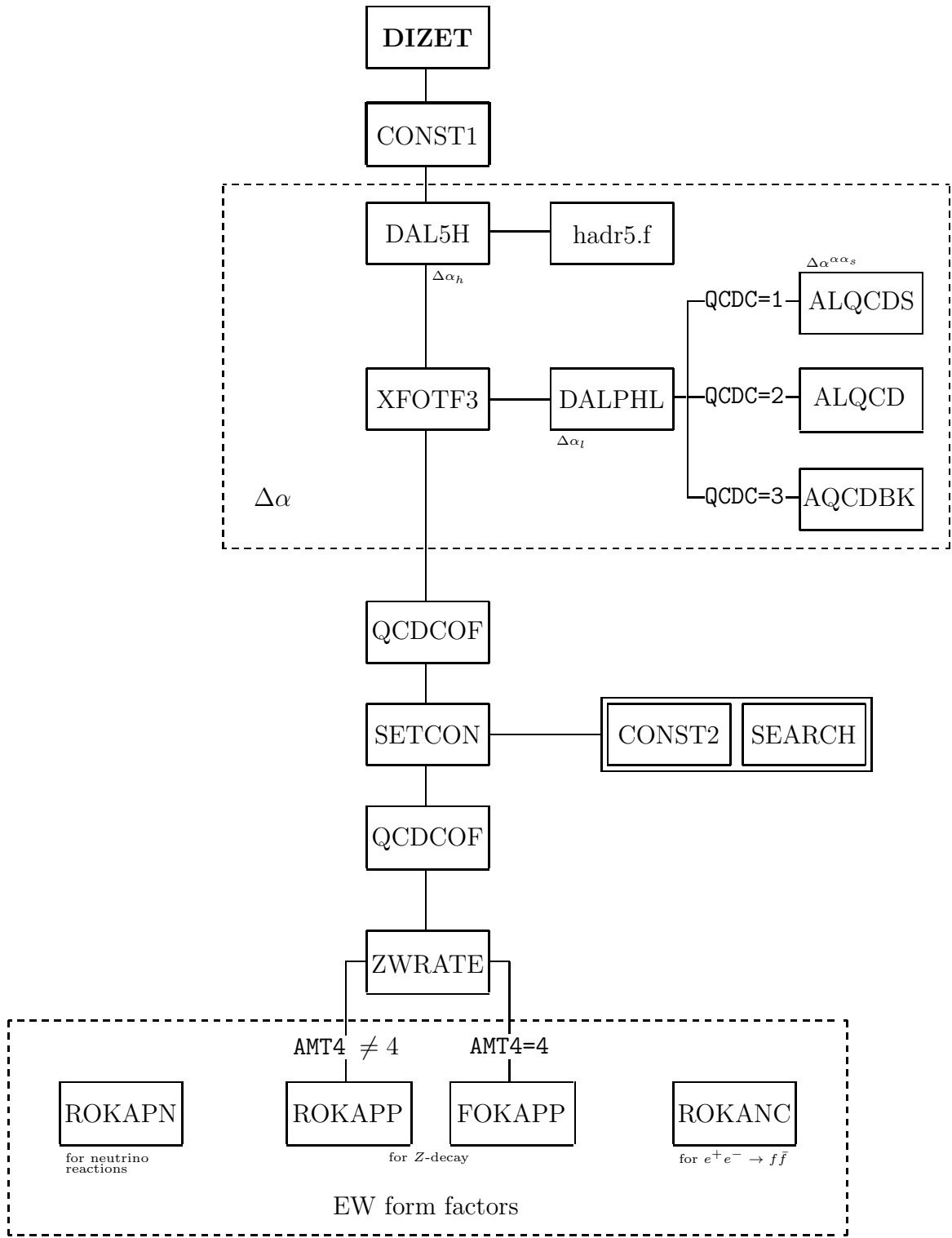
\clearpage 
\vspace*{5.0cm}

\begin{table}[bthp]\centering
\begin{tabular}{|c|c|c|c|c|c|c|c|c|}  \hline
{\tt    CHFLAG}&{\tt IVALUE}&
{\tt    CHFLAG}&{\tt IVALUE}&
\mbox
{\tt    CHFLAG}&{\tt IVALUE}\\
\hline
  {\tt AFBC} & 1
& {\tt AFMT} & 1 
& {\tt ALEM} & 3 \\
  {\tt ALE2} & 3
& {\tt AMT4} & 4
& {\tt BARB} & 2 \\
  {\tt BORN} & 0
& {\tt BOXD} & 1
& {\tt CONV} & 1 \\
  {\tt CZAK} & 1
& {\tt DIAG} & 1
& {\tt EXPF} & 0 \\
  {\tt EXPR} & 0
& {\tt FINR} & 1
& {\tt FOT2} & 3 \\
  {\tt FSRS} & 1
& {\tt FTJR} & 1
& {\tt GAMS} & 1 \\
  {\tt GFER} & 2
& {\tt HIGS} & 0
& {\tt HIG2} & 0 \\
  {\tt INTF} & 1
& {\tt ISPP} & 2
& {\tt IPFC} & 5 \\
  {\tt IPSC} & 0
& {\tt IPTO} & 3
& {\tt MISC} & 0 \\
  {\tt MISD} & 1
& {\tt PART} & 0
& {\tt POWR} & 1 \\
  {\tt PREC} & 10
& {\tt PRNT} & 0
& {\tt QCDC} & 3 \\
  {\tt SCAL} & 0
& {\tt SCRE} & 0
& {\tt VPOL} & 1 \\
  {\tt WEAK} & 1  & & & & \\
\hline
\end{tabular}
\caption[Flag settings for \zf]
{\it Flag settings for \zf;
     the flags are listed in alphabetical order. The numerical values
     are the default settings
}
\label{ta7}
\end{table}

\vfill

\begin{table}[bthp]
\begin{center}
\begin{tabular}{|l|cccccccccccc|}
\hline
 Final- & & & & & & & & & & & &  \\
 state  & $\nu{\bar{\nu}}$ & $e^+e^-$  & $\mu^+\mu^-$ &
          $\tau^+\tau^-$   & $u\bar{u}$& $d\bar{d}$   & 
          $c\bar{c}$       & $s\bar{s}$& $t\bar{t}$   &
          $b\bar{b}$      &hadrons&Bhabha\\
 fermions & & & & & & & & & & & &  \\ \hline  \hline
{\tt INDF}  &~~0~~&~~1~~&~~2~~&~~3~~&~~4~~&~~5~~&~~6~~&~~7~~&~~8~~&
         ~~9~~&~~10~~&~~11~~\\
\hline
\end{tabular}
\end{center}
\caption[Indices for the selection of final states]
{\it
Indices for the selection of final states.
Note that \mbox{ {\tt INDF} = 0,1} returns only s-channel observables,
\mbox{{\tt INDF} = 8} always returns zero,  and
\mbox{{\tt INDF} = 10} indicates a sum over all open quark channels.
\label{indf}
}
\end{table}

\bigskip

\clearpage

\begin{table}[bthp]\centering
\begin{tabular}{|c|c|c|c|c|c|}  \hline
I  &{\tt 'FLAG'}&  name in  & Position in & Position in & default \\
     &          & programs  &{\tt NPAR(1:21)}&{\tt NPAR(1:30)}
     &    \\  & & &({\tt DIZET})&({\tt ZFITTER})& \\
\hline
\hline
 1 &{\tt AFBC} &{\tt IAFB}  &    & 13 & 1 \\
 2 &{\tt SCAL} &{\tt ISCAL} &  9 & 15 & 0 \\
 3 &{\tt SCRE} &{\tt ISCRE} &  6 &    & 0 \\
 4 &{\tt AMT4} &{\tt IAMT4} &  2 & 16 & 4 \\
 5 &{\tt BORN} &{\tt IBORN} &    & 14 & 0 \\
 6 &{\tt BOXD} &{\tt IBOX}  &    &  4 & 1 \\
 7 &{\tt CONV} &            &    &    & 1 \\
 8 &{\tt FINR} &{\tt IFINAL}&    &  9 & 1 \\
 9 &{\tt FOT2} &{\tt IPHOT2}&    & 10 & 3 \\
10 &{\tt GAMS} &            &    &  5 & 1 \\
11 &{\tt DIAG} &            &    &  7 & 1 \\
12 &{\tt INTF} &{\tt INTERF}&    &  8 & 1 \\
13 &{\tt BARB} &{\tt IBARB} & 10 &    & 2 \\
14 &{\tt PART} &{\tt IPART} &    & 17 & 0 \\
15 &{\tt POWR} &            &    &    & 1 \\
16 &{\tt PRNT} &            &    &    & 0 \\
17 &{\tt ALEM} &{\tt IALEM} &  7 & 20 & 3 \\
18 &{\tt QCDC} &{\tt IQCD}  &  3 &  3 & 3 \\
19 &{\tt VPOL} &{\tt IHVP}  &  1 &  2 & 1 \\
20 &{\tt WEAK} &{\tt IWEAK} &    &  1 & 1 \\
21 &{\tt FTJR} &{\tt IFTJR} & 11 &    & 1 \\
22 &{\tt EXPR} &{\tt IFACR} & 12 &    & 0 \\
23 &{\tt EXPF} &{\tt IFACT} & 13 & 19 & 0 \\
24 &{\tt HIGS} &{\tt IHIGS} & 14 &    & 0 \\
25 &{\tt AFMT} &{\tt IAFMT} & 15 &    & 1 \\
26 &{\tt CZAK} &{\tt ICZAK} & 17 &    & 1 \\
27 &{\tt PREC} &            &    &    &10 \\
28 &{\tt HIG2} &{\tt IHIG2} & 18 &    & 0 \\
29 &{\tt ALE2} &{\tt IALE2} & 19 & 21 & 3 \\
30 &{\tt GFER} &{\tt IGFER} & 20 &    & 2 \\
31 &{\tt ISPP} &{\tt ISRPPR}&    &    & 1 \\
32 &{\tt FSRS} &            &    &    & 1 \\
33 &{\tt MISC} &{\tt IMISC} &    &    & 0 \\
34 &{\tt MISD} &{\tt IMISD} &    &    & 1 \\
35 &{\tt IPFC} &{\tt IPFC}  &    &    & 5 \\
36 &{\tt IPSC} &{\tt IPSC}  &    &    & 0 \\
37 &{\tt IPTO} &{\tt IPTO}  &    &    & 3 \\
   &           &{\tt IMOMS} &  4 &    & 1 \\
   &           &{\tt IMASS} &  5 &    & 0 \\
   &           &{\tt IMASK} &  8 &    & 0 \\
   &           &{\tt IEWLC} & 16 &    & 1 \\
   &           &{\tt IDDZZ} & 21 &    & 1 \\
\hline
\end{tabular}
\caption[Flags used in {\tt DIZET} and {\tt ZFITTER}]
{\it Flag settings for \zf\ and {\tt DIZET};
     the flags are listed in the order of vector {\tt IFLAGS}.  
The
     corresponding names used internally in the programs, the position of
     the flags in vector {\tt NPAR(1:21)} of {\tt DIZET} and 
     {\tt NPAR(1:30)} of {\tt ZFITTER} and the default values 
     are given
}
\label{tab:xxxx}
\end{table}

\def\theequation{\arabic{chapter}.\arabic{section}.\arabic{equation}}
\def\thefigure{\arabic{chapter}.\arabic{figure}}
\def\thetable{\arabic{chapter}.\arabic{table}}
\chapter{Photonic Corrections\label{ch-phot}}
\section{Born Cross-Sections\label{i.born}}
\eqnzero
Born cross-sections are calculated in \zf\ for two purposes:
\begin{itemize}
\item
If the user wants to calculate a Born cross-section;
\item
In order to be convoluted during the calculation of a QED
corrected cross-section.
\end{itemize}
Born approximations are calculated with \zf\ by choosing flag {\tt BORN}=1 
in subroutine {\tt ZUFLAG}\footnote{The
logical structure of \zf\ is shown in \fig{zf_flow_main}. 
}.
The fermion production channel is chosen with flag {\tt INDF}. 
For a list of the fermion indices see \tbn{indf}\footnote{A 
systematic presentation of the initialization of \zf\ is given in
\appendx{zfguide}.  
}.
There are two different cases:
\begin{itemize}
\item 
Calculation of a differential cross-section 
$d\sigma^{\rm{Born}}/d\cos\vartheta$.
This is foreseen within the Standard Model only and initiated by the
user with the interface {\tt ZUATSM}\footnote{A systematic
presentation of the interfaces of \zf\ is given in 
\appendx{interfaces}.  
}.
Subroutine {\tt ZANCUT} is called, which calls  subroutines {\tt EWCOUP} 
and {\tt COSCUT}. The latter calls   subroutine {\tt BORN}.
The Born cross-section $d\sigma^{\rm{Born}}/d\cos\vartheta$ is
calculated in subroutine {\tt COSCUT}.
\item 
Calculation of an integrated cross-section $\sigma_{\sss{T}}^{\rm{Born}}$
(or angular asymmetry $A_{\sss{FB}}^{\rm{Born}}$):
This may be initialized by any of the other interfaces.
For the Standard Model, e.g., subroutine {\tt ZUTHSM} calls subroutine
{\tt ZCUT}, which calls subroutines {\tt EWCOUP} and {\tt SFAST} (without
angular acceptance cut) or 
subroutines {\tt EWCOUP} and {\tt SCUT} (with angular acceptance cut).
Then subroutine {\tt  SFAST} or  subroutine {\tt SCUT} calls
subroutine {\tt BORN} and calculates 
the integrated Born cross-section $\sigma_{\sss{T}}^{\rm{Born}}$ and angular
asymmetry $A_{\sss{FB}}^{\rm{Born}}$.  
\end{itemize}
In all cases it is subroutine {\tt EWCOUP} (see \sect{ewcoup}) where
the coupling constants of the 
cross-section are determined. 
Polarization effects are also taken into account there.
Then, in subroutine {\tt BORN}  (see \sect{IBORNA}) the angular
independent pieces of the Born cross-section are composed.

In this Section we will describe the Born formulae for the simplest
case of photon and $\zb$ exchange with constant, real-valued vector and
axial vector couplings.
This corresponds e.g. to the tree level Standard Model (interfaces 
{\tt ZUATSM} or {\tt ZUTHSM}) or to a 
semi-model-independent ansatz (e.g. interface {\tt ZUXSA}).

The differential Born cross-section {\tt SIGBRN} is:
\bqa
\frac{d\sigma^{\rm{Born}}}{d\cos\vartheta}
&=& 
D_{\sss{T}}(\cos\vartheta)\, \sigma_{\sss{T}}^0(\sman)+\frac{4\mfs}{\sman}\,
D_{\sss{T}}^m(\cos\vartheta)\,\sigma_{\sss{T}}^{0,m}(\sman) 
+D_{\sss{FB}}(\cos\vartheta)\,\sigma_{\sss{FB}}^0(\sman),
\label{sigdiff}
\eqa

\begin{figure}[th]
\vspace{-17mm}
\[
\begin{array}{ccc}
\begin{picture}(125,86)(0,40)
  \Photon(25,43)(100,43){3}{25}
  \ArrowLine(125,86)(100,43)
  \Vertex(100,43){2.5}
     \ArrowLine(100,43)(125,0)
     \ArrowLine(0,0)(25,43)
     \Vertex(25,43){2.5}
     \ArrowLine(25,43)(0,86)
\Text(14,74)[lb]{$\fbe$}
\Text(108,74)[lb]{$\fbf$}
\Text(62.5,50)[bc]{$\ph$}
\Text(14,12)[lt]{$\fe$}
\Text(108,12)[lt]{$\ff$}
\end{picture}
\qquad
&+&
\qquad
\begin{picture}(125,86)(0,40)
  \Photon(25,43)(100,43){3}{15}
  \ArrowLine(125,86)(100,43)
  \Vertex(100,43){2.5}
     \ArrowLine(100,43)(125,0)
     \ArrowLine(0,0)(25,43)
     \Vertex(25,43){2.5}
     \ArrowLine(25,43)(0,86)
\Text(14,74)[lb]{$\fbe$}
\Text(108,74)[lb]{$\fbf$}
\Text(62.5,50)[bc]{$\zb$}
\Text(14,12)[lt]{$\fe$}
\Text(108,12)[lt]{$\ff$}
\end{picture}
\end{array}
\]
\vspace*{10mm}
\caption[Born approximation for $\fe\fbe\to(\zb,\ph)\to\ff\fbf$]
{\it
Born approximation for $\fe\fbe\to(\zb,\ph)\to\ff\fbf$ 
\label{zavert2}
}
\end{figure}
\vspace*{3mm}

with

\bqa 
\sigma_{\sss{T}}^0(\sman) &=&  
\frac{\pi \alpha^2}{\sman} \beta_f(\sman)
\left[ K_{\sss{T}}(\ph) +K_{\sss{T}}(I) \,\Re e\chi_{\sss{Z}}(\sman)
+K_{\sss{T}}(Z)\,|\chi_{\sss{Z}}(\sman)|^2 \right],
\\ \nl 
\sigma_{\sss{T}}^{0,m}(\sman) &=&  
\frac{\pi \alpha^2}{\sman} \beta_f(\sman)
\left[ K^{m}_{\sss{T}}(\ph) 
      +K^{m}_{\sss{T}}(I) \,\Re e\chi_{\sss{Z}}(\sman)
      +K^{m}_{\sss{T}}(Z) \, |\chi_{\sss{Z}}(\sman)|^2 \right],
\\ \nl 
\sigma_{\sss{FB}}^0(\sman) 
&=& 
\frac{\pi \alpha^2}{\sman} \beta_f(\sman) 
\left[ K_{\sss{FB}}(I)
\Re e \chi_{\sss{Z}}(\sman)
+K_{\sss{FB}}(Z)\left|\chi_{\sss{Z}}(\sman)\right|^2\right],
\label{sigdiff1}
\eqa
\noindent and
\bqa
D_{\sss{T}}(\cos\vartheta)  &=& \frac{1}{2} \left( 1+\cos^2\vartheta \right),
\label{sigdiff2}
\\ \nl
D_{\sss{T}}^m(\cos\vartheta)&=& \frac{1}{2} \sin^2\vartheta,
\\ \nl
D_{\sss{FB}}(\cos\vartheta) &=& \cos\vartheta, 
\label{sigdiff3}
\\ \nl
\beta_f(\sman) &=& \sqrt{1-\frac{4\mfs}{\sman}}\,.
\label{mus}
\eqa

For hadrons, we sum over $\fu,\fd,\fc,\fs,\fb$ quarks (above their production
thresholds). 
Masses of quarks are usually treated together with QCD corrections,
see \sect{qcdrun}.
Some deviating choice of inclusion of final state masses is possible for
hadronic final states with flags {\tt FINR}=-1 and {\tt POWR}, 
see \appendx{zuflag}.

The $\zb$ propagator is contained in the factor:
\ba
\chi_Z(\sman) &=& 
\frac{\gf}{\sqrt{2}}\frac{\mzs}{8\pi\alpha}\frac{\sman}{\sman-m^2_{\sss{Z}}},
\label{chiZ}
\\ \nl
m^2_{\sss{Z}} &=& \mzs - \ib\mzl\gz(\sman).
\label{mZ2}
\ea
An $\sman$ independent width may be taken into account with the
transformation of constants as introduced in~\cite{Bardin:1988xt}.
This is realized in \zf\ as described in \sect{Zpropagator}.
For the calculation of $\gz$ in the Standard Model we refer to
\sect{subgamma}. 
We also define accordingly for the photon propagator:
\ba
\chi_{\ph}(\sman) &=& 1,
\label{kg}
\ea
and use the following conventions:
\bqa
\label{kg1}
|Q_e|&=&1, 
\\
\label{kg2}
a_f&=&1, 
\\
\label{kg3}
v_f&=& 1 - 4 |Q_f| \siws.
\eqa
Further, the coupling functions are:
\ba
\label{kgizsplit}
K_{\sss{T}}(\ph) &=& K^{m}_{\sss{T}}(\ph)=\qes\qfs\cf, 
\\ \nl
\label{kgfb}
K_{\sss{FB}}(\ph)&=& 0,
\\ \nl
K_{\sss{T}}(I)  &=& K^{m}_{\sss{T}}(I)  =2|Q_eQ_f|v_ev_f\cf, 
\\ \nl
K_{\sss{FB}}(I) &=& 2|Q_eQ_f| a_ea_f\beta_f\cf,
\\ \nl
K_{\sss{T}}(Z)  &=&\bigl(v_e^2+a_e^2\bigr)
                    \bigl(v_f^2+a_f^2\beta_f^2\bigr)\cf, 
\\ \nl
K_{\sss{FB}}(Z) &=& 4v_ea_ev_fa_f\beta_f\cf, 
\\ \nl
K_{\sss{T}}^m(Z)&=&\bigl(v_e^2+a_e^2\bigr)v_f^2\cf, 
\\ \nl
{\bar{K}}_{\sss{T}}^m(Z)&=&\bigl(v_e^2+a_e^2\bigr)a_f^2\cf,
\label{kgiz}
\ea
where $\cf$ is the color factor 1(3) for leptons (quarks).
We added the definition of ${\bar{K}}_{\sss{T}}^m(Z)$ for later use.

Later on, in \chaptsc{ch-PO}{ch-IBA} the corrections
from weak loop insertions, from the running  
QED coupling
and, for quark production, from QCD will be absorbed in the couplings $K_{\sss{A}}$ 
and in the prediction for the $\zb$ width\footnote{Note however that
corrections from $\zb\zb$ and $\wb\wb$ box diagrams have a 
complicated angular dependence; see \subsect{ew_boxes}. }.
In model independent approaches, the $K_{\sss{A}}$ will be, together
with $\mzl$ and $\gz$, the parameters to be determined; see
\sect{ewcoup}. 

The most general expressions for four
different polarizations of the fermions are given explicitely
in \sect{helipol}.
Weak higher order corrections influence also exclusively the
combinations $K_{\sss{A}}$.

QED corrections depend
only on $\mzl, \gz, \cos\vartheta, \sman$ and, due to mass
singularities, on fermion masses.
The explicit expressions for the QED corrections will
contain {\em effective} Born cross-section factors with a reduced invariant
mass $\smanp$: 
\ba
\label{sa0}
\sigma_{\sss{A}}^0(\smanp)
&=& \sum_{m,n=\ph,Z} \sigma_{\sss{A}}^0(s',s';m,n), 
\\  \nl   
\sigma_{\sss{A}}^0(s,s')
&=&
\sum_{m,n=\ph,Z}
\sigma_{\sss{A}}^0(s,s';m,n),
\\ \nl
\sigma_{\sss{A}}^0(s,s';m,n)
&=&
\frac{\pi \alpha^2}{s'}
K_{\sss{A}}(m,n) 
\frac{1}{2} 
\left[\chi_m(\sman)\chi_n^*(\smanp) + \chi_m(\smanp)\chi_n^*(\sman)\right], 
\hspace{.5cm} A=T,FB.
\nl
\label{bornss'}
\ea
The $K_{\sss{A}}(m,n)$ are those defined in \eqns{kgizsplit}{kgiz}.
Further,  we use the
correspondences $(\ph\ph, \ph Z+Z\ph,ZZ) \sim (\ph, I, Z)$. 
The functions $\sigma_{\sss{A}}^0(s,s')$ are needed when we treat the
interference of initial state radiation (with scale $\smanp$) and final
state radiation (with scale $\sman$).

After angular integration, the total Born cross-section 
{\tt SBORN} = $\sigma_{\sss{T}}^{\rm{Born}} (s,c)$
and
forward-backward asymmetry 
{\tt ABORN} = $A_{\sss{FB}}^{\rm{Born}} (s,c)$ become:
\ba
\sigma_{\sss{T}}^{\rm{Born}} (s,c)
&=&  C_{\sss{T}}(c) \, \sigma_{\sss{T}}^0(\sman) 
+ \frac{4\mfs}{\sman} \,
 C_{\sss{T}}^m(c) \, \sigma_{\sss{T}}^{0,m}(\sman),
\label{sigtot}
\\ \nl
A_{\sss{FB}}^{\rm{Born}} (s,c)&=&
\frac{\sigma_{\sss{FB}}^{\rm{Born}}(s,c)}{\sigma_{\sss{T}}^{\rm{Born}}(s,c)},
\\ \nl
\sigma_{\sss{FB}}^{\rm{Born}}(s,c) &=& 
C_{\sss{FB}}(c) \, \sigma_{\sss{FB}}^{0}(\sman).
\nl
\ea
Here, we allow for an angular acceptance cut,
\bqa
C_{\sss{T}}(c)&=&
\int_{-c}^c d\cos \vartheta \, D_{\sss{T}}(\cos \vartheta)
= c\left(1+\frac{c^2}{3}\right),
\label{ct}
\\
C_{\sss{T}}^m(c)&=&
\int_{-c}^c d\cos \vartheta \, D_{\sss{T}}^m(\cos \vartheta)
= c\left(1-\frac{c^2}{3}\right),
\label{ctm}
\\
\label{cfb}
C_{\sss{FB}}(c) &=& \left\{ \int_0^c - \int_{-c}^0 \right\}  d\cos
\vartheta \, D_{\sss{FB}}(\cos\vartheta) =  c^2. 
\eqa
The usual normalization factors $C_{\sss{T}}(1)=4/3, C_{\sss{FB}}=1$ 
are obtained if the full scattering region is explored. 
In the program, we often use the abbreviations 
{\tt COPL3} = $c + c^3/3$ and {\tt COPL2} = $1 + c^2$.  
 
Finally we should mention that mass effects in the improved Born 
cross-sections of \zf\ are contained in only three factors: 
\bqa
{\tt THRESH}&=& \beta_f(s),
\label{thresh}
\\
{\tt CORF2} &=& c_1(\mfl) = 1+2\frac{\mfs}{\sman},   
\label{corf2}
\\
{\tt CORF3} &=& c_2(\mfl) =  -6\frac{\mfs}{\sman}.
\label{corf3}
\eqa
Further, there are mass dependent QCD corrections, see \subsect{R-factors}.

The mass factors \eqns{thresh}{corf3} are valid for integrated
cross-sections without cuts. 
For differental cross-sections and for those with cuts applied, they have to 
be considered as approximations.
This seems to be no numerical problem for final state fermions foreseen 
in \zf.  
We just remind that \zf\ does not yet cover $\ft$-quark production.

\section{Photonic Corrections. Overview\label{chains}}
\setcounter{equation}{0}
The branches with different treatments of photonic cuts are chosen with flag
{\tt ICUT} in subroutine {\tt ZUINIT} when calling subroutine {\tt ZUCUTS}: 
the values $-1, +1, +2, +3$ correspond to the 
application of: no cuts but an $s'$ cut, the $s'$ cut plus acceptance
cut, the acollinearity cut without/with acceptance cut.   
In the case of the angular distribution $d\sigma/d\cos\vartheta$,
the flag values are reduced to $+1, +2, +3$ where ${\tt ICUT}=+1$ 
means an $s'$ cut, and ${\tt ICUT}=+2$ (and for checking purposes 
equivalently $+3$) treats an acollinearity cut. 

The numerical values of the cut variables are set by the user when
calling subroutine {\tt ZUCUTS}, see \appendx{zucuts}.
Real photonic corrections are influenced by the following additional flags:
{\tt INTF}, {\tt FINR}, {\tt FOT2}.
 
\begin{figure}[t]
\begin{center}
\vfill
\setlength{\unitlength}{1pt}
\SetWidth{0.8}
\begin{picture}(180,120)(0,0)
\thicklines
\ArrowLine(10,110)(40,80)
\Vertex(30,90){1.8}
\Photon(30,90)(70,110){2}{7}
\ArrowLine(30,90)(60,60)
\Vertex(60,60){1.8}
\ArrowLine(60,60)(10,10)
\Photon(60,60)(120,60){3}{8}
\Vertex(120,60){1.8}
\ArrowLine(120,60)(170,110)
\ArrowLine(170,10)(120,60)
\Text(5,117)[]{$e^-$}
\Text(5,5)[]{$e^+$}
\Text(90,50)[]{$\gamma\,,\,Z$}
\Text(178,8)[]{$\bar{f}$}
\Text(178,113)[]{$f$}
\Text(77,114)[]{$\gamma$}
\end{picture}
\setlength{\unitlength}{1pt}
\SetWidth{0.8}
\begin{picture}(200,120)(0,0)
\thicklines
\ArrowLine(10,110)(60,60)
\Vertex(60,60){1.8}
\ArrowLine(60,60)(10,10)
\Photon(60,60)(120,60){3}{8}
\Vertex(120,60){1.8}
\ArrowLine(170,10)(150,30)
\Vertex(150,30){1.8}
\Photon(150,30)(190,50){2}{7}
\ArrowLine(150,30)(120,60)
\ArrowLine(120,60)(170,110)
\Text(5,117)[]{$e^-$}
\Text(5,5)[]{$e^+$}
\Text(90,50)[]{$\gamma\,,\,Z$} 
\Text(178,8)[]{$\bar{f}$}
\Text(178,113)[]{$f$}
\Text(197,53)[]{$\gamma$}
\end{picture}

\vspace*{0.5cm}
\caption
[Real photon emission]
{\it
Examples for real photon emission 
\label{fig.realbr}
}
\end{center}
\end{figure}
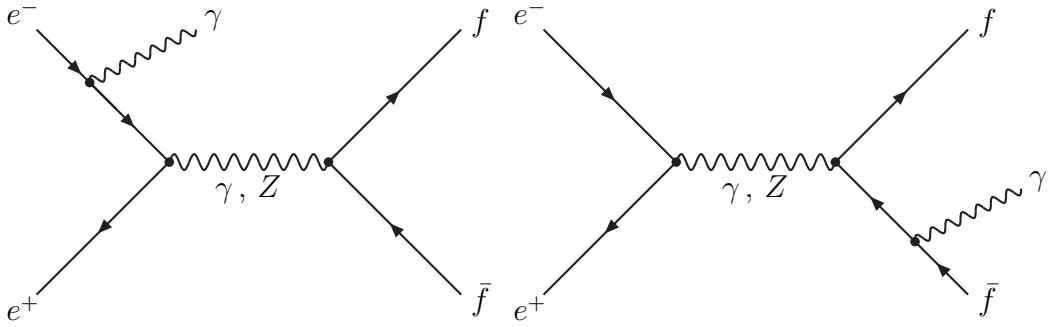

\begin{figure}[tbhp]
\begin{center}
\vfill
\setlength{\unitlength}{1pt}
\SetWidth{0.8}
\begin{picture}(180,120)(0,0)
\thicklines
\ArrowLine(10,110)(40,80)
\Vertex(40,80){1.8}
\ArrowLine(40,80)(60,60)
\Photon(40,80)(40,40){2}{5}
\ArrowLine(60,60)(40,40)
\Vertex(40,40){1.8}
\ArrowLine(40,40)(10,10)
\Vertex(60,60){1.8}
\Photon(60,60)(120,60){3}{8}
\Vertex(120,60){1.8}
\ArrowLine(120,60)(170,110)
\ArrowLine(170,10)(120,60)
\Text(5,117)[]{$e^-$}
\Text(5,5)[]{$e^+$}
\Text(90,50)[]{$\gamma\,,\,Z$}
\Text(178,8)[]{$\bar{f}$}
\Text(178,113)[]{$f$}
\Text(32,60)[]{$\gamma$}
\end{picture}
\setlength{\unitlength}{1pt}
\SetWidth{0.8}
\begin{picture}(180,120)(0,0)
\thicklines
\ArrowLine(10,110)(60,60)
\Vertex(60,60){1.8}
\ArrowLine(60,60)(10,10)
\Photon(60,60)(120,60){3}{8}
\Vertex(120,60){1.8}
\ArrowLine(170,10)(140,40)
\Vertex(140,40){1.8}
\ArrowLine(140,40)(120,60)
\Photon(140,40)(140,80){2}{5}
\ArrowLine(120,60)(140,80)
\Vertex(140,80){1.8}
\ArrowLine(140,80)(170,110)
\Text(5,117)[]{$e^-$}
\Text(5,5)[]{$e^+$}
\Text(90,50)[]{$\gamma\,,\,Z$}
\Text(178,8)[]{$\bar{f}$}
\Text(178,113)[]{$f$}
\Text(149,60)[]{$\gamma$}
\end{picture}

\vspace*{0.5cm}
\caption[
Photonic vertex corrections]
{\label{fig2paw}
{\it
The photonic vertex corrections 
}}
\label{fig.virtph}
\end{center}
\end{figure}
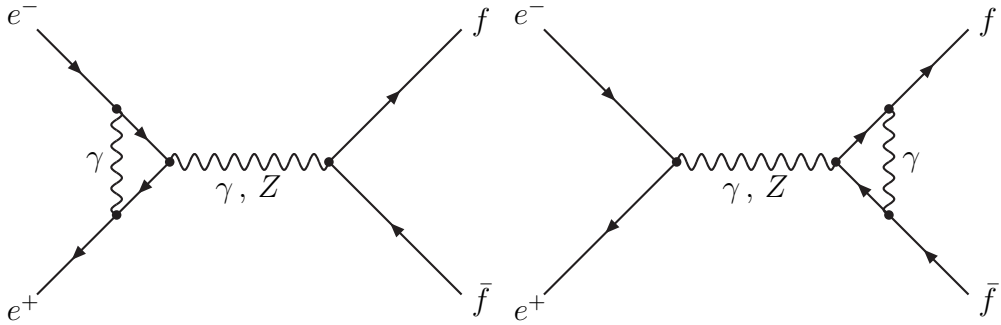

In this Section, we will describe real photonic corrections to 
$d\sigma/d\cos\vartheta$, $\sigma_{\sss{T}}(c)$ and $\sigma_{\sss{FB}}(c)$.
A complete treatment of photonic corrections also includes
the running of the electromagnetic coupling constant.
This will be discussed in \sect{IBORNA}.
For the calculation of reaction \eqn{firsteq} to order \oalf, 
the cross-sections $\sigma$ are split into contributions
from initial state radiation, $\sigma^{{ini}}$,
final state radiation, $\sigma^{{fin}}$, and their
interference, $\sigma^{{int}}$.
In order to get a finite, gauge invariant result,
real photon bremsstrahlung from the diagrams of
\fig{fig.realbr} is combined with  
photonic vertex corrections of \fig{fig.virtph}
for initial or final state radiation, and
with the box diagram corrections of \fig{fig.virtbx} for their
interference.

In \zf, initial and final state corrections may be combined in
different ways thus reaching different numerical accuracy.
The simplest choice, with flag {\tt FINR}=0, is to take final
state corrections into account by a numerical factor ($A=T,FB$):
\ba
\sigma_{\sss{A}} &=&
\sigma_{\sss{A}}^{{ ini}}\left(1+\delta_{\sss{A}}^{fin}\right)  
+ \sigma_{\sss{A}}^{{int}},
\label{inifin1}
\\
\delta_{\sss{T}}^{fin} &=& \frac{3}{4}\frac{\alpha}{\pi} Q_f^2,
\label{}
\\
\delta_{\sss{FB}}^{fin} &=& 0.
\label{inifin2}
\ea
The factor $\delta_{\sss{A}}^{fin}$ has been given here for the massless case
without cuts.
The contribution $\sigma^{{  int}}$ comes from initial-final state
interferences and may be switched on or off with flag {\tt INTF}.
The calculation of complete \oalf\  corrections,
$\sigma =
\sigma^{{ini}}  + \sigma^{{fin}} + \sigma^{{int}}$,
is not foreseen in
\zf\ since initial state radiation contains necessarily soft photon
exponentiation.
The default approach to photonic bremsstrahlung for leptons is a {\em
common} soft 
photon exponentiation for initial and 
final state radiation (with flag {\tt NPAR(9)} = {\tt FINR}=1):
\bq
\sigma =
\sigma^{{ ini + fin}}  + \sigma^{{  int}}.
\label{inifin3}
\eq
In such an approach higher order
non-factorizing QCD and QED corrections 
can be treated only approximately, since they are known only for 
an inclusive setup.

An alternative, of some practical importance for the production of $b$
quark pairs e.g., is the common treatment of final state QED and QCD
corrections with account of running mass effects.
For this case, higher order results are available and replace successfully
the exponentiation of final state soft photonic corrections.
For details see \sect{qcdrun}.

In the other cases, the recommended cross-section is \eqn{inifin3}
and we give here the generic formula as an instructive example: 
\ba
\sigma_{\sss{T}}(c) &=& 
\int_0^{1-4m_f^2/s}dv\,
\Biggl[\sigma_{\sss{T}}^0(s')\,R_{\sss{T}}^{ini+fin}(v,c) 
\nl \nl
&& 
+~\sum_{m,n=\gamma,Z} \sigma_{\sss{FB}}^0(s,s',m,n) R_{\sss{T}}^{int}(v,c,m,n)
\Biggr] .
\label{generic1}
\ea
In \eqn{generic1}, we assume for definiteness no cuts on the final state.
The variable $c$ is to be understood as either $\cos\vartheta$ (then
$\sigma_{\sss{T}}$ is the differential cross-section part,
symmetric in the scattering angle) or as the integration limit for the cut
angular integration (for $c=1$, $\sigma_{\sss{T}}$ is the total cross-section).
When replacing index $T$ by $FB$, \eqn{generic1} expresses a
forward-backward anti-symmetric combination. 
 
The photonic corrections consist of Born-like virtual+soft parts $S$,
photonic box parts $B$, and hard parts $H$:

\ba
\sigma_{\sss{T}}(c) &=&  \int_0^{1-4m_f^2/s} dv \,
\Biggl\{
\left[ \sigma_{\sss{T}}^{Born}(s',c)\left(1+{\bar S}^{ini}\right)\beta_e
v^{\beta_e-1} + \sigma_{\sss{T}}^0(s') {\bar H}_{\sss{T}}^{ini} (v,c) \right]
{\bar R}_{\sss{T}}^{fin}(v) 
\nl \nl
&&+~ \sigma_{\sss{FB}}^0(s,s') \left[ H_{\sss{T}}^{int}(v,c) -
\frac{\sigma_{\sss{FB}}^0(s)}{\sigma_{\sss{FB}}^0(s,s')} 
H_{\sss{T}}^{int,sing}(v,c)\right]
\Biggr\} 
\nl \nl &&
+~ \sigma_{\sss{FB}}^0(s,c) {\bar S}_{\sss{T}}^{int}  + \sum_{m,n=\gamma,Z}
\sigma_{\sss{FB}}^0(s,s,m,n) B_{\sss{T}}(c,m,n) .
\label{generic2}
\ea
Here, the $\sigma_{\sss{T}}^{Born}(s,c), \sigma_{\sss{T}}^{0}(s)$ etc. 
are generic expressions
denoting either effective Born angular distributions or integrated effective
Born cross-sections.
They are defined in \sect{i.born}, \sect{ewcoup},
and \sect{IBORNA}.   
The other expressions will be explained in the next Sections.

Technically, the integration is performed as a numerical integration
over variable $R$:
\ba
R= \frac{s'}{s} = 1-v,
\label{rg1mv}
\ea
and the variable $v$ was defined in~\eqn{defv}.
The expression to be calculated with \zf\ is (see also~\eqn{generic3} below):

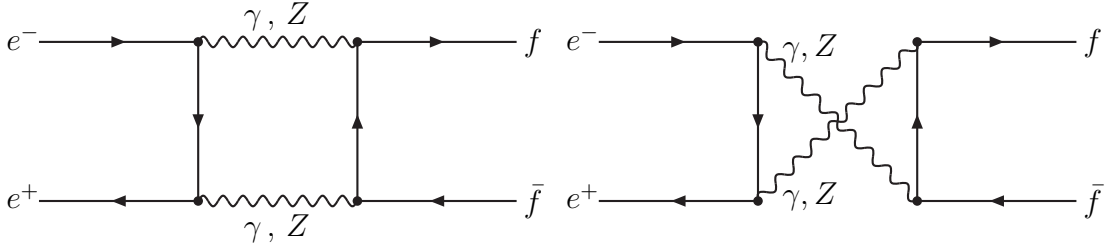
\begin{figure}[thbp]
\begin{center}
\vfill
\setlength{\unitlength}{1pt}
\SetWidth{0.8}
\begin{picture}(200,120)(0,0)
\thicklines
\ArrowLine(10,90)(70,90)
\Vertex(70,90){1.8}
\ArrowLine(70,90)(70,30)
\Vertex(70,30){1.8}
\ArrowLine(70,30)(10,30)
\Photon(70,90)(130,90){2}{8}
\Photon(70,30)(130,30){2}{8}
\ArrowLine(190,30)(130,30)
\Vertex(130,30){1.8}
\ArrowLine(130,30)(130,90)
\Vertex(130,90){1.8}
\ArrowLine(130,90)(190,90)
\Text(4,92)[]{$e^-$}
\Text(4,32)[]{$e^+$}
\Text(197,30)[]{$\bar{f}$}
\Text(197,90)[]{$f$}
\Text(100,100)[]{$\gamma\,,\,Z$}
\Text(100,20)[]{$\gamma\,,\,Z$}
\end{picture}
\setlength{\unitlength}{1pt}
\SetWidth{0.8}
\begin{picture}(200,120)(0,0)
\thicklines
\ArrowLine(10,90)(70,90)
\Vertex(70,90){1.8}
\ArrowLine(70,90)(70,30)
\Vertex(70,30){1.8}
\ArrowLine(70,30)(10,30)
\Photon(70,90)(130,30){2}{9}
\Photon(70,30)(130,90){2}{9}
\ArrowLine(190,30)(130,30)
\Vertex(130,30){1.8}
\ArrowLine(130,30)(130,90)
\Vertex(130,90){1.8}
\ArrowLine(130,90)(190,90)
\Text(4,92)[]{$e^-$}
\Text(4,32)[]{$e^+$}
\Text(197,30)[]{$\bar{f}$}
\Text(197,90)[]{$f$}
\Text(90,88)[]{$\gamma,Z$}
\Text(90,32)[]{$\gamma,Z$}
\end{picture}

\caption[
Box diagrams with virtual photons
]{\label{fig3paw}
{\it
The $\gamma\gamma$ and $\gamma Z$ box diagrams 
}}
\label{fig.virtbx}
\end{center}
\end{figure}

\ba
\sigma_{\sss{T}}(c) &=&  \int_{4m_f^2/s}^{1} dR \,
\left[\beta_e(1-R)^{\beta_e-1}\right] \, {\cal R}(R,c) 
\nl \nl
&&+~\sigma_{\sss{FB}}^0(s,c) {\bar S}_{\sss{T}}^{int}  + \sum_{m,n=\gamma,Z}
\sigma_{\sss{FB}}^0(s,s,m,n) B_{\sss{T}}(c,m,n) ,
\label{sigTnum}
\ea
with
\ba
 {\cal R}(R,c) &=&
\left[ \sigma_{\sss{T}}^{Born}(s',c)\left(1+{\bar S}^{ini}\right)
+ \sigma_{\sss{T}}^0(s') \, {\bar H}_{\sss{T}}^{ini} (v,c)\, {\cal Y} \right]
{\bar R}_{\sss{T}}^{fin}(v) 
\nl \nl
&&+~ \sigma_{\sss{FB}}^0(s,s') \left[ H_{\sss{T}}^{int}(v,c) -
\frac{\sigma_{\sss{FB}}^0(s)}{\sigma_{\sss{FB}}^0(s,s')} 
H_{\sss{T}}^{int,sing}(v,c)\right]
{\cal Y},
\label{RRc}
\ea
and
\ba
{\cal Y} &=&  \left[ \beta_e (1-R)^{\beta_e-1}\right]^{-1} .
\label{Ymap}
\ea

The integrand is smoothened with the following variable transformation
(mapping), using functions {\tt FACT(R)} = $F$ and {\tt FACINV(F)} = $R$:
\ba
dF &=&  \beta_e (1-R)^{\beta_e-1} \, dR,
\\ \nl
F &=& - (1-R)^{\beta_e},
\label{dfun}
\\ \nl
R &=& 1-(-F)^{1/\beta_e}.
\label{dfunin}
\ea

Further, at $R \to 1$ the function $ {\cal R}$ is smooth:
\ba
{\cal R}(1,c) &=& \sigma_{\sss{T}}^{Born}(s,c) \left(1+{\bar S}^{ini}\right)
{\bar R}_{\sss{T}}^{fin}(0) .
\label{R1c}
\ea

This allows to perform the integral explicitly in the neighborhood
of that limit.
In subroutine {\tt SFAST}, the following expression is calculated numerically:
\ba
\sigma_{\sss{T}}(c) &=& 
{\cal R}(1,c) \, \epsilon_s^{\beta_e} 
- \int_{F(1-\epsilon_s)}^{F(4m_f^2/s)} dF \, {\cal R} \left(
1-(-F)^{1/\beta_e}, c \right) 
\nl \nl
&&+~ \sigma_{\sss{FB}}^0(s,c) {\bar S}_{\sss{T}}^{int}  + \sum_{m,n=\gamma,Z}
\sigma_{\sss{FB}}^0(s,s,m,n) B_{\sss{T}}(c,m,n) ,
\label{generic3}
\ea
with typically $\epsilon_s=10^{-12}$.
The integration is done this way with subroutine {\tt FDSIMP} for the
integrated QED corrections without cuts.

Subroutine {\tt FDSIMP} has, among others, the arguments 
{\tt FUNCT, DFUN, DFUNIN}. 
They denote the integrand function and functions {\tt FACT, FACINV} in \zf.  

Alternatively, one may split the integrand into two pieces; a soft one with
the soft photon factor, and the hard rest.
This is done in subroutines {\tt COSCUT} and {\tt SCUT}.
It suffices to map the soft part and to directly integrate the
rest (hard corrections)\footnote{The choice of an $s'$ cut ({\tt ICUT}=1)
or cuts on acollinearity and 
fermion energy ({\tt ICUT}=2,3) is of no influence for the presentation
here. 
The decision between these two sets of cuts leads only to different
expressions for the hard corrections $H$. 
}: 
\ba
\sigma_{\sss{T}}(c) &=& 
{\cal R}(1,c) \, \epsilon_s^{\beta_e} 
- \int_{F(1-\epsilon_s)}^{F(R_{min})} dF \, {\cal R}^{soft} \left(
1-(-F)^{1/\beta_e} , c\right) 
\nl \nl
&&+ \int_{R_{min}}^{1-\epsilon_h} dR \,  
\Biggl\{
\sigma_{\sss{T}}^0(s') {\bar H}_{\sss{T}}^{ini} (v,c)
{\bar R}_{\sss{T}}^{fin}(v) 
\nl \nl &&
+~ \sigma_{\sss{FB}}^0(s,s') \left[ H_{\sss{T}}^{int}(v,c) -
\frac{\sigma_{\sss{FB}}^0(s)}{\sigma_{\sss{FB}}^0(s,s')} 
H_{\sss{T}}^{int,sing}(v,c)\right]
\Biggr\}
\nl \nl
&&+~ \sigma_{\sss{FB}}^0(s,c) {\bar S}_{\sss{T}}^{int}  + \sum_{m,n=\gamma,Z}
\sigma_{\sss{FB}}^0(s,s,m,n) B_{\sss{T}}(c,m,n) ,
\label{generic4}
\ea
with
\ba
 {\cal R}^{soft}(R,c) &=&
\sigma_{\sss{T}}^{Born}(s',c)\left(1+{\bar S}^{ini}\right)
{\bar R}_{\sss{T}}^{fin}(v) .
\label{Rsoft}
\ea
Further, it is typically $\epsilon_h=10^{-5}$ and the second, direct
integration over $dR$ in~\eqn{generic4} is done with subroutine {\tt
  SIMPS} \cite{Silin:19xy}.

From the above it is seen that \zf \ performs {\em one-fold numerical
  integrations}  for the contributions introduced
in~\eqns{inifin1}{generic2}:
\ba
\frac{d\sigma_{\sss{A}}^{{ini}}}{d\cos\vartheta} &=&    \int_{\Omega} dv 
 \, \sigma_{\sss{A}}^0(s')  \, R_{\sss{A}}^{{ ini}}(v,\cos\vartheta) ,
\label{siginic}
\\  \nl
\frac{d\sigma_{\sss{A}}^{{ini + fin}}}{d\cos\vartheta} &=&   
\int_{\Omega} dv 
 \,   \sigma_{\sss{A}}^0(s') R_{\sss{A}}^{{ ini}}(v,\cos\vartheta)  \,  
{\bar R}_{\sss{A}}^{{ fin}}(v), 
\label{siginific}
\\ \nl
\frac{d\sigma_{\sss{A}}^{{ int}}}{d\cos\vartheta} &=& 
   \int_{\Omega} dv 
         \sum_{m,n}  
\sigma_{\bar{\sss{A}}}^0(s,s';m,n)
  \,       R_{\sss{A}}^{{int}}(v,\cos\vartheta;m,n),
\label{sigintc}
\\ \nl
\sigma_{\sss{A}}^{{ini}}(c) &=&    
\int_ {\Omega}    
dv   \,  \sigma_{\sss{A}}^0(s')  \,  R_{\sss{A}}^{{ ini}}(v,c) ,
\label{sigini}
\\  \nl
\sigma_{\sss{A}}^{{ini+fin}}(c) &=&    
\int_{\Omega} dv 
\, \sigma_{\sss{A}}^0(s') \, R_{\sss{A}}^{{ini}}(v,c) \, 
{\bar R}_{\sss{A}}^{{ fin}}(v),
\label{siginifin}
\\  \nl
\sigma_{\sss{A}}^{{ int}}(c) &=&     \int_{\Omega}
dv 
     \sum_{m,n} \sigma_{\bar{\sss{A}}}^0(s,s';m,n)
   \,        R_{\sss{A}}^{{int}}(v,c;m,n),
\label{sigint}
\ea
where $A=T,FB$, ${\bar A}=FB,T$, and $m,n = \gamma,Z$.
The initial-final state interference contributions from $\gamma$ and
\Z \, exchange and 
from their interference have to be separated.
This is unavoidable since the $\gamma\gamma$ and $\gamma Z$ boxes differ.
We further would like to stress that for the
initial-final state interferences the C even and the C odd properties
of the basic cross-section (see the couplings' labels) are opposite to
those of Born terms or of initial or final state radiation.

The integral over $s'$, or, equivalently, over $R$, has to be
performed numerically since soft  
photon exponentiation makes a complicated integrand.
The radiator functions 
$R_{\sss{A}}^a(v, \cos\vartheta)$ 
and $R_{\sss{A}}^a (v, c)$, $A = T, FB$, 
are obtained
from two-fold and three-fold analytic  phase space integration, respectively:
\ba
R_{\sss{A}}^a (v, \cos\vartheta [,m,n]) &=&  ~~~~~~~~~~~~~\int dv_2  \,  
\int d\phi_{\gamma} \, 
\left| \chi^a_{\sss{A}}(s,v,v_2, \cos\vartheta, \phi_{\gamma})\right|^2,
\label{symb1}
\\ \nl
R_{\sss{A}}^a (v, c[,m,n]) &=& \int_{-c}^c d\cos\vartheta \,
 \int dv_2  \,  \int d \phi_{\gamma} \,
\left| \chi^a_{\sss{A}}(s,v,v_2, \cos\vartheta, \phi_{\gamma})\right|^2,
\label{symb}
\ea
where the squared matrix elements $\left|\chi_{\sss{A}}^a\right|^2$ are the 
result of a
Feynman diagram calculation. 
Further, $s'=M_{{\bar f} f}^2$, $v_2=M_{{f}\gamma}^2/s$,
and $\phi_{\gamma}$ is one of the photon angles in the
($\gamma,{f}$) rest system.
More details on the definition of the phase space parameterization used are given
in \sect{subsec:acol}.  

The $s'$ integration region is indicated by the symbol $\Omega$.
Several different cases have been prepared in \zf:

\begin{itemize}
\item   Born cross-sections (flag {\tt BORN}=1); option is described in
        \sect{i.born} and \sect{IBORNA}; 
\item   no cut but a simple cut on $s'$ (flag {\tt ICUT}=--1);
        option is available for $\sigma_{\sss{T,FB}}$~\cite{Bardin:1989cwt} 
and is
        described in \sect{subsec:nocut};
\item   cut on $s'$ for $d\sigma / d\cos\vartheta$~\cite{Bardin:1991fut} and on
        $s'$ and $\cos\vartheta$ for 
        $\sigma_{\sss{T,FB}}$~\cite{Bardin:1991det}
        (flag {\tt ICUT}=1); option is described in \sect{somana};
\item  cut on $s'$ in the course of numerical integration, 
       combined with cuts on $v_2$ applied in 
        the analytical integration  (flag {\tt ICUT}=2,3); their combination
       creates combined 
        cuts on acollinearity and minimal energy of the
       fermions~\cite{Christova:1999cct,Christova:1999gh}; for the $s'$
integration this means in 
       effect that in different $s'$ regions different analytical expressions
       are integrand; option is described in \sect{subsec:acol}.
\end{itemize}

The angular acceptance cut, $c_1 \leq \cos \vartheta \leq c_2$,
limits the scattering angle $\vartheta$ of the final-state
{\em anti}fermions (see \fig{fig.theta}).
The
scattering angle of {\em fermions}   remains unrestricted if the other
cut(s) do not imply an implicit restriction (see \sect{subsec:acol}).
 In \zf, the QED contributions include the complete \oalf cor\-rect\-ions,
plus soft photon exponentiation, plus selected higher order photonic
corrections (chosen with flag {\tt FOT2}). 
As a matter of fact, we mention that the radiator functions (flux functions),
$R_{\sss{A}}^a$, {\em differ} for different 
observables (i.e. different index $A$) and for different bremsstrahlung types
($a={{ini,fin,int}}$). 
Only close around the \Z\ resonance where hard photon emission is
suppressed and for loose cuts some of the radiator functions
agree approximately~\cite{Riemann:1989b,Bardin:1989cwt}.

Finally we mention that it is foreseen in \zf\ to calculate the following
contributions: 
\ba
\frac{d\sigma_{\sss{A}}^{{fin}}}{d\cos\vartheta} &=&   
 \sigma_{\sss{A}}^0(s) \int_{\Omega} dv 
  R_{\sss{A}}^{{fin}}(v,\cos\vartheta) ,
\label{sigfic}
\\  \nl
\sigma_{\sss{A}}^{{fin}}(c) &=&    \sigma_{\sss{A}}^0(s) 
\int_{\Omega}   dv \, R_{\sss{A}}^{{fin}}(v),
\label{sigfin}
\ea
i.e. {\em singly deconvoluted} observables --- with FSR but without ISR 
corrections. This is achieved by setting {\tt FOT2}=-1.

\section{Photonic Corrections with $\smanp$ Cut
\label{subsec:nocut} }
\setcounter{equation}{0}
%
The calculational chain with $\smanp$ cut is chosen when calling
subroutine {\tt ZUCUTS} with flag {\tt ICUT}=--1. (The internal 
flag {\tt IFAST} is set equal to 1.)
The treatment of photonic ${\cal O} (\alpha)$ corrections is based on
\cite{Bardin:1989cwt}. 
Higher order QED corrections depend on the flaggs {\tt FOT2} and {\tt ISPP}
For references for this, see \sect{zuflag}.

In this calculational chain,
the cross-sections $\sigma_{\sss{T}}$ and $\sigma_{\sss{FB}}$ are calculated 
with formulae that assume 
{\em no angular cuts} being  applied to the final state phase space.
It is computationally fast.
The integrations are performed with subroutine {\tt FDSIMP} as described in 
\eqn{generic3} and \eqns{RRc}{R1c}.

In the notations introduced, the phase space is bounded by the values:
\ba
c &=& 1,
\\ \nl
E_{\gamma}^{\max} &=& \frac{\sqrt{s}}{2} \Delta.
\label{egam}
\ea
Equivalently to~\eqn{egam},  
\ba
\Delta &\leq& 1- \frac{4 m_f^2}{s},
\label{cuts2}
\\
s'^{\min} &=& 4 m_f^2.
\ea
Thus, the radiative corrections may depend on fermion masses even for 
light quarks and leptons.
For their choice, see \sect{IPS1}.
This dependence can be important when total cross-sections
are determined from experimental data, and {\it is} often of special importance
when comparing results from other programs.

\zf\ returns the cross-section \eqn{inifin1}
with QED corrections ({\tt SIGQED}, asymmetry {\tt AFBQED}) and without 
({\tt SIGBRN, AFBBRN}) from
subroutine {\tt ZCUT}.
Technicalities of the calculation are described in \sect{chains}.
Subroutines {\tt FCROS} and {\tt FASYM} are used for the calculation
of the symmetric and anti-symmetric cross-sections, respectively. 
\subsection{Initial state corrections with soft photon exponentiation
and higher order corrections\label{isr}}
Cross-sections with initial state QED corrections as introduced
in~\eqn{sigini} are 
understood to include soft  
photon exponentiation plus, optionally, further  higher order
corrections and are calculated   
as follows~\cite{Bardin:1989cwt}: 
\ba
\sigma_{\sss{A}}^{\rm{ini}}(1) &=&    
\int_
 0^{\Delta} 
dv  \,  \sigma_{\sss{A}}^0(s') \,  R_{\sss{A}}^{\rm{ ini}}(v,1) ,
\label{sigini2}
\\ \nl
\label{saini0}
R_{\sss{A}}^{\rm{ ini}}(v,1) 
&=& 
 C_{\sss{A}}(1) \left[ \left(1+ {\bar S}^{ini}\right) \beta_ev^{\beta_e-1}
+ {\bar H}_{\sss{A}}^{ini}(v,1)\right],   \hspace{.7cm} A=T,FB,
\label{R0}
\ea 
with 
\ba
\beta_e&=&\frac{2\alpha}{\pi}Q_e^2\left( L_e-1\right),
\label{betae}
\\ \nl
L_e &=& \ln\frac{s}{m_e^2}.
\label{le}
\ea
The soft plus virtual corrections ${\bar S}^{ini}$ = {\tt SOFTER} are
calculated according to Eq.~(50) of~\cite{Bardin:1991fut} in
subroutine {\tt SETFUN} in the course of initialization of \zf: 
\ba
{\bar S}^{ini}
&=&
\frac{\alpha}{\pi} Q_e^2
\left[ \frac{3}{2} \left(L_e-1\right) + \frac{\pi^2}{3} - \frac{1}{2} \right]
+  S^{(2)} + S^{pairs}.
\label{softer}
\ea
The higher order virtual and soft photonic corrections, $S^{(2,3)}$, and
pair production corrections, $S^{pairs}$, are described in \sect{sosv}.
They are calculated depending on flag {\tt FOT2} and {\tt ISPP}
The hard photon corrections ${\bar H}_{\sss{T}}^{ini}$ = {\tt H0} in
subroutine {\tt FCROS} and ${\bar H}_{\sss{FB}}^{ini}$ =
{\tt H3} in subroutine {\tt FASYM} are, respectively:  
\ba
{\bar H}_{\sss{T}}^{ini}(v,1)
&=&
\frac{\alpha}{\pi} Q_{e}^{2} (L_{e}-1)
\left[\frac{1+(1-v)^2}{v} \right] - \frac{\beta_e}{v} + H_{\sss{T}}^{(2)}(v), 
\label{htht1}
\\ \nl 
{\bar H}_{\sss{FB}}^{ini}(v,1)
&=&
\frac{\alpha}{\pi} Q_{e}^{2} 
\left[ (L_{e}-1)- \ln\frac{1-v}{(1-\frac{1}{2}v)^2} \right] 
\left[ \frac{1+(1-v)^2}{v} 
\frac{1-v}{(1-\frac{1}{2}v)^2}
\right]  
\nl \nl
&& -~\frac{\beta_e}{v} +~ H_{\sss{FB}}^{(2)}(v).
\label{hthfb1}
\ea
The higher order corrections $H_{\sss{T}}^{(2,3)}$ and
$H_{\sss{FB}}^{(2,3)}$ also depend on
flaggs {\tt FOT2} and {\tt ISPP} and are described in \sect{soh}.
\subsection{Final state radiation\label{common}}
The final state photonic corrections are calculated in dependence on
flag {\tt IFINAL} = {\tt FINR}.
All options of flag {\tt FINR} are possible.

Details may be found  in \sect{commexp}.
\subsection{Initial-final state interference corrections \label{inifi}} 
The QED corrections from the interference of initial and final state
radiation are taken into account for flag setting {\tt INTF}=1.
Combined with the interference of the $\gamma \gamma$ and $\gamma Z$ box
diagrams with the Born matrix element, they give the following cross-section
contributions: 
\ba
\sigma_{\sss{A}}^{\rm{ int}}(1) &=& \int_0^{\Delta} dv 
         \sum_{m,n}  \sigma_{\bar A}^0(s,s';m,n)
         R_{\sss{A}}^{\rm{int}}(v,1;m,n),
\label{sigint2}
\ea
with
\ba
R_{\sss{A}}^{int}(v,1,m,n)
&=&
\delta(v) \left[  {\bar S}_{\sss{A}}^{int}(1) + B_{\sss{A}}(1,m,n) \right] 
+ {\bar H}_{\sss{A}}^{int}(v,1) 
.
\label{Rint}
\ea
We use 
$\sigma_{\sss{A}}^0(s,s';m,n)$ from~\eqn{bornss'},
${\bar S}_{\sss{A}}^{int}(c)$ from~\eqn{sdelt} and~\eqn{sdelfb},
$B_{\sss{A}}(c,m,n)$ from~\eqn{bpm}.
When $A=T$, it is ${\bar A} = FB$ and vice versa.

In \zf, the soft photon corrections {\tt SFTIS, SFTIA} from subroutine
{\tt SFTINT} as well as the
photonic box corrections {\tt BOXIS, BOXIA} from subroutine {\tt BOXINT}
for integrated cross-sections are called in subroutine {\tt SFAST} and
calculated as functions of $c$ taken 
by fixing there $c=1$. 
So, for details we may refer to \sect{intc}.  
The hard correction to the total cross-section is calculated in the
combination ($\sigma^0_{\sss{FB}} {\bar H}_{\sss{T}}^{int}) \sim $ 
 {\tt (H4*ABORN+8/v*ABORN0)} 
and is added in subroutine {\tt SFAST}
after an integration over the function {\tt FCROS}.
Analogously,  
($\sigma^0_{\sss{T}} {\bar H}_{\sss{FB}}^{int}) \sim $ 
{\tt [H1*SBORN+4*(8*AL2+1)/(3v)*SBORN0]}
results from the integration over the function {\tt FASYM} and contributes to
the forward-backward asymmetry.

The ${H}_{\sss{T}}^{int}(v,1)~\sim$ {\tt H4} is given in Eq.~(29) and
${H}_{\sss{FB}}^{int}(v,1)~\sim$ {\tt H1} in 
Eq.~(22) of~\cite{Bardin:1991fut}.
From these functions, the barred ones are derived:
\ba
\sigma_{\sss{FB}}^0(s,s') {\bar H}_{\sss{T}}^{int}(v,1) 
&=&
\frac{\alpha}{\pi} Q_e Q_f
\left[  \frac{3}{v} (1-v)(v-2)  \sigma_{\sss{FB}}^0(s,s')
+ \frac{6}{v} \sigma_{\sss{FB}}^0(s) \right],
\\ \nl 
\sigma_{\sss{T}}^0(s,s')
 {\bar H}_{\sss{FB}}^{int}(v,1)
&=&
\frac{\alpha}{\pi} Q_e Q_f
\Biggl\{
\frac{2}{3v}   \Biggl[
2 \frac{1-v}{2-v} \left( v^2+2v-2 \right) 
\nl \nl && +~ (1-v) \left( 5v^2-10v+8
\right) \ln (1-v) 
\nl &&-~
\left(5v^2-2v+8 \right)(2-v) \ln (2-v)
\Biggr] \sigma_{\sss{T}}^0(s,s')
\nl \nl
&& +~
\frac{4}{3v}\left(8 \ln2 +1\right) 
\sigma_{\sss{T}}^0(s) \Biggr\} .
\label{hisa}
\ea
These functions are constructed such that they are finite at $v \to 0$.
The subtracted pieces were integrated analytically and combined with the
soft photon corrections. 
This makes the soft corrections (artificially) dependent on integration
boundaries (if any) but cancels the dependence on the infrared
cut-off. 

%
\section{Photonic Corrections with Cuts on {$s'$} and {$\cos\vartheta$}
\label{somana}
}
\setcounter{equation}{0}
Cross-sections with $s'$ cut and cut on the acceptance angle are
chosen by flag {\tt ICUT}=1. 

The calculation of differential cross-sections is also possible.
Differential cross-sections are calculable with the interface {\tt ZUATSM}.
The calculation is organized by subroutine {\tt ZANCUT} and the cross-section
is determined in subroutine {\tt COSCUT}.
Integrated cross-sections may have an acceptance cut in the production angle.
They are accessible by all the other interfaces, e.g. {\tt ZUTHSM} for Standard
Model calculations,  and go via subroutine {\tt ZCUT} with call of 
subroutine {\tt SCUT}

Integrations are done with subroutines {\tt SIMPS} and {\tt FDSIMP} 
following \eqn{generic4} with \eqn{R1c} and \eqn{Rsoft}.

\begin{figure}[thbp]
\begin{center} \mbox{
\epsfysize=8.0cm
\epsffile{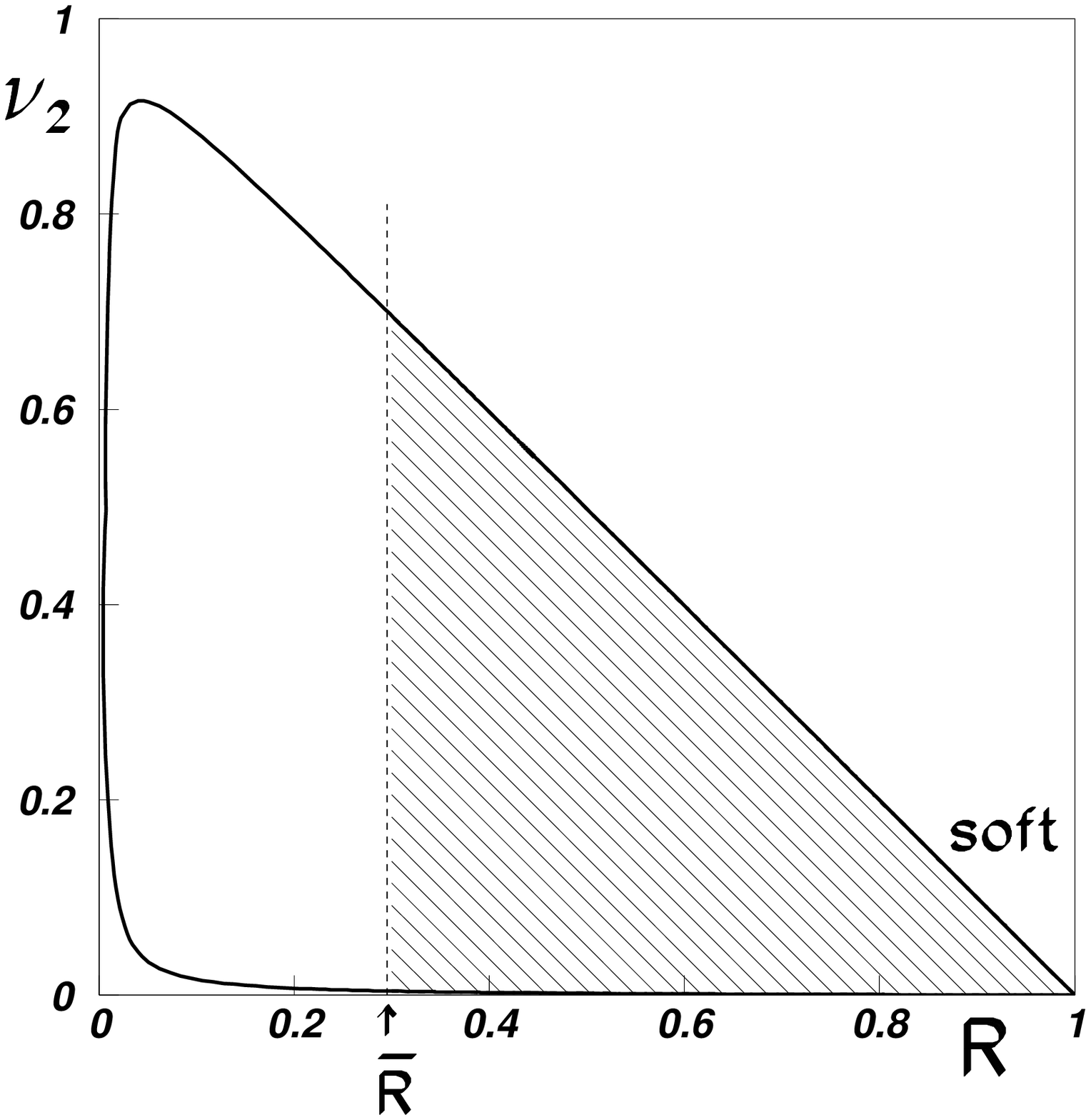}
}
\end{center}
\caption[
Dalitz plot with a cut on $s'$
]{\it
Dalitz plot with a cut on $s'$
\label{dalitz0}
}
\end{figure}

The minimal invariant mass of the final state fermion pair,
$R_{\min}$, may be restricted by choosing the cut variable $\Delta$:
\ba
\Delta = 1-R_{\min} &=& v_{\max},
\\
4m_f^2 \leq s'_{\min} &=& R_{\min} s \leq s.
\label{spmin}
\ea
For single photon emission, this may be re-interpreted as a cut on the
maximum allowed photon energy of the bremsstrahlung photon:
\ba
\Delta= \frac{E_{\gamma}^{\max}}{E_{beam}}.
\ea 
The Dalitz plot in \fig{dalitz0} shows this cut in the variable
$R$.

In the ultra-relativistic limit,
the allowed region in the $v_2, R$ plane is a triangle. 
It is independent of the scattering angle.
 
For the integrated cross-sections, the $s'$ cut may be combined with an angular
acceptance cut: 
\bq
 c_1 \leq \cos \vartheta \leq c_2.
\label{cuts1}
\eq

\vspace{.9cm}

For a symmetric choice of the angular cuts, $c_1 = - c_2$, we use the
abbreviation \ba
c = |c_1| = c_2.
\ea

As a matter of convention this cut is imposed on the antifermion.
Because of CP invariance the cut could equally well be applied on the
fermion instead.
However, it may not be understood as being applied to both fermions at once.

In the notion of \zf, for the $s'$ cut case, it is $R > R_{min}$ =
{\tt RECUT2}. 
This is used for some branchings in the course of calculation.
More details on the notion of cut defining variables will be given in
\sect{subsec:acol}.

At the end of these introductory remarks, we should comment on the 
normalizations.
This is easiest studied with the Born cross-sections.
From subroutine {\tt SCUT}, we get e.g.:

\ba
\sigma_{\sss{T}}^0(c) = \frac{3}{8} \, \,{\tt CSIGNB} \cdot {\tt SBORN0} \cdot 
{\tt COPL3}. 
\label{copl3}
\ea

The corresponding definitions are chosen such that  {\tt CSIGNB} contains a 
factor 4/3 (and the conversion factor),  {\tt SBORN0} contains no numerical 
constant (i.e. equals one for muon pair production with photon exchange),
and {\tt COPL3} gets 4/3 if no angular acceptance cut is applied.
If we write in this article, e.g.:
\ba
\sigma_{\sss{T}}^0(c) = \sigma_{\sss{T}}^0(s) \, C_{\sss{T}}(c),
\label{ctc}
\ea
then there is a difference by a factor of 2 
between \eqn{copl3} and \eqn{ctc} since $\sigma_{\sss{T}}^0(s)$ has no 
numerical factor, and $C_{\sss{T}}(c)$ = {\tt COPL3}.
This is exactly the numerical factor mentioned in \sect{chains}.
For symmetrical acceptance cut, the normalization is correct in
subroutine {\tt   ZCUT} where  
the cross-section finally is prepared for output.
In the following, this kind of numerical overall factor will be 
assumed to be controlled when we say, e.g., $\sigma_{\sss{T}}^0(c)$ = {\tt
SBRN} in subroutines  
{\tt SCUT} and {\tt SFAST}.
The program returns correct quantities.

Another remark concerns contributions of the form $\ln c_{\pm}$ etc, which 
get singular if $\cos\vartheta=\pm 1$ or $c= \pm 1$.
This happens already in subroutine {\tt SFAST} with simple $s'$ cut. 
A careful analysis of the corresponding expressions shows that they usually 
result from expressions like e.g.
\ba
c_{\pm} = \frac{1}{2} 
\left( 1 \pm \sqrt{1-\frac{4m_e^2}{s}} \cos \vartheta 
\right)
&\approx& \frac{1}{2} \left( 1   \pm \cos \vartheta\right)
\mp \frac{m_e^2}{s} \cos \vartheta .
\ea
These expressions never get smaller than $m_e^2/s$ and thus their
logarithms are well-defined.
In subroutine {\tt ZETCOS} this is specially arranged to be 
fulfilled.
\subsection{Initial state corrections with soft photon exponentiation
  and higher order corrections
\label{isp}
} 
\subsubsection{Differential cross-sections
\label{ispdiff}
}
The differential cross-sections are determined with subroutine {\tt COSCUT}.
Common exponentiation of initial and final state radiation is described in 
\sect{commexp}. 
Here, we restrict ourselves to the case of only initial state radiation.  
According to~\eqn{siginic}, it is: 
\ba
\frac{d\sigma^{{ini}}}{d\cos\vartheta} &=&  \sum_{A=T,FB}
~  \int_
0^{\Delta} dv
\sigma_{\sss{A}}^0(s') R_{\sss{A}}^{{ ini}}(v,\cos\vartheta) ,
\label{siginic1}
\ea
with
\ba
   {R}_{\sss{A}}^{ini}(v, \cos\vartheta) = 
   D_{\sss{A}}(\cos\vartheta) \left[ 1 + \bar{S}^{ini} \right]
  \beta_{e} v^{\beta_{e}-1} + \bar{H}_{\sss{A}}^{ini}(v,\cos\vartheta),
\ea
with $\bar{S}^{ini}$ from~\eqn{softer} and 
\ba
 \bar{H}_{\sss{A}}^{ini}(v,\cos\vartheta) = H_{\sss{A}}^{ini}(v,\cos\vartheta) 
+  D_{\sss{A}}(\cos\vartheta)  
\left[- \frac{\beta_{e}}{v} + H_{\sss{A}}^{(2)}(v) \right]. 
\ea
The functions $D_{\sss{A}}$ are defined in \eqn{sigdiff2} and
\eqn{sigdiff3}, functions $H_{\sss{A}}^{(2)}$ 
in \eqn{soh},
and the symmetrized (anti-symmetrized) hard  photon parts 
$\bar{H}_{\sss{T}}^{ini}$ 
= {\tt H0} and $\bar{H}_{\sss{FB}}^{ini}$ = {\tt H3} are:
\begin{equation}
   H_{\sss{T,FB}}^{ini}(v,\cos\vartheta) = \frac{\alpha}{2\pi} Q_{e}^{2}
\left[ h_{\sss{T,FB}}^{ini}(v,\cos\vartheta) \pm
  h_{\sss{T,FB}}^{ini}(v,-\cos\vartheta) \right ] .
\label{eq:s36}
\end{equation}
They are calculated in  function {\tt HARD}.
The functions $H_{\sss{T,FB}}^{ini}(v,\cos\vartheta)$ 
(with same variable names) 
are calculated in
subroutine {\tt HCUT}, and 
$h_{\sss{T}}^{ini}(v,\cos\vartheta)$ = {\tt H0M} and 
$h_{\sss{FB}}^{ini}(v,\cos\vartheta)$ = {\tt H3M}
in subroutine {\tt HINIM}, region $(-,+)$:
\ba
\frac{v^3}{R}
   h_{\sss{T}}^{ini}(v,\cos\vartheta) &=& 
   \frac{L_{c}}{\gamma^{2}} 
    r_{2}\left(r_{2} - \frac{2R}{\gamma} r_{1}
    + \frac{2R^{2}}{\gamma^{2}}\right) +
    \left(-\frac{2}{3R} r_{4} + \frac{10}{3} r_{2} - 4R + \frac{2}{3\gamma
    R} r_{2} r_{3}\right)  
\nonumber \\ 
&&
 -~  \frac{1}{\gamma^{2}} \left(3r_{4} + 8 R r_{2} + \frac{26}{3} R^{2}\right)
   + \frac{R}{\gamma^{3}} \left( 8 r_{3} + \frac{44}{3} R r_{1}\right) 
\nonumber \\
&&
  -~  \frac{R^{2}}{\gamma^{4}} \left(
    \frac{22}{3} r_{2} + 4 R\right) ,
\\ \nl 
\frac{v^2}{R}
    h_{\sss{FB}}^{ini}(v,\cos \vartheta) 
&=&
 r_{2}
    \frac{L_{c}}{\gamma^{2
   }} \left(r_{1} - \frac{2R}{\gamma}\right) + \frac{2}{\gamma} r_{2}
    - \frac{4}{\gamma^{2}} r_{1} (r_{2} + R) + \frac{2R}{\gamma^{3}}
   ( 3 r_{2} + 2 R).
\ea
Functions $h_{\sss{A}}^{ini}(v,\cos \vartheta)$ 
were given in Eqs. (37) and (38) of 
\cite{Bardin:1991fut}\footnote{There are printing errors in the
corresponding Eqs. (13) and (14) of 
 \cite{Bilenkii:1989zg}.
}.
The following abbreviations are used:
\ba
    L_{c} &=& \ln\frac{\gamma^{2}}{R} + L_{e},
\\[2mm]
    R &=& 1 - v, 
\\[2mm]
r_{n} &=& 1 + R^{n},
\label{6.14}
\\[2mm]
    C_{\pm} &=& \frac{1}{2}(1 \pm \cos\vartheta),
\label{cpm}
\\[2mm] 
\gamma &=&
    C_{+} + RC _{-}.
\label{eq:c312}
\ea
\subsubsection{Integrated cross-sections delta cut \label{ispint}}
The Born plus initial state corrections are:
\ba
\label{saini}
\sigma_{\sss{A}}^{ini} (c)
&=&
\int_0^{\Delta} dv ~ \sigma_{\sss{A}}^0(s') R_{\sss{A}}^{ini}(v,c),
\hspace{.7cm} A=T,FB,
\\ \nl 
R_{\sss{A}}^{{ ini}}(v,c) 
&=&
 C_{\sss{A}}(c) \left(1+ {\bar S}^{ini}\right) \beta_ev^{\beta_e-1}
+ {\bar H}_{\sss{A}}^{ini}(v,c).
\label{R02}
\ea 
The cross-section is calculated in subroutine {\tt SCUT}.
The function 
${\bar S}^{ini}$ is defined in \eqn{softer}.
The hard photon corrections ${\bar H}_{\sss{T}}^{ini}$={\tt H0} and 
${\bar H}_{\sss{FB}}^{ini}$={\tt H3} 
are calculated by a numerical integration of functions {\tt SHARD} and 
{\tt AHARD}, respectively:
\ba
{\bar H}_{\sss{A}}^{ini}(v,c)
&=&
\frac{\alpha}{\pi}Q_e^2 \left[
\frac{h_{\sss{A}}^{ini}(v,c)}{v} - C_{\sss{A}}(c) \frac{\beta_e}{v}\right]
+  C_{\sss{A}}(c)  H_{\sss{A}}^{(2)}, 
\hspace{.7cm} A=T,FB. 
\label{6.19}
\ea
The second order hard photonic corrections $H_{\sss{A}}^{(2)}, A=T,FB$,
are described in \sect{soh} and the $C_{\sss{A}}(c)$ are defined in \eqn{ct} 
and \eqn{cfb}.
Further, it is:
\begin{eqnarray}
 h_{\sss{T}}^{ini}(v,c) &=&
\frac{4}{3}r_{2}
  \left[ \frac{\ln\gamma_{+}}{\gamma^{3}_{+}}\left(c^{3}_{+}-R^{3}c^{3}_{-}
\right)
-        \frac{\ln\gamma_{-}}{\gamma^{3}_{-}}
\left(c^{3}_{-}-R^{3}c^{3}_{+}\right)
\right]
\nonumber \\
&&+~  \frac{2c}{\gamma^{3}_{-}\gamma^{3}_{+}}
\Biggl\{ R r_{2}\left[\left(L_{e}-1\right)-\ln R\right] 
\left[\frac{2}{3}R^{2}(1-c_{+}c_{-})
+  v^{2}c_{+}c_{-}(R+r_{2}c_{+}c_{-})\right]
\nonumber \\
&&+~2 v^{2}R(c_{+}c_{-})^{2}\left(r_{2}R-r_{4}
   + \frac{4}{3} \frac{R^{2}}{c_{+}c_{-}} - \frac{22}{3}R^{2}\right)
\nl &&
+~   2 v^{4}(c_{+}c_{-})^{3} \left(
    \frac{5}{3} r_{2}R - 2 R^{2} - \frac{1}{3}r_{4}\right)\Biggr\} ,
\label{eq.2.5}
\\ \nl
h_{\sss{FB}}^{ini}(v,c) &=&
8R\frac{r_{2}}{r_{1}^{2}}\ln\frac{r_{1}}{2}
  +  \left.\frac{2R}{\gamma^{2}_{-}\gamma^{2}_{+}}\right\{
  \left[\left(L_{e}-1\right)-\ln R\right]\Biggl[2r_{2}\frac{R^{2}}{r_{1}^{2}}
 \nonumber \\
&& -~r_{2}^{2} c_{+}c_{-} + 4r_{2}c_{+}c_{-}\frac{v^{2}}{r_{1}^{2}}
   ( R + r_{2}c_{+}c_{-})\Biggr]
 \nonumber \\
&&  +~ v^{2}c_{+}c_{-}\left[4R - 4c_{+}c_{-}r_{1}^{2} + v^{2}\right]
  \left\} 
  - 4Rr_{2}c_{+}c_{-}\left(\frac{\ln \gamma_{-}}{\gamma_{-}^{2}}\right.
 +\frac{\ln \gamma_{+}}{\gamma_{+}^{2}}\right) .
\label{eq.2.6}
\end{eqnarray}
For the $s'$ cut, the square brackets in \eqn{6.19}
are calculated in the functions {\tt SHFULL} and  {\tt AHFULL}.

The following abbreviations are used:
\ba
c_{\pm} &=& \frac{1}{2}(1 \pm c),
\label{6.22}
\\ \nl
\gamma_{\pm} &=& c_{\pm} + Rc_{\mp} .
\label{eq:c311}
\ea
\subsection{Final state radiation \label{sprfinal} }
Final state radiation may be treated in quite different ways in \zf.
The treatment is chosen by flag {\tt IFINAL}={\tt FINR}
and described in detail in \sect{IBORNA}.
The default value is {\tt FINR}=1 for {\it leptons}.
A common exponentiation of initial and final state radiation will be 
performed when  {\tt FINR}=1 is chosen.
This situation is described in \sect{commexp}.
For  {\tt FINR}=0, final state photonic corrections are treated as a simple 
factors to $\sigma_{\sss{A}}^0(s')$ in \eqn{siginic1} and in \eqn{saini} 
 according to \eqns{inifin2}{inifin1}.
This situation is described in  \sect{qcdrun} 
and \sect{ewcoup}.
Here we mention only that for the two  
effective Born cross-sections $\sigma_{\sss{T}}^0$ and $\sigma_{\sss{FB}}^0$ 
the factors are different.  
For  {\tt FINR}=--1,  no final state QED or QCD corrections are 
applied.
For {\it neutrinos}, 
{\tt FINR}=0 is fixed and in effect no final state radiation contributes.
For details see subroutine {\tt EWCOUP} where the generalized weak couplings 
are set. 
%
%
\subsubsection{Common exponentiation of initial and final state radiation
\label{commexp} 
}
%
Common exponentation of initial and final state soft photon radiation is done 
for flag {\tt FINR}=1 and follows Section 5.3 of \cite{Bardin:1991fut}.
 
The cross-section contributions are:
\ba
\frac{d\sigma^{ini + fin}}{d\cos\vartheta} &=&  \sum_{A=T,FB}  
\int_0^{\Delta} dv 
\sigma_{\sss{A}}^0(s')R_{\sss{A}}^{ini}(v,\cos\vartheta)  
{\bar R}_{\sss{A}}^{fin}(v),
\label{siginific2}
\\  \nl
\sigma_{\sss{A}}^{ini + fin}(c) &=&   
\int_0^{\Delta} dv
         \sigma_{\sss{A}}^0(s')
         R_{\sss{A}}^{ini}(v,c)
        {\bar R}_{\sss{A}}^{fin}(v)
,  \hspace{1.5cm} A=T,FB.
\label{siginifin2}
\ea
For the definitions of 
$\sigma_{\sss{A}}^0(s'), R_{\sss{A}}^{ini}(v,\cos\vartheta),
R_{\sss{A}}^{ini}(v,c)$ see \sect{i.born} and \sect{isp}.

We mention here that the case of no cuts (but a simple $s'$ cut) is
chosen with $c=1$. In this case the subroutine {\tt SCUT} returns numbers 
identical to subroutine to {\tt SFAST}.
\footnote{Subroutine {\tt SFAST} is called in subroutine
{\tt ZCUT} when flag {\tt IFAST}=1.
For further comments see \sect{common}.}.

The integrations in \eqns{siginific2}{siginifin2} are performed 
in subroutines {\tt COSCUT} and {\tt SCUT}.
The final state factors ${\bar R}_{\sss{T}}^{fin}(v)$ = {\tt SFIN} and 
${\bar R}_{\sss{FB}}^{fin}(v)$ = {\tt AFIN}
are calculated by calls of subroutine {\tt FUNFIN} from subroutine 
{\tt BORN}:
\ba
\label{barrfinI}
{\bar R}_{\sss{A}}^{fin}(v)
&=& \Delta '^{\beta_{f}'}\left[1+\bar{S}^{fin}(\beta_{f}')\right] + 
{\bar G}_{\sss{A}}(v),
\ea
%
%
The virtual and soft photon part $\bar{S}^{fin}(\beta_{f}')$ = 
{\tt (SOFTFR + 3/2*ADD)} has a rescaled argument $\beta_{f}'$:
\ba
\label{barsfin}
\bar{S}^{fin}(\beta_{f}')
&=&
\frac{3}{4}\beta_{f}'
+
\frac{\alpha}{\pi} Q_f^2
\left( \frac{\pi^2}{3} - \frac{1}{2} \right),
\\ \nl  
\label{barbetfin}
\beta_{f}' &=& \frac{2\alpha}{\pi}Q_{f}^{2}\left(\ln\frac{s'}{m_{f}^{2}}
   -1\right).
\ea
The effective $s'$ cut $\Delta'$  = {\tt ALIM = A} is also rescaled:
\ba
  \Delta ' &\equiv& 1 - \frac{s_{\min}'}{s'} =
\frac{\Delta-v}{1-v}.
\ea
Due to this rescaling, the soft corrections for the final state 
$\bar{S}^{fin}$ have
to be calculated in parallel to the hard corrections and may not be
predetermined in the initialization phase as is done for the initial
state corrections  $\bar{S}^{ini}$.

The hard parts are:
\ba
 {\bar G}_{\sss{A}}(v) = \frac{1}{4}\beta_{f}'\Delta '\left(\Delta
'-4\right) 
 +\frac{\alpha}{\pi}Q_{f}^{2}\left[- 2\litwo(\Delta ') + g_{\sss{A}}
\left(\Delta '\right)\right],
\ea
where  
\ba 
g_{\sss{T}}(\Delta ') &=& \frac{1}{2}\left(1-\Delta '\right)\left(3-\Delta '
\right)\ln\left(1-\Delta '\right)
  - \frac{1}{4}\Delta '\left(\Delta '-6\right) 
-   
\frac{\Delta '^{2}}{2}\, {\cal G}   ,
\label{gtdel}
\\ \nl
g_{\sss{FB}}(\Delta ') &=& \frac{1}{2}\Delta '^{2}.
\ea
In subroutine {\tt FUNFIN}, 
it is ${\bar G}_{\sss{A}}+(\alpha/\pi)Q_{f}^{2} g_{\sss{A}}$ = {\tt
FIN - FINS}. 
The nonuniversal hard final state correction $g_{\sss{T}}(\Delta')$ contains an
approximation since it contains a term, which 
arises from the angular integration of the product of
the (angular dependent) initial state factor with
\ba 
{\cal G}(\cos\vartheta)=\left(3-\frac{4}{1+\cos^2\vartheta}\right),
\ea
and which has been approximated in~\eqn{gtdel} by
\ba
{\cal G}  = 0. 
\ea
\subsection{Initial-final state interference corrections
\label{s:inifin}
} 
%
The initial-final state interference corrections are added when flag 
{\tt INTERF}=1 is chosen. 

The corresponding cross-sections have the following generic structure: 
\ba
\frac{d \sigma^{int}}{d \cos\vartheta}
&=&
\sum_{m,n} \int_{0}^{\Delta} dv 
\, \Re e 
\Biggl[
\sigma_{\sss{FB}}^0(s,s';m,n) \, R_{\sss{T}}^{int}(v,\cos\vartheta;m,n) 
\nl &&+~
\sigma_{\sss{T}}^0(s,s';m,n)  \, R_{\sss{FB}}^{int}(v,\cos\vartheta;m,n) 
\Biggr],
\label{int1}
\\
\sigma_{\sss{T}}^{int} &=&
\sum_{m,n} \int_{0}^{\Delta} dv  \, \Re e 
\Biggl[
\sigma_{\sss{FB}}^0(s,s';m,n)  \, R_{\sss{T}}^{int}(v,c;m,n) \Biggr],
\label{int2}
\\
\sigma_{\sss{FB}}^{int} &=&
\sum_{m,n} \int_{0}^{\Delta} dv 
 \, \Re e 
\Biggl[
\sigma_{\sss{T}}^0(s,s';m,n)  \, R_{\sss{FB}}^{int}(v,c;m,n) \Biggr] .
\label{int3}
\ea
In all these formulae, $m,n$ take the values $\gamma,Z$ and
we use $\sigma_{\sss{A}}^0(s,s';m,n)$ from~\eqn{bornss'}.

The  initial-final state interference corrections have the following
property (for a proof see preprint version of \cite{Riemann:1988gy}):
\ba
 R_{\sss{A}}^{int}(v,C;m,n) =
 \frac{1}{2}\left[R_{\sss{A}}^{int}(v,C;m,m) +
 R_{\sss{A}}^{int}(v,C;n,n)^{*}\right], ~~~m,n = \gamma,Z. 
\label{eq:r41}
\ea
Here, the $^*$ means complex conjugation and
$C$ may be either $c$ or $\cos\vartheta$ and $A=T,FB$.
This property allows to determine the $\gamma Z$ interference from the
$\gamma$ and $Z$ exchange corrections.
In \zf, the photon functions are the conjugated ones. It is important
to use \eqn{eq:r41} in accordance with \eqn{bornss'}.
\subsubsection{Differential cross-sections
\label{intcos}
}
%
Differential cross-sections are calculated in subroutine {\tt COSCUT}.
Up to some normalizations to be discussed yet, 
it is calculated for \eqn{int1}:
\ba
R_{\sss{A}}^{int}(v,\cos\vartheta;m,n) 
 &=&
 \frac{\alpha}{\pi}Q_{e}Q_{f}
\Bigl\{
\delta(v) \left[ D_{\bar A}(\cos\vartheta) {\bar S}^{int}
(\Delta,\cos\vartheta) + B_{\sss{A}}(\cos\vartheta,m,n) \right]
\nl && +~ 
{\bar H}_{\sss{A}}^{int}(v,\cos\vartheta) \Bigr\} .
\label{rint0}
\ea
Here, $D_{\bar{A}}(\cos\vartheta)$ is defined 
in \eqn{sigdiff2} and \eqn{sigdiff3}.
The functions $R^{int}$ consist of soft bremsstrahlung $S$, hard 
bremsstrahlung $H$,
and $\gamma\gamma$ and $\gamma Z$ box terms $B$.
The soft correction factor 
${\bar S}^{int} (\Delta,\cos\vartheta)$ is independent of $m,n$ and also of 
$A=T,FB$ and ${\bar A} = FB,T$.
It does depend on the cut $\Delta$. 
This is an artefact of a regularization 
of the hard photonic corrections $H_{\sss{A}}^{int}(v,\cos\vartheta)$ 
near $v=0$:
\ba
  \sigma_{\bar A}^0(s,s') {\bar H}_{\sss{A}}^{int}     (v,\cos\vartheta) 
&=& 
  \sigma_{\bar A}^0(s,s')        H_{\sss{A}}^{int}     (v,\cos\vartheta) 
- \frac{\sigma_{\bar A}^0(s)}{v} D_{\bar A}(\cos\vartheta)
 H^{int}_{sing}(\cos\vartheta), 
\nl
\label{rint1}
\ea
with:
\ba
{H}^{int}_{sing}(\cos\vartheta) &=& 
4 \ln \frac{C_-}{C_+}.
\label{singA}
\ea
The subtraction in \eqn{rint1} makes the hard corrections finite at $v=0$.
It has to be added (after integration) to the soft photonic correction
$S_{\sss{A}}(\cos\vartheta,\epsilon)$ of Eq. (93) in \cite{Bardin:1991fut}, 
which has the same singularity with opposite sign:
\ba
{D}_{\bar A}(\cos\vartheta)
  {\bar S}^{int}(\Delta, \cos\vartheta) &=& 
S_{\sss{A}}(\cos\vartheta,\epsilon) + {D}_{\bar A}(\cos\vartheta)
H^{int}_{sing}(\cos\vartheta)
 \int_{\epsilon}^{\Delta} 
  \frac{dv}{v}.
\label{IntSing}
\ea
The subroutine {\tt COSCUT}
calls subroutine {\tt SOFTIN} with argument {\tt SFTI} for the
calculation of the soft photon corrections, subroutine {\tt BOXIN} with
argument {\tt BOXI} for the box corrections, and performs a numerical
integration over function {\tt HARD} (with variable $R=1-v$) in order to
determine the hard photonic corrections {\tt HRD}.   
Their sum is then transferred to subroutine {\tt ZANCUT}.
In terms of the variables of \zf, the differential cross-section contribution
from the initial-final state interference is:
\ba
\frac{d\sigma^{int}}{d \cos\vartheta}  
&=&
{\tt SIGQED} 
\nl &=& \frac{3}{8} 
\frac{\tt CSIGNB}{s} 
[( {\tt SFTI} + {\tt BOXI})  {\tt CORINT} - {\tt HRD} ] .
\ea
The factor 
\ba
{\tt CORINT} = {\tt RQCDV} 
\ea
is taken from 
subroutine {\tt EWCOUP}, see \sect{ewcoup}.
Further, we have to notice that the Born factors in the box terms are 
composed not in subroutine {\tt BORN} but in subroutine {\tt BOXIN}.
 
Now we give the explicit expressions for the various contributions and
begin with ${\bar S}^{int}$ = {\tt SFTI}\footnote{A dependence on the
infrared pole has been already cancelled between the box  
and soft photon functions used.
To control them, consult Section 4.1 of \cite{Bardin:1991fut}.
}:
\ba
 {\bar S}^{int}(\Delta, \cos\vartheta)
&=&
2 
~\left[ 2\ln\Delta
\ln\frac{C_{-}}{C_{+}} + \litwo(C_{+}) - \litwo(C_{-})
 -\frac{1}{2}\left( \ln^{2}C_{+} - \ln^{2}C_{-} \right)\right] ,
\nl
\label{sbarc}
\ea
with $C_{\pm}$ from ~\eqn{cpm}.

We now come to the calculations of the photonic box contributions 
from Section 4.1 of~\cite{Bardin:1991fut}, used in \eqn{int1} and 
\eqn{rint0}.

Up to the normalization factor {\tt (3/8)$\cdot$CSIGNB/S}, they are calculated 
completely inside the subroutine {\tt BOXIN}.
It suffices to derive the pure $\gamma$ and the pure $Z$ exchange 
function; then the $\gamma Z$ interference is also known from \eqn{eq:r41}. 
The box functions for $A=T$ and $A=FB$ are related:
\ba
 B_{\sss{T,FB}}(\cos\vartheta,n,n) = 
\left[ b(\cos\vartheta,n,n)
 \pm b(-\cos\vartheta,n,n) \right] . 
\label{BTFB}
\ea
Thus, only two different box functions remain to be
calculated\footnote{In \cite{Bardin:1991fut}, the overall signs in
Eqs. (95) and (96) are wrong. 
To control the divergent terms, consult Section 4.1 of \cite{Bardin:1991fut}.
There, the divergent term in Eq. (96) is by a factor 2 too small.
}:
\ba
b(\cos\vartheta,\gamma,\gamma) &=&
   C_{+}^{2}\ln\frac{C_{+}}{C_{-}}
\left[
- 2\pi i \right] 
- \frac{1}{2}\ln C_{-} \Bigl[-2C_{+} +\cos\vartheta(\ln C_{-} 
   + 2\pi i)\Bigr] 
\nl && 
+~ \pi i C_{+},
\\ \nl
 b(\cos\vartheta,Z,Z) &=& 
 2C_{+}\left(1-R_{Z}\right)
\Biggl\{ \ln\frac{C_{-}}{R_{Z}}
 - (1-R_{Z})L_{Z} +
   \nonumber \\
&& \frac{1}{C_{+}}(1-R_{Z} - 2C_{+})
\left[ l(1) - l(C_{-}) -L_{Z}\ln C_{-} \right] \Biggr\}
   \nonumber \\
&&  
+~C_{+}^{2} \left\{
\left[ 
4L_{Z} + \ln(C_{+}C_{-}) \right]
   \ln\frac{C_{+}}{C_{-}} + 2l(C_{+}) - 2l(C_{-}) \right\} .
\ea
The following abbreviations are used:
\ba
 l(a) &=& \litwo\left(1 - aR_{Z}^{-1}\right),
\label{la}
\\ \nl
   L_{Z} &=& \ln\left(1 - R_{Z}^{-1}\right) ,
\label{LZ}
\ea
and $R_{Z}$ = {\tt XR}  is:
\ba
R_Z &=& \frac{m_Z^2}{s}.
\label{Rz}
\ea
The complex mass $m_Z$ is defined in \eqn{mZ2}.
 
To state it correctly: In \zf\ the programming of the box contributions to 
the angular distribution deviates slightly from the conventions used in this 
description so far.
In subroutine {\tt BOXIN}, the real and imaginary parts of $B_{\sss{A}}(n,n)$ 
in \eqn{BTFB} are calculated explicitely:
\ba 
\Re e \, B_{\sss{T}}(\cos\vartheta;\gamma,\gamma) &=& - {\tt REF4BX},
\label{rf4bx}
\\
\Im m \,  B_{\sss{T}}(\cos\vartheta;\gamma,\gamma) &=& - {\tt AMF4BX},
\label{af4bx}
\\
\Re e \,  B_{\sss{FB}}(\cos\vartheta;\gamma,\gamma) &=& {\tt REF1BX},
\label{rf1bx}
\\
 \Im m \, B_{\sss{FB}}(\cos\vartheta;\gamma,\gamma) &=& {\tt AMF1BX},
\label{af1bx}
\\
B_{\sss{T}}(\cos\vartheta;Z,Z) &=& - {\tt XH4BX},
\label{h4bx}
\\
B_{\sss{FB}}(\cos\vartheta;Z,Z) &=& {\tt XH1BX} .
\label{h1bx}
\ea

Further, the box corrections do not take the overall factors with
coupling constants, $\sigma_{\sss{A}}^0(s,s')$, from subroutine {\tt BORN}.

These factors are constructed in {\tt BOXIN} directly out of the
building blocks of subroutine {\tt EWCOUP}.
As an example, we give the cross-section contribution from the pure photonic 
boxes: 
\ba
\frac{d \sigma^{int,box}}{d \cos\vartheta}(\gamma,\gamma)
&=&
\frac{\alpha}{\pi}Q_{e}Q_{f} \left[
\sigma_{\sss{FB}}^0(s,s;\gamma,\gamma)\, 
B_{\sss{T}}(\cos\vartheta,\gamma,\gamma)
+
\sigma_{\sss{T}}^0(s,s;\gamma,\gamma)\, 
B_{\sss{FB}}(\cos\vartheta,\gamma,\gamma) 
\right] .
\nl
\label{boxggc}
\ea 
This is the $(\gamma,\gamma)$ part of the variable {\tt BOXI} = 
{\tt BOXIS + BOXIA}. 
In terms of the variables used in \zf\footnote{The imaginary parts of
the pure photonic boxes contribute through  
the $\gamma Z$ interference.}:
\ba
\frac{d \sigma^{int,box}}{d \cos\vartheta}(\gamma,\gamma)
&=&
- {\tt ALQEF} \cdot  {\tt VPOL2}  \cdot {\tt (AEFA} \cdot {\tt REF4BX}
 +{\tt VEFA} \cdot {\tt REF1BX}) 
 \cdot {\tt CORINT} .
\label{bxg}
\nl
\ea
We see that besides the box functions there is an explicit dependence on the 
combinations of couplings like {\tt AEFA, VEFA, VPOL2}.  
 

Explicitely, 
\ba
{\tt REF4BX} &=& 
- \left[ 
\cos \vartheta \ln \frac{C_+}{C_-} 
\left[ \ln (C_+  C_-) -1 \right]  
+ \ln (C_+  C_-) \right],
\\
{\tt AMF4BX} &=& 2 \pi \left( \cos \vartheta \ln \frac{C_+}{C_-} -1 \right),
%
%
%
%
%
%
%
%
%
%
%
%
%
\\
    {\tt  REF1BX} &=& -\cos\vartheta \left[ \ln^2 C_+
 +              \ln^2 C_- - \ln \left(C_+C_-\right)  \right]
 - \ln\frac{C_+}{C_-},
\\
  {\tt  AMF1BX} &=& - 2\pi \left[ \cos\vartheta \left[\ln
\left(C_+C_-\right) -1 \right] 
                         +(1+\cos^2\vartheta) \ln\frac{C_+}{C_-}\right] .
\ea
%
%
%
%
%
%
%
%
%
%
%
%

The  hard  radiator parts $H_{\sss{T}}^{int}$ =  
{\tt H4} and $H_{\sss{FB}}^{int}$ =  
{\tt H1} are independent of the  gauge  boson exchanged.
They are calculated as arguments of subroutine {\tt HCUT}, called by function 
{\tt HARD}:  
\ba
   H_{\sss{T,FB}}^{int}(v,\cos\vartheta) = 
    \left[h_{\sss{T,FB}}^{int}(v,\cos\vartheta) \pm
    h_{\sss{T,FB}}^{int}(v,-\cos\vartheta) \right] . 
\ea
While the box terms for $A = FB$ and $A = T$ are expressed by
one and the same function, this is not the case for 
$H_{\sss{T}}^{int}$ and $H_{\sss{FB}}^{int}$.
They are composed as symmetric and anti-symmetric combinations of variables
$h_{\sss{T}}^{int}$ = {\tt H4M}, 
$h_{\sss{FB}}^{int}$ = {\tt H1M}
(Eqs. (100) and (101), respectively,
in \cite{Bardin:1991fut}). 
This is done in \zf\ by two subsequent calls of subroutine
{\tt HINTFM}: 
\ba
h_{\sss{T}}^{int}(v,\cos\vartheta) &=& 
 2 C_{+} \left\{ \left[ \frac{4}{v} - 3
 - R(2+R)\right] \ln \frac{C_{-}}{C_{+}} - (1+R)^{2} \ln
 \frac{\bar \gamma}{\gamma} \right\}
 \nonumber \\
&& + 2\left[ -\frac{4}{v}+4+R(2+\ln R)\right]
  + \frac{4}{\gamma}\left[\frac{2}{v}
  -2-R(1+R)\right],
\ea
\ba
 h_{\sss{FB}}^{int}(v,\cos\vartheta) &=& 
 2\left( 1+\cos^2\vartheta \right) \ln C_{-} \left( \frac{2}{v} - 1 -
 R - R^{2}\right) 
\\
&&    +~ 4C_{+}\left[ -\frac{4}{v} + 4 + 2R - R(1-R)\ln R \right]
+\frac{4}{\gamma} 
   \left( -\frac{2}{v^{2}}+ \frac{5}{v}-3-R \right)
\nonumber \\
&& +~\frac{2}{\gamma^{2}}\left(-\frac{2}{v^{2}} +
\frac{6}{v}-4-2R-R^{2} \right) 
   +2(R^{2}-1)C_{+}\ln C_{+}C_{-} 
\nonumber  \\
\nonumber  
&& +~
2\left[
(1-R+R^{2})+\cos\vartheta(1-R^{2})+\cos^2\vartheta(1+R+R^{2})\right]
\ln\gamma .
\ea
Here, $\gamma$ is defined in \eqn{eq:c312}) and 
${\bar \gamma} = \gamma (C_{\pm} \to C_{\mp})  $.
\subsubsection{Integrated cross-sections
\label{intc}
}
The radiator functions in \eqn{int2} and \eqn{int3} are:
\ba
R_{\sss{A}}^{int}(v,c,m,n)
&=&
\frac{\alpha}{\pi} Q_e Q_f 
\left\{
\delta(v) \left[{\bar S}_{\sss{A}}^{int}(c) + B_{\sss{A}}(c,m,n) \right] +
{\bar H}_{\sss{A}}^{int}(v,c) 
\right\}.
\label{Rint2}
\ea
In \zf, the soft photon corrections are
${\bar S}_{\sss{T}}^{int}(c)$ = - {\tt SFTI4}
and 
${\bar S}_{\sss{FB}}^{int}(c)$ = {\tt SFTI1}, and their contributions to
cross-sections are then {\tt SFTIS} = - {\tt SFTI4 $\cdot$ ABORN0 $\cdot$
  ALQEF} and   {\tt SFTIA = SFTI1 $\cdot$ 
SBORN0 $\cdot$ ALQEF}.
They are  used in all calculational chains and are
calculated in subroutine {\tt SFTINT}, which is
called by subroutine {\tt SFAST} or, alternatively, by {\tt SCUT}.
Their explicit expressions are derived from the functions
$S_{\sss{A}}^{int}(c,\epsilon)$, Eqs. (20) and (25)
of~\cite{Bardin:1991det}, with a regularization:
\ba
{\bar S}_{\sss{A}}^{int}(c)
&=&
{S}_{\sss{A}}^{int}(c,\epsilon) + H^{int}_{sing,\sss{A}}(c)
\int_{\epsilon}^{\Delta} \frac{dv}{v} ,
\label{sintdel}
\ea
while at the same time compensating terms contribute to functions {\tt
  SHARD} and {\tt AHARD}:
\ba
\sigma_{\bar A}^0(s,s')
{\bar H}_{\sss{A}}^{int}(v,c)
&=&
\sigma_{\bar A}^0(s,s')
{H}_{\sss{A}}^{int}(v,c)-\frac{\sigma_{\bar A}^0(s)}{v}
H^{int}_{sing,\sss{A}}(c).
\label{hintdel}
\ea
Functions ${H}^{int}_{sing,\sss{A}}(c)$ are the angular integrals of functions
$D_{\bar A}(\cos\vartheta) {H}^{int}_{sing}(\cos\vartheta)$:
\ba
H^{int}_{sing,\sss{T}}(c) &=& -4 \left[
\left(c^{2}-1\right)\ln\frac{c_{+}}{c_{-}} + 2c\right],
\label{intsingT}
\\
H^{int}_{sing,\sss{FB}}(c) &=& \frac{4}{3}
\left[-8\ln 2 - 4 \ln(c_{+}c_{-})
  - 3 C_{\sss{T}}(c) \ln\frac{c_{+}}{c_{-}} - c^{2}\right].
\label{intsingFB}
\ea
Then, we may write with $\ln \Delta $ = {\tt ALDEL}:
\begin{eqnarray}
  {\bar S}_{\sss{T}}^{int}(c) 
&=&
H^{int}_{sing,\sss{T}}(c)  \ln \Delta 
 + 2\left(c^{2}-1\right)\left[\litwo(c_{+}) - \litwo(c_{-})\right]
 \nonumber \\
&& -~\left(c^{2}-1\right)\ln\left(c_{+}c_{-}\right)\ln\frac{c_{+}}{c_{-}}
 - 4c\ln(c_{+}c_{-}) - 4 \ln\frac{c_{+}}{c_{-}} + 8c,
\label{sdelt}
\\ \nl
  \frac{3}{4}{\bar S}_{\sss{FB}}^{int}(c) 
&=& 
H^{int}_{sing,\sss{FB}}(c) \ln\Delta
 \nonumber \\
&&  +~ \frac{3}{2}C_{\sss{T}}(c)\left[\litwo(c_{+}) -
\litwo(c_{-})\right] 
  + 2\left[\litwo(c_{+}) + \litwo(c_{-})\right]
\nl &&
  -~
  \frac{3}{4}C_{\sss{T}}(c)\ln(c_{+}c_{-})\ln\frac{c_{+}}{c_{-}} 
- \left(\ln^{2}c_{+} + \ln^{2}c_{-}\right)
  + 4\ln^{2}2 + \ln 2 
\nl &&
- \frac{1}{2}\ln(c_{+}c_{-})(c^{2}-1) 
 + \frac{c^{2}}{2}  - 2 \litwo(1).
\label{sdelfb}
\end{eqnarray}
For $c=1$, the expressions simplify considerably: 
${\bar S}_{\sss{T}}^{int}(1) = -8(\ln \Delta -1)$ 
and ${\bar S}_{\sss{FB}}^{int}(1) =
-(1 + 8 \ln 2 ) \ln \Delta + 2 \litwo(1) + 4 \ln^2 2 + \ln 2 + 1/2$.
Though, this is not used in \zf.

The net box contributions to the integrated cross-sections are called 
{\tt BOXIS} and  {\tt BOXIA}.  
They 
are calculated in subroutine {\tt BOXINT}, which itself is 
called by either subroutine {\tt SFAST} or subroutine {\tt SCUT}. 
For the integrated box functions apply the correspondences \eqn{rf4bx} to 
\eqn{h1bx}.   
The box functions 
have an explicit dependence on the kind of exchanged vector 
boson and are written in (anti-)symmetrised form (see Eqns.~(21),
(22), (26), (27) of~\cite{Bardin:1991det})\footnote{\zf\ calculates
the expressions $B_{\sss{T,FB}}$. 
We remark here that in Eq. (19) of~\cite{Bardin:1991det} the $\pm$ has to 
be a $\mp$ and that in Eq. (22) the first term in the curly bracket has a 
wrong sign. Further, in Eq. (18) the factor $\sigma^0$ is superfluous.
}:
\ba
 B_{\sss{T,FB}}(c;n,n) &=& b_{\sss{T,FB}}(c;n,n) 
                             \mp b_{\sss{T,FB}}(-c;n,n),
\label{bpm}
\ea
with
\ba
b_{\sss{T}}(c;\gamma,\gamma)  &=&
\frac{1}{2}\left(c^{2}-1\right)\ln^{2}c_{+} 
+ \left(- c^{2}+ 2c  -3    \right) \ln c_{+}  - 3c
 -~i \pi (c^{2}-1) \ln c_{+},
\label{bt00}
\\ \nl
b_{\sss{FB}}(c;\gamma,\gamma) 
 &=& -\frac{1}{2}\left(c^{2}-1\right)\ln^{2}c_{+} 
- \left(- c^{2}+ 2c  -3    \right) \ln c_{+}
- \frac{1}{2}\left(\ln^{2}2 + 6\ln 2 +
      c^{2}\right) 
 \nonumber \\
&&   -~ \frac{i \pi}{3} 
\left[5 \ln 2 + \left(2c^3+3c^{2}+6c+5 \right)\ln c_{+} 
- \frac{2}{3}c^{2}\right], 
\label{bfb00}
\\
\nl
 b_{\sss{T}}(c;Z,Z) &=&
 -2
\Biggl\{ -2c R_Z (1-R_Z)\ln c_{+}
\\
&& -~ \left[-2R_Z^{2} + R_Z\left(c^{2}+1\right)+c^{2}-1\right]\ln c_{+} -
c R_Z (R_Z+1)\Biggr\} 
L_Z 
 \nonumber \\
&& +~ 4  c R_Z (R_Z-1) l(1) 
- 2c\ln R_Z +
\left(c^{2}-1\right)\ln^{2}c_{+} 
 \nonumber \\
&&  +~ 2\left[R_Z\left(c^{2}-1\right)-c^{2}+2c+3\right]\ln c_{+} 
 \nonumber \\  \nonumber 
&&-~ 2 \left[2R_Z^{2}+2c R_Z(R_Z-1)-R_Z\left(c^{2}+1\right)\right]
l(c_+) 
+ 2R_Zc - 6c,
\hspace{1cm}     n \neq 0,
\label{btnn}
\\
 b_{\sss{FB}}(c;Z,Z) &=& 
\left[C_{\sss{T}}(c) + \frac{4}{3} \right] \ln^{2}c_{+} 
\\
&&+~ 
\Biggl\{ \left(4R_Z^{2} - 2R_Z + \frac{10}{3}\right) \ln 2
\nl
&&+~\left[4R_Z^{2}-2R_Z\left(c^{2}+1\right)+2c^{2}+\frac{10}{3}\right]\ln c_{+}
 \nonumber \\
&&  +~ 4 \left[ cR_Z(R_Z-1)+C_{\sss{T}}(c) \right] \ln c_{+} +
c^{2}\left(-R_Z^{2}+3R_Z-\frac{4}{3}\right)\Biggr\} L_Z
 \nonumber \\
&& +~ c^{2}\left(\frac{4}{3}R_Z-\frac{5}{3}\right)\ln R_Z  
+   \left(\frac{16}{3}R_Z^{3} - 4R_Z^{2} + 2R_Z -\frac{2}{3}\right)
l \left( \frac{1}{2} \right)  
 \nonumber \\
&& +~ 2c^{2}(R_Z-1) l(1) 
 - \frac{4}{3}\ln^{2}2 + \left(\frac{8}{3}R_Z^{2}+\frac{8}{3}R_Z - 6\right)
\ln 2
 \nonumber \\
&&+~\left[\frac{8}{3}R_Z^{2} + R_Z\left(-\frac{4}{3}c^{2}+\frac{8}{3}\right)
+ 2\left(c^{2}-3\right)\right] \ln c_{+} 
\nl && +~\left(\frac{8}{3}R_Z^{2}+\frac{4}{3}R_Z-4\right)c\ln c_{+} 
 \nonumber \\
&& +~ 2\left[-\frac{8}{3}R_Z^{3} + 2R_Z^{2}-R_Z\left(c^{2}+1\right)+c^{2}
 +\frac{1}{3}\right] l(c_+) 
 \nonumber \\ \nonumber
&& +~ 2\left[
2c\left(R_Z^{2}-R_Z\right) +
C_{\sss{T}}(c)\right]  l(c_+) 
 + \left(\frac{2}{3}R_Z-1\right)c^{2}
,\hspace{0.2cm}
     n \neq 0.
\label{bfbnn}
\ea
For $l(a)$, $L_Z$, and $R_Z$ we use \eqns{la}{Rz}.
Both the soft photon and the box corrections simplify considerably if no
acceptance cut is applied.
For this case, the latter are given in~\cite{Bardin:1987hv,Bardin:1988ze},
where also the completely 
integrated corrections $\sigma_{\sss{T,FB}}^{QED,int}(1)$ (i.e. with no cut) 
may be found in a rather compact form which is, though, not realized 
in subroutine {\tt SFAST} of \zf.

The hard radiator functions ${\bar H}_{\sss{A}}^{int}$ in \eqn{hintdel} are 
calculated
in functions {\tt SHARD} and {\tt AHARD}.
They are explicitely regulated there at $v \to 0$, while the functions 
$H_{\sss{A}}^{int}$ = {\tt H4, H1}  are taken from Eqs. (24) and (28) of 
\cite{Bardin:1991det} and
called from subroutines {\tt SHFULL, AHFULL}:
\begin{eqnarray}
{H}_{\sss{T}}^{int}(v,c) 
&=&
 4c_{+}c_{-}\left[\frac{r_{1} r_{2}}{v}\ln\frac{c_{+}}{c_{-}}
  + \right(\frac{c^{2}}{c_{+}c_{-}}R - v^{2}\left)
\ln\frac{\gamma_{+}}{\gamma_{-}}\right]
+ 4cR\left(-\frac{r_{1}}{v} + \ln R\right)
,
\nl
\label{htint}
\\ \nl
{H}_{\sss{FB}}^{int}(v,c) 
&=&
 \frac{2c^{2}R}{\gamma_{+}\gamma_{-}}
\left(\frac{1}{3} v^{2} - \frac{2}{3 v} + \frac{2 R}{r_{1}}
+\frac{1}{3}c^{2} v r_{1} \right)
\nonumber  \\
&& -~ 2c^{2}R v\ln R -4\frac{r_{1}}{v}c_{+}c_{-}r_{2}\ln (c_{+}c_{-})
\nonumber  \\
&& -~\frac{16}{3}\frac{r_{3}}{v}\left(c_{-}^{3}\ln c_{-} +
c_{+}^{3}\ln c_{+}\right)
 + \frac{2}{3}\left(2R-5r_{2}\right)\frac{r_{1}}{v}\ln r_{1}
\nonumber  \\
&& -~ 4\ln\gamma_{-} \left[ 2cc_{-}^{2} - \gamma_{-}
\left(v + \frac{4}{3}
 \frac{\gamma_{-}^{2}}{v} - 4c_{-}^{2} + \gamma_{-}\right) \right]
\nonumber  \\
&& +~ 4\ln\gamma_{+} \left[ 2cc_{+}^{2} + \gamma_{+}\left(v + \frac{4}{3}
 \frac{\gamma_{+}^{2}}{v} - 4c_{+}^{2} + \gamma_{+}\right) \right], 
\end{eqnarray}
with $r_n, c_{\pm}, \gamma_{\pm}$ defined in \eqn{6.14}, \eqn{6.22},
\eqn{eq:c311}. 

\section{
\label{subsec:acol}
Photonic Corrections with Acollinearity Cut
}
\setcounter{equation}{0}
The calculational chain with cuts on angular acceptance, 
acollinearity, and minimal energy of the final state fermions
was originally implemented in \zf\ by M. Bilenky and A. Sazonov
(1989) \cite{Bilenkii:1989zg} (flag {\tt ICUT}=0).
The present default treatment of photonic ${\cal O} (\alpha)$ corrections with 
the package {\tt acol.f} is chosen with  
flag {\tt ICUT}=2 or  {\tt ICUT}=3 in the call of subroutine {\tt
ZUCUTS}; see also \sect{zucuts}.
It is based on
\cite{Christova:1999cct,Christova:1999gh,Christova:2000x1}.  
The phase space parameterisation derived in \cite{Passarino:1982zp} is applied.
Higher order QED corrections depend on the flag {\tt FOT2}, see
\sect{secondo} and \appendx{zuflag}. 

In this approach, the cross-sections $\sigma_{\sss{T}}$ and
$\sigma_{\sss{FB}}$ are calculated with formulae that assume 
the following cuts on the production angle $\cos\vartheta$ of one fermion 
({\it acceptance cut}), on the final state fermions' energies
$E_{\bar f}, E_{f}$, and 
on the fermions' acollinearity angle $\xi$:
\bq
\label{cuts}
\begin{tabular}{ccccc}
 $c_1$ &$\leq$& $\cos \vartheta$ &$\leq$& $c_2$,
\\
\nl
$E_{f}^{min}$ &$\leq$& $E_{\bar f} ,E_{f}$ &$\leq$&
$\frac{\displaystyle \sqrt{s}}{\displaystyle 2}$ ,
\\
\nl
$0$ &$\leq$& $\xi$ &$\leq$& $\xi^{max}$.
\end{tabular}
\eq
The equality of the lower fermion energy limits is not essential but
simplifies some formulae and is assumed to be valid in \zf.
Some technical restrictions of the cut boundaries will be mentioned
later. 
The cut conditions do not explicitly depend on fermion masses in the
approximations assumed, but
there is an influence of the acollinearity cut on the scattering
angle, see remark after \eqn{reblim}.

\zf\ returns the cross-section with QED corrections 
{\tt SIGQED}, the asymmetry {\tt AFBQED}, and the
effective Born quantities {\tt SIGBRN} and {\tt AFBBRN} 
from subroutine {\tt ZCUT}. 
The angular distributions over $\cos\vartheta$ are calculated 
by subroutine {\tt ZANCUT} (see e.g.~\eqn{generic2} 
and~\eqns{siginic}{sigint} and the flowcharts 
in \figs{structZCUT}{structZANCUT}). 
Technicalities of the calculation are described in \sect{chains}.

%
\begin{figure}[thbp]  
\vspace{7.0cm}
\hspace{2.0cm}
\begin{minipage}[bht]{10.5cm}
{\begin{center}
  \vspace{0.0cm}
  \hspace{0.0cm}
  \mbox{
  \epsfysize=16cm
  \epsffile[0 0 500 500]{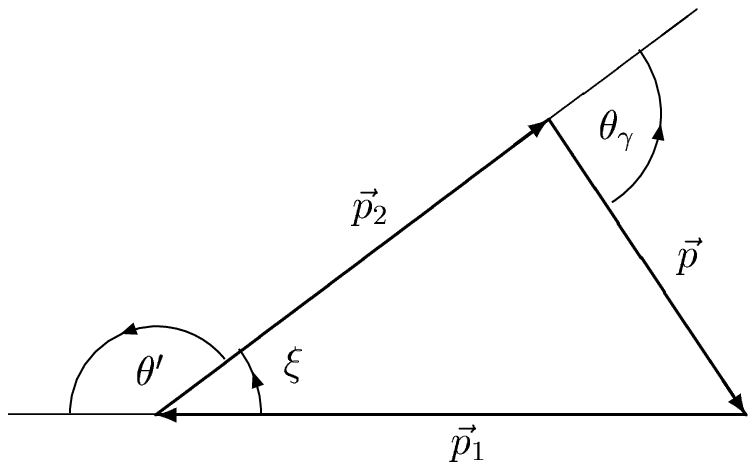}
  }

\end{center}}
\end{minipage}
\vspace{-18cm}
\caption[Acollinearity angle $\xi$]
{\it Acollinearity angle $\xi$
\label{acolang}}
\end{figure}
%

\subsection{Kinematics
\label{acolkin}
}
A three-fold analytical integration of the squared matrix elements
had to be performed over three angles of phase space:
$\varphi_{\gamma}, v_2(\cos\theta_{\gamma})$ and $\cos\vartheta$,
with $\varphi_{\gamma}, \theta_{\gamma}$ as photon angles (in the
fermion-photon restframe) and $v_2$ as invariant mass of 
photon and antifermion. The last integration, that over $R=s'/s$, 
is then performed numerically by \zf. 

The three variables $\cos\theta_{\gamma}$, $v_2$, and $R$ 
may be related to the three final state momenta 
$p_f, p_{\bar f}, p_{\gamma}$ in the centre of mass system 
(cms) where the three-momenta form a triangle:
${\vec p}_f + {\vec p}_{\bar f} + {\vec p}_{\gamma} =0$
with $v_2=2 p_{\gamma} p_{\bar{f}}$ and $s'=(p_f+p_{\bar{f}})^2$
(see \fig{acolang} with ${\vec p}_1 = {\vec p}_f,
{\vec p}_2 = {\vec p}_{\bar f}, {\vec p} = {\vec p}_{\gamma}$).
This triangle is characterized by two angles, besides
$\theta_{\gamma}$ also by $\theta'=\pi-\xi$, where $\xi$ is the acollinearity
angle (the deviation of the fermions' opening angle from $\pi$).

Conditions on the two angles, 
\ba
\sin^2\theta_\gamma &\geq& 0,
\\
\sin^2(\xi/2) &\leq& \sin^2(\xi^{max}/2),
\ea
are related to the moduli of the three-momenta in the triangle in the cms
\cite{Byckling:1973}.  
These moduli may be expressed by $s$, $R=s'/s$ and $v_2$
so that the photon angle $\theta_\gamma$ as variable can be substituted 
by $v_2$
which simplifies the analytical integrations substantially.
This chain of relations gives an access to the boundary conditions of
the physically allowed region:
\ba
v_{2_m}^{min}(R)\quad\leq &v_2&\leq\quad v_{2_m}^{max}(R),
\label{vminmax1}
\\
R_E - R \quad\leq &v_2&\leq\quad 1-R_E,
\label{rminmax}
\ea
and
\ba
&v_2& \leq v_{2_\xi}^{min}(R),
\ea
or
\ba
v_{2_\xi}^{max}(R)
\leq 
&v_2,& 
\label{raco}  
\ea
where
\ba
v_{2_m}^{max,min}(R) &=& \frac{1}{2}(1-R)
\left(1\pm\sqrt{1-\frac{4 m_f^2}{s'}}\right), 
\label{vminmax2}
\\ 
v_{2_\xi}^{max,min}(R) &=& \frac{1}{2}\,(1-R)\left[1\pm
\sqrt{1-\frac{R}{R_\xi}\frac{(1-R_\xi)^2}{(1-R)^2}}\right],
\label{vximinmax2}
\\ 
R_E &=& \frac{2 E_{f}^{min}}{\sqrt{s}}, 
\label{rvv1}
\\
R_{\xi} &=& \frac{1-\sin(\xi^{max}/2)} {1+\sin (\xi^{max}/2)}.
\label{rxi}
\ea

\begin{figure}[tbhp]
  \begin{center}
\vspace*{-1.2cm}
  \mbox{%
  \epsfig{file=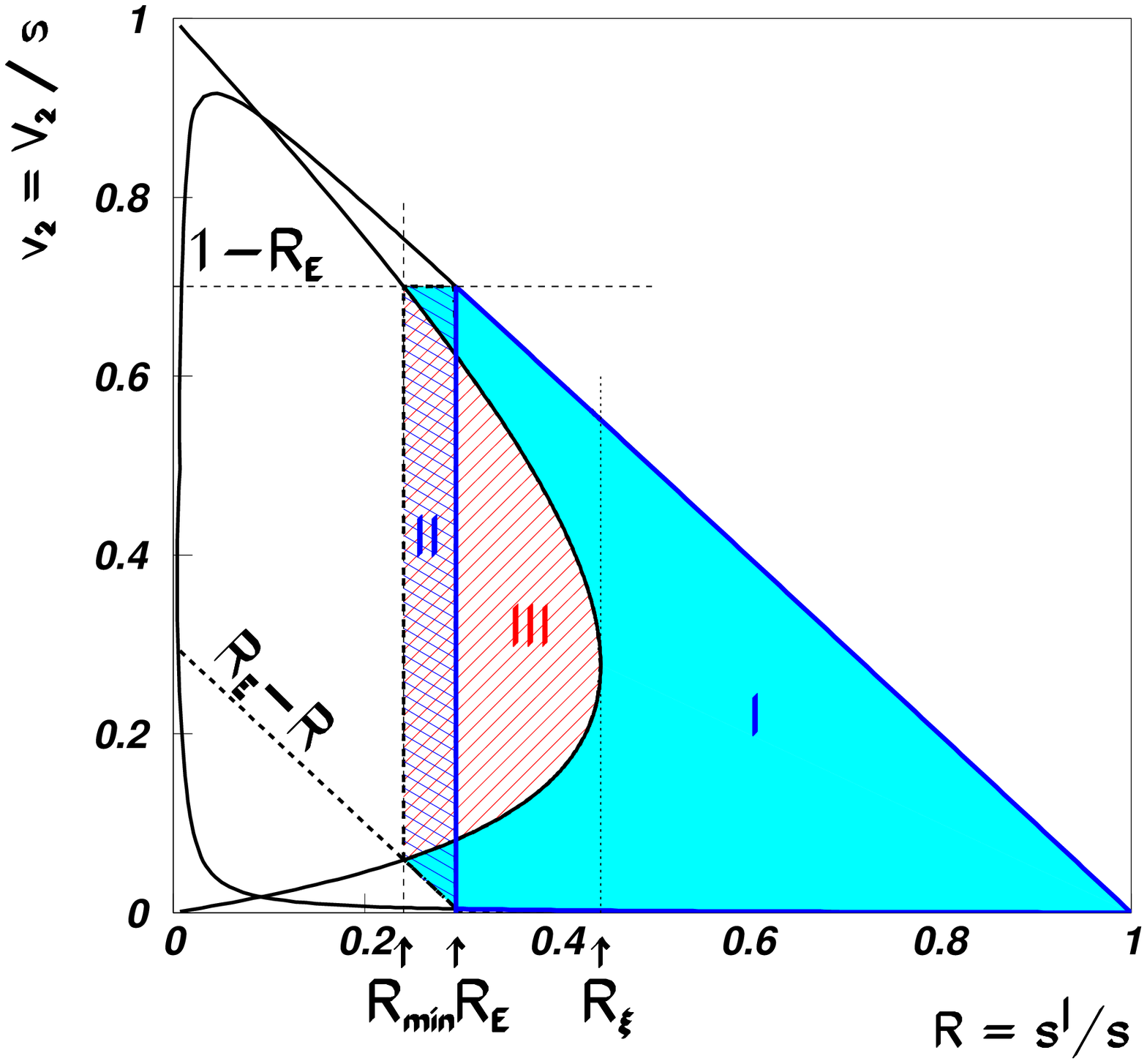
          ,height=9cm  
          ,width=9cm   
         }%
  }
\caption[Dalitz plot with cuts on $\xi$ and $E_{f}$ (I)]
{\label{dalitz1}\it 
Dalitz plot with cuts on  $\xi$ and $E_{f}$ (I)
}
  \end{center}
\end{figure}

The Dalitz plot given in \fig{dalitz1} may help to understand the 
relation between a simple $s'$ cut (just region I) and 
the cuts applied here (regions I with II and III).

\begin{figure}[htb]
\begin{minipage}[bht]{9.6cm}
{\begin{center}
  \vspace{0cm}
  \hspace{-3cm}  
  \mbox{
    \epsfxsize=9.5cm
    \epsffile{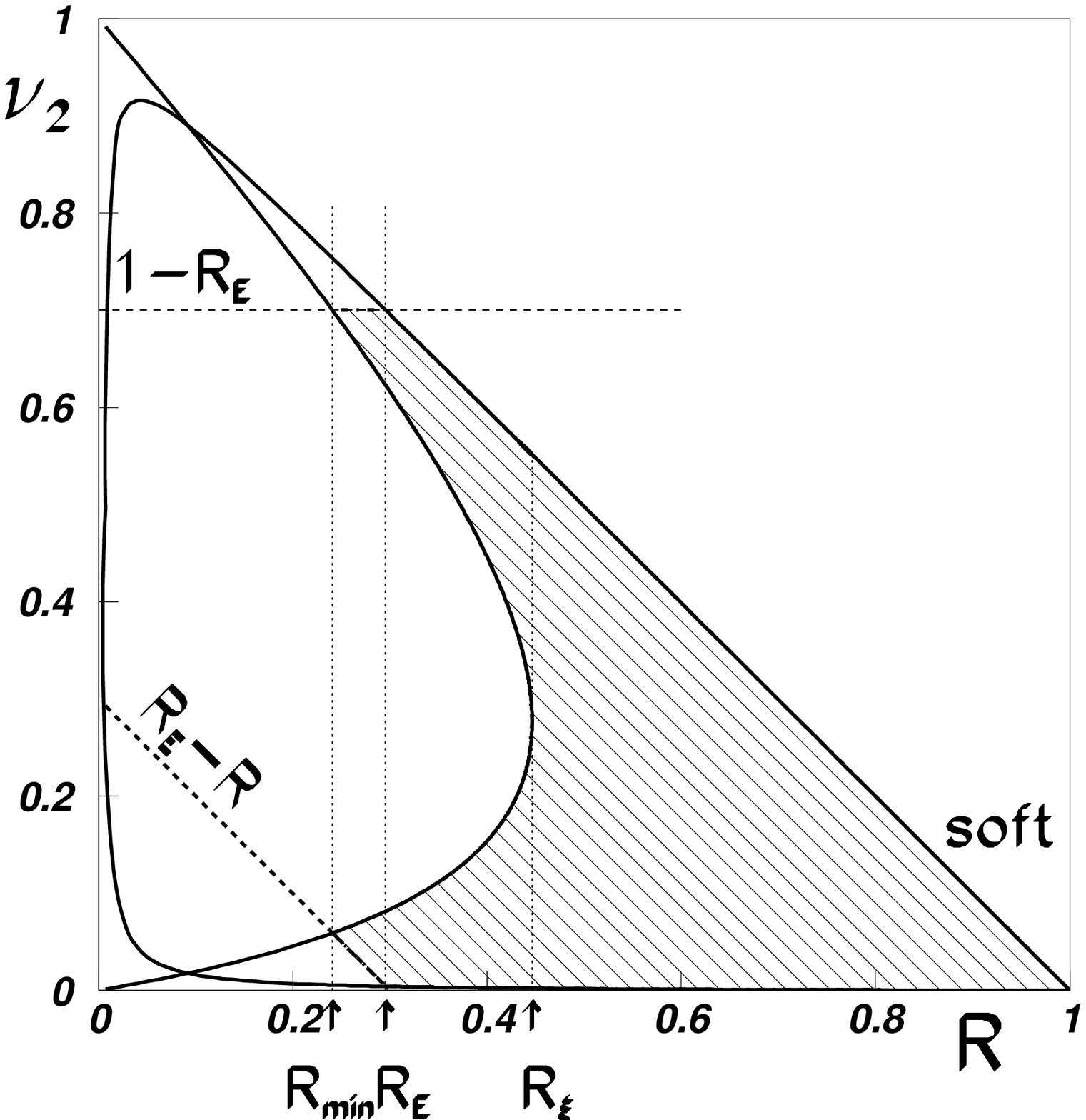}
  } 
 \end{center}}
\end{minipage} 
\begin{minipage}[bht]{9.6cm}
{\begin{center}
  \vspace{0cm}
  \hspace{-6cm}  
  \mbox{
    \epsfxsize=9.5cm
    \epsffile{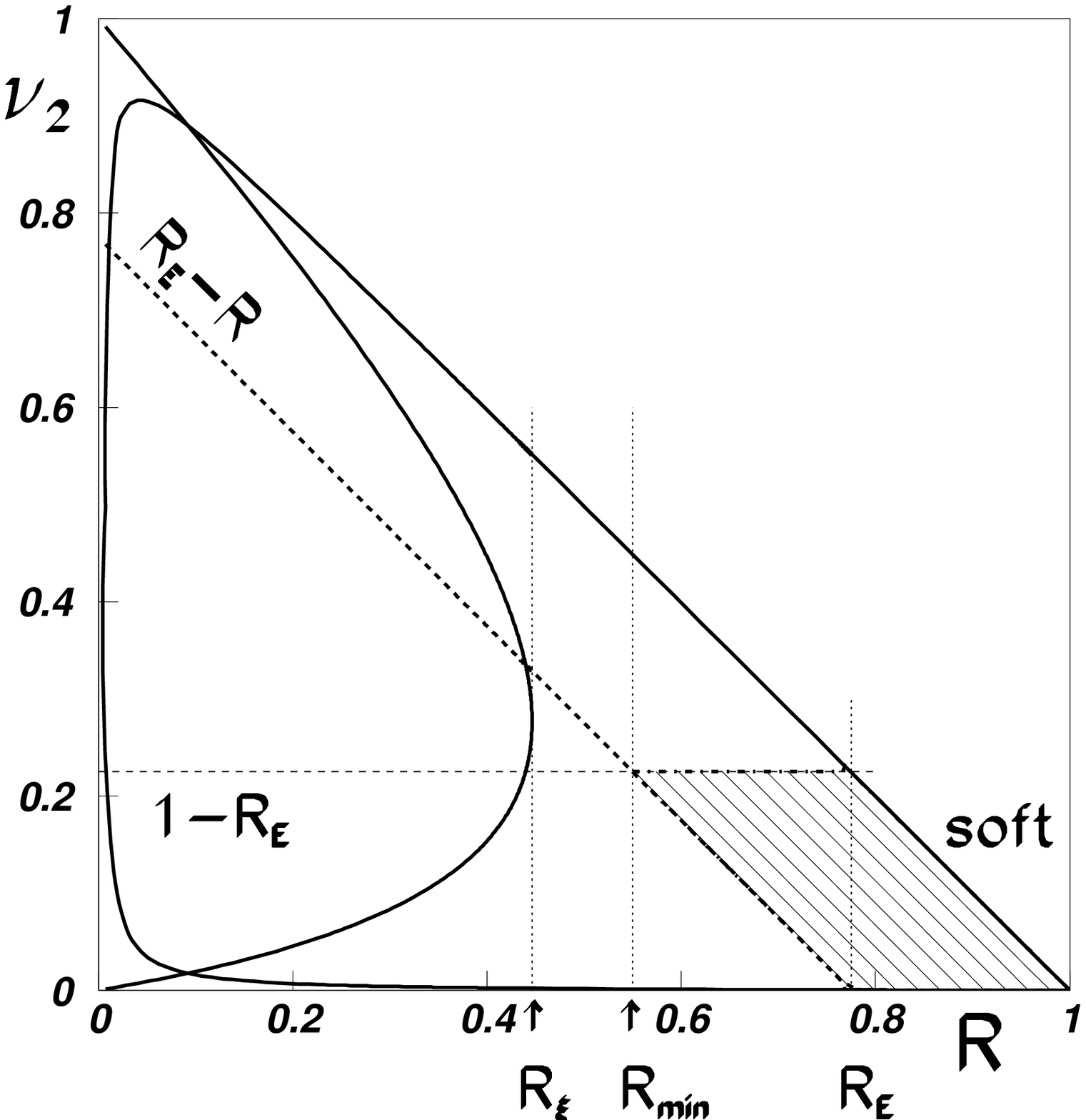} 
  }
 \end{center}}
\end{minipage}
\caption[Dalitz plots with cuts on  $\xi$ and $E_{f}$ (II)]
{\it Dalitz plots with cuts on $\xi$ and $E_{f}$ (II);
the two cases are discussed in the text
\label{dalitz2}}
\end{figure}

As \fig{dalitz1} shows, we have to determine the cross-sections 
in three phase-space regions with different boundary values of $v_2$ at
given $R$:  
\ba
\frac{d\sigma}{d\cos\vartheta} = \left[ \int_{\mathrm{I}} +
\int_{\mathrm{II}}   
- \int_{\mathrm{III}} \right]~ dR ~ d{v_2} 
\frac{d\sigma}{dR d{v_2} d\cos\vartheta} .
\label{sig}
\ea
Region I corresponds to the simple $s'$ cut.
The integration over $R$ extends from $R_{min}$ to 1:
\ba
\label{rmin}
R_{min} &=& R_E \left(1 - \frac{\sin^2(\xi^{max}/2)}
{1-R_E\cos^2(\xi^{max}/2)} \right) . 
\ea
The soft-photon corner of the phase space is at $R=1$.
Thus, the additional contributions related to the acollinearity cut are
exclusively due to hard photons.
The boundaries for the integration over $v_2$ can now be summarized 
with one general parameter $A=A(R)$ to:
\ba
\label{ibound}
{v_2}^{max,min}(R) &=& \frac{1}{2} (1-R)~ \left[1 \pm A(R)\right],
\ea
where in every region $A=A(R)$ depends on only one of
the applied cuts:
\ba
\label{arI}
A_{\mathrm{I}}(R) &=& \sqrt{1-\frac{R_{m}}{R}} \approx 1,
\label{A1}
\\
\label{arII}
A_{\mathrm{II}}(R) &=& \frac{1+R-2R_E}{1-R}, 
\label{A2}
\\
\label{arIII}
A_{\mathrm{III}}(R) &=& \sqrt{1 - \frac{R(1-R_{\xi})^2}{R_{\xi}(1-R)^2}},  
\label{A3}
\ea
with
\ba
  \label{eq:rs}
R_{m}  &=& \frac{4m_f^2}{s},
\\
  \label{eq:re}
R_E &=& \frac{2E_{min}}{\sqrt{s}}, 
\\
  \label{eq:rx}
R_{\xi} &=&
\frac{1-\sin(\xi^{max}/2)}{1+\sin(\xi^{max}/2)}.     
\ea
The turning point $P_t$ of the acollinearity bound is given by
\bq
P_t \equiv
[R_t;v_{2,t}] =
\left[ R_\xi; \frac{1}{2}(1-R_\xi) \right].
\label{pt}
\eq
The upper and lower fermion energy cuts of \eqn{rminmax} are straight
lines meeting at a value $\tilde{R}_E$ of $R$:
\bq
\label{intsec}
\tilde{R}_E = R^f_E+R^{\bar{f}}_E-1. 
\eq
Two qualitatively different cases may arise.
In the first case, $\tilde{R}_E < R_\xi$, the acollinearity cut affects
the integration region and the absolute minimum of $R$ is given by
\bq
R_{min} = R_{E}\left(1-\frac{\sin^2 (\xi^{max}/2)}
{1-R_E\cos^2 (\xi^{max}/2)}\right).
\label{Rmin}
\eq
The complete integration region to be used in \eqn{symb1} and
\eqn{symb} can then be split into three different parts where
the kinematical bound by $m_f$, the linear bounds by the 
energy cuts, and the curved acollinearity bound are considered
separately:
\ba
\Gamma &=& \Gamma_{I}\quad +\quad\Gamma_{II}\quad -\quad\Gamma_{III}\nl
&=&\int_{\bar{R}_E}^{1} d{R} 
\int_{v_{2_m}^{min}(R)}^{v_{2_m}^{max}(R)} d{v_2}
+\int_{R_{min}}^{\bar{R}_E} d{R} \int_{R_E-R}^{1-R_E} d{v_2}
-\int_{R_{min}}^{R_\xi} d{R} \int_{v_{2_\xi}^{min}(R)}
^{v_{2_\xi}^{max}(R)} d{v_2}.
\label{regions1}
\ea
In the other case, $\tilde{R}_E\geq R_\xi$, the energy cuts are so stringent
that the acollinearity cut has no effect. The minimum value of $R$ is
$R_{min}=\tilde{R}_E$ and the integration region is simplified to a trapezoid: 
\bq
\Gamma\quad =\quad\Gamma_{I}\,+\,\Gamma_{II}
\quad =\quad\int_{\bar{R}_E}^{1} d{R}
\int_{v_{2_m}^{min}(R)}^{v_{2_m}^{max}(R)} d{v_2} 
+\int_{R_{min}}^{\bar{R}_E} d{R} \int_{R_E-R}^{1-R_E} d{v_2}.
\label{regions2}
\eq
The cut value $\bar{R}_E$ in ~\eqn{regions1} and \eqn{regions2}, 
\ba
\bar{R}_E &=&  \frac{2\frac{m_f^2}{s}+(1-R_E)
\left(R_E+\sqrt{R_E^2-4\frac{m_f^2}{s}}\right)}
{2(1-R_E+\frac{m_f^2}{s})},
\label{reblim}
\ea
can safely be set to $R_E$ neglecting $m_f$ here, 
as is done in \zf\ (see also~\eqn{limemin}).
The two corresponding Dalitz plots are shown in
\fig{dalitz2}.

%
%
%
%

\bigskip

In subroutine {\tt ZCUT} the distinction 
of the two situations~\eqn{regions1} and
\eqn{regions2} is organized in a straightforward manner where the hard 
acollinearity cut corrections are calculated by subtractions 
from the $s'$ cut result, in subroutine {\tt SCUT} including all
angular cuts (by functions {\tt SHARD} and {\tt AHARD}), 
and in {\tt SFAST} with acollinearity, but without acceptance cut
(by functions  {\tt FCROS} and {\tt FASYM}).
The corresponding hard corrections to the angular distributions 
are treated in an analogous way in {\tt ZANCUT} 
(see subroutines {\tt COSCUT} and {\tt HARD}).

\bigskip

The above relations are independent of the scattering angle and thus,
as mentioned, compatible with the angular acceptance cut defined in 
\eqn{cuts1}. Though, we would like to point out that the acollinearity cut
has an indirect influence on the acceptance cut.
It is easy to see that the maximal scattering
angle of the second fermion (which is unrestricted by the user's
acceptance cut) becomes limited by an acollinearity cut, i.e.
the scattering angle of the second fermion is limited to
$[-(\xi^{max}+\vartheta^{max}),(\xi^{max}+\vartheta^{max})]$.

The user may apply
a sometimes reasonable approximation of the acollinearity
cut in terms of the  simpler $\Delta$ cut ($\equiv$ $s'$ cut);
this can be achieved by using  $\Delta_\xi$ for the definition of the
integration limit~\eqn{spmin} in \sect{somana}:
\bq
\Delta_{\xi}  \equiv 1 - R_\xi =
\frac {2 \sin(\xi^{max}/2) }
                          { 1 + \sin (\xi^{max}/2)} \approx \xi^{max}.
\label{delxi}
\eq
The quality of such an approximation depends critically on the values
of the
$E_{min}$ cut and the $\xi^{max}$ cut; for loose cuts it improves.
Because of the approximations that have been implemented in the code
for this calculational chain, the user must be cautious when applying severe
cuts. Since the approximation is ultra-relativistic one should restrict oneself
to the following region of the phase space:
\ba
E^{min} &\gg& m_f, 
\label{limemin}
\\ \nl
 \xi^{max} &\ll& \left( 1 - \frac{8m_f}{ \sqrt{s}} \right) \pi.
\label{limxi}
\ea
On the other hand, the validity of inclusive soft-photon exponentiation
(see \sect{sprfinal} and \sect{commexp})
comes into question if the kinematical region shrinks to the soft
photon corner either due to too stringent energy cuts or if the 
turning point $P_t$ introduced in \eqn{pt} is moved too far to the right:
\ba
E^{min} &<& 0.95 \, E_{\rm{beam}},
\\
\xi^{max}  &>& 2^{\circ}.
\label{lime}
\ea

In this Section, we discussed the {\em final state kinematics} and the
resulting physical phase space boundaries.
In \sect{radiacol}, we will discuss yet the influence of the {\em initial
state kinematics} on the singularity structure of the squared radiative
matrix elements and the resulting peculiarities of the integrated
radiator functions; see discussion around \eqn{iniintprop}.
\subsection{Cross-sections
\label{crosacol}}
The cross-sections with acollinearity cut calculated finally in \zf\
may be written generically in close analogy to those given in  
\sect{somana}, e.g. \eqn{siginic1} and \eqn{saini}:
\ba
\frac{d\sigma^{a}}{d\cos\vartheta} &=&  
\sum_{B=T,FB}  
\left[
 \int_{R_E}^{1}  
+\int_{R_{min}}^{R_E} 
-\theta_R
\int_{R_{min}}^{R_{\xi}} 
\right]dR 
~  \sigma_{\sss{B}}^0(s') R_{\sss{B},reg}^{a}(R,\cos\vartheta),
\label{siginifica}
\\  \nl
\sigma_{\sss{B}}^{a}(c) &=&   
\left[
 \int_{R_E}^{1}  
+\int_{R_{min}}^{R_E} 
-\theta_R
\int_{R_{min}}^{R_{\xi}} 
\right]dR 
        ~ \sigma_{\sss{B}}^0(s')
         R_{\sss{B},reg}^{a}(R,c)
,  \hspace{0.5cm} B=T,FB,
\nl
\label{siginifina}
\ea
with
\ba
\label{thetar}
\theta_R &=&
\theta\left(R_{\xi}-(2R_E-1)\right).
\ea
The subindex {\it reg} indicates that the expressions for the radiators
$R$ depend on the region, {\it reg = I,II,III}.
In region $I$, the usual decomposition of the radiators into hard and
soft+virtual components holds as was described in much detail in \sect{somana}.
Further, soft photon exponentiation may be performed there.
If it is performed, then not necessarily from $R_E$ to the upper
integration bound in $R$, but starting from
$\max(R_E,R_{\xi})$.
In the other two regions, there are only hard radiatior parts to be
considered.  

For later reference, we give here also the generic formula for common
exponentiation of initial and final state photons:
\ba
\frac{d\sigma^{ini + fin}}{d\cos\vartheta} &=&  
\sum_{B=T,FB}  
\left[
 \int_{R_E}^{1}  
+\int_{R_{min}}^{R_E} 
-\theta_R
\int_{R_{min}}^{R_{\xi}} 
\right]dR 
  \sigma_{\sss{B}}^0(s')
 R_{\sss{B},reg}^{ini}(R,\cos\vartheta)  {\bar R}_{\sss{B}}^{fin}(R),
\label{siginificb}
\\  \nl
\sigma_{\sss{B}}^{ini + fin}(c) &=&   
\left[
 \int_{R_E}^{1}  
+\int_{R_{min}}^{R_E} 
-\theta_R
\int_{R_{min}}^{R_{\xi}} 
\right]dR 
         \sigma_{\sss{B}}^0(s')
         R_{\sss{B},reg}^{ini}(R,c)
        {\bar R}_{\sss{B}}^{fin}(R)
,  \hspace{.5cm} B=T,FB.
\nonumber  \nl
\label{siginifinb}
\ea
We will not give here the various hard radiator expressions to be used 
in \zf.
Instead, after some introduction to their treatment we will quote as
an instrucive 
example the rather compact formulae for the integrated cross-sections 
without acceptance cut ($c=1$).
Further, the explicit expressions for ${\bar R}_{\sss{B}}^{fin}(R)$ will 
be given in \sect{commexa}. 
With these two examples at hand, the expert user will be able to find
any further information needed from the Fortran code itself.
\subsection{Hard radiator parts. Package {\tt acol.f}
\label{radiacol}}
In presence of an acollinearity cut, the hard radiator parts have to
be calculated differently in three regions of the Dalitz shown in
\fig{dalitz1}.
For regions II and III, they are calculated with the package {\tt acol.f}. 
\subsubsection{Flags and subroutines
}
\label{flagsacol}
In \zf\ the calculation of cross-sections and asymmetries is 
done through subroutine {\tt ZCUT} which either calls subroutine
{\tt SCUT} (formulae with all cuts) or {\tt SFAST} 
(formulae for special cases). The corresponding
angular distributions can be calculated by subroutine 
{\tt ZANCUT} which calls {\tt COSCUT} for the hard flux functions. 
Flag {\tt ICUT} from subroutine {\tt ZUCUTS} decides in both cases
which cuts are being applied:
\begin{itemize}
 \item
{\tt ICUT}=2 :\,
Cuts on acollinearity, minimal energies, but no acceptance cut.
Subroutine {\tt SFAST} calls
{\tt FCROS} and {\tt FUNFIN} for $\sigma_{\sss{T}}$ and  
{\tt FASYM} and {\tt FUNFIN} for $\sigma_{\sss{FB}}$  
\item
{\tt ICUT}=3 :\,
Cuts on acollinearity, minimal energies, and acceptance.
Subroutine {\tt SCUT} calls 
{\tt SHARD} and {\tt FUNFIN} for $\sigma_{\sss{T}}$ and  
{\tt AHARD} and {\tt FUNFIN} for $\sigma_{\sss{FB}}$ 
\end{itemize}
The flag value {\tt ICUT}=2 also sets the flag 
{\tt IFAST}=1 in {\tt ZUCUTS}, so subroutine
{\tt SFAST} is called in {\tt ZCUT}.
We call {\tt ICUT}=2 in the calculational chain related to subroutine
{\tt SFAST} since it is a generalization of the numerically
fastest case with no cuts but on $s'$.
In all other cases, the calculations are done by subroutine
{\tt SCUT} (with {\tt IFAST}=0)\footnote{In order to maintain
compatibility with earlier releases, the branch 
{\tt ICUT}=0 (corresponding to the old coding for general cuts)
also calculates with the old coding of the final state corrections
(flag {\tt IFUNFIN}=0), while all other branches use the newly 
corrected final state contributions
(flag {\tt IFUNFIN}=1).
}.

The hard flux functions $H_{\sss{B}}^a$ for the acollinearity cut are
called from essentially five different functions in \zf: 
{\tt SHARD}, {\tt AHARD}, {\tt FCROS}, {\tt FASYM}, and {\tt FUNFIN}. 
%
Functions {\tt SHARD} and {\tt AHARD} call the hard corrections for
the integrated cross-sections 
{from} subroutines {\tt SHFULL, AHFULL} (integration region I) and  
{\tt SHCUTACOL, AHCUTACOL} (integration regions II and III) from the
linked package {\tt acol.f}.
The complete hard corrections with all cuts are then further 
defined by subroutines {\tt RADTIN, RTFUN0, RTFUN1}
and {\tt RADFBIN, RFBFUN0, RFBFUN1}.
Similarly, functions {\tt FCROS} and {\tt FASYM}
call {\tt SHACOL, AHACOL} for the shorter expressions 
without acceptance cut ($c=1$).

The angular distributions $d{\sigma_{\sss{T,FB}}}/d{\cos\vartheta}$
are calculated via subroutine {\tt ZANCUT},
and the acollinearity cut is chosen with flag {\tt ICUT}=2 (or
equivalently, {\tt ICUT}=3).
The hard flux functions $H_{\sss{B}}^a(R,A(R),\cos\vartheta)$
are derived by function {\tt HARD}.
In {\tt HARD}, subroutine {\tt HCUT} is called which itself
uses subroutines {\tt HINIM} for the initial state contributions 
$H_{\sss{B}}^{ini}$
and {\tt HINTFM} for the initial-final state interference contributions
$H_{\sss{B}}^{int}$. Final state corrections are again provided by 
subroutine {\tt FUNFIN}. One should mention that {\tt FUNFIN} 
is always called in subroutine {\tt BORN} so final state
corrections are always considered together with the (effective) Born 
observables. 
\subsubsection{Phase space splitting with acollinearity cut}
\label{phspacol}
At this point one has to note that the treatment of mass
singularities in the hard initial state and initial-final
state interference contributions (when setting the initial state
mass $m_e$ to zero) necessitates a splitting of the remaining 
phase space of $\cos\vartheta$ and $s'$ into
different regions of phase space with different analytical expressions.
This is due to the fact that 
squared matrix elements contain, from initial state radiation
and the initial-final state interference, 
the electron (positron) propagator after radiation of a photon, and
these terms are 
proportional to 
\ba
\label{iniintprop}
\frac{1}{Z_{1(2)}} &=& - \frac{1}{(k_{1(2)}-p)^2-m_e^2} 
    = \frac{1}{2k_{1(2)}p}
\ea
up to the second power of this expression.
In the course of the analytical integration over the first
two  angles of phase space ($\varphi_\gamma$ and $\theta_{\gamma}$)
they generate logarithmic expressions with vanishing arguments 
when neglecting the electron mass $m_e$.
This happens for certain values of $\cos\vartheta$ depending 
on the remaining kinematical variable $R$ and the applied cuts.
One has to distinguish four different regions
of phase space for $\cos\vartheta$ and $R$, each region with 
different analytical expressions for the radiators 
$H^{ini,int}_{\sss{B}}(R,A(R),\cos\vartheta)$ 
of the angular distribution.
In turn,  four (or respectively 
six) different expressions for the integrated results
$H^{ini,int}_{\sss{T}}(R,A(R),c)$ ($H^{ini,int}_{\sss{FB}}$) have to be used 
in \zf.
For a detailed analysis on this issue see~\cite{Christova:1999gh}.
The phase space splitting for the initial state and interference
terms is organized by subroutine {\tt PHASEREGC}.
{\tt PHASEREGC} decides which phase space region and which
correponding formulae have to be chosen depending on $c$ 
(or $\cos\vartheta$), $R$, $\xi^{max}$, 
and $E_{min}$.

Without acceptance cut ($c=1$), the number 
of different cases and analytical expressions for 
$H^{ini,int}_{\sss{T}}(R,A(R),C)$ in phase space boil down to one, 
and for $H^{ini,int}_{\sss{FB}}(R,A(R),C)$
down to two ($C=\cos\vartheta,c$) \cite{Christova:1999cct}. 
These results are very compact and attached
to the {\tt SFAST} branch of {\tt ZFITTER}.
\subsubsection{Hard radiators without acceptance cut
  ($c=1$)} 
\label{hardshortacol}
The hard corrections $H_{\sss{B}}^a$ for $c=1$ 
to the integrated flux functions \cite{Christova:1999cct} are written below. 
They are defined -- as stated above -- in subroutines 
{\tt SHACOL} and {\tt AHACOL}.
For brevity we set everywhere $A\equiv A(R)$.
The hard contributions to $\sigma_{\sss{T}}(R)$ are:
\ba
H^{ini}_{\sss{T}} (R;A)\quad &=& 
\frac{3\alpha}{4\pi} Q_e^2\cdot\left[ \left(A+\frac{A^3}{3}\right)
\frac{1+R^2}{1-R} 
\left( \ln\frac{s}{m_e^2}-1\right) + (A-A^3) 
\frac{2 R}{1-R}\right],
\label{hardtA1}
\nonumber\\
\\
H^{int}_{\sss{T}}(R;A)\quad &=&  
-\frac{\alpha}{\pi} Q_eQ_f\cdot
\frac{4 A R (1+R)}{1-R},
\label{hardtA2}
\\
H^{fin}_{\sss{T}}(R;s,A)&=& 
\frac{\alpha}{\pi} Q_f^2\cdot
\Biggl[
\frac{1+R^2}{1-R} \ln\frac{1+A}{1-A}
-\frac{8 A\cdot m_f^2/s}{(1-A^2)(1-R)}-A(1-R)
\Biggr].
\label{hardtA3}
\ea
For $\sigma_{\sss{FB}}^{ini,int}(R)$, 
two cases have to be
distinguished depending on the parameter $A_0$:
\ba
\label{funyz3}
A_0(R) &=& \frac{1-R}{1+R}.
\ea
For $A>A_0$ (as is the case in the entire region I with $A=1$):
\ba
H^{ini}_{\sss{FB}}(R;A\geq A_0)&=&
\frac{\alpha}{\pi}Q_e^2\cdot\frac{1+R^2}{1-R}\left[
\frac{4 R}{(1+R)^2} \left( \ln\frac{s(1+R)^2}{4m_e^2R} - 1 \right)
\right.
\nonumber\\
&&\left.
-~\frac{1}{(1+R)^2}
\left[
y_+ y_- \ln\left|y_+ y_-\right| 
+{4 R}\ln(4 R)
\right]\right.
\nonumber\\
&&\left.
-~(1-A^2) \left( \ln\frac{s}{4m_e^2(1+A)^2R} - 1\right)
\right]
\nonumber\\
&&
+~\frac{4A(1-A)R}{1-R},
\label{hardfbA1}
\\
H^{int}_{\sss{FB}}(R,A\geq A_0)&=&
\frac{\alpha}{\pi} Q_eQ_f\cdot 
\Biggl\{
\frac{3R}{2}\left[\ln\frac{z_+}{z_-}
+\frac{2-R+\frac{5}{3}R^2}{1-R}\ln R 
\right]
\nonumber\\
&&-~\frac{1+R}{2(1-R)}
(5-2 R+5 R^2)
\ln\frac{(1+R)(1+A)}{2} 
\nonumber\\
&&+~\frac{1}{4(1-R)}\Biggl[
\frac{(1-4R+R^2)[A(1+R)^2-(1-R)^2]}{1+R}
\nl &&+~ 2 A (1-A) (1+R^3)\Biggr]
\Biggr\} .
\label{hardfbA2}
\ea
While, for $A<A_0$:
\ba
H^{ini}_{\sss{FB}}(R;A<A_0)&=&
\frac{\alpha}{\pi}Q_e^2\cdot\frac{1+R^2}{1-R}
\left[
-\frac{y_+ y_-}{(1+R)^2}
\ln\left|\frac{y_+}{y_-}\right|
+(1-A^2)\ln\frac{1+A}{1-A}
\right]
\nonumber\\
&&
+~\frac{8 A R^2}{(1+R)(1-R)},
\label{hardfbA3}
\\
H^{int}_{\sss{FB}}(R,A<A_0)&=&
\frac{3\alpha}{2\pi} Q_eQ_f\cdot ~ R
\Biggl\{
\ln\frac{z_+}{z_-}
-\frac{2-R+\frac{5}{3}R^2}{1-R}
+ A (1-R)  
 \Biggr\}.
\label{hardfbA4}
\ea
The hard contribution to $\sigma_{\sss{FB}}^{fin}$ is in both cases:
\ba
H^{fin}_{\sss{FB}}(R;s,A)\quad
&=& 
H^{fin}_{\sss{T}}(R;s,A)
+
\frac{\alpha}{\pi} Q_f^2\cdot
\left[
A(1-R)
-(1+R)\ln\frac{z_+}{z_-}
\right].
\label{hardfbA5}
\ea
The abbreviations used are
\ba
\label{funyz1}
y_{\pm} &=& (1-R) \pm A(1+R),
\\
\label{funyz2}
z_{\pm} &=&(1+R) \pm A(1-R).
\ea
We just mention that the final state mass $m_f$ 
must be included for the final state 
contributions $H^{fin}_{\sss{B}}$
in region~I, where $A=\sqrt{1-4 m_f^2/s'}$. 
For this case,
the formulae from~\eqn{hardtA1} to~\eqn{funyz3} deliver the already 
well known results {from}~\cite{Bardin:1991det,Bardin:1989cwt,Bardin:1991fut} 
(with acceptance cut) or~\cite{Bonneau:1971mkx} 
(no cuts) and are therefore calculated by subroutines {\tt SHFULL} and 
{\tt AHFULL} for region I in {\tt SHARD} or are directly inserted 
in {\tt FCROS} and {\tt FASYM} for region I in {\tt SFAST}.

Finally, soft and virtual corrections $S^{ini,int}_{\sss{T,FB}}$ are  
taken care of as was the case for the $s'$ cut (see~\sect{somana}).
Subtractions from the hard
corrections $H^{ini,int}_{\sss{T,FB}}$,  
so that the overall radiators $R^{ini,int}_{\sss{B}}$
in {\tt SHARD}, {\tt AHARD}, {\tt FCROS}, and {\tt FASYM}
are finite at $v \to 0$, affect only region~I. 

The complete hard radiator functions 
${\bar H}_{\sss{T,FB}}^{ini,fin}(R,A(R),\cos\vartheta)$ for the
angular distributions to order \oalf\ with all cuts, as well as their
integrals over $\cos\vartheta$, 
will be given elsewhere~\cite{Christova:2000x1}.
%
\subsection{Common soft photon exponentiation of initial and final
state radiation\label{commexa}}
The final state correction factor is needed both for the calculation
of the soft+virtual initial state corrections (integration of functions 
{\tt SSOFT, ASOFT} which delivers the terms {\tt SSFT, ASFT} in {\tt SCUT})
and the hard one (attached to {\tt SHINI, AHINI} and calculated in 
{\tt SHARD, AHARD} in {\tt SCUT}).
This treatment of final state effects is done in a completely 
analogous way in subroutine {\tt COSCUT} (angular distributions) 
and similarly in subroutine {\tt SFAST} (special cuts). 
Thus, the final state factors are calculated by several calls to
subroutine {\tt FUNFIN}, which itself is called from subroutine {\tt
BORN}:
\ba
\label{defgt}
 {\tt SFIN} &=&{\bar R}_{\sss{T}}^{fin}(R),
\\
{\tt AFIN} &=&{\bar R}_{\sss{FB}}^{fin}(R).
\label{defgfb}
\ea
In case of common initial
and final state soft photon exponentiation, 
final state radiation is taken into account in \eqn{siginificb} and
\eqn{siginifinb} as follows \cite{Bardin:1991fut}:  
\ba
\label{barrfin}
{\bar R}_{\sss{B}}^{fin}(R)
&=& 
\int_{R_{min}/R}^1 du
\left[[1+\bar{S}^{fin}(\beta_{f}')]\beta_{f}'(1-u)^{\beta_{f}'-1}
\theta(u-R_{max}) + {\bar H}_{\sss{B}}^{fin}(u) \right]
\nl
&=&
[1+\bar{S}^{fin}(\beta_{f}')]
(1-R_{max})^{\beta_{f}'} +  {\bar G}_{\sss{B}}(R) ,
\ea
with
\ba
R_{max}&=& max\left(\frac{R_{min}}{R},R_E\right),
\\
 {\bar G}_{\sss{B}}(R) &=& \int_{R_{min}/R}^1 du ~ {\bar H}_{\sss{B}}^{fin}(u).
\ea 
The $\bar{S}^{fin}(\beta_{f}')$
 and $\beta_{f}'$ are defined in \eqn{barsfin} and \eqn{barbetfin}.
For region I, the hard radiator part ${\bar G}_{\sss{B}}(v)$ was introduced
in \eqn{barrfinI} and an analytical expression was given there.
For the other two regions, one has to perform the integration of
$H_{\sss{B}}^{fin}$ over $u$ 
taking into account the dependence of $A(u)$ on the integration
variable $u$.
So the hard contributions ${\bar G}_{\sss{B}}(R)$ are explicitly integrated 
in the following way:
\ba
\label{finalhard}
{\bar G}_{\sss{B}}(R)
&=& 
\Biggl\{
\int_{R_{min}/R}^1 du
\theta\left(u-R_E\right)
+~
\theta\left(R_E-\frac{R_{min}}{R}\right)
\int_{R_{min}/R}^{R_E} du
\nl
&&-~
\theta\left(R_{\xi}-\frac{R_{min}}{R}\right)
\int_{R_{min}/R}^{R_{\xi}} du
\Biggr\} {\bar H}_{\sss{B}}^{fin}(u)
\nl
&=& 
\Biggl\{
\int_{R_{min}/R}^1 du
\theta\left(u-R_E\right)
+~
\theta\left(R-\frac{R_{min}}{R_E}\right)
\int_{R_{min}/R}^{R_E} du
\nl
&&-~\theta\left(R-\frac{R_{min}}{R_{\xi}}\right)
\int_{R_{min}/R}^{R_{\xi}} du
\Biggr\} {\bar H}_{\sss{B}}^{fin}(u)
.
\nonumber\\
\ea
Further, for regions II and III -- with the soft-photon corner being
in region I -- \zf\ adds or subtracts the integrals over the
unregularized hard flux functions:
\ba
 {\bar H}_{\sss{B}}^{fin} (u) = H_{\sss{B}}^{fin}(u),
\ea
while in region I the regularized hard radiator is integrated for
$R>R_{max}$ and added. 

These results are included in the terms {\tt SHRD} and {\tt AHRD1}
which arise from integation of functions {\tt SHARD}, {\tt AHARD} 
over $R$ in subroutine {\tt SCUT}, as indicated in Eq.~(\ref{finalhard}).
These final state corections {\tt SFIN}, {\tt AFIN} are contained in 
subroutine {\tt BORN} (which is called in {\tt SHARD,AHARD}) 
as factors to the effective Born cross-sections.
In a similar fashion the acollinearity options of subroutines {\tt
  SFAST} and {\tt COSCUT} are organized with terms {\tt SHRD}, {\tt AHRD1} 
substituted by term {\tt HRD}, or by terms {\tt CROS,ASYM} respectively, 
and functions {\tt SHARD}, {\tt AHARD} replaced by function {\tt HARD}, 
or by functions {\tt FCROS}, {\tt FASYM} respectively. 
This dependence of $A(u)$ on the integration variable $u$ presents 
an additional complication compared to region~I with $A=1$.
Nevertheless, the integrations have been done
analytically with exclusion of two logarithms.
The results of the hard integration over $u$ for region II are found
in subroutine {\tt FUNFIN} ($R_{II}\equiv R_E$):
\ba
{\bar G}_{\sss{T,II}}(R)
&=& 
\frac{\alpha Q_f^2}{\pi}
\int_{R_{min}/R}^{R_{II}} du 
\left[
\frac{1+u^2}{1-u} \ln \frac{1+A_{II}(u)}{1-A_{II}(u)} - A_{II}(u)(1-u)
\right]
\nl
&=&
\frac{\alpha Q_f^2}{\pi}
\left[ {\tt ALE1I} + D_1 \left( \frac{1}{2}D_1 -D_2 \right)\right],
\\
{\bar G}_{\sss{FB,II}}(R)
&=& \frac{\alpha Q_f^2}{\pi}
\int_{R_{min}/R}^{R_{II}} du
 \Biggl[
\frac{1+u^2}{1-u} \ln \frac{1+A_{II}(u)}{1-A_{II}(u)} 
\nl &&
+~ (1+u) \ln \frac{(1+u)+A_{II}(u)(1-u)}{(1+u)-A_{II}(u)(1-u)}  \Biggr]
\nl
&=&
\frac{\alpha Q_f^2}{\pi}
\left[ {\tt ALE1I} + {\tt ALE2I} \right],
\ea
with
\ba
D_1 &=& {\tt ALIM} - {\tt RECUT2} = \frac{R_{min}}{R} - R_{II},
\\
D_2 &=& {\tt RECUT2} - 1 = R_{II}-1,
\ea
and $A_{II}(u)$ defined in \eqn{arII}.
The electromagnetic coupling is defined in \sect{ALQF2Z}.
Functions  {\tt ALE1I} and {\tt ALE2I} are the integrals over the
logarithms. 
In \zf, they are obtained from an interpolation 
of tabulated values of these integrals (over region II) of:
\ba
\label{fal1}
{\tt FAL1(u)} &=& \frac{1+u^2}{1-u}\ln \frac{1+A(u)}{1-A(u)},
\\
\label{fal2}
{\tt FAL2(u)} &=& -(1+u) \ln \frac{N_+(u)}{N_-(u)},
\\
N_{\pm}(u)    &=&  (1+u) \pm A(u) (1-u).
\ea
In fact, the two logarithms are integrated numerically.
This is done using {\tt SIMPS}, for pre-defined small, and rising
intervals of $u$ inside the integration region, thus forming a table
with 20 entries ({\tt NP}=20), the fields being called {\tt ALEi}
and {\tt ALAi} ($i=1,2$) for regions II and III, repectively. 
This table then is used for the truly needed integrations 
{\tt ALAiI},  {\tt ALEiI}, $i=1,2$,
by
interpolation with subroutine {\tt INTERP}, called in {\tt FUNFIN} for
the calculation of factors {\tt SFIN} and {\tt AFIN}.
The factors are calculated in three steps, in the following order: II
+ I -- III.

For equidistant points, fulfilling $x_1 < x_2 < x_3$ and distance
$\Delta$, the interpolation is done with the Lagrange interpolation formula:
\ba
f(x) &=& \frac{1}{2\Delta^2}
\left[
(x-x_1)(x-x_2)y_3 - 2~ (x-x_1)(x-x_3)y_2 + (x-x_2)(x-x_3)y_1\right].
\nl
\ea
This formula is realized in {\tt INTERP} and
makes a quadratic 
interpolation of the integral over $u$ of the two logarithms mentioned
using tables.
The tables are prepared by {\tt SETFIN} using {\tt SIMPS} for selected
values of $R_{min}/R$, covering the full regions II and III.

Analogously, for region III (to be subtracted) ($R_{III}\equiv R_{\xi}$):
\ba
{\tt SFIN}^{III} &=& 
\frac{\alpha Q_f^2}{\pi}
\int_{R_{min}/R}^{R_{III}} du 
\left[
\frac{1+u^2}{1-u} \ln \frac{1+A_{III}(u)}{1-A_{III}(u)} - A_{III}(u)(1-u)
\right]
\\
&=&
\frac{\alpha Q_f^2}{\pi}
\left[ {\tt ALA1I} 
+ \frac{1}{2}
\left( (R_{min}/R +B ){\tt SQR}-(1-B^2)\ln
\left|\frac{{\tt RACUT}+B}{{\tt SQR}+ R_{min}/R +B}\right| \right)
\right],
\nl
{\tt AFIN}^{III} &=& \frac{\alpha Q_f^2}{\pi}
\int_{R_{min}/R}^{R_{III}} du
\Biggl[
\frac{1+u^2}{1-u} \ln \frac{1+A_{III}(u)}{1-A_{III}(u)} 
\nl &&
+~ (1+u) \ln \frac{(1+u)+A_{III}(u)(1-u)}{(1+u)-A_{III}(u)(1-u)}  \Biggr]
\nl
&=&
\frac{\alpha Q_f^2}{\pi}
\left[ {\tt ALA1I} + {\tt ALA2I} \right],
\ea
with $A_{III}(u)$ defined in \eqn{arIII} and  
\ba
{\tt SQR}
&=& \sqrt{
\max\left[
0,(R_{min}/R)^2+2BR_{min}/R+1\right]},
\\
B &=& - \frac{1+\sin^2 (\xi^{max}/2)}{\cos^2 (\xi^{max}/2)
},
\\
{\tt RACUT}&=& R_{III}.
\ea
Functions  {\tt ALE1I} and {\tt ALE2I} are again the integrals over the
logarithms, 
obtained from an interpolation 
of tabulated values of these integrals (over region III now) of \eqn{fal1} 
and \eqn{fal2}.

\section{Higher Order QED Corrections\label{secondo}}
\eqnzero
\subsection{Virtual and soft photonic and fermion pair production
corrections\label{sosv}} 
There are two classes of ISR QED corrections: {\em photonic} and {\em
fermion pair production} corrections.
Photonic corrections  are governed by flag {\tt FOT2}
and pair production corrections  by flag {\tt ISPP}.

For the photonic corrections in {\tt ZFITTER} three different radiators 
are accessible:
\begin{itemize}
\item {\tt FOT2}=0,3 -- 
additive radiator of the Kuraev-Fadin type \cite{Kuraev:1985}
\item {\tt FOT2}=4 -- the QED-E radiator, Eqs. (3.31) and (3.32) of
\cite{Bardin:1989qr} (F. Berends and W. van Neerven, unpublished)
\item {\tt FOT2}=5 -- ``pragmatic'' factorized radiator, Eq. (63)
of \cite{Skrzypek:1992vk}
\end{itemize}

For fermion pair production corrections the following options are accessible:
\begin{itemize}
\item {\tt ISPP}=--1 -- a ``fudge'' implementation using formulae of  
                        \cite{Kniehl:1988id}
\item {\tt ISPP}=0 -- pair corrections are switched off
\item {\tt ISPP}=1 -- an implementation fully consistent with
                   \cite{Kniehl:1988id}, supplemented by  
                        a naive reweighting procedure; option used 
                        in \cite{Bardin:1999gt}\footnote{This option
produces reasonable numbers below and at the $\zb$ 
resonance and begins 
to quickly deteriorate above the resonance; 
for this reason it is considered now as obsolete option.
}
\item {\tt ISPP}=2 -- allows to include the singlet channel and
higher orders according to \cite{Arbuzov:1999uq}
\item {\tt ISPP}=3 -- according to \cite{Jadach:1992aa}
\item {\tt ISPP}=4 -- as {\tt ISPP}=3, but supplemented by an
     extended treatment of hadronic pairs \cite{Arbuzov:1999uq}
\end{itemize}
\subsubsection{Photonic corrections}
Higher order photonic corrections are applied up to the third order 
leading log contribution of order $\ord{(\alpha L)^3}$.

All the higher order virtual and soft photonic corrections are
determined in subroutine {\tt SETFUN}.

The second order virtual and soft pure photonic corrections $S^{(2)}$ in
\eqn{softer} are calculated following Eq. (2.30) of \cite{Berends:1988ab}; 
for the third order correction we follow \cite{Montagna:1997jv}:
\ba
S^{(2+3)} &=&
 \left(\frac{\alpha}{\pi}\right)^2\delta_2^{V+S}
+\left(\frac{\alpha}{\pi}\right)^3\delta_3^{V+S}
\\ \nl
\delta_2^{V+S}
&=&
 \left( \frac{9}{8}-2 \zetaz \right) L_e^2
+\left[ -\frac{45}{16} + \frac{11}{2} \zetaz + 3 \zetad \right] L_e
\nl
&&
+\left[
-\frac{6}{5} \zetaz^2 - \frac{9}{2}\zetad - 6\zetaz\ln 2 +\frac{3}{8}\zetaz
+\frac{19}{4}\right],
\\
\delta^{\rm V+S}_3&=&\lpar L_e-1\rpar^3
\lrbr \frac{9}{16}-3\ztwo+\frac{8}{3}\ztri\rrbr.
\label{soft2}
\ea
The terms proportional to $L_e^2$ are taken into account for setting
{\tt FOT2}=0 (or higher), those to $L_e$ for {\tt FOT2}=1 (or higher), 
the non-logarithmic ones  for  {\tt FOT2}=2 (or higher).
\subsubsection{Pair production corrections}
The virtual and soft corrections from initial
state fermion pair production $\delta^{V+S,p}$ are included
following Eq. (21) of \cite{Kniehl:1988id}:   
\ba
\delta^{V+S,p}&=&
\left(\frac{\alpha}{\pi}\right)^2
\sum_{n=e,\mu,\tau,\had} P_n^{V+S},
\label{spairs}
\\ \nl 
P_n^{V+S}&=&
\Biggl[ R(\infty) \left(\frac{1}{2}{L}_n^2 
-  \zetaz \right) + R_0 {L}_n 
+ R_1 \Biggr] \left( \frac{2}{3} l + \frac{1}{2} \right)
\nl &&
+\bigl[ R(\infty) l + R_0 \bigr] \left( \frac{2}{3} l^2  
-\frac{7}{12}\right)
+ R(\infty)\left[\frac{4}{9} {L}_n^3 + \frac{2}{3}\zetad + \frac{5}{8}\right],
\label{pairso}
\ea
where:
\ba
{L}_{n} &=&  \ln \frac{s}{m_{n}^2}~~~\mbox{for~leptons},
\label{spairs2}
\\ \nl
L_{\had} &=& \ln \frac{s}{4m_{\pi}^2}~~\mbox{for~hadrons},
\\ \nl
m_{\pi} &=& 0.1396~~\mbox{GeV},
\\ \nl
l &=& \ln \frac{2\delta}{\sqrt{s}},
\ea
\bq
\frac{\delta}{\sqrt{s}} = \left\{
\begin{array}{ll}
0.0055 & {\tt ISPP}=-1 \\ \\
10^{-6}& {\tt ISPP}=1
\end{array}
.\right.
\label{softfac}
\eq
The constant {\tt PAIRDL}=$\delta/\sqrt{s}$ is an auxiliary parameter used to
adjust a smooth connection of soft and hard radiation parts, with
$2m/\sqrt{s}\ll\delta/\sqrt{s}\ll (1-\sqrt{R})$.

The additional constants are for hadrons:
\ba
R(\infty)&=&4.0, 
\\
 R_0 &=& -8.31,
\\
 R_1&=&13.1 ,
\label{ri012}
\ea
and for leptons: 
\ba
R(\infty)&=&1, 
\\
R_0 &=& -\frac{5}{3},
\\
R_1&=&\frac{28}{9} - \zetaz. 
\label{ri012a}
\ea
 
Further abbreviations are:
\ba
\zetaz &=& \frac{\pi^2}{6},
\\ \nl 
\zetad &=& 1.202\,056\,903\,159\,6.
\label{abbrev1}
\ea
For {\tt ISPP} =--1, pairs are treated like photons, i.e. their $V+S$
contributions are added to the photonic ones. 
In this way an interplay of pairs with photons is realized. 
\subsection{Hard corrections for the total cross-section\label{soh}}
The higher order hard corrections consist of hard photonic
corrections and of hard contributions due to real initial state
fermion pair production.
The influence of flags {\tt FOT2} and {\tt ISPP}
is described in \sect{sosv}.

For $\sigma_{\sss{T}}$ and for {\tt ISPP} = --1, both photonic and
pair production 
contributions are calculated in function {\tt SH2}, which is called by 
function {\tt FCROS} for the cross-sections without cuts, by function 
{\tt SHFULL} for the integrated cross-sections, and by function 
{\tt HARD} for the differential cross-section:
\ba
H_{\sss{T}}^{(2+3)}(v)&=& h_{\sss{T}}^{(2+3)} + h_{\sss{T}}^{p}.
\label{t2svp}
\ea
We remind that $v=1-z,\;z=R$.

For {\tt ISPP} $\geq$ 1, the pair production contribution is
calculated in function {\tt PH2}. 
For {\tt ISPP} = 1, instead of \eqn{t2svp} one applies a ``reweighting'' 
procedure, which may be described as follows:
\bq
\sigma^{\sss{\rm{QED}}}=\sigma^{\rm{photonic}}
\lpar 1+\delta^{V+S,p}+\delta^{H,p} \rpar,
\eq
where $\delta^{V+S,p}$ is given by \eqn{pairso} and 
\bq
\delta^{H,p}=\frac{\sigma^{H,p}}{\sigma^{\rm{Born}}}\,.
\eq
Finally, for the implementation of pairs 
for {\tt ISPP} $\geq$ 2 we refer the reader to \cite{Arbuzov:1999uq}.
\subsubsection{Photonic corrections}
The hard photonic corrections are
calculated according to ${\tilde \delta}_2^{H}$ (Eq. (2.30)
of~\cite{Berends:1988ab}): 
\ba
h_{\sss{T}}^{(2+3)}
&=&
 \left(\frac{\alpha}{\pi}\right)^2{\tilde \delta}_2^{H}
+\left(\frac{\alpha}{\pi}\right)^3{\tilde \delta}_3^{H},
\\ \nl 
{\tilde \delta}_2^{H}
&=&
h_2 L_e^2 + h_1 L_e + h_0,
\\ \nl  
h_2 &=&
-\frac{1+z^2}{1-z}  \ln z  
+(1+z) \left[ -2 \ln (1-z) + \frac{1}{2} \ln z \right]
-\frac{5}{2} -\frac{z}{2}\,,
\\ \nl 
h_1 &=&
 \frac{1+z^2}{1-z}  \left[
\litwo(1-z)+\ln z \ln (1-z) +\frac{7}{2}\ln z - \frac{1}{2} \ln^2
z\right]
\nl && 
+(1+z) \left[ \frac{1}{4} \ln^2 z +4 \ln (1-z) -2\zetaz \right] -\ln z
+7 +\frac{z}{2}\,,
\\ \nl 
h_0 &=&
\frac{1+z^2}{1-z}
\Biggl[
-\frac{1}{6} \ln^3z +\frac{1}{2}\ln z \litwo(1-z)
+\frac{1}{2}\ln^2z\ln(1-z) 
\nl &&-\frac{3}{2}\litwo(1-z)- \frac{3}{2}\ln z \ln(1-z) 
+\zetaz \ln z - \frac{17}{6}\ln z - \ln^2 z 
\Biggr] 
\nl && 
+(1+z) \left[ \frac{3}{2}\lithree(1-z) -2 S_{1,2}(1-z)
-\ln(1-z)\litwo(1-z)-\frac{5}{2} \right]
\nl && -\frac{1}{4}(1-5z)\ln^2(1-z) +\frac{1}{2}(1-7z)\ln z \ln(1-z)
\nl && -\frac{25}{6}z\litwo(1-z)+\left(1+\frac{19}{3}z\right) \zetaz 
-\left(\frac{1}{2}+3z\right)\ln(1-z) 
\nl && +\frac{1}{6}(11+10z)\ln z + \frac{2}{(1-z)^2}\ln^2 z 
-\frac{25}{11}z\ln^2 z
\nl && 
-\frac{2z}{3(1-z)}\left[1+\frac{2}{1-z}\ln z +\frac{1}{(1-z)^2} \ln^2 z
\right] 
\\ \nl
\delta^{H}_3&=&\lpar L_e-1\rpar^3
\frac{1}{6}
\biggl\{
-\frac{27}{2}+\frac{15}{4}v
+4\lpar 1-\frac{v}{2}\rpar
\Bigl[6\ztwo-6\ln^2 v+3\litwo{2}\lpar v\rpar\Bigr]
\nl &&
 +3\lpar-\frac{6}{v}+7-\frac{3}{2}v\rpar\ln(1-v)
 +\lpar\frac{4}{v}-7+\frac{7}{2}v\rpar\ln^2(1-v)
\nl &&
 -6\lpar 6-v\rpar\ln v
 +6\lpar-\frac{4}{v}+6-3v\rpar\ln(1-v)\ln v
\biggr\}.
\label{d2h}
\ea
\subsubsection{Pair production corrections}
The hard part of the initial state corrections due to fermion pair
creation is according to Eqs.~(22) and (23) of~\cite{Kniehl:1988id}:
\ba
h_{\sss{T}}^{p}&=&
\theta(R-z_{\min})
\theta\left[ (1-\sqrt{R})^2 - \delta^2 \right]
\left[ 
h_e^{p}+h_{\mu}^{p}+h_{\tau}^{p}+h_{had}^{p}
\right],
\\ \nl 
h_{had}^{p}
&=&
\left( \frac{\alpha}{\pi} \right)^2
\frac{1}{3}
\Biggl\{
\frac{1+z^2}{(1-z)}
\left[ R(\infty)
\left( 
\frac{1}{2}\ln^2 \frac{s(1-z)^2}{4m_{\pi}^2z}-\zetaz\right)
+R_0
\ln \frac{s(1-z)^2}{4m_{\pi}^2z} 
+R_1\right]
\nl &&-~
(1-z)\left[
R(\infty)\left(2\ln \frac{s(1-z)^2}{4m_{\pi}^2z}-3\right)+2R_0\right]
\nl &&-~
R(\infty)\left[
\frac{z^2}{1-z} \left( \frac{1}{2}\ln^2z + \litwo(1-z)\right) +\ln z\right]
\Biggr\} , 
\\ \nl 
h_{\mu}^{p}
&=&
\left( \frac{\alpha}{\pi} \right)^2
\frac{1}{3}
\Biggl\{
\frac{1+z^2}{2(1-z)}\bar{L}_{\mu}^2 +\left[ \frac{1+z^2}{1-z} \left(
\ln\frac{(1-z)^2}{z} -\frac{5}{3}\right) -2(1-z) \right] \bar{L}_{\mu}
\nl && +~
\frac{1+z^2}{1-z}\left(\frac{1}{2}\ln^2 \frac{(1-z)^2}{z}-\frac{5}{3}
\ln \frac{(1-z)^2}{z} -2\zetaz +\frac{28}{9}\right) 
\nl && -~
(1-z) \left(2\ln \frac{(1-z)^2}{z} - \frac{19}{3}\right) -
\frac{z^2}{1-z} \left[ \frac{1}{2}\ln^2z + \litwo(1-z)\right] -\ln z
\Biggr\} ,
\ea
with $\bar{L}_{\mu}=\ln(s/m_{\mu}^2)$, etc., and 
\ba
z_{\min} &=& \frac{\smanp_{\min}}{\sman}\,.
\label{zmin}
\ea

We use further:
\ba
\lithree(y) 
\equiv 
S_{2,1}(y) &=& \int_0^1 \frac{dx}{x}\ln x \ln(1-xy) =
\int_0^y \frac{dx}{x} \litwo(x),
\\ \nl 
S_{1,2}(y) &=& \frac{1}{2}\int_0^1 \frac{dx}{x} \ln^2 (1-xy) 
= \frac{1}{2} \int_0^y \frac{dx}{x}\ln^2(1-x).
\ea
For their numerical calculation in the routines {\tt TRILOG} and {\tt S12} we
use \cite{Matsuura:1987}.

In order to get a better agreement with the pair production correction
shown in Figure 4, bottom dotted curve of \cite{Jadach:1992aa}, we somewhat
arbitrarily multiplied the soft and hard muonic 
corrections by a correction factor,
\ba
{\tt CORFAC} = 2.6 .
\ea
This factor is set in subroutine {\tt SETFUN} and used for {\tt
ISPP}=--1 only; an outdated option, which is retained for backward
compatiblity. 
\subsection{Hard corrections for the forward-backward asymmetry\label{sohfb}}
The corresponding corrections for $\sigma_{\sss{FB}}$
are calculated in function {\tt AH2}, which is called by 
function {\tt FASYM} for the cross-sections without cuts, by function 
{\tt AHFULL} for the integrated cross-sections, and by function 
{\tt HARD} for the differential cross-section:
\ba
H_{\sss{FB}}^{(2)}(v) &=& h_{\sss{T}}^{(2)} + \delta h_{\sss{FB}}^{(2)}\,, 
\ea
where the $h_{\sss{T}}^{(2)}$ is taken over from~\eqn{t2svp}, and 
the additional correction in leading logarithmic approximation is:
\ba 
\delta h_{\sss{FB}}^{(2)}&=&
\frac{1}{4}
\left( \frac{\alpha}{\pi} \right)^2
L_e^2
\Biggl[\frac{(1-z)^3}{2z}
+\frac{(1-z)^2}{\sqrt{z}}\left(\arctan\frac{1}{\sqrt{z}} 
-\arctan\sqrt{z} \right) 
\nl && 
- (1+z) \ln z + 2(1-z) \Biggr] + {\cal O}(\alpha^3).
\label{a2h}
\ea
The correction $\delta h_{\sss{FB}}^{(2)}$ is 
coded according to ${\tilde \Phi}_{\sss{FB}}$ from Eq.~(2.35)
of~\cite{Beenakker:1989km}, 
where it was calculated for the total forward-backward asymmetry
without cuts.

The initial state fermion pair production correction $h_{\sss{FB}}^{p}$
is unknown and set equal to zero.

\chapter{Pseudo-Observables\label{ch-PO}} 
\section{Introduction\label{introdPOs}}
\setcounter{equation}{0}
In this Section we describe the calculation of 
the so-called {\em pseudo-observables}, PO,
which are used for the combination of LEP data from the four LEP
Collaborations 
\cite{Vanc-Gruenewald:1998,home-LEPEWWG}, 
and which are also needed for the calculation of the cross-sections 
within the Standard Model. 
The various cross-sections,
asymmetries, or polarizations 
are called {\em realistic observables}.

Pseudo-observables, which may derived from them, are usually
definition dependent quantities. 
Typical examples of pseudo-observables are partial $\zb$-decay widths, 
$\gff$, effective sinuses of weak mixing angles, $\seffsf{\ff}$, the
parameter $\dr$, or more general --- amplitude form factors. 
In some renormalization schemes the $\wb$-boson mass, $\mwl$, 
may be also considered as a PO.

For the calculations of POs {\tt ZFITTER} uses the {\tt DIZET} package
which relies on the so-called {\em on-mass-shell (OMS)} renormalization 
scheme \cite{Bardin:1980fet,Bardin:1982svt}.
\section{Input Parameters\label{input}}
\eqnzero
\subsection{Input parameter set \label{IPS1}}
A very important notion is the notion of an {\em input parameter set},
IPS, 
i.e. the choice of the input parameters of a given renormalization scheme.

Our OMS uses the masses of all fundamental particles, both fermions 
and bosons, and two coupling constants: $\alpha(0)$
and $\als(\mzs)$.
There is the exception of the ill-defined masses of light quarks
$\fu,\fd,\fc,\fs$ and $\fb$: they are  
excluded and replaced by $\alpha(\mzs)$ by making use of a dispersion relation
\eqn{disprel}
between the real and imaginary parts of the hadronic vacuum polarization 
$\pir{\sman}$ 
\eqn{rgamrat}
and the optical theorem relating the imaginary part 
with the total cross-section 
$\sigma_{\rm tot}\lpar\fep\fem\to\ph^*\to{\hbox{\rm hadrons}}\rpar$ 
\cite{Eidelman:1995ny}.
More details are given in \subsect{qedrun}. 

More rigorously, in this way we exclude mass singular logarithmic terms like
$\ln({\mqs}/s)$, while power suppressed terms of order
$\ord{\mqs/\mzs}$ are either  
totally ignored for light quarks ($\fu,\fd$ and $\fs$), or are kept only in 
FSR QCD corrections for heavy quarks ($\fc$ and $\ffb$), or are treated within 
the language of effective quark masses (see \subsect{runnmass}).
For the description of the FSR QCD corrections we use the
notion of running masses in the $\MSB$ scheme. This introduces into the IPS 
the pole masses of $\fc$ and $\ffb$ quarks. They are set inside 
subroutine {\tt QCDCOF} and are not supposed to be varied by the user.

The knowledge about the hadronic vacuum polarization is contained in
the quantity 
$\dalhv$, which is treated as one of the input parameters and
used in parallel to $\alpha(0)$ since the latter is also needed
at least for the ISR corrections. 

The $\dalhv$ can be either computed from quark masses 
or, preferrably, fitted to experimental annihilation data. 
In the {\tt DIZET} package both treatments
are foreseen; see the description of user option {\tt ALEM} in
\appendx{dizetflags}.   

Nowadays, the mass of the $\wb$ boson is experimentally known with 
a precision of about 60 MeV both from $\prot\aprot$ and $e^+e^-$ measurements,
that is, with a combined precision of about 40 MeV.
Although this precision is supposed to be reduced further with future
\LEPII\ data taking and analysis 
and with forthcoming improvements from FNAL, still $\mwl$
can be calculated with a better theoretical error exploiting the very precise 
knowledge of the Fermi constant in $\flm$-decay, $\gf$. 
For this reason, $\mwl$ is usually replaced in the IPS by $\gf$, see
\subsect{imoms}. 

The following parameters are passed to subroutine {\tt DIZET} by its
argument list: 
\ba
\label{amwi}
{\tt AMW}&=&\mwl,
\\
{\tt AMZ}&=&\mzl,
\\
{\tt AMT}&=&\mtl,
\\
{\tt DAL5H}&=&\dalhv,
\\
{\tt ALSTR}&=& \als(\mzs).
\label{sminput}
\ea
$\gf$ is set inside subroutine {\tt CONST1} for {\tt DIZET} together with
all the other numerical input for the calculation of POs.
Two important constants are:
\ba
\mbox{\tt ALFAI} &=& 1/\alpha(0) = 137.0359895, 
\\
\alpha(0) &=& \alpha,
\\
\mbox{\tt GMU} &=& \gf = 1.16639\, 10^{-5}\;\mbox{GeV}^{-2};
\ea
see also the description of flag {\tt GFER}, see \subsect{dizetflags}. 
At present they both are also set in subroutine {\tt EWINIT}.
\noindent
A special auxiliary constant $A_0$,  
\bqa
\mbox{\tt A0}&=& A_0 = \sqrt{\frac{\pi\alpha(0)}{\sqrt{2}\gf}}
= 37.28052\;,
\label{ao_est}
\eqa
is also computed in subroutine {\tt CONST1}. 

Note that \eqns{amwi}{sminput} together with $\gf$ is over-complete; 
only two of the three parameters
\bq
\gf,\quad \mwl,\quad\mbox{and}\quad\mzl
\eq
are independent. 
This is exploited by option {\tt IMOMS} = {\tt NPAR(4)}, see \subsect{imoms}.

The definition of the {\em effective quark masses} is initialized 
in subroutine {\tt CONST1} by  flag {\tt MQ}.
Effective quark masses are some fitted values which allow to reproduce 
in the one-loop order the quantity $\dalhv$. They are also used in 
power suppressed mass dependent terms of self-energies 
and in the calculation of various imaginary parts.
The power dependence of POs on effective quark masses is very weak
(in contrast to the logarithmic mass singularities) and as far as the
effective masses  
are compatible with a chosen value of $\dalhv$ the residual theoretical 
uncertainty is totally ignorable.
 
For imaginary parts the situation is even better since we
usually are well above the thresholds of light quarks where the
imaginary parts are  
equal to $\ib\pi$ and the precise values of effective masses are irrelevant. 

The preferred setting is {\tt MQ}=1 ({\tt MQ} = {\tt ITQ}=2--{\tt{IHVP}}),
and for {\tt IHVP}=1 one uses a set of effective masses, which is compatible 
with the fit of~\cite{Eidelman:1995ny}.

All lepton and effective quark masses are set in subroutine {\tt CONST1}.

The lepton masses are taken from ~\cite{Caso:1998aa}:
\ba
m_{\nu} &=& 0,
\\
m_e &=& 0.510\,999\,07\,\times 10^{-3}\, \mathrm{GeV},
\\
m_{\mu} &=& 0.105\,658\,389\,\mathrm{GeV},
\\
m_{\tau} &=& 1.777\,05\,\mathrm{GeV} .
\ea
 The scheme of determination of quark masses for all applications but QCD 
final state corrections is as follows.
We describe it in some detail since an expert user could like to change the 
preset values for some reasons.
One of three foreseen sets {\tt AMQ(I)} of quark masses is
selected. The default setting is:
\ba
m_u &=& 0.062\,\mathrm{GeV},\\
m_d &=& 0.083\,\mathrm{GeV},\\
m_c &=& 1.50\,\mathrm{GeV}, \\
m_s &=& 0.215\,\mathrm{GeV},\\
m_t &=& {\tt TMASS},        \\
m_b &=& 4.7\,\mathrm{GeV}.
\ea 
For the alternative choices of the flag, {\tt IHVP}=2,3, it is chosen: 
\ba
m_u &=& 0.04145\,\mathrm{GeV}, \\
m_d &=& 0.04146\,\mathrm{GeV}, \\ 
m_c &=& 1.5 \,\mathrm{GeV},    \\
m_s &=& 0.15\,\mathrm{GeV},    \\
m_t &=& {\tt TMASS},           \\
m_b &=& 4.7 \,\mathrm{GeV}.
\ea
{\tt DIZET} returns 
the following quantities which are described in this section:
\ba
\mbox{\tt ALQED}=\alpha(\mzs),
\ea
the running QED coupling at $\mzs$; further:
\ba
\mbox{\tt ALSTRT}=\als(\mts),
\ea
the running QCD coupling at $\mts$; further:
\ba
\nonumber
\mbox{\tt ZPAR}  &-& \mbox{array of useful quantities},
\\ \nonumber
\mbox{\tt PARTZ} &-& \mbox{array of partial $\zb$-widths}, 
\\ \nonumber
\mbox{\tt PARTW} &-& \mbox{array of partial $\wb$-widths}.
\eqa
The last three arrays are described in \sect{dizetug}.

 The conversion factor, making cross-sections to be
measured in nanobarn when energies are used in GeV, is {\tt CONHC}.
The settings in subroutine {\tt EWINIT} are:
\ba
{\tt CONHC} &=& 0.389\,379\,66\,\times 10^{6}\,\mathrm{GeV}^2 \mathrm{nbarn},
\ea
and cross-sections are normalized with the factor 
\ba
{\tt CSIGNB} &=& \frac{4}{3} \pi \alpha^2 {\tt CONHC} .
\label{csignb}
\ea

  In the OMS renormalization scheme the weak mixing angle is defined uniquely 
through the gauge-boson masses:
\bq
\sin^2\theta_{\sss{W}} \equiv \siws = 1 - \frac{\mws}{\mzs}\;.
\label{defsw2}
\eq
 We should mention here that the calculations of electroweak
corrections have been done in {\tt DIZET} in the unitary gauge.
The pseudo-observables (the weak form factors in particular)
are independent of the gauge since they are  (if only in principle)
measurable quantities.
\subsection{Further specification of input quantities \label{imoms}}
As soon as the IPS is specified, one may calculate some potentially measurable
quantity taking into account as many orders of the perturbative series
as are available in the literature. 

First of all, after a call of {\tt CONST1}, {\tt DIZET} computes
by function {\tt XFOTF3} the running QED coupling $\alpha(\mzs)$.
For the leptonic part, $\dall$, up to three-loop QED corrections are taken.
For $\dalhv$ the parameterization depends on flag {\tt VPOL}, 
see \subsect{qedrun}. By call of subroutine {\tt QCDCOF},
FSR QCD corrections include up to four loops, see \subsect{qcdrun}.

Then, {\tt DIZET} offers three opportunities in subroutine {\tt SETCON}, 
depending on flag {\tt IMOMS}:

\begin{itemize} 
\item
 The default is {\tt IMOMS}=1; the $\wb$ boson mass is calculated 
iteratively from the equation:
\bq
\mwl=
 \frac{\mzl}{\sqrt{2}}\sqrt{1+\sqrt{1-\frac{4 A^2_0 }
            {\mzs \lpar 1-\Delta r \rpar }}}\;,
\label{wmass}
\eq
using the standard IPS \eqns{amwi}{sminput}.
\item
The alternative, chosen with {\tt IMOMS}=2 and foreseen for expert
users only, calculates iteratively
$\mzl$ from $\gf,\,\mwl$ and other input parameters:
\bq
\mzl=\frac{\mwl}{\sqrt{1-\frac{\displaystyle
        A^2_0 }{\displaystyle \mws \lpar1-\Delta r \rpar }}}\;.
\label{zmass}
\eq
\item
A third setting,  {\tt IMOMS}=3, was foreseen for the calculation of $\gf$
from $\mwl,\,\mzl$ by means of iterations of the equation 
\bq
\gf=\frac{\alpha \pi}{\sqrt{2} \mws (1-\mws/\mzs) (1-\dr)}\;
\label{Gmu}
\eq
with respect to $\gf$. 
However, this is not realized in the code. 
\end{itemize}

\eqn{Gmu} follows from the calculation of one-loop EWRC
for $\flm$-decay, followed by the renormalization group re-summation
of $\dr$, governed by $\dalpha$:
\bq
\dr=\dalpha(\mzs)+\dr_{\EW}.
\eq
The re-summation of $\dr_{\EW}$ is not justified by renormalization group
arguments, it simply `accompanies' $\dalpha$ in \eqn{Gmu}. The only 
correct generalization of \eqn{Gmu} 
is to compute the next perturbative order and to write 
an expression which is consistent with higher order calculations.
More details about $\dr$ may be found in \subsect{deltar}.

\eqns{wmass}{zmass} are nothing but algebraic solutions of \eqn{Gmu} with
respect to $\mwl$ and $\mzl$, correspondingly. However, since $\dr$ also
depends implicitely on $\mwl$ and $\mzl$, one has to
solve them 
iteratively.
 
These equations establish interdependences between input parameters 
using the precisely known quantity $\gf$. 
Actually only {\tt IMOMS}=1 
is meaningful for the analysis of $\zb$-resonance data. 
That's why {\tt IMOMS}
is fixed in subroutine {\tt ZDIZET} (a CPU time saving interface to 
{\tt DIZET}) and is not accessible to changes by users.

After execution of {\tt SETCON}, the $\mwl$ is computed and the IPS becomes
fully specified.
\subsection{The running QED coupling\label{qedrun}}
The running QED coupling $\alpha(\sman)$ appears in many places in 
calculations.

 Fermion one-loop insertions to the photon propagator are summed together 
 with the photonic Born diagram 
to form the matrix element $A^{\sss{OLA}}_{\ph}$:
\bq
A^{\sss{OLA}}_{\ph} = 
   \ib \chi_{\ph}(\sman) \alpha(\sman) \gadu{\mu} \otimes \gadu{\mu}\,.
\label{Born_modulo}
\eq
Dyson summation leads to the change
of $\alpha$ into $\alpha(\sman)$:
\bq
\alpha(\sman) = \frac{\alpha(0)}{1-\dalpha^{\fer}(\sman)}=
                \frac{\alpha(0)}{1-\dalpha^{(5)}(\sman)
                                  -\dalpha^{\ft}(\sman)
                           -\dalpha^{\alpha\als}(\sman)}\,,
\label{dysfa}
\eq
where we explicitly disentangled the one-loop top-quark contribution
$\dalpha^{\ft}(\sman)$ and the two-loop irreducible higher-order 
correction $\dalpha^{\alpha\als}(\sman)$.

 The function {\tt XFOTF3}, belonging to the {\tt DIZET} package, is used 
to calculate $\alpha(\sman)$.

In function {\tt XFOTF3('ALEM','ALE2','VPOL','QCDC',ITOP,DAL5H,Q2)} 
we use the convention that {\tt Q2}=$-\sman$ in the $\sman$ channel
and {\tt Q2}=$\tman$ in the $\tman$ channel. 

Important user options are {\tt ALEM}, {\tt ALE2}, {\tt VPOL}, {\tt
QCDC}, described in \sect{zuflag}), while {\tt ITOP}
is an internal flag switching on/off for {\tt ITOP=1/0}  the two last terms
in \eqn{dysfa}.

The 5-flavor part is a sum of leptonic and hadronic contributions:
\bqa
\dalpha^{(5)} = \dalpha_{\lep} + \dalpha^{(5)}_{\had}.
\eqa 

The leptonic part is assigned to the variable {\tt XCHQ21}, 
and the hadronic part to {\tt UDCSB}.
$\dalpha^{(5)}_{\had}$ might be also treated as an input parameter for the fit.

The $\dalpha^{\alpha\als}$ correction is given by $\alpha/(4\pi)$ 
{\tt ALFQCD}.

 The result for $\dalpha^{(5)}_{\had}(\sman)$ was obtained in 
\cite{Eidelman:1995ny} by making use of a dispersion relation: 
\bqa
\label{disprel}
\dalpha^{(5)}_{\had}(\sman) =
 -\frac{\alpha}{3\pi}\sman\,\Reb\int_{4m^2_\pi}^{\infty} d\smanp
  \frac{R_{\ph}\lpar \smanp\rpar}{\smanp\lpar\smanp-\sman-\ib\ep\rpar}\,,
\eqa
with 
\bqa
\label{rgamrat}
R_{\ph}(\sman)=
    \frac{\sigma\lpar\fep\fem\to\ph^{*}\to\mbox{hadrons}\rpar}
         {\sigma\lpar\fep\fem\to\ph^{*}\to\flmp\flmm    \rpar}
\eqa
as an experimental input. 
For the hadronic contribution at $\mzl$ it gives:
\bq 
\dalpha^{(5)}_{\had}(\mzs) = 0.0280398. 
\eq
The parameterization of \cite{Eidelman:1995ny}
is chosen with {\tt VPOL}=1 and $\dalpha^{(5)}_{\had}(\sman)$
is calculated by a call to subroutine {\tt hadr5} 
\cite{Jegerlehner:1995ZZ}. 

 The leptonic one-loop contribution, $\dalpha_{\lep}(\sman)$, 
is defined by 
\bq
\alpha(\sman)=\frac{\alpha}
{\ds{1-\frac{\alpha}{4\pi}\Bigl[\Pgg^{\fer}(\sman)-\Pgg^{\fer}(0)\Bigr]}}\,,
\label{alpha_fer}
\eq
with
\bqa
\Pgg^{\fer}(\sman) &=& 4 \sum_f c_f  Q^2_f  \bff{f}{-\sman}{\mfl}{\mfl},
\label{pigg_fer}
\\
\bff{f}{\pmoms}{\Mind{1}}{\Mind{2}}
&=&
2\lrbr\bff{21}{\pmoms}{\Mind{1}}{\Mind{2}}+\bff{1}{\pmoms}{\Mind{1}}{\Mind{2}}
 \rrbr
\nll &=& -\frac{1}{\epsb}
+2\int_0^1 dx x(1-x)\ln\frac{\pmoms x(1-x)+\Mind{1}(1-x)+\Mind{2} x}{\tHss}\,.
\qquad\quad
\label{Bf_function}
\eqa
For the two-loop corrections we use \cite{Kallen:1955ks} 
and for the three-loop terms \cite{Steinhauser:1998rq}.
The contributions in {\tt ZFITTER} are:
\bq
\dalpha_{\lep} = 
        314.97637\cdot 10^{-4} 
=\bigl[ 314.18942_{\rm 1-loop}
        + 0.77616_{\rm 2-loop}
        + 0.01079_{\rm 3-loop} \bigr] \cdot 10^{-4}.
\eq
These numbers were derived with leptonic masses taken
from~\cite{Caso:1998aa} and with $\mzl=91.1867$ GeV.
The leptonic contribution is calculated by function {\tt DALPHL}.

 The top contribution is given by \eqn{dalphat}, it depends on the mass 
of the top quark, and we present its numerical value for $\mtl=173.8$ GeV:
\bq
\dalpha^{\ft}(\mzs) = -0.585844\cdot 10^{-4}.
\eq

The mixed two-loop $\ord{\alpha\als}$ correction 
arising from $\ft\bart$ loops with gluon exchange is calculated 
using formulae from~\cite{Kniehl:1990yc}. For more details see
\subsect{mixed_bkqcd}. Its numerical value at $\mtl=173.8$ and $\als=0.119$ 
is
\bq
\dalpha^{\alpha\als}(\mzs) = - 0.103962 \cdot 10^{-4}.
\eq

All numerical results presented in \chapt{ch-PO} are derived with 
{\tt DIZET} v.6.05 as were those shown in \cite{Bardin:1999gt}.
 
\section{Electroweak Renormalization, Parameters $\drho$ and $\dr$
\label{drhodr}}
\eqnzero
In this Subsection we discuss two important ingredients of the  
renormalization scheme: parameters $\dr$ and $\drho$. 

\subsection{$\dr$ at one loop\label{deltar}}
Sirlin's parameter $\dr$~\cite{Sirlin:1980nh} 
is calculated in subroutine {\tt SEARCH}.
There are several user options, depending on the  
flags {\tt BARB}, {\tt QCDC}, {\tt VPOL}, {\tt MASS}, {\tt ALEM}, {\tt
AFMT}, {\tt AMT4}, {\tt HIGS}, {\tt SCRE},
which influence its calculation; 
see the description of flags in \sect{zuflag}.

The one-loop expression for $\dr$ is given by 
\bqa
\deltar&=&\gspi\Biggl\{ 
         -\frac{2}{3}-\Pgg^{\fer,F}(0)+\frac{\ctws}{\stws}
          \delrho{F} + \frac{1}{\stws} \Biggl[
          \delrho{F}_{\sss{\wb}}
         +\frac{11}{2}-\frac{5}{8} \ctws (1+\ctws)
         +\frac{9}{4} \frac{\ctws}{\stws} \ln \ctws \Biggr]
               \Biggr\},
\nll
\label{delta_r}
\eqa
The expression given in our old, short description in \cite{Bardin:1995a2}
and coded in {\tt DIZET} differs from \eqn{delta_r} only by notations. 
We give the dictionary: 
\bqa
\ctws &=&\mbox{\tt R},
\\ \Pgg^{\fer,F}(0)&=&
\frac{4}{3}\sum_f\qfs\ln\frac{\mfs}{\mws}=\frac{4}{3}\mbox{\tt (SL2+SQ2)},
\\
\Sigma^{F}_{\sss{VV}}(x)&=&V(x),\qquad\mbox{with}\quad V=W,Z,
\\[3mm]
\delrho{F}&=&{W}(\mws)-{Z}(\mzs)=\siws\mbox{\tt(XDWZ1F+XWZ1R1)}
\\[3mm]
\label{dict_dr}
\delrho{F}_{\sss{\wb}}&=&{W}(0)-{W}(\mws)
                        =\mbox{\tt W0F+W0-DREAL(XWM1F+XWM1)}
\nll
&=&\frac{1}{\mws}
\lrbr\Sigma^{F}_{\sss{WW}}(0) 
    -\Sigma^{F}_{\sss{WW}}(\mws)\rrbr.
\label{delta_rhoFW-old}
\eqa
The one-loop $\dr$ in this form was introduced in appendices (E.8) and
(F.3) of \cite{Bardin:1980fet,Bardin:1982svt}, with the $t$ mass dependent 
terms being given in the appendix of~\cite{Bardin:1989dit}. 

In \eqn{delta_r} are certain combinations of self-energy
functions.
One of them, appearing in every
amplitude form factors, is a gauge invariant quantity which plays a special
role in all calculations:
\bqa
\Delta \rho &=&
\frac{1}{\mws}
\lrbr \Sigma_{_{WW}}(\mws) - \Sigma_{_{ZZ}}(\mzs) \rrbr,
\label{rhodef}
\eqa
\subsection{The parameter $\Delta\rho$ \label{drho}}
$\Delta\rho$ is closely related to the well-known Veltman parameter
$\rho$ defined in~\cite{Ross:1975fq,Veltman:1977kh} 
as the ratio of neutral and charged current effective coupling 
constants in neutrino scattering for very low transferredd momentum
squares $\imoms$: 
\bq
\rho=\frac{\mws}{\mzs\cos^2\theta^{0}_{\sss{W}}}\,,
\label{Veltman_rho}
\eq
where 
\bq
\cos^2\theta^{0}_{\sss{W}}
\label{mixinglow}
\eq
depends on a weak mixing angle $\theta^{0}_{\sss{W}}$ measured at low
$\imoms$. 
If all radiative corrections (RC) are ignored, then it is equal to 
\bq
\cows=\mws/\mzs\,. 
\label{mixingmass}
\eq
RCs induce some difference between $\cows$ and $\cos^2\theta^{0}_{\sss{W}}$. 
To evaluate this difference, let us consider both masses and 
$\cos^2\theta^{0}_{\sss{W}}$ to be the running quantities (because of RCs) 
and differentiate \eqn{Veltman_rho} in the vicinity of 
$\imoms=0$:
\bq
\delta\rho(0)=\frac{\delta\mws(0)}{\mzs\cos^2\theta^{0}_{\sss{W}}}
             -\frac{\delta\mzs(0)\mws}{\mzq\cos^2\theta^{0}_{\sss{W}}}\,,
\eq
where we made use of the fact that $\delta\cos^2\theta^{0}_{\sss{W}}(0)=0$. 
Noting that all {\em derivatives} $\delta$ are of order $\ord{\alpha}$
and that the difference between \eqnsc{mixinglow}{mixingmass} is also
of order $\ord{\alpha}$, we have in the one-loop approximation
\bqa
\drhov(0)=\frac{\delta\mws(0)}{\mws}
         -\frac{\delta\mzs(0)}{\mzs}=\frac{\gbs}{16\pi^2\mws}
\Bigl[\Sigma_{\sss{WW}}(0)-\Sigma_{\sss{ZZ}}(0)\Bigr].
\eqa
For $\zb$ resonance observables more relevant is the quantity defined 
as the difference of two complete one-loop self-energies taken at
their physical masses, 
\eqn{rhodef}: 
\ba
\delta\rho&=&\frac{\alpha}{4\pi\siws}\drho,
\\
\Delta\rho&=&\frac{1}{\mws}
\Bigl[\Sigma_{\sss{WW}}(\mws)-\Sigma_{\sss{ZZ}}(\mzs)\Bigr],
\label{rho_1l}
\ea
which is closely related to \eqn{Veltman_rho} and 
which arises naturally within the OMS scheme.

Although \eqn{rho_1l} implies the use of self-energies computed at 
one-loop level,
it may be considered as 
generalized to higher perturbative order in coupling constants $\alpha$ 
and $\als$.
For this reason it is convenient to include the coupling constant in
the superscript of $\delta{\rho}$. At the one-loop order we introduce:
\bqa
\delta{\rho}^{\alpha}=\frac{\alpha}{4\pi\siws}\drho.
\label{insert3}
\eqa
Now we proceed by presenting $\delta\rho$ as a sum of terms which are 
known in the literature:
\bq
\delta\rho = \delta\rho^{\alpha}+\delta\rho^{\alpha\als}
            +\delta\rho^{\alpha^2}+\delta\rho^{\alpha\alpha^2_S}_{\sss{L}}.
\label{deltarho_s}
\eq

  We note that only for the first two terms the complete calculations are
available~\cite{Djouadi:1988di,Bardin:1989aa,Kniehl:1990yc,Halzen:1991je}. 
For the two-loop electroweak 
correction, $\delta\rho^{\alpha^2}$, only the two first terms in an expansion 
in $\mts$ are known (\cite{Degrassi:1996mg,Degrassi:1997ps}): 
\bq
\delta\rho^{\alpha^2}=\delta\rho^{\alpha^2}_{\sss{L}}  
                     +\delta\rho^{\alpha^2}_{\sss{NL}}.
\label{deltarho_d}
\eq
For the three-loop correction {\tt AFMT} $\sim
\delta\rho^{\alpha\als^2}$ \cite{Avdeev:1994db,Chetyrkin:1995ix},
only the leading in $\mts$ term is known.
The flag {\tt AFMT} is explained in \subsect{dizetflags}.

Below we discuss briefly each term. 

By examining the one-loop $\delta\rho^{\alpha}$ it is easy to verify that it
is enhanced {\it quadratically} with the top quark mass, i.e. if one expands 
the complete one-loop expression for the fermionic component of
$\delta\rho^{\alpha}$ into a series and retains
only the first term one gets:
\bqa
\delta{\rho}^{\alpha}&\approx&\delta{\rho}^{\alpha}_{_{L}}\equiv
- N_c \frac{\alpha}{16\pi\siws\cows}\frac{\mts}{\mzs}\,= -{\tt DRHO1}.
\label{leading_1l}
\eqa
\eqn{leading_1l} introduces on passing the notion of {\em leading} term.
Actually, in many cases a radiative correction may be split into 
two parts: {\it leading} part and {\it remainder}. \\

This splitting should satisfy the following requirements: 
\begin{enumerate}
\item
There is a guiding principle of splitting: Usually a splitting
is undertaken in order to re-sum the leading term to all orders 
in perturbation theory.
\item
The leading term should be bigger than the remainder.
\item Splitting should respect gauge invariance. For instance, if we define
\bqa
\delta\rho = \delta\rho_{\sss{L}} + \delta\rho_{\rm{rem}}\,,
\label{splitting}
\eqa  
both $\delta\rho_{\sss{L}}$ and $\delta\rho_{\rm{rem}}$ must be 
separately gauge invariant. 
The $\delta\rho_{_{L}}$ is re-summed to all orders, 
while $\delta\rho_{\rm{rem}}$ is  treating 
in a fixed order of perturbation theory.
\end{enumerate}

The splitting (\ref{splitting}) is gauge invariant because 
$\Delta{\rho}$, as defined by \eqn{rho_1l}, is a gauge invariant object,
and the leading term (\ref{leading_1l}) is apparently gauge-invariant.

However, since the top quark mass is only two times bigger than the $\zb$
boson mass, the quadratic enhancement is not so pronounced and the second 
condition is not fulfilled in a strict sense. 
A way out could be found observing that the complete $\Delta{\rho}$ is a good 
candidate for re-summation because it is gauge invariant. 
In \cite{Fanchiotti:1991kc} the following quantity was introduced: 
\bqa
\delta\hat{\rho}^{\alpha} & \equiv &
\frac{\alpha}{4\pi\siws}\frac{1}{\mws}
\biggl[\Sigma_{\sss{WW}}(\mws)-\Sigma_{\sss{ZZ}}(\mzs)\biggr]
\Bigg|_{\overline{MS},\;\tHs=\mzl},
\\
\delta\hat{\rho}^{\alpha} & = &
 \frac{\siws}{\cows}
\lrbr -{\tt AL4PI}*{\tt DWZ1AL} + {\tt SCALE}\rrbr.
\label{insert7}
\eqa
Since this quantity is divergent it is understood as regularized in 
the $\MSB$ sense, that means dropping out the pole term $1/\epsb$. 
All divergences cancel for physical observables, therefore we may simply 
drop them all for every ingredient of calculations.
After that we are left with the residual 
dependence on the t'Hooft scale parameter $\tHs$ 
in the individual components, again cancelling in the sum
\footnote{In fact,
every divergence $1/\epsb$ is accompanied by
$-\ln{\Minds{}}/{\tHss}$.}.
It became customary to define all scale dependent quantities 
at $\tHs=\mzl$, which is conventionally considered to be the natural scale 
not only for the  $\zb$-resonance, but also for all electroweak
observables. 

We note that the two separately gauge invariant terms in \eqn{splitting} 
may be considered with different coupling constants: $\alpha$ or $\gf$. 
As was proven in~\cite{Consoli:1989pc}, 
the leading term should be considered with coupling constant 
$\gf$ 
, i.e. one
should perform a {\em conversion} of couplings:
\bq
\alpha\to\gf\,.
\eq
This may be achieved by multiplying \eqn{leading_1l} by the conversion 
factor
\bq
f=\frac{\sqrt{2}\gf\mzs\siws\cows}{\pi\alpha}\,.
\label{conversionf}
\eq
As a result,  $\delta{\rho}_{\sss{L}}$ becomes:
\ba
\delta{\rho}^{\sss{G}}_{\sss{L}}= - N_c x_t 
 = -{\tt DRIRR} + 3 x^2_t {\tt AMT4C}\,,
 \ea
with
\ba
 x_t = \frac{\gf\mts}{8\sqrt{2}\pi^2}={\tt AXF}\,.
\label{xt}
\ea
Here the superscript $G$ emphasizes the use of the scale $\gf$.

For a discussion about $\delta\rho^{\alpha\als}$ we refer to 
\subsect{mixed_bkqcd} and  
about $\delta\rho^{\alpha^2}$ to \subsect{EW_twoloops}. 

The sum of the two leading terms is:
\bq
-\lpar\delta\rho^{\sss{G}\als}_{\sss{L}}
+\delta\rho^{\sss{G}\als^2}_{\sss{L}}\rpar=3 x_t {\tt TBQCD0},
\label{TBQCD0}
\eq
where the second term is the {\tt AFMT} correction:
\ba
\delta\rho^{\sss{G}\als^2}_{\sss{L}} = {\tt AFMT}.
\ea
It
is accessed through variable {\tt TBQCD0}.
\subsection{$\dr$ beyond one loop\label{dr_beyond}}
We begin with a rearrangement of terms in \eqn{delta_r}:
\bqa
\dr&=&\dalphav+\frac{\alpha}{4\pi\siws}
\Biggl\{\siws
\Biggl[
-\frac{2}{3}-\Pgg^{\ft,F}\lpar 0\rpar-\Pgg^{\fl+5\fq,F}\lpar\mzs\rpar
\Biggr]
+\frac{\cows}{\siws}\Delta\rho^{F}
\nll &&
+\Delta\rho^{F}_{\sss{W}}
+\frac{11}{2}-\frac{5}{8}\cows\lpar 1+\cows\rpar
+\frac{9\cows}{4\siws}\ln{\cows}
\Biggr\}.
\label{drdalpha}
\eqa
Here the superscript $\fl+5\fq$ denotes a summation over leptons and
five light quarks.  
Remember now that any self-energy function in \eqn{drdalpha} is defined 
at the scale $\tHs=\mwl$ as an artifact of the definition:
\bqa
\bff{0}{\pmoms}{\Mind{1}}{\Mind{2}} &=&
 \dlt-\ln\frac{\Mind{1}\Mind{2}}{\tHss}
+\frac{\Lambda}{\pmoms}
 \ln\frac{-\pmoms-\ib\epsilon+\Minds{1}+\Minds{2}-\Lambda}{2\Mind{1}\Mind{2}}
\nll &&
 +~\frac{\Minds{1}-\Minds{2}}{2\pmoms}\ln\frac{\Minds{1}}{\Minds{2}}+2
\label{b0answer}
\\
&=& 
\dlt-\ln\frac{\mws}{\tHss} + \fbff{0}{\pmoms}{\Mind{1}}{\Mind{2}},
\label{b0m1m2}
\eqa
where $\Lambda^2=\lkall{\pmoms}{\Minds{1}}{\Minds{2}}$ is the K\"allen 
$\lambda$-function:
\ba
\lkall{x}{y}{z}=x^2+y^2+z^2-2xy-2xz-2yz.
\ea
However, we need them being defined at scale $\tHs=\mzl$ in the $\MSB$ spirit.
For instance, $\Delta\rho^{F}$ re-scales as follows:
\bqa
\frac{\cows}{\siwf}\Delta\rho^{F}\Big|_{\tHs=\mwl} &=&
\frac{\cows}{\siwf}\Delta\rho^{F}\Big|_{\tHs=\mzl} 
+\frac{1}{\siws}
\biggl[
 \frac{1}{6}\Nf-\frac{4}{3}\siws\asums{\ff}\cf\qfs
-\frac{1}{6}-7\cows
\biggr]
\ln{\cows}.
\eqa
That explains the definition of variable {\tt SCALE}:
\ba
\label{scale}
{\tt SCALE} = - \frac{\alpha}{4\pi\siws} 
\biggl[
 \frac{1}{6}\Nf-\frac{4}{3}\siws\asums{\ff}\cf\qfs
 -\frac{1}{6}-7\cows
 \biggr]
 \ln{\cows}.
\ea
  
After rescaling to $\tHs=\mzl$ one gets:
\bqa
\dr&=&\dalphav+\dalpha^{\ft}(\mzs)+\frac{\alpha}{4\pi\siws}
\Biggl\{\Biggl[\frac{\cows}{\siws}\Delta\rho^{F}+\Delta\rho^{F}_{\sss{W}}
-\siws\Pgg\lpar\mzs\rpar\Biggr]\Bigg|_{\tHs=\mzl}
\nll&&
-\frac{2}{3}\siws
+\lpar\frac{1}{6}\Nf-\frac{1}{6}-7\cows\rpar\ln{\cows}
+\frac{11}{2}-\frac{5}{8}\cows\lpar 1+\cows\rpar
+\frac{9\cows}{4\siws}\ln{\cows}
\Biggr\},
\label{dr_scaled}
\eqa
with
\bq
\dalpha^{\ft}(\mzs)=\frac{\alpha}{4\pi}
\Bigl[\Pgg^{\ft,F}\lpar\mzs\rpar-\Pgg^{\ft,F}\lpar 0\rpar\Bigr].
\label{dalphat}
\eq

In \eqn{dr_scaled} we disentangled the pure QED contributions
$\dalphav$ and $\dalpha^{\ft}(\mzs)$, the parameter $\drho$, and the rest. 

The one-loop result for $\dr$ contains two {\it large} terms: 
the running coupling $\dalphav$ 
({\em logarithmically} enhanced by fermionic mass singularities,
$\ln(\mfl/s)$),  
and the parameter $\drho$ (enhanced {\em quadratically} by $\mts$).
In the spirit of leading/remainder splittings it may be presented
in two ways: 
\bq
\dr=\dalpha(\mzs)+\frac{\cows}{\siws}\drho^{\alpha}_{\sss{L}}
+\dr_{\rm{rem}}
\equiv\dalpha(\mzs)+\frac{\cows}{\siws}\Delta{\hat{\rho}}^{\alpha}
+\drh_{\rm{rem}}.
\label{insert2}
\eq
The first term is:
\bq
\dalpha(\mzs)=\dalphav+\dalpha^{\ft}(\mzs),
\label{dal_first}
\eq
and the second term is the {\em leading} 
electroweak part of $\dr$.

In \eqn{insert2} all three terms are separately gauge invariant.
The full $\dr$ is gauge invariant, being the physical amplitude of
$\flm$-decay.  
The $\dalphav$ is gauge invariant since 
it contains only fermionic loops. Then, the second terms (\eqn{leading_1l}
for the first choice or \eqn{insert7} for the second choice)
are gauge invariant as discussed above.
As a consequence, both $\dr_{\rm{rem}}$ and
$\drh_{\rm{rem}}$ are also gauge invariant.

Different variants of leading/remainder splitting are governed by
the user option {\tt AMT4}, see \subsect{dizetflags}. We define:
\bqa
\dr = \dalpha(\mzs)-\dr_{\sss{L}}+\dr_{\rm{rem}}\,,
\label{insert9}
\eqa
where $\dr_{\sss{L}}$ (and consequently $\dr_{\rm{rem}})$ depends on 
{\tt AMT4}:
\bq
\dr^{(\alpha)}_{\sss{L}}=-\frac{\cows}{\siws}\times\left\{
\begin{array}{ll}
 \delta\rho^{\alpha}_{\sss{L}}
+\delta\rho^{\alpha^2}_{\sss{L}}
&\quad\mbox{for}\quad\mbox{\tt AMT4}=1,
{}\\
 \delta\rho^{\alpha}_{\sss{L}}
+\delta\rho^{\alpha^2}_{\sss{L}}
+\delta\rho^{\alpha\als}_{\sss{L}}
&\quad\mbox{for}\quad\mbox{\tt AMT4}=2,
{}\\
 \delta\hat{\rho}^{\alpha}
+\delta\rho^{\alpha^2}_{\sss{L}}
+\delta\rho^{\alpha\als}_{\sss{L}}
&\quad\mbox{for}\quad\mbox{\tt AMT4}=3.
\end{array}
\right. 
\label{dr_amt4}
\eq
Moreover, superscript $\alpha$ emphasizes the use of scale $\alpha$.  
Here we indicated also some higher order terms
in order to show the difference between different settings of {\tt AMT4}.
The {\it re-summed} version used in {\tt ZFITTER} for {\tt AMT4}=1,2,3 is
\bqa
\label{romus}
\frac{1}{1-\dr}&=&
\frac{1}{\Bigl[1-\dalpha(\mzs)\Bigr]
         \Bigl[1+f\dr^{(\alpha)}_{\sss{L}}\Bigr]
-\dr^{(\alpha)}_{\rm{rem}}} \,,
\eqa
where $f$ is the conversion factor of \eqn{conversionf}.
This old implementation (before the advent of next-to-leading
two-loop electroweak corrections) was described in~\cite{Bardin:1995a2}.
These options are not supported beginning with \zf\ v.5.10. 
We remind that the re-summation of $\dalpha(\mzs)$ is dictated by the
renormalization group. 
The re-summation of $\mtl$ enhanced terms comprises 
to special factorization in \eqn{romus} and
to {\it conversion} of the scale of $\dr^{(\alpha)}_{\sss{L}}$
from $\alpha$ to $\gf$ with the conversion factor (\ref{conversionf}).
An alternative notion is useful:
\bqa
\label{romusG}
\frac{1}{1-\dr}&=&
\frac{1}{\Bigl[1-\dalpha(\mzs)\Bigr]
   {\ds{\Biggl[1-\frac{\cows}{\siws}\drhovb^{(G)}\Biggr]}}
-\dr^{(\alpha)}_{\rm{rem}} }\,,
\eqa
where the superscript $G$ denotes the conversion to the scale $\gf$
in all terms, see \eqn{dr_amt4}.

 When \cite{Bardin:1995a2} was written it was not 
clear where to place $\dr_{\rm{rem}}$, either as in \eqn{romusG} 
or as in rows of Eq.~(99) of~\cite{Bardin:1995a2},
and which scale to use for it. 
That's why various {\it options} were considered, depending on flag
{\tt EXPR}. 
It may be easily seen that these options differ by terms of order 
${\cal{O}}(\gfs\mts\mzs)$.

This uncertainty is removed now by a calculation of order
${\cal{O}}(\gfs\mts\mzs)$
two-loop electroweak corrections enhanced by the top quark mass 
\cite{Degrassi:1996mg,Degrassi:1997ps}.
The leading term of order ${\cal{O}}(\gfs\mtq)$, which was known
since 1992 \cite{Barbieri:1992nz,Fleischer:1993ub}
was also re-calculated in \cite{Degrassi:1996mg,Degrassi:1997ps}.

The implementation of terms of ${\cal{O}}(\gfs\mts\mzs)$ into {\tt ZFITTER}
is realized by option {\tt AMT4}=4 which is by now the default.
The other options became obsolete.

In \cite{Degrassi:1996mg,Degrassi:1997ps}
it was shown that one should place the one-loop $\dr_{\rm{rem}}$ as follows:
\bqa 
\frac{1}{1-\dr}&=&\frac{1}{\Bigl[1-\dalpha(\mzs)-\dr^{(G)}_{\rm{rem}}\Bigr]
   {\ds{\Biggl[1-\frac{\cows}{\siws}\drhovb^{(G)}\Biggr]}}}\,.
\label{romus_deg}
\eqa

Irreducible higher order corrections are applied by means of a simple 
modification of the leading and remainder terms
\bqa
\drhovb&\rightarrow&\drhovb + \drhovb^{ho}, 
\\
\dr_{\rm{rem}}& \rightarrow & \dr_{\rm{rem}}+\dr^{ho}_{\rm{rem}}\,.
\label{ho_rem_corr}
\eqa
A higher order ({\it ho}) term is a sum of known corrections:
\bqa
\drhovb^{ho}&=&\drhovb^{G\als}+\drhovb^{G\als^2}+\drhovb^{G^2},
\\
\dr^{ho}_{\rm{rem}}&=&\dr^{G\als}_{\rm{rem}}+\dr^{G^2}_{\rm{rem}}\,.
\label{ho_avail}
\eqa

Terms of the orders ${\cal{O}}(\gfs\mtq)$ and ${\cal{O}}(\gfs\mts\mzs)$
are calculated in subroutine {\tt GDEGNL}, an interface to the package
{\tt m2tcor.f}~\cite{Degrassi:1996ZZ,Degrassi:1998oo}; 
see \subsect{EW_twoloops} for more details.
Complete two-loop mixed EW$\otimes$QCD corrections may be calculated by 
two packages: with
{\tt bcqcdl5$\_$14.f} for {\tt QCDC=1,2} \cite{Bardin:1989aa}, and 
with {\tt bkqcdc5$\_$14.f} for {\tt QCDC=3} (based on formulae of 
\cite{Kniehl:1990yc}, presented in \subsect{mixed_bkqcd}).
They produce equal results within the accuracy of integration. 
However, {\tt QCDC=3} is much faster 
since it uses analytically integrated expressions, while {\tt QCDC=2}
 relies on a numerical integration. 

We supply a dictionary:

for $\drho$:
\bqa
-\drhovb^{G} 
&=&{\tt DROBLO},
\\
-\bigl(\drhovb^{G\als}_{\sss L}+\drhovb^{G\als^2}_{\sss L}\bigr) 
&=&
 3 x_t {\tt TBQCD0},
\\
-\drhovb^{G^2}
&=&{\tt DRHOD},
\\
-\drhovb^{(G)} 
&=&{\tt DROBAR}; 
\label{dict_rho}
\eqa

for $\dr_{\rm{rem}}$:
\bqa
\dr^{\alpha}_{\rm{rem}}+\dr^{\alpha\als}_{\rm{rem}}
 +\dalpha^{\alpha\als}(\mzs) 
&=& {\tt DRREM}-{\tt DRHIG1},
\\
\dr^{\alpha\als}_{\rm{rem}}
&=& {\tt TBQCD} + 2 {\tt CLQQCD} - {\tt TBQCDL},
\\
\dr^{\alpha^2}_{\rm{rem}}  
&=&{\tt DRDREM},
\\
\dalphav+\dalphat &=& {\tt DALFA},
\\
\dalpha^{\alpha\als}(\mzs) &=& \frac{\alpha}{4\pi} {\tt ALQCD};
\label{dict_rem}
\eqa

and finally for the complete $\dr$:
\bq
\dr = {\tt DRBIG} = 1-\left(1+\frac{R}{1-R} {\tt DROBAR}-{\tt DRHIGS}\right)
      \bigl( 1-{\tt DALFA}-{\tt DRREM}-{\tt DRDREM}\bigr).
\label{dict_dr2}
\eq
\subsection{Simulation of theoretical uncertainties\label{options_tu}}
After the advent of next-to-leading two-loop electroweak corrections,
the options developed to simulate the theoretical uncertainties
as described in \cite{Bardin:1995a2} became obsolete and were critically
revisited. 
In this subsection we discuss new realizations.

With the aid of the option {\tt EXPR} we select different {\tt EXP}ansions
of $\dr$:  
\bqa
\deltar
= \left\{
\begin{array}{ll}
{\ds{ 1 - \lpar 1 - \frac{\ctws}{\stws}\delta\bar\rho-\drhigsg \rpar
          \lrbr 1 - \Delta\alpha-\drrem
             - \lpar\deltarremho + \dr^{\hb,\alpha^2}_L\rpar\Ksc \rrbr}},\\
{}\\
{\ds{ 1 - \lpar 1 - \frac{\ctws}{\stws}\delta\bar\rho-\drhigsg \rpar
                    \lpar 1-\dalpha-\drrem \rpar
             + \lpar\deltarremho + \dr^{\hb,\alpha^2}_L \rpar\Ksc}}, \\
{}\\
{\ds{ 1 - \lpar 1 - \frac{\ctws}{\stws}\delta\bar\rho-\drhigsg \rpar
          \lpar 1 - \Delta\alpha \rpar}}
{}\\ \hspace{3.8cm}
{\ds{        + \lpar 1 - \frac{\ctws}{\stws}\delta\rho^{G} \rpar\drrem
             + \lpar \dr^{\hb,\alpha^2}_L+\drrem \rpar \Ksc }}.
\end{array}
\right.
\label{dr_facr}
\eqa

We note that the first row ({\tt EXPR}=0) realizes the so-called 
OMS-I renormalization scheme 
\cite{Degrassi:1996mg,Degrassi:1997ps}, while the third row 
approaches the spirit of the OMS-II renormalization scheme -- a fully
expanded option -- with the second row being an intermediate step on 
the road from OMS-I to OMS-II. 

In \eqnst{dict_rho}{dict_rem}{dr_facr}
are two terms, which optionally affect
the re-summation of the leading Higgs contribution to $\dr$: 
\bqa
\drhigsg&=&\frac{\sqrt{2}\gf\mws}{4\pi^2}
\frac{11}{12}\lpar\ln \frac{\mhs}{\mws} - \frac{5}{6}\rpar
={\tt DRHIGS},
\\
\drhigsa&=&\frac{\alpha}{4\pi\siws}
\frac{11}{12}\lpar\ln \frac{\mhs}{\mws} - \frac{5}{6}\rpar
={\tt DRHIG1}. 
\label{drhiggs}
\eqa
We assume here:
\bqa
\ln \frac{\mhs}{\mws} - \frac{5}{6} \geq 0.
\eqa
For flag {\tt IHIGS}=1, the $\drhigs$
is extracted from the remainder with scale $\alpha/\siws$
and added to the leading term with scale $\gf$, as in \eqn{drhiggs}.
We observe that $10/12$ of $\drhigs$ is contained in $\drhovb$. 
Therefore, for {\tt AMT4}=4 only $1/12$ of it is additionally
re-summed. The influence of this option on theoretical errors was
found to be tiny. For this reason, this re-summation has been
implemented only for $\dr$ and not for other POs.

With the options {\tt HIG2}=1/0 it is possible to switch on/off the 
quadratically enhanced two-loop Higgs contribution
\cite{vanderBij:1984bw,vanderBij:1984aj}:
\bq
{\tt  DRHHS} = 
\deltar^{\hb,\alpha^2}_L
=-0.005832 \frac{\alpha^2}{\pi^2\stwf} \frac{\mhs}{\mws}\,.
\label{dr_h}
\eq

With the choices {\tt SCRE}=0,1,2 one changes the 
scale in the two-loop remainder terms as in the following three rows 
\bq
 K=\left\{
\begin{array}{ll}
    1  ,    \\
    f^2 ,   \\ 
    f^{-2} .   
\end{array}
\right. 
\label{scaler-d} 
\eq
The default is {\tt SCRE}=0. Then one uses the scale as suggested by
the author of {\tt m2tcor.f}. 
For more details see \subsect{EW_twoloops}.
With conversion factor $f^{-2}$ it is converted back to the 
scale $\alpha/\siws$ and this operation is symmetrized in order to have
symmetric errors due to change of scale of remainders.

The option {\tt SCAL}=0,1,2,3,4 is the only QCD option.
At the default, {\tt SCAL}=0, we implemented the exact {\tt AFMT} correction.
For {\tt SCAL}=1,2,3 we implemented the $\xi$-factor as given
in~\cite{Kniehl:1995yr}. 
Finally, for {\tt SCAL}=4,
Sirlin's scale $\xi=0.248$ (see~\cite{Sirlin:1995yr})
was implemented. 
Only {\tt SCAL}=0,4 are left among the working options.

The last step in subroutine {\tt SEARCH} is the calculation of a factor $A$,
\bqa
A=\frac{A_0}{ \sqrt{1-\dr}}={\tt AAFAC},
\label{dr_facrr}
\eqa
that is used for various options {\tt IMOMS},
see the iterations in~\eqnsc{wmass}{zmass}.

\section{Partial and Total $\zb$ Decay Widths\label{subgamma}}
\eqnzero
The one-loop electroweak corrections to the $Z$ boson width are
calculated in {\tt ZFITTER} following \cite{Akhundov:1986fct},
higher orders are treated as described in \cite{Bardin:1995a2}, and  the QCD
corrections follow \cite{Chetyrkin:1994js3}, and references therein. 
\subsection{Notations \label{notation}}
We recall the amplitude of the decay $\zb\to\ff\fbf$ in one-loop 
approximation and in the approximation of vanishing external fermion 
masses~\cite{Akhundov:1986fct,Bardin:1989dit}, given by:
\bq
V^{\zb\ff\fbf}_{\mu}\lpar\mzs\rpar = 
\lpar 2\pi\rpar^4\ib\,\ib\,\sqrt{\sqrt{2}\gf\mzs} 
\sqrt{\rho^{\ff}_{\sss{\zb}}}\tcif\gamma_\mu
\lrbr\lpar 1+\gfd \rpar - 4|\qd|\stws\kappa^{\ff}_{\sss{\zb}}\rrbr,
\label{decayrhokappadef}
\eq
where $\rZf$ and $\kZf$ are called {\em effective couplings} of 
$\zb$-decay; they are some complex-valued constants.

  The two effective couplings for each fermionic partial width of the $\zb$ 
boson are computed in subroutine {\tt FOKAPP}.
The relations between variables of the code and the
notations used in \eqn{decayrhokappadef} are:
\bqa
{\tt XRO1}&=&\rZdf{\ff}\equiv\rZdf{\ff,\alpha},
\\
{\tt XAK1}&=&\kZdf{\ff}\equiv\kZdf{\ff,\alpha},
\eqa
where the superscript $\alpha$ is introduced
in order to emphasize the perturbative order and scale.
For their ingredients (on top of dictionary \eqn{dict_dr}) we have:
\bqa
\mbox{\tt XZFM1A}&=&\zmz,
\\
\mbox{\tt XZM1A-W0A}&=& -\delrho{F}_{\sss{Z}},
\\
\mbox{\tt XAMM1A}&=&\Pzg^{F}(\mzs),
\\
{\tt XV1ZZ}&=&\cvetril{}{\sss{\zb}}{\mzs},
\\
{\tt UFF} &=& u_{\ff}.
\eqa

We use two expressions for the partial decay widths of 
the $\zb$ boson. One for decays into a pair of leptons, $\fl=\fe,\flm,\flt$: 
\bqa
\gll\equiv\Gamma\lpar\zb\to\fl\barl\rpar&=&\Gamma_0\big|\rZl\big|
\sqrt{1-\frac{4\mls}{\mzs}}
\lrbr\lpar 1+\frac{2\mls}{\mzs}\rpar\bigl(|\Rvaz{\fl}|^2 + 1)
-\frac{6\mls}{\mzs}\rrbr
\nll && \times
\lrbr 1+\frac{3}{4}\frac{\alpha(\mzs)}{\pi}\qfs\rrbr
\label{defzwidthl},
\eqa
and another one for decays into a pair of quarks,
$\ff=\fu,\fd,\fc,\fs,\fb$: 
\bq
\gff\equiv\Gamma\lpar\zb\to\ff\fbf\rpar=\Gamma_0\,\cf\,\big|\rZf\big|
\Bigl[
|\Rvaz{\ff}|^2\,R^{\ff}_{\sss{V}}(\mzs) + R^{\ff}_{\sss{A}}(\mzs)
\Bigr] 
+\Delta_{_{\rm EW/QCD}}\;.
\label{defzwidthq}
\eq
Here,
\bq
\Gamma_0 = \frac{\gf\mzc}{24\srt\,\pi} = 82.945(7)\,\mbox{MeV}
\eq
is the `standard' partial width. The complex-valued variable
\bq
\Rvaz{\ff} = \frac{\vc{\ff}}{\ac{\ff}}
=1-4|\qf|\bigl(\kZdf{\ff}\siws+\Imsi{\ff}\bigr)
\label{varatiorez}
\eq
is the ratio of effective vector and axial couplings\footnote{Beware
of two different definitions of $\vc{\ff}$ and $\ac{\ff}$ 
used in this description. 
We use here
the definitions 
\eqn{ratio}, while in \zf\ and everywhere else in the description we use
instead \eqns{kg1}{kg3}, 
i.e. quantities normalized by $\tcif$.
},
\bqa
\vc{\ff}&=&\tcif-2\qf\bigl(\kZdf{\ff}\siws+\Imsi{\ff}\bigr)\,,
\\
\ac{\ff}\,=\,\tcif.
\label{ratio}
\eqa
Finally, $\cf=3$ is the color factor. 

One should note the appearance of the order $\ord{\alpha^2}$ term $\Imsi{\ff}$ 
which denotes a real-valued contribution originating from 
the product of two imaginary parts of finite parts of polarization 
operators:
\bq
\Imsi{\ff}=\lpar\frac{\alpha(\sman)}{4\,\pi}\rpar^2
\Imb\Pgg(\sman)\;\Imb\Pzg(\sman)
=\alpha^2(\sman)\frac{35}{18}
\lrbr 1-\frac{8}{3}\Reb\bigl(\kZdf{\ff}\bigr)\siws\rrbr,
\label{addseffim}
\eq
giving a sometimes non-negligible contribution due to the enhancement
by the factor $\pi^2$. 
Note that it bears the channel index $\ff$.

The radiator factors
\bq
R^{\ff}_{\sss{V,A}}(\mzs)
\label{QCDradiators}
\eq
describe the final state QED and QCD vector and axial vector corrections 
for quarkonic decay modes. 
They are described in \subsect{qcdrun}.

The combinations
\bq
\seffsf{\ff} = \Reb\bigl(\kZdf{\ff}\bigr)\siws+\Imsi{\ff}
\label{sw2eff}
\eq
define the flavor dependent {\em effective weak mixing angles}. 

The \eqnss{defzwidthl}{defzwidthq} differ by the
treatment of final 
state corrections and finite mass effects originating from the Born amplitude
and from the final state QCD corrections. 
For leptonic decay modes, 
\eqn{defzwidthl}, the mass effects are separated from the final state QED 
correction factor since only Born-induced mass effects are taken into account. 
For quarkonic decay modes, QED, QCD, mixed QED$\otimes$QCD
corrections and finite mass corrections (the latter are accounted for in terms
of {\em running} masses) are included all together by means of radiator 
factors $R^{\ff}_{\sss{V,A}}$.
Another important feature of quarkonic widths, $\gff$, is the presence of 
the {\em non-factorizable} EW$\otimes$QCD corrections 
$\Delta_{_{\rm EW/QCD}}\,$
\cite{Czarnecki:1996ei,Harlander:1998zb}, 
which will be referred to as {\tt CKHSS} corrections.
Their handling is governed by the flag {\tt CZAK}.

Since these corrections have a relative order of magnitude of less than one
per mill, they are implemented in {\tt ZFITTER} as fixed numbers
taken from \cite{Czarnecki:1996ei,Harlander:1998zb}:
\bqa 
\Delta_{_{\rm EW/QCD}} = {\tt ARCZAK} = \left\{
\begin{array}{rl}
-0.113&\mbox{MeV, ~for $\fu$ and $\fc$ quarks,} \\
-0.160&\mbox{MeV, ~for $\fd$ and $\fs$ quarks,} \\
-0.040&\mbox{MeV, ~for $\fb$ quark. }
\end{array}
\right.
\label{ck_hhs}
\eqa

 The form factors comprise one-loop and higher order virtual
electroweak and {\em internal} QCD corrections, see \subsect{mixed}.
Their implementation relies on the notion of leading and remainder terms. 

 Here we present only the realization for {\tt AMT4}=4, i.e. that with 
next-to-leading two-loop electroweak corrections. 
First of all we have to
rewrite identically the one-loop expressions 
in \eqn{decayrhokappadef} in order to introduce a {\em
leading-remainder splitting}:  
\bqa
\rZf&=&1-\delta\hat{\rho}^{\alpha}+\drhov^{\ff,\alpha}_{\rm{rem}}\,,
\\
\kZf&=&1-\frac{\cows}{\siws}\delta\hat{\rho}^{\alpha}
                                  +\dkapv^{\ff,\alpha}_{\rm{rem}}\,.
\eqa
 These expressions uniquely define leading terms containing
$\delta\hat{\rho}^{\alpha}$  and remainder terms containing the rest.
\subsection{Simulation of theoretical uncertainties}
In a complete analogy with \eqn{dr_facr} one may develop options to simulate
theoretical errors for $\rZf$ and $\kZf$ due to non-calculated orders.
They are governed by the flag {\tt EXPF}=0,1,2, correspondingly.
For $\rZf$ one has
\bqa
\rZf
= \left\{
\begin{array}{ll}
{\ds{\frac{1+\drhov^{\ff,[G]}_{\rm{rem}}+\drhov^{\ff,G^2}_{\rm{rem}}\Ksc}
          {1+\drhovb^{(G)}\,\lpar 1-\Delta{\bar{r}}^{[G]}_{\rm{rem}}\rpar}}}\;,
\\{}\\
{\ds{\frac{1+\drhov^{\ff,[G]}_{\rm{rem}}}
          {1+\drhovb^{(G)}\,\lpar 1-\Delta{\bar{r}}^{[G]}_{\rm{rem}}\rpar}
            +\drhov^{\ff,G^2}_{\rm{rem}}\Ksc }}\,,
\\{}\\
{\ds{1-\drhovb^{(G)}+\lpar\drhovb^{G}\rpar^2
            +\drhovb^{G}\Delta{\bar{r}}^{[G]}_{\rm{rem}}
            +\drhov^{\ff,[G]}_{\rm{rem}}
      \lpar 1-\drhovb^{G}\rpar+\drhov^{\ff,G^2}_{\rm{rem}}\Ksc }}\,.    
\end{array}
\right.
\label{drho_fact}
\eqa
Here superscript $(G)$ stands for the inclusion of all known terms,
as in the first \eqn{ho_rem_corr}, while superscript $[G]$ denotes the
inclusion of one-loop EW corrections together with  
all known orders in $\als$, i.e. $[G]=G+G\als+\dots$
We also note that the $\rho$ and $\kappa$-remainders are channel-dependent
quantities ($\ff$-dependent).

Moreover, it is:
\bq
\Delta{\bar{r}}^{[G]}_{\rm{rem}}=\Delta{\bar{r}}^{G}_{\rm{rem}}
                                +\dr^{G\als}_{\rm{rem}},
\eq
with the one-loop remainder \cite{Degrassi:1998oo}:
\bqa
\Delta{\bar{r}}^{G}_{\rm{rem}}&=&
\frac{\srt\gf\mzs\siws\cows}{4\pi^2}
\biggl\{
-\frac{2}{3}
+\frac{1}{\siws}
\biggl[
 \frac{1}{6}\Nf-\frac{1}{6}-7\cows
\biggr]
\ln{\cows}
\nll &&
+\frac{1}{\siws}
\biggl[
\Delta\rho^{\sss{F}}_{\sss{W}}
+\frac{11}{2}-\frac{5}{8}\cows\lpar 1+\cows\rpar
+\frac{9\cows}{4\siws}\ln{\cows}
\biggr]
\biggr\}.
\label{drrembar}
\eqa
Finally, we note that all rows for user options {\tt EXPF}
are analogously as for {\tt EXPR}, \eqn{dr_facr}.

Similarly, for $\kZf$ (realized also by flag {\tt EXPF}):
\bqa
\kZf
= \left\{
\begin{array}{ll}
{\ds{\lpar 1+\dkapv^{\ff,[G]}_{\rm{rem}}+\dkapv^{\ff,G^2}_{\rm{rem}}\Ksc\rpar
\lrbr
1-\frac{\cows}{\siws}\drhovb^{(G)}\lpar 1-\Delta{\bar{r}}^{[G]}_{\rm{rem}}\rpar
\rrbr}},
\\{}\\
{\ds{\lpar 1+\dkapv^{\ff,[G]}_{\rm{rem}}\rpar
\lrbr
1-\frac{\cows}{\siws}\drhovb^{(G)}\lpar 1-\Delta{\bar{r}}^{[G]}_{\rm{rem}}\rpar
\rrbr
+\dkapv^{\ff,G^2}_{\rm{rem}}\Ksc\,,}}
\\{}\\
{\ds{1-\frac{\cows}{\siws}\drhovb^{(G)}
 +\frac{\cows}{\siws}\drhovb^{G}\Delta{\bar{r}}^{[G]}_{\rm{rem}}
+\dkapv^{\ff,[G]}_{\rm{rem}}
\lpar 1-\frac{\cows}{\siws}\drhovb^{G}\rpar
+\dkapv^{\ff,G^2}_{\rm{rem}}\Ksc\,.}}
\\
\end{array}
\right.
\label{dkap_fact}
\eqa
In \subsect{EW_twoloops} we will show how the quantities introduced in
\eqnsc{drho_fact}{dkap_fact} are related to variables supplied by the package
{\tt m2tcor.f} \cite{Degrassi:1996ZZ}. 

Since the next-to-leading two-loop EW corrections are not known for the 
$\zb\to\fb\barb$ channel, the code internally realizes {\tt AMT4}=3 for this
channel. 
That's why we refer the reader to old publications for the
description of the implementation of higher orders and theoretical options
\cite{Bardin:1995a2,Bardin:1992jc}.
The one-loop corrections for the $\zb\to\fb\barb$ channel are fully described
elsewhere.

 The $\zb$ decay rate is calculated in subroutine {\tt ZWRATE}, where
also the $\wb$ decay rate is calculated \cite{Bardin:1986fi}.
\subsection{Weak form factors for $\zb\to\fb\barb$\label{zbbff}}
Due to the large mass splitting between the $\ft$ and $\fb$ quarks, there
are two one-loop vertex diagrams for the $\zb$ decay into $\fb$ quarks
and also higher order vertex
corrections, which 
contribute additional {\em non-universal} $\mtl$-dependent terms that are 
absent in the cases of light quarks ~\cite{Akhundov:1986fct,Mann:1984} 
(see also \cite{Beenakker:1988pv,Bernabeu:1988me,Bernabeu:1991ws}).
Their leading term and 
the corresponding higher order corrections are taken
into account with the correction $\tau_{\fb}$:
\bq
\tau_{\fb} = -2 x_t\left[1-\frac{\pi}{3}\als(\mts) 
+ x_t \tau^{(2)}\left(\frac{\mts}{\mhs}\right) \right] 
= {\tt TAUBB1} + {\tt TAUBB2}.
\label{taubb}
\eq
For the first term, $tau_{\fb} = -2 x_t$,
defined in \eqn{xt}, we refer to
\cite{Bardin:1992jc}, 
for the second term -- which will be called {\tt FTJR} correction --
to \cite{Buchalla:1993zm,Fleischer:1992fq,Degrassi:1993ij,Chetyrkin:1993jp},
and for the last one to~\cite{Barbieri:1992nz}.

Since the first term represents a one-loop correction with scale $\gf$,
one has to subtract it with scale $\alpha$ from the $\zb\fb\barb$ 
decay amplitude form factor $F_{\sss{L}}$:
\bqa
V^{\zb\ff\fbf}_{\mu}\lpar\sman\rpar
 &=& \lpar 2\pi\rpar^4\ib\frac{\ib g^3}{16\pi^2 2\cow}\gamma_\mu
\lrbr I^{(3)}_f
 F_{\ZL}\lpar\sman\rpar \lpar 1+\gfd \rpar - 2\qd\stws
 F_{\ZQ}\lpar\sman\rpar \rrbr,
\label{fourformdef}
\\
F_{\sss{L}} &\to& F_{\sss{L}}
-\lpar -\frac{\alpha}{8\pi\stws}\frac{\mts}{\mws}\rpar.
\label{subt_zbb}
\eqa
If $\rho_{\fb}$ and $\kappa_{\fb}$ denote couplings based on the 
subtracted quantity \eqn{subt_zbb}, then $\tau_b$ has to be accounted
for by means of the following replacements:
\bqa
\rho_{\fb} &\to& \rho_{\fb}\lpar 1 + \tau_{\fb} \rpar^2,
\\
\kappa_{\fb}&\to& \frac{\kappa_{\fb}}{\lpar 1+\tau_{\fb}\rpar}\,,
\label{barbierizationofbb}
\eqa
as was proven in~\cite{Barbieri:1992nz}.

A compact form of the function $\tau^{(2)}$ may be taken 
from~\cite{Fleischer:1993ub}.

At the end we give the two leading terms of the unsubtracted $F_{\sss{L}}$ in  
the limit of large $\ft$-quark masses \cite{Akhundov:1986fct}:
\bqa
 F_{\sss{L}}\approx-\frac{\alpha}{8\pi\stws}\left|V_{\ft\fb}\right|^2
 \lrbr \frac{\mts}{\mws} + \lpar\frac{8}{3}+\frac{1}{6\ctws}\rpar
 \ln\frac{\mts}{\mws} \rrbr.
\eqa
\section{Table of Pseudo-Observables\label{fullPOs}}
\eqnzero
We complete the section about pseudo-observables 
with a full list of their {\em definitions}, 
with indication of their locations in {\tt COMMON} blocks, and
with a Table of all POs accessible in {\tt ZFITTER}.

In order to compute POs, only a part of the {\tt ZFITTER} package is needed.
POs may be computed with a main routine {\tt zfmain$\_$PO.f} which has to be
compiled together with the following packages: {\tt dizet6$\_$21.f},
{\tt zf514$\_$aux.f}, {\tt m2tcor5$\_$11.f}, {\tt bkqcdl5$\_$14.f}, and {\tt
bcqcdl5$\_$14.f}. 

All accessible POs are listed in \tabn{tabPOs}. 
In the first two rows we see the $\wb$ boson mass and $\siws=1-\mws/\mzs$.
Given the discussion in the previous Section about the calculation of $\mwl$
in {\tt ZFITTER}, these quantities need not be defined. 
Further, all partial $Z$ widths are shown, then the invisible width,
simply equal to $3\gn$, then the total hadronic width, equal to the
sum of all quarkonic widths, and finally the total width.

Then there is a group of quantities straightforwardly derived from widths:
first the ratios
\ba 
R_{\fl}&=&\frac{\gh}{\gel}\,,
\\
R^0_{\ff}&=&\frac{\gff}{\gh}\,,
\ea
and then the hadronic and leptonic pole cross-sections 
\ba
\sigma^0_{\had} &=& 12\pi\,\frac{\gel\gh}{\mzs\gzs}\,,
\\
\sigma^0_{\ell} &=& 12\pi\,\frac{\gel\gl}{\mzs\gzs}\,.
\ea

Next there follow three effective weak mixing angles defined by \eqn{sw2eff}
and three $\rZdf{\ff}$ parameters ($\ff=\fe,\fb,\fc$) 
defined by the first of \eqn{drho_fact}.
The so-called {\em coupling factors} are trivially made out of the {\em
real parts}  of the ratios of effective couplings \eqn{ratio}: 
\bq
{\cal A}_{\ff}=\frac{2\,\Reb\,\Rvaz{\ff}}{\Bigl(\Reb\,\Rvaz{\ff}\Bigr)^2+1}\,.
\label{def_coupling}
\eq
They are also known as {\em left-right flavor asymmetries}.
Finally, {\em forward-backward peak asymmetries} are mere combinations of these
coupling factors:
\bq
\afba{0 f}=\frac{3}{4}\,{\cal A}_{\fe}\,{\cal A}_{\ff}.
\label{def_afbs}
\eq
\clearpage
\begin{table}[ht]
\vspace*{1cm}
\begin{center}
\begin{tabular}{|c||r|r|r|r|}
\hline
Observable             &  minimal  & preferred &  maximal  &  interval  \\
\hline
\hline
$\mwl\,$[GeV]          &  80.3669  &  80.3738  &  80.3742  &  0.73  MeV \\
$\siws$                &  0.22309  &  0.22310  &  0.22323  &  0.00013   \\
\hline
\hline
$\gn\,$[MeV]           &  167.215  &  167.234  &  167.239  &  0.024 MeV \\ 
$\gel\,$[MeV]          &   83.983  &   83.995  &   83.999  &  0.016 MeV \\
$\gmu\,$[MeV]          &   83.982  &   83.995  &   83.998  &  0.016 MeV \\
$\gt\,$[MeV]           &   83.793  &   83.805  &   83.808  &  0.015 MeV \\
$\gu\,$[MeV]           &  300.092  &  300.154  &  300.168  &  0.076 MeV \\
$\gd\,$[MeV]           &  382.928  &  382.996  &  383.012  &  0.084 MeV \\ 
$\gc\,$[MeV]           &  300.031  &  300.092  &  300.106  &  0.075 MeV \\ 
$\gbq\,$[MeV]          &  375.886  &  375.993  &  375.995  &  0.109 MeV \\  
\hline
$\gi\,$[GeV]           &  0.50165  &  0.50170  &  0.50172  &  0.071 MeV \\
$\gh\,$[GeV]           &  1.74187  &  1.74223  &  1.74227  &  0.40  MeV \\
$\gz\,$[GeV]           &  2.49527  &  2.49573  &  2.49578  &  0.51  MeV \\ 
\hline
\hline
$R_{\fl}$              &  20.7406  &  20.7420  &  20.7421  &  0.0015    \\
$R^0_{\fb}$            &  0.215786 &  0.215811 &  0.215813 &  0.000027  \\
$R^0_{\fc}$            &  0.172243 &  0.172246 &  0.172252 &  0.000009  \\
\hline
$\sigma^0_{\had}\,$[nb]&  41.4777  &  41.4777  &  41.4785  &  0.8   pb  \\
$\sigma^0_{\lep}\,$[nb]&   1.9997  &   1.9997  &   1.9999  &  0.2   pb  \\
\hline
\hline
$\seffsf{\rm{lept}}$   &  0.231594 &  0.231601 &  0.231653 &  0.000059  \\
$\seffsf{\fb}$         &  0.232941 &  0.232950 &  0.233004 &  0.000063  \\ 
$\seffsf{\fc}$         &  0.231488 &  0.231495 &  0.231547 &  0.000059  \\
\hline
$\rhoe$                &  1.00516  &  1.00528  &  1.00532  &  0.00016   \\
$\rhoi{\fb}$           &  0.99403  &  0.99424  &  0.99424  &  0.00021   \\
$\rhoi{\fc}$           &  1.00586  &  1.00598  &  1.00601  &  0.00015   \\
\hline
\hline
${\cal A}_{\fe}$       &  0.145988 &  0.146396 &  0.146452 &  0.000464  \\  
${\cal A}_{\fb}$       &  0.934573 &  0.934607 &  0.934613 &  0.000040  \\ 
${\cal A}_{\fc}$       &  0.667416 &  0.667595 &  0.667620 &  0.000204  \\
\hline
$\afba{0,\fl}$         &  0.015984 &  0.016074 &  0.016086 &  0.000102  \\
$\afba{0,\fb}$         &  0.102327 &  0.102617 &  0.102657 &  0.000330  \\
$\afba{0,\fc}$         &  0.073076 &  0.073300 &  0.073331 &  0.000255  \\
\hline
\end{tabular}
\caption[Predictions for pseudo-observables from {\tt ZFITTER}]
{\it
Predictions for pseudo-observables from {\tt ZFITTER}
\label{tabPOs}}
\end{center}
\end{table}
\clearpage
\subsection{Pseudo-Observables in common blocks of {\tt DIZET} \label{POCOMM}} 
Channel dependent quantities are stored in the common block
{\tt COMMON/CDZRKZ/} and 
in the array {\tt PARTZ(0:11)}; we quote from the code:
\begin{verbatim}
*
      COMMON/CDZRKZ/ARROFZ(0:10),ARKAFZ(0:10),ARVEFZ(0:10),ARSEFZ(0:10)
     &             ,AROTFZ(0:10),AIROFZ(0:10),AIKAFZ(0:10),AIVEFZ(0:10)
*---
      DIMENSION NPAR(21),ZPAR(31),PARTZ(0:11),PARTW(3)
*---
      CALL DIZET
     &(NPAR,AMW,AMZ,AMT,AMH,DAL5H,ALQED,ALPHST,ALPHTT,ZPAR,PARTZ,PARTW)
*
\end{verbatim}

\noindent The correspondences are as follows:
\bqa
\mbox{\tt ARROFZ(0:10)    }&=&(\rZdf{\ff})^{'}\,,
\\
\mbox{\tt AROTFZ(0:10)    }&=&\Reb\,\rZdf{\ff}\,,
\\
\mbox{\tt ARKAFZ(0:10)    }&=&\Reb\,\kZdf{\ff}\,,
\\
\mbox{\tt ARVEFZ(0:10)    }&=&\Reb\,\Rvaz{\ff}\,,
\\
\mbox{\tt AIROFZ(0:10)    }&=&\Imb\,\rZdf{\ff}\,,
\\
\mbox{\tt AIKAFZ(0:10)    }&=&\Imb\,\kZdf{\ff}\,,
\\
\mbox{\tt AIVEFZ(0:10)    }&=&\Imb\,\Rvaz{\ff}\,,
\\
\mbox{\tt ARSEFZ(0:10)    }&=&\seffsf{\ff}\,,
\\
\quad\mbox{\tt PARTZ(0:11)}&=&\gff\,,
\label{CDZRKZ}
\eqa
with the usual {\tt ZFITTER} channel assignments (see \fig{interfEWCOUP}),
and with all quantities but $(\rZdf{\ff})^{'}$ already defined.
The quantity $(\rZdf{\ff})^{'}$ absorbs imaginary 
parts of $\Rvaz{\ff}$. 
Namely, if one defines the partial widths \eqnsc{defzwidthl}{defzwidthq} using 
$(\Rvaz{\ff})^2$ instead of $|\Rvaz{\ff}|^2$ one has to use the 
pre-factor $(\rZdf{\ff})^{'}$ instead of $|\rZdf{\ff}|$. 
Both options {\tt MISC=1/0} (with $(\rZdf{\ff})^{'}/|\rZdf{\ff}|$) 
are used in the so-called {\em Model Independent interfaces} of {\tt ZFITTER},
see \appendx{interfaces}.

In view of the potential use of quantities $(\rZdf{\ff})^{'}$, another
common block, {\tt  COMMON /CDZAUX/}, deserves an explanation.
We again quote from the code:
\begin{verbatim}
* 
      COMMON /CDZAUX/PARTZA(0:10),PARTZI(0:10),RENFAC(0:10),SRENFC(0:10)
*
\end{verbatim}

It is:
\bqa
\reni{\ff}&=&{\tt RENFAC(0:10)}=
\frac{\gz \mbox{~computed with~~~~ Im parts of effective couplings}}
     {\gz \mbox{~computed without  Im parts of effective couplings}}\,,
\nll\\
\sreni{\ff}&=&{\tt SRENFC(0:10)}.
\eqa
The vectors {\tt PARTZA(0:10)} and {\tt PARTZI(0:10)} are for expert
use only. 
                        
\chapter{Improved Born Cross-Section\label{ch-IBA}}
\section{Introduction\label{improvedborn}}
\eqnzero
Having discussed POs, we have to make the next step towards 
{\em realistic observables}, i.e. various cross-sections and asymmetries
with experimental cuts. 
We have to construct the so-called 
{\em improved Born approximations}, IBAs.
They are also called sometimes doubly deconvoluted
observables~\cite{Bardin:1999gt} since 
they are made free of initial state QED 
and final state QED$\otimes$QCD corrections.
The latter two groups of corrections form two separately gauge invariant 
sub-sets of diagrams. 
All remaining diagrams contribute to the IBA: purely EW and {\em
internal} QCD corrections. 

In the first three Sections of this Chapter 
we introduce some of the ingredients needed to construct the IBA.
First we discuss the $\zb$ boson parameter transformation, then the one-loop 
EW form factors, and finally the two classes of higher order corrections:
internal, mixed EW$\otimes$QCD corrections and two-loop EW corrections.

The calculation of the IBA from these ingredients does not belong to
the {\tt DIZET} 
package and will be discussed in \sects{ewcoup}{IBORNA}.
\section{The $\zb$ Propagator\label{Zpropagator}}
\eqnzero
The $\zb$ boson propagator in the Breit-Wigner form appears already at the 
level of the Born cross-sections.
It ensures the finiteness of the Born cross-section.

In general, the imaginary part depends on $\sman$:
\bqa
{\cal K}_{\sss{Z}}(\sman) &=& \frac{\sman}{\sman-{m^{2}_{\sss{Z}}}}\,,
\label{propagn}
\\
{{m}^{2}_{\sss{Z}}} &=& \mzs - \ib\mzl\gz(\sman).
\label{Gamzs}
\eqa
At LEP, it is common to use the following definition for the $Z$ width 
function (default, for flag {\tt GAMS}=1):
\bq
\gz(\sman) = \frac{\sman}{\mzs}\gz\,.
\label{mg}
\eq
This is an approximation, valid far away from production thresholds.
In the Standard Model the $\zb$ boson width $\gz$ is predicted as a
result of quantum corrections and calculated in {\tt ZFITTER} according to the
formulae of \sect{subgamma}.

In another approach the $\zb$ width function is treated as a constant
(chosen with flag {\tt GAMS}=0):
\bqa
{\bar \Gamma}_{\sss{Z}}(\sman) = {\bar \Gamma}_{\sss{Z}},
\\
{\bar {\cal K}}_{\sss{Z}}(\sman)&=&\frac{\sman}{\sman-{\bar m}_{\sss{Z}}^2}\,,
\label{propagb}
\\
{\bar m}_{\sss{Z}}^2 &=& 
{\bar M}_{\sss{Z}}^2 - \ib{\bar M}_{\sss{Z}} \Gamma_{\sss{Z}}.
\label{Gamzb}
\eqa
The following equality holds \cite{Bardin:1988xt}:
\bqa
G_{\mu} {\cal K}_{\sss{Z}}(\sman) \equiv  
    {\bar G}_{\mu} {\bar {\cal K}}_{\sss{Z}}(\sman),
\label{equivmzmz}
\eqa
with: 
\bqa
{\bar M}_{\sss{Z}} &=&
\frac{1}{\sqrt {1 + \gzs / \mzs}} \mzl
      \approx 
\mzl - \frac{1}{2} \frac{\Gamma^2_{\sss{Z}}}{\mzl} 
      \approx  
\mzl - 34 \; {\rm{MeV}},
\label{mbar}
\\
{\bar \Gamma}_{\sss{Z}} &=&
\frac{1}{\sqrt{ 1 + \Gamma^2_{\sss{Z}}/\mzs}}\gz
      \approx 
\gz - \frac{1}{2} \frac{\Gamma^3_{\sss{Z}}}{\mzl}
      \approx 
\gz - 1\ {\rm{MeV}},
\label{gbar}
\\
{\bar G}_{\mu} &=&\frac {G_{\mu}} {1+\ib\gz/\mzl}\,.
\label{gmub}
\eqa
The choice of the definition of the $\zb$ mass and width is
left to the user, see flag {\tt GAMS}, defined in \appendx{zuflag}.
Additional literature on the definition of the $\zb$ boson parameters
may be found in \cite{Bardin:1995a2,Riemann:1997tj}.  
\section{One-Loop Electroweak Form Factors\label{OLA_ff}}
\eqnzero
The amplitude of the process $\fep\fem\to\ff\fbf$ in the Standard Model 
gets in the one-loop approximation contributions from self-energy insertions, 
vertex corrections, box diagrams, and bremsstrahlung diagrams.
We divide these diagrams into several gauge invariant subsets.

We recall that first of all we disentangle a QED subset: QED-vertices
and fermionic self-energies, $\ph\ph$ and $\zb\ph$ boxes and bremsstrahlung,
see \sect{chains}.

Then, 
we divide the 
remaining diagrams into two more gauge-invariant subsets, giving rise to
two {\em improved (or dressed)} amplitudes: 
i) improved $\ph$ exchange amplitude with running QED-coupling 
where only fermion loops contribute (see \eqn{Born_modulo}), 
and 
ii) improved $\zb$ exchange amplitude with four, in general 
complex-valued {\em EW form factors}
$\rho_{ef},\kappa_{e},\kappa_{f},\kappa_{ef}$: 
\bqa
{\cal A}^{\sss{OLA}}_{\sss{\zb}}(\sman,\tman)&=& 
        \ib\,e^2\,4\,\tcie\tcif\frac{\chi_{\sss{Z}}(\sman)}{\sman} 
        \rho_{ef}(\sman,\tman)
        \biggl\{
        \gadu{\mu}{\lpar 1+\gfd \rpar }
        \otimes \gadu{\mu} { \lpar 1+\gfd \rpar}    
\nll\nll &&
-4 |\qe | \stws \kappa_e(\sman,\tman)
        \gadu{\mu} \otimes \gadu{\mu}{\lpar1+\gfd\rpar}
       -4 |\qf | \stws \kappa_f(\sman,\tman)
        \gadu{\mu} {\lpar 1+\gfd \rpar } 
        \otimes \gadu{\mu}                              
\nll\nll &&
+16 |\qe \qf| \stwf \kappa_{e,f}(\sman,\tman)
        \gadu{\mu} \otimes \gadu{\mu} \biggr\}.
\label{processrhokappadef}
\eqa
\newpage

\noindent
The form factors are simply related to the one-loop form factors
\bqa
\label{form1}
\rho_{ef}  &=& 1+\vvertil{}{\sss{LL}}{\sman,\tman}-\siws\Delta r,       
\\
\kappa_{e} &=& 1+\vvertil{}{\sss{QL}}{\sman,\tman}
                -\vvertil{}{\sss{LL}}{\sman,\tman},  
\\
\kappa_{f} &=& 1+\vvertil{}{\sss{LQ}}{\sman,\tman}
                -\vvertil{}{\sss{LL}}{\sman,\tman},  
\\
\label{form4}
\kappa_{ef}&=& 1+\vvertil{}{\sss{QQ}}{\sman,\tman}
                -\vvertil{}{\sss{LL}}{\sman,\tman}.
\eqa
This set of form factors corresponds to an equivalent Born-like ansatz:
\bqa
{\cal A}^{\sss{OLA}}_{\sss{\zb}}&=&
\ib \gspi \,e^2\,4\,\tcie\tcif\frac{\chi_{\sss{Z}}(\sman)}{\sman} 
\nll &&
\times \Biggl\{\gadu{\mu} {\lpar 1+\gfd \rpar } \otimes
       \gadu{\mu} {\lpar 1+\gfd \rpar } \vvertil{}{\sss{LL}}{\sman,\tman}     
-4 |\qe | \stws \gadu{\mu}     
      \otimes \gadu{\mu} {\lpar 1+\gfd \rpar}\vvertil{}{\sss{QL}}{\sman,\tman} 
\nll &&
-4 |\qf | \stws
\gadu{\mu}{\lpar 1+\gfd\rpar}\otimes\gadu{\mu}\vvertil{}{\sss{LQ}}{\sman,\tman}
+16|\qe\qf|\stwf\gadu{\mu}\otimes\gadu{\mu}
\vvertil{}{\sss{QQ}}{\sman,\tman}\Biggr\}.\qquad\quad
\label{structures}
\eqa
These form factors are functions of two Mandelstam invariants 
$(\sman,\tman)$ due to the remaining $\wb\wb$ and $\zb\zb$ box
contributions.
The Mandelstam variables are defined such that they 
satisfy the identity
:
\bq
\sman+\tman+\uman=0, \qquad\tman=-\frac{\sman}{2}(1-\cos\vartheta). 
\eq
\subsection{Subroutine {\tt ROKANC}\label{rokanc}}
The weak form factors in the one-loop approximation are calculated 
in subroutine {\tt ROKANC}. 
The four 
\eqns{form1}{form4}
are assigned to the complex-valued array {\tt XROK(4)}:
\bqa
{\tt XROK(1)}&=& \rho^{\alpha}_{ef}(\sman,\tman),
\\
{\tt XROK(2)}&=&\kappa^{\alpha}_{e}(\sman,\tman),
\\
{\tt XROK(3)}&=&\kappa^{\alpha}_{f}(\sman,\tman),
\\
{\tt XROK(4)}&=&\kappa^{\alpha}_{ef}(\sman,\tman).
\eqa
The four form factors \eqns{form1}{form4} are split into leading
and remainder parts following exactly the same philosophy as described in
\subsect{subgamma}. The analogs of first rows of \eqnsc{drho_fact}{dkap_fact}
are used, for instance:
\bq
\rho_{ef}={\ds{\frac{1+\drhov^{ef,(G)}_{\rm{rem}}}
     {1+\drhovb^{(G)}\,\lpar 1-\Delta{\bar{r}}^{[G]}_{\rm{rem}}\rpar}}}\;,
\eq
etc., with a trivial generalization concerning the treatment of 
next-to-leading order two-loop EW remainder terms, 
$\drhov^{\ff,G^2}_{\rm{rem}}$ and $\dkapv^{\ff,G^2}_{\rm{rem}}$.
For $\kappa_{e}$ one uses the $\kappa$ remainder for the
$\zb\to\fep\fem$ channel,  
for $\kappa_{f}$ the $\kappa$ remainder for the $\zb\to\ff\fbf$ channel, 
and for $\kappa_{ef}$ both. 
For $\rho_{ef}$ one uses 
$1/2(\drhov^{\fe,G^2}_{\rm{rem}}+\drhov^{\ff,G^2}_{\rm{rem}})$.
However, an emulation of the theoretical uncertainties as it is done
in the second and third rows of \eqnsc{drho_fact}{dkap_fact} was not
realized.  
The imaginary parts of EW form factors are attributed to remainders.

When \cite{Bardin:1997xq} was written, 
{\tt ZFITTER} had yet a simplified treatment of the 
additional terms in $\mtl$ for the $\fep\fem\to\fb\barb$ channel.
They were contained in three constant terms contributing
to the $\zb\fb\barb$-vertex \cite{Akhundov:1986fct}
and vanishing in the limit $\mtl\to 0$: 
${V}^{\ft}_{1\sss{W}}(\mzs),\;{V}^{\ft}_{2\sss{W}}(\mzs)$, and
$\delta^{\ft}_{ct,\fb}$ (see also Eqs.(268)--(271) of \cite{Bardin:1997xq}
and the description in \cite{Bardin:1992jc}).
Now, we take into account in these terms the complete kinematical
dependencies, arising from diagrams with virtual $\wb$ boson exchange
(see for details \cite{LKreptoLEP98}):   
$\cvetri{}{\wb a}{\sman},\;\avetri{}{\wb a}{\sman},\;$
$\hvetri{}{\wb n}{\sman},\;\hvetri{}{\wb  }{\sman}\;$ 
and $\hboxc{\sss{\wb\wb}}{d}{\sman,\tman}$.
These {\em additions} form a gauge invariant subset of terms.
They are supplied with a call to subroutine 
{\tt VTBANA(NUNI,S,WWv2,WWv11,WWv12)}:
\bqa
\cvetri{}{\wb a}{\sman}&=&{\tt WWv11},
\\
\avetri{}{\wb a}{\sman}&=&{\tt WWv12},
\\
\hvetri{}{\wb n}{\sman}&=&{\tt WWv2},
\\
\hvetri{}{\wb  }{\sman}&=&{\tt VTBX1}.
\eqa
The last variable and another one, 
\bq
{\tt VTBX2}=\lpar\frac{\mzs}{\sman}-1\rpar\cows
\Bigl[\hvetri{}{\wb n}{\sman}-\lpar 1-|\qb|\rpar\cvetri{}{\wb a}{\sman}\Bigr],
\eq
are used in the calculation of two corrections: 
\bqa
{\tt DVTBB1}&=&\frac{\alpha}{4\pi\siws}
\lrbr {\tt VTBX1}-\lpar-\frac{\mts}{2\mws}\rpar\rrbr,
\\
{\tt DVTBB2}&=&\frac{\alpha}{4\pi\siws}{\tt VTBX2}.
\label{DVTBB12}
\eqa
In the first one we subtracted the leading asymptotic term of
${\tt VTBX1}=-\mts/\mws/2$, (see \eqn{subt_zbb}), which is shifted to 
the leading term with the scale $\gf$ and accounted for by the substitutions
\eqn{barbierizationofbb}. 
The non-box $\mtl$-additions are applied by 
\bqa
\drhov^{ef}_{\rm{rem}}&\to&
\drhov^{ef}_{\rm{rem}}+{\tt DVTBB1},
\\
\kappa^{e }_{\rm{rem}}&\to&
\kappa^{e }_{\rm{rem}}\;+{\tt DVTBB2},
\\
\kappa^{f }_{\rm{rem}}&\to&
\kappa^{f }_{\rm{rem}}\;-{\tt DVTBB1},
\\
\kappa^{ef}_{\rm{rem}}&\to&
\kappa^{ef}_{\rm{rem}}\;-{\tt DVTBB1}.
\eqa

The $\wb\wb$ box contributes to the EW form factor
$\vvertil{}{\sss{LL}}{\sman,\tman}$.
It is written 
as a sum of two terms, 
corresponding to a zero $\ft$
mass part and an additional one vanishing at $\mtl\to 0$. 
The additional terms also contain non-unitary contributions,
which have to be cancelled analytically.
Otherwise the treatment with {\tt BOXD=1}  
may induce heavy gauge violations, see \cite{LKreptoLEP98}. 
After this cancellation some remnant arises in \eqns{form1}{form4}:
\bq
\frac{\rt}{4}\Bigl[\fbff{0}{-\sman}{\mwl}{\mwl}+1\Bigr]={\tt XROBT}.
\eq
It is present only in the $\zb\to\fb\barb$ channel.

In \subsect{ew_boxes} we will describe the implementation of
the EW boxes in {\tt ROKANC} in more detail.
\subsection{Treatment of electroweak boxes in {\tt DIZET}\label{ew_boxes}}
Every weak form factor gets a contribution from the purely weak boxes, 
the $\wb\wb$ and $\zb\zb$ diagrams.  
 They are vanishingly small at the $\zb$ peak but
very important at higher energies. 
For the \LEPII\ case see~\cite{Boudjema:1996qg}.

Calculation and treatment of these boxes deserve a dedicated discussion. 
There are three options governed by the flag {\tt BOXD}:

\begin{itemize}
\item{} {\tt BOXD}=0, the box contributions are ignored.
Strictly speaking such a {\em truncation} of form factors is not 
a gauge invariant procedure and it is more or less justified only 
in the vicinity of the $Z$ resonance where the net box contribution is
small anyway since it is suppressed by a factor $(\sman-\mzs)/\sman$.

\item{} {\tt BOXD}=1, the boxes are calculated as additive separate
contribution 
to the cross-section. 
Since box contributions depend on the scattering angle 
it should be integrated over an angular interval. 
This is done by a numerical integration of the differential
distribution supplied by the function {\tt DZEWBX(RACOS)}, where {\tt RACOS}
is the cms scattering angle.
This treatment is gauge invariant only for deconvoluted quantities.
If one subsequently convolutes an IBA part with ISR, but without 
boxes, and adds the boxes without convolution instead, then an
apparent violation of gauge invariance is being introduced.
This procedure is applicable only near the resonance.

\item{} {\tt BOXD}=2, box contributions are added 
to the four form factors in subroutine {\tt ROKANC}.
This is the only gauge invariant treatment.
However, it may be done only via interface {\tt ZANCUT}, which is
differential in the scattering angle, 
followed by a subsequent numerical integration over the scattering
angle if integrated quantities are needed. 
\end{itemize}

The box additions
$\rho^{\rm{box}}_{ef}(\sman,\tman)\;,\kappa^{\rm{box}}_{e,f}(\sman,\tman)$
and $\kappa^{\rm{box}}_{ef}(\sman,\tman)$ to the form factors are functions
of two Mandelstam variables: 
\bqa
\rho^{\rm{box}}_{ef}(\sman,\tman)   &=& 
\rho^{\rm{box}}_{_{ZZ}}(\sman,\tman)+\rho^{\rm{box}}_{_{WW}}(\sman,\tman),
\\
\kappa^{\rm{box}}_{e,f}(\sman,\tman) &=&
\kappa^{\rm{box}}_{e,f_{ZZ}}(\sman,\tman)  
-\rho^{\rm{box}}_{_{ZZ}}(\sman,\tman)-\rho^{\rm{box}}_{_{WW}}(\sman,\tman),
\\
\kappa^{\rm{box}}_{ef}(\sman,\tman) &=&
\kappa^{\rm{box}}_{ef_{ZZ}}(\sman,\tman)  
-\rho^{\rm{box}}_{_{ZZ}}(\sman,\tman)-\rho^{\rm{box}}_{_{WW}}(\sman,\tman).
\eqa
For the $\wb\wb$ contributions the dictionary is:
\bqa
\rho^{\rm{box}}_{_{WW}}(\sman,\tman)&=&\frac{\alpha}{4\pi\siws}{\tt XWWRO}
\\
&=&
- s_e s_f \cows \lpar\frac{\mzs}{\sman}-1\rpar
\left(\frac{\sman}{\sman-\tman}\right)^2
\left\{
\begin{array}{l}
{\cal B}^{(+)}_{_{WW}}={\tt XWWP} 
\\ \\
{\cal B}^{(-)}_{_{WW}}={\tt XWWM}
\end{array}\right.
\nll
&=&
\left\{
\begin{array}{l}
-\cows\lpar\Rz-1\rpar\sman\hboxc{\sss{\wb\wb}}{d}{\sman,\tman}
\\ \\
+\cows\lpar\Rz-1\rpar\sman\hboxc{\sss{\wb\wb}}{c}{\sman,\tman}
\end{array}\right.,
\nonumber
\eqa
with $s_{\ff}=2\tcif$,  
and $\hboxc{\sss{\wb\wb}}{d}{\sman,\tman}$ given by the ``unitarized''
direct box 
and 
$\hboxc{\sss{\wb\wb}}{c}{\sman,\tman}$ given by 
the crossed one,
which has to be ``unitarized'' by dropping out two terms with $1/\Rw$.
\subsubsection{Contributions to the weak form factors from $\zb\zb$ boxes
\label{zz_boxes}}
The $\zb\zb$ box contribution forms a separate gauge invariant sub-set
of terms: 
\bqa
\rho^{\rm{box}}_{_{ZZ}}(\sman,\tman)
&=&{\tt XZZRO}=
 \frac{s_e s_f}{32R}\frac{s - \mzs}{\sman} \Biggl\{
 \left[ (v^2_e+a^2_e)(v^2_f+a^2_f) + 4 v_e v_f \right] 
\left(\frac{\sman}{I_1}\right)^2
{\cal B}^{(+)}_{_{ZZ}}
\nll && \hspace{2.1cm}
-\left[ (v^2_e+a^2_e)(v^2_f+a^2_f) - 4 v_e v_f \right] 
\left(\frac{\sman}{I_2}\right)^2
{\cal B}^{(-)}_{_{ZZ}}
\Biggr\},
\nll
\\
\kappa^{\rm{box}}_{e,f_{ZZ}}(\sman,\tman){}
&=& {\tt XZZJ} =
 \frac{s_e s_f}{32R} \frac{s - \mzs}{\sman}(v_{e,f}-a_{e,f})
 \Biggl[(v_{e,f}+a_{e,f})^2
\left(\frac{\sman}{I_2}\right)^2
{\cal B}^{(-)}_{_{ZZ}}
\nll && \hspace{4.2cm}
      -(v_{e,f}-a_{e,f})^2
\left(\frac{\sman}{I_1}\right)^2
{\cal B}^{(+)}_{_{ZZ}}
\Biggr],
\\
\kappa^{\rm{box}}_{ef_{ZZ}}(\sman,\tman) 
&=&{\tt XZZIJ}   =
 \frac{s_e s_f}{16R}\frac{s - \mzs}{\sman}(v_e-a_e)(v_f-a_f)
\left(\frac{\sman}{I_1}\right)^2
{\cal B}^{(+)}_{_{ZZ}},
\eqa
where 
\bqa
{\cal B}^{(+)}_{_{ZZ}} = {\tt XZZP} &=&
   + {\cal B}(s,I_1,I_2,\mzs)
   - 2\left(\frac{I_1}{\sman}\right)^3 {\cal F}_4(s,I_1,\mzs),
\\
{\cal B}^{(-)}_{_{ZZ}} = {\tt XZZM} &=&
 - {\cal B}(s,I_2,I_1,\mzs)
 - 2\left(\frac{I_2}{\sman}\right)^3 {\cal F}_4(s,I_2,\mzs),
\eqa
with 
the angular dependent invariants: 
\bq
I_1= -\uman,\qquad I_2= -\tman,\qquad\tman=-\frac{\sman}{2}(1-\cos\vartheta).
\eq
In this and the next Subsections the usual {\tt ZFITTER} conventions 
\eqns{kg1}{kg3} are used.
The genuine box-function
${\cal B}(s,I_1,I_2,M^2_{_V}) $ is calculated by function {\tt XBOX}:
\bqa
{\cal B}(s,I_1,I_2,\mvs) &=& {\tt XBOX} =
2\frac{I_1}{\sman}\ln\frac{I_2}{\mvs}
+\frac{I_1}{\sman}\left(1-4\frac{\mvs}{\sman}\right)s
{\cal F}(-s,M^2_{_V},M^2_{_V})
\\ &&
+2\frac{I_2-I_1-2\mvs}{\sman}
\left[\Litwo(1)-\Litwo\left(1-\frac{I_2}{\mvs}\right)
+{\cal F}_3(s,\mvs)\right]
\nll &&
+\left\{2\frac{\mvs}{\sman}\left[\mvs\left(2s-I_1\right)-2I^2_2\right]
+\frac{I_2}{\sman}\left(I^2_1+I^2_2\right)\right\}\frac{1}{s^2}
{\cal F}_4(s,I_1,\mvs).
\nonumber
\label{boxgenuine}
\eqa
The functions ${\cal F}(-s,M^2_{_V},M^2_{_V})$, ${\cal F}_3(s,\mvs)$
and ${\cal F}_4(s,I_1,\mvs)$ were introduced in \cite{Bardin:1989dit}.
\subsubsection{The box-contribution as an additive part of the cross-section}
For {\tt BOXD}=1 the box-contribution is treated differently -- as an 
extra piece of the cross-section. 
The corresponding formulae were presented in
~\cite{Bardin:1989dit}. 
Here we repeat them again\footnote{Actually the expressions for the
differential weak box terms in \cite{Bardin:1989dit} contained some
misprints.}.  
The angular distribution is returned by function
{\tt DZEWBX}$(\cos\vartheta)$:
\bqa
\frac{d\sigma^{\rm{box}} }{d\cos\vartheta} &=&
\frac{\alpha G^2_{\mu} M^4_{_W}}{8s\pi^2}
\Biggl\{
 \lambda^L h^L {\cal R}e \Biggl[
\left(Q_e Q_f F^*_{\sss{A}} + s_e s_f \chi v^{^L}_e v^{^L}_f \right)
\left({\cal B}^{(+)}_{_{WW}}+{\cal B}^{(-)}_{_{WW}}\right)
\Biggr]
\nll &&
+\frac{1}{64}{\cal R}e \Biggl[
 \Biggl(Q_e Q_f F^*_{\sss{A}}
\left[ \lambda^L h^L (v^{L}_e)^2 (v^{L}_f)^2
     + \lambda^R h^R (v^{R}_e)^2 (v^{R}_f)^2 \right]
\nll &&\hspace{1.3cm}
+ s_e s_f \chi^*
\left[ \lambda^L h^L (v^{L}_e)^3 (v^{L}_f)^3
     + \lambda^R h^R (v^{R}_e)^3 (v^{R}_f)^3 \right]
\Biggr)
{\cal B}^{(+)}_{_{ZZ}}
\nll &&\hspace{1.3cm}
+\Biggl(Q_e Q_f F^*_{\sss{A}}
\left[ \lambda^L h^R (v^{L}_e)^2 (v^{R}_f)^2
     + \lambda^R h^L (v^{R}_e)^2 (v^{L}_f)^2 \right]
\nll &&\hspace{1.3cm}
+ s_e s_f \chi^*
\left[ \lambda^L h^R (v^{L}_e)^3 (v^{R}_f)^3
     + \lambda^R h^L (v^{R}_e)^3 (v^{L}_f)^3 \right]
\Biggr)
{\cal B}^{(-)}_{_{ZZ}}
\Biggr]
\Biggr\}.
\eqa
Here we use:
\ba
\label{f-a}
F_{\sss{A}} = \alpha(s)/\alpha,
\ea
with $\alpha(s)$ defined by \eqn{alpha_fer}), and:
\bqa
v^{L}_e &=& v_e + a_e,
\\
v^{R}_e &=& v_e - a_e,
\\
v^{L}_f &=& v_f + a_f,
\\
v^{R}_f &=& v_f - a_f,
\\
\lambda^L &=& \lambda_1 + \lambda_2\;=\;(1+\lambda_+)(1-\lambda_-),
\\
\lambda^R &=& \lambda_1 - \lambda_2\;=\;(1-\lambda_+)(1+\lambda_-),
\\
h^L &=& h_1 + h_2\;=\;\frac{1}{4}(1+h_+)(1-h_-),
\\
h^R &=& h_1 - h_2\;=\;\frac{1}{4}(1-h_+)(1+h_-).
\eqa
\section{Mixed Electroweak and QCD Corrections\label{mixed}}
\eqnzero
Mixed $\ord{\alpha\als}$ corrections originate from gluon insertions
to the fermionic components of bosonic self-energies. 
It is easy to derive the $\ord{\alpha\als}$ contributions
to $\dr$, $\rho$, and $\kappa$ 
simply by replacing every
self-energy contribution of order $\ord{\alpha}$ by a contribution of order
$\ord{\alpha\als}$. 
In this way we get:
\bqa
\Delta\dr&=&\frac{\alpha\als}{12\pi^2}
\biggl[
-\Pgg^{\ft,F}\lpar 0\rpar
-\Pgg^{\fl+5\fq,F}\lpar\mzs\rpar
+\frac{\cows}{\siwf}\Delta\rho^{{\fer},F}
+\frac{1}{\siws}\Delta\rho^{{\fer},F}_{\sss{W}}
\biggr],
\label{drmQCD}
\\
\Delta\rho_{\fe\ff}&=&\frac{\alpha\als}{12\pi^2}
\frac{1}{\siws}
\Bigl[
-\Delta\rho^{{\fer},F}_{\sss{Z}}+\Dz{{\fer},F}{\sman}
\Bigr],
\\
\Delta\kappa_{\fe}&=&\Delta\kappa_{\ff}=\frac{1}{2}\kappa_{\fe\ff}=
\frac{\alpha\als}{12\pi^2}\frac{1}{\siws}
\biggl[
-\frac{\cows}{\siws}\Delta\rho^{{\fer},F}-\Pzga{{\fer},F}{\sman}
\biggr].
\label{rokamQCD}
\eqa
\subsection{Package {\tt bcqcdl.f}\label{mixed_bcqcd}}
The implementation of internal QCD corrections is governed by flag {\tt QCDC}.
If {\tt QCDC}=0, they are not implemented.
The options {\tt QCDC}=1,2 are based on \cite{Djouadi:1987gn} 
and~\cite{Djouadi:1988di} where the two-loop self-energy functions of order
$\ord{\alpha\als}$ were reduced to two numerically treated integrals.
In \cite{Bardin:1989aa} all these corrections,
\eqns{drmQCD}{rokamQCD}, were 
computed in terms of three self-energy functions $\Pi^{\sss{V,A,W}}$ and
Taylor expansions in powers of $\sman/\mts$ were developed.
These expansions, accessible with {\tt QCDC}=1, ensured sufficient
precision in the  
beginning of \LEPI\ running and allowed a very fast computing compared
to the exact, but CPU time consuming option {\tt QCDC}=2. 
The relevant functions are supplied by the package {\tt bcqcdl.f}.
Both options are fully described
in~\cite{Bardin:1997xq,Bardin:1989aa}. We do not repeat their
description here since a better option is available.
\subsection{Package {\tt bkqcdl.f}\label{mixed_bkqcd}}
With the option {\tt QCDC}=3 we implemented analytic results for
$\Pi^{\sss{V,A,W}}$ derived in~\cite{Kniehl:1990yc} in the dimensional
regularization scheme. 
A {\tt FORTRAN} code for their calculations was provided by B. Kniehl.

From \eqns{drmQCD}{rokamQCD} it is easy to derive two sets of equations:
\begin{itemize}
\item[(i)]  for a light quark $\fu$--$\fd$ doublet;
\item[(ii)] for the heavy quark $\ft$--$\ffb$ doublet.
\end{itemize}

\noindent The expressions for a light quark doublet are very short:
\bqa
\dr^{\fu\fd}&=&{\tt CLQQCD}=
-\frac{\alpha\als}{\pi^2}\frac{1}{4}
\frac{\cows-\siws}{\siwf}\ln\cows,
\\
\Delta\rho^{\fu\fd}&=&{\tt ROQCD}=
-\frac{\alpha\als}{\pi^2}\frac{1}{16}
\frac{1}{\siws\cows}\Bigl[2+\vfwsi{\ft}+\vfwsi{\ffb}\Bigr]
\frac{\sman}{\sman-\mzs}\ln\Rz,
\\
\Delta\kappa^{\fu\fd}&=&{\tt AKQCD}=
\frac{\alpha\als}{\pi^2}\frac{1}{4\siwf}
\biggl\{\cows\ln\cows
+\siws\Bigl[1-4\siws\lpar\qus+\qds\rpar\Bigr]\ln\Rz
\biggr\}.
\eqa
Here and below we use the ratios: 
\ba
\cows&=&\frac{\mws}{\mzs},
\\
\siws&=&1-\cows,
\\
\rw&=&\frac{\mhs}{\mws},
\\
\rz&=&\frac{\mhs}{\mzs},
\\
\Rw&=&\frac{\mws}{\sman},
\\
\Rz&=&\frac{\mzs}{\sman},
\label{abbrev}
\ea
and:
\ba
\vfwi{\ff}=|\vc{\ff}+\ac{\ff}|=1-4|\qf|\siws.
\ea

For a heavy quark doublet one derives:
\bqa
\dr^{\ft\ffb}&=&{\tt XTBQCD}=
\frac{\alpha\als}{\pi^2}
\biggl\{
\qts V'_1\lpar 0\rpar   
+\frac{\cows}{\siwf}\frac{\rt}{4}\lrbr\ztwo+\frac{1}{2}\rrbr
-\frac{\zt}{4\siwf}
\Reb\Bigl[\vfwsi{\ft} V_1\lpar \rvari{\sss{Z}}\rpar 
+ A_1\lpar \rvari{\sss{Z}}\rpar 
\nll &&
- A_1\lpar 0\rpar 
\Bigr]
+\frac{\cows-\siws}{\siwf} \rt 
\Bigl[
\Reb F_1\lpar\xvar\rpar -\Reb F_1\lpar 0\rpar 
\Bigr]
-\frac{1}{8\siwf}\vfwi{\ffb}\ln \zt
\biggr\},
\\
\Delta\rho^{\ft\ffb}&=&{\tt ROQCD}=
\frac{\alpha\als}{\pi^2}
\frac{1}{4\siws\cows}
\biggl\{
\zt
\Bigl[
\vfwsi{\ft} V_1\lpar\rvari{\sss{Z}}\rpar + A_1\lpar\rvari{\sss{Z}}\rpar 
\Bigr]
\nll &&
+\frac{\mts}{\mzs-\sman}
\biggl\{
\vfwsi{\ft}
\Bigl[
V_1\lpar\rvar\rpar -V_1\lpar\rvari{\sss{Z}}\rpar 
\Bigr] 
+A_1\lpar\rvar\rpar-A_1\lpar\rvari{\sss{Z}}\rpar 
\biggr\}
\nll &&
-\frac{1}{4}
\Bigl[ 
1+\vfwsi{\ffb}
\Bigr] 
\frac{\sman}{\sman-\mzs}\ln\Rz
-2z_f\lrbr\frac{23}{8}-\ztwo-3\ztri\rrbr 
\biggr\},  
\\
\Delta\kappa^{\ft\ffb}&=&{\tt AKQCD}=
\frac{\alpha\als}{\pi^2}
\biggl\{
\frac{\cows}{4\siwf} \rt 
\Bigl[
\vfwsi{\ft} V_1\lpar \rvari{\sss{Z}}\rpar+A_1\lpar \rvari{\sss{Z}}\rpar 
\Bigr]
-\frac{\cows}{\siwf} \rt F_1\lpar\xvar\rpar 
\nll &&
+\frac{\vfwi{\ft}}{\siws}|\qt| \rt V_1\lpar\rvar\rpar 
-\frac{1}{16\siwf}
\Bigl[
-\lpar 1+\vfwi{\ffb}\rpar\ln\zt-4\vfwi{\ffb}|\qb|\siws\ln\Rz
\Bigr]
\biggr\},
\eqa
where
\ba
\rt&=&\frac{\mts}{\mws}\,,
\\
\rvar&=&\frac{\sman + \ib\ep}{4\mts}\,,
\\
\rvari{\sss{Z}}&=&\frac{\mzs + \ib\ep}{4\mts}\,,
\\
\xvar&=&\frac{\mws + \ib\ep}{4\mts}\,.
\ea
They are returned by complex-valued functions:
\bqa
V_1 \lpar r \rpar&=&{\tt CV1(CR)},
\\
A_1 \lpar r \rpar&=&{\tt CA1(CR)},
\\
F_1 \lpar x \rpar&=&{\tt CF1(CX)}.
\eqa
\section{Two-Loop Electroweak Corrections. Package {\tt
m2tcor.f}\label{EW_twoloops}}  
\eqnzero
The two-loop EW corrections are supplied by the package {\tt m2tcor.f} 
\cite{Degrassi:1996ZZ}.
It is based on~\cite{Degrassi:1996mg,Degrassi:1997ps,Degrassi:1999jd}.
The package is interfaced to {\tt ZFITTER} by subroutine {\tt GDEGNL}:

\vspace*{2mm}

{\fbox{\tt GDEGNL(GMU,MZ,MT,AMH,MW,PI3QF,SMAN,DRDREM,DRHOD,DKDREM,DROREM)}}

\vspace*{2mm}

In the argument list, the following variables are {\tt INPUT}:
\bqa
{\tt GMU}&=&\gf,
\\
{\tt MZ}&=&\mzl,
\\
{\tt MT}&=&\mtl,
\\
{\tt AMH}&=&\mhl,
\\
{\tt MW}&=&\mwl,
\\
{\tt PI3QF}&=&\qf,
\\
{\tt SMAN }&=&\sman;
\eqa
and the following ones are {\tt OUTPUT}:
\bqa
{\tt DRDREM}&=&\dr^{\alpha^2}_{\rm{rem}}\,,
\\
{\tt DRHOD }&=&-\drhovb^{G^2}\,,
\\
{\tt DKDREM}&=&\dkapv^{\ff,G^2}_{\rm{rem}}\,,
\\
{\tt DROREM}&=&\drhov^{\ff,G^2}_{\rm{rem}}\,.
\eqa
The various two-loop corrections  consist of two pieces. 
One is derived in the $\MSB$ framework and the other one, with the 
superscript ${\rm{OMS}}$, contains the additional corrections coming from 
the expansion of the OMS sine of weak mixing angle in the one-loop result. 
The OMS result is a sum: {\tt something} + {\tt something}$_{\rm{OMS}}$, 
with the coupling constant being expressed in terms of OMS quantities.

Typically, every correction has two branches: $\Ht < 1/4$ --- 
the {\it light Higgs case}, and $\Ht > 4$ --- the {\it heavy Higgs case},
with an interpolation in between. 
The expressions are rather lengthy in spite of expansions in positive powers
of $\mts$. 
The two-loop level is quite involved and since the
Higgs mass is unknown we would like to have an expression valid
for any mass of the Higgs.

 In what follows we use the short hand notations:
\bqa
\wt  &=& \frac{\mws}{\mts}\,,
\\
\zt  &=& \frac{\mzs}{\mts}\,,
\\
\Ht  &=& \frac{\mhs}{\mts}\,.
\eqa

The leading two-loop term in the calculation of $\dr$
is accounted for by the following substitutions:
\bqa
f\dr_{_L} &\to& f\dr_{_L}+\frac{\cows}{\siws} N_c\xts
\left(\Delta\rho^{(2)}+\Delta\rho^{(2)}_{_{\rm{OMS}}}\right),
\\
-\Delta{\hat{\rho}} &=& \Delta\rho^{(2)}+\Delta\rho^{(2)}_{_{\rm{OMS}}}\,.
\label{drhosubrepl}
\eqa
For the leading contributions to $\rho^f_{\sss{L}}$ and
$\kappa^f_{\sss{L}}$ they are~\cite{Degrassi:1999jd}:
\bqa
f\rho^f_{\sss{L}}  &\to&
\Delta{\hat{\rho}}\left(1-\Delta{\bar{r}}_{\rm{rem}}\right),
\\
f\kappa^f_{\sss{L}}&\to&-\frac{\cows}{\siws}\Delta{\hat{\rho}}
\left(1-\Delta{\bar{r}}_{\rm{rem}}\right),
\eqa
where $f$ is the conversion factor \eqn{conversionf} and 
$\Delta{\bar{r}}_{\rm{rem}}$ =  {\tt DF1BAR}
 is defined by \eqn{drrembar}.

 The $\Delta\rho^{(2)}$ is the two-loop irreducible
contribution to Veltman's $\rho$-parameter, as defined by \eqn{insert7}:
\bq
\Delta{\hat{\rho}}=
\frac{\alpha}{4\pi\siws}\frac{1}{\mws}
\biggl[\Sigma_{\sss{WW}}(\mws)-\Sigma_{\sss{ZZ}}(\mzs)\biggr]
\Bigg|_{\MSB,\;\tHs=\mzl},
\label{Vrhos}
\eq
but now to orders $\ord{\gfs\mtq}$ and $\ord{\gfs\mts\mzs}$.
 
  All ~results ~for ~$\Delta\rho^{(2)}$ ~are ~returned ~by ~the  
~subroutines ~{\tt ew2ltobf} ~and ~{\tt ew2ltobfOS}.   
 The  ~subroutine ~{\tt ew2ltobf} ~computes ~the ~two-loop ~expression ~of
~\eqn{Vrhos} ~in ~units ~of
$N_c \alpha^2 / (16\pi\stws\zt\ctws)^2$:
\bqa
\Delta\rho^{(2)}&=&{\tt tobf2lew}.
\eqa
 The subroutine {\tt ew2ltobfOS} computes the additional term to be added
to the $\MSB$ two-loop expression of \eqn{Vrhos} in units of 
$N_c \alpha^2 / [(4\pi\stws)^2 4\zt\ctws]$:
\bqa
\Delta\rho^{(2)}_{_{\rm{OMS}}} &=&4\zt \cows \cdot {\tt tobf2ewOS}.
\eqa

The three {\it remainders} are modified as follows:
\bqa
\dr_{\rm{rem}}&\to&\dr_{\rm{rem}}+\dr^{(2)}_{\rm{rem}},
\\
\rho^f_{\rm{rem}}  &\to&f \rho^f_{\rm{rem}}+\rho^{f(2)}_{\rm{rem}},
\\
\kappa^f_{\rm{rem}}&\to&f\kappa^f_{\rm{rem}}+\kappa^{f(2)}_{\rm{rem}}.
\label{remmodtwoloop}
\eqa

 However, contrary to the leading term \eqn{drhosubrepl}, 
the remainders contain also contributions from vertex
diagrams.  
That's why they are channel dependent and bear the superscript $f$.
Now we list all terms entering \eqn{remmodtwoloop}:
\begin{itemize}
\item[$\bullet$] $\dr^{(2)}_{\rm{rem}}$:
\end{itemize}
\bqa
\dr^{(2)}_{\rm{rem}}
&=& {\tt DRDREM}=
N_c\left(\frac{\alpha}{4\pi\siw}\right)^2\left(\frac{\mts}{\mws}\right)^2
\left[
 \Delta r^{(2)}_{\sss{W}}+\Delta r^{(2)}_{_{W,{\rm{OMS}}}}
-\left(\frac{\delta e}{e}\right)^{(2)}
\right].\qquad
\label{drdrem}
\eqa

 The $\Delta r^{(2)}_{\sss{W}}$ is made of the two-loop contribution to 
\bq
\Delta{\hat{\rho}}_{\sss{W}}=\frac{1}{\mws}
\biggl[
\Sww^{F}\left( 0\right)-\Sww^{F}\left(\mws\right)
\biggr]\Bigg|_{\MSB,\;\tHs=\mzl},
\eq
and a vertex correction $V_{r}$:
\bq
\Delta r^{(2)}_{\sss{W}} = -\Delta\rho^{(2)}_{\sss{W}} + V_{r}\,.
\eq
 The $\left({\delta e}/{e}\right)^{(2)}$
is the two-loop contribution to the electric charge renormalization. 

\noindent
The Subroutine {\tt ew2ldeltarw} returns these two-loop expressions 
in the variable {\tt drs2lew}:
\bqa
\Delta r^{(2)}_{\sss{W}}&=&{\tt drs2lew},
\\
-\Delta\rho^{(2)}_{\sss{W}}&=& {\tt aww},
\\
V_{r}&=& {\tt vertex1}.
\eqa
The ~({\tt aww} ~and ~{\tt vertex1} ~are ~internal ~variables.
~Subroutine ~{\tt ew2ldeltarwOS} ~computes ~the ~term 
~to ~be ~added ~to ~the ~$\MSB$ ~two-loop ~expression ~of 
~$\Delta\rho^{(2)}_{\sss{W}}$
~in ~units ~of
$N_c \alpha^2 / [(4\pi\stws)^2 4\zt\ctws]$:
\bq
\Delta\rho^{(2)}_{_{W,{\rm{OMS}}}}=\frac{1}{4}\zt\cow{\tt drs2ewOS}.
\eq
The subroutine {\tt ew2ltwodel}
computes the two-loop contribution to
the electric charge renormalization in units of
$N_c \alpha^2 / (4\pi\stws\zt\ctws)^2$:
\bq
\left(\frac{\delta e}{e}\right)^{(2)} = {\tt deleoe2lew}.
\eq
\begin{itemize}
\item{ $ \rho^{f(2)}_{\rm{rem}}(s)$ }:
\end{itemize}
\bq
\rho^{f(2)}_{\rm{rem}}(s)={\tt DROREM}= N_c\left(4 x_t \right)^2
\Bigl[
\wt\Bigl(\eta^{(2)}+\eta^{f(2)}_{_{\rm{OMS}}}(s)-2\cows\log\cows\Bigr)
-\Bigl(\Delta r^{(2)}_{\sss{W}}+\Delta r^{(2)}_{_{W,\rm{OMS}}}\Bigr)
\Bigr].
\eq

 The $\eta^{(2)}$ is the two-loop contribution to 
\bq 
{\rm Re}
\left[
\frac{\Szz(s)-\Szz(\mzs)}{s-\mzs}
\right]
\Bigg|_{s=\mzs}.
\label{deriv}
\eq
 The subroutine {\tt ew2leta} returns its two-loop expression
in units of $N_c \alpha^2 / [(4\pi\stws)^2\zt\ctws]$:
\bq
\eta^{(2)} = {\tt eta2lew}.
\eq
The additional piece to be added to the $\MSB$ two-loop contribution
{\tt eta2lewOS}, \eqn{deriv}, 
is returned by subroutine {\tt ew2letaOS} 
in units of $N_c \alpha^2 / [(4\pi\stws)^2 4\zt\ctws]$:
\bq
\eta^{f(2)}_{_{\rm{OMS}}}(s)=\frac{1}{4}{\tt eta2lewOS}.
\eq

\begin{itemize}
\item{ $ \kappa^{f(2)}_{\rm{rem}}(s) $ }:
\end{itemize}
\bq
\kappa^{f(2)}_{\rm{rem}}(s)= {\tt DKDREM} = 
N_c\xts
\Bigl[ k^{(2)}+k^{f(2)}_{_{\rm{OMS}}}(s) \Bigr].
\label{DKDREM}
\eq

 The $k^{(2)}$ is the two-loop contribution to
\bq
 \Pzg(\mzs)+V_{m}\,,
\eq
i.e. the two-loop expressions for the $\zb\gamma$-mixing 
and the relevant vertex correction.

 The subroutine {\tt kappacur2l} returns it 
in units of $N_c \alpha^2 / (16\pi\stws \zt \ctws)^2$:
\bq
k^{(2)} = {\tt k2lew}.
\eq
 The additional piece to be added to the $\MSB$
two-loop contribution is calculated in subroutine 
{\tt kappacur2lOS} in units of
$N_c \alpha^2 / [(4\pi\stws)^2 4 \zt]$:
\bqa
k^{f(2)}_{_{\rm{OMS}}}(s) &=& -4 \cows \zt {\tt k2lewOS}.
\eqa

\section{Final State QCD and QED Corrections\label{qcdrun}}
\eqnzero
This subsection is devoted to the detailed description of the
corrections $R^{\ff}_{\sss{V}}(s)$ and $R^{\ff}_{\sss{A}}(s)$
introduced in the expressions for the $Z$ width \eqn{defzwidthl} and
\eqn{defzwidthq}. 

\subsection{Case of leptons \label{FSR-leptons}}
In \eqn{defzwidthl}, for the $\zb$ decays into leptons 
the QED FSR for an inclusive setup are described by a simple factor
$R_{\rm{\sss{QED}}}(\mzs)$:
\bqa
R_{\rm{\sss{QED}}}(\sman) =
1 + \frac{3}{4} \frac{\alpha(\sman)}{\pi}\qfs\,,
\\
R_{\rm{\sss{QED}}}(\mzs) = 1 + 0.0017\qfs\,.
\label{ALQF2Z}
\eqa
The same factor is used for the total cross-section of 
lepton pair production for the case {\tt FINR}=0.
For lepton pair production with a proper treatment of final state radiation 
with cuts ({\tt FINR}=1) this factor is replaced by different formulae,
see Chapter \ref{ch-phot}.

\subsection{Case of quarks \label{R-factors}}
In \eqn{defzwidthq} for
the $\zb$ decays into quarks the 
radiator factors $R^{\ff}_{\sss{V,A}}(\mzs)$ introduced in \eqn{defzwidthq}
are used.
For quark production cross-sections these radiator factors get
$\sman$-dependent.
They will be given in this section.
When cuts are applied, the QED is treated properly. 
This means that the one-loop QED pieces of the radiator factors
\eqns{rvfact}{rafact} are left out.
Then, the higher order QCD and mixed QCD$\otimes$QED
pieces are applied in an approximate way since they are known only for 
the total cross-section.

For $A_{\sss{FB}}$, the one-loop QED and QCD
corrections vanish in massless approximation.
These factors are applied only for the $\fc\barc$ and $\fb\barb$ channels
to first order in $\ord{\mq/\sqrt{\sman}}$~\cite{Arbuzov:1992pr}.

Subroutine {\tt QCDCOF} computes the inclusive $R_{\sss{V,A}}$-factors:
\bqa
R^{\fq}_{\sss{V}}(\sman)&=& 1 + \frac{3}{4} Q^2_q \frac{\alpha(\sman)}{\pi}
              +\frac{\als(\sman)}{\pi}
              -\frac{1}{4}Q^2_q\frac{\alpha(\sman)}{\pi}\frac{\als(\sman)}{\pi}
\nll
        & &   + \left[C_{02}+C^t_2\left(\frac{s}{\mts}\right)\right]
                \left(\frac{\als(\sman)}{\pi}\right)^2         
              + C_{03}\left(\frac{\als(\sman)}{\pi}\right)^3
\nll
        & &   + \frac{\mcS(\sman)+\mbS(\sman)}{s} C_{23}
                                \left(\frac{\als(\sman)}{\pi}\right)^3
\nll
        & &   + \frac{\mqS(\sman)}{s} \Biggl[ 
                        C^V_{21}      \frac{\als(\sman)}{\pi}
                      + C^V_{22}\left(\frac{\als(\sman)}{\pi}\right)^2
                      + C^V_{23}\left(\frac{\als(\sman)}{\pi}\right)^3
                                  \Biggr]
\nll
   & &+\frac{\mcQ(\sman)}{\smans}\left[ C_{42}-\ln\frac{\mcS(\sman)}{s}\right]
                                \left(\frac{\als(\sman)}{\pi}\right)^2 
      +\frac{\mbQ(\sman)}{\smans}\left[ C_{42}-\ln\frac{\mbS(\sman)}{s}\right]
                                \left(\frac{\als(\sman)}{\pi}\right)^2 
\nll 
        & &   + \frac{\mqQ(\sman)}{\smans} \Biggl\{
                       C^V_{41}       \frac{\als(\sman)}{\pi}
                 + \left[C^V_{42}+C^{V,L}_{42}\ln\frac{\mqS(\sman)}{s}\right]
                                \left(\frac{\als(\sman)}{\pi}\right)^2
                                    \Biggr\}
\nll
        & &   +12\frac{\mqpQ(\sman)}{\smans}
                                \left(\frac{\als(\sman)}{\pi}\right)^2
              -\frac{\mqX(\sman)}{s^3}\Biggl\{8+\frac{16}{27}
               \left[155+6\ln\frac{\mqS(\sman)}{s}\right]
                                      \frac{\als(\sman)}{\pi}\Biggr\},
\label{rvfact}
\\
\nll
R^{\fq}_{\sss{A}}(\sman)&=&1 + \frac{3}{4} Q^2_q \frac{\alpha(\sman)}{\pi}
             + \frac{\als(\sman)}{\pi}
             - \frac{1}{4}Q^2_q\frac{\alpha(\sman)}{\pi}\frac{\als(\sman)}{\pi}
\nll
        & & + \left[C_{02}+C^t_2\left(\frac{s}{\mts}\right)
            - \left(2I^{(3)}_q\right)
              {\cal I}^{(2)}\left(\frac{s}{\mts}\right)\right]
              \left(\frac{\als(\sman)}{\pi}\right)^2         
\nll
        & & + \left[C_{03} - 
              \left(2I^{(3)}_q\right) 
              {\cal I}^{(3)}\left(\frac{s}{\mts}\right)\right]
              \left(\frac{\als(\sman)}{\pi}\right)^3
\nll
        & & + \frac{\mcS(\sman)+\mbS(\sman)}{s} C_{23}
                                \left(\frac{\als(\sman)}{\pi}\right)^3
            + \frac{\mqS(\sman)}{s} \Biggl[C^A_{20} 
                      + C^A_{21}      \frac{\als(\sman)}{\pi}
                      + C^A_{22}\left(\frac{\als(\sman)}{\pi}\right)^2
\nll
        & & + 6\left(3 + \ln\frac{\mts}{s}\right)
                                \left(\frac{\als(\sman)}{\pi}\right)^2 
                      + C^A_{23}\left(\frac{\als(\sman)}{\pi}\right)^3
                                \Biggr]
\nll
        & & - 10\frac{\mqS(\sman)}{\mts}
              \left[\frac{8}{81}+\frac{1}{54}\ln\frac{\mts}{s}\right]
                                \left(\frac{\als(\sman)}{\pi}\right)^2
\nll
& & + \frac{\mcQ(\sman)}{\smans}\left[ C_{42}-\ln\frac{\mcS(\sman)}{s} \right]
                                \left(\frac{\als(\sman)}{\pi}\right)^2
    + \frac{\mbQ(\sman)}{\smans}\left[ C_{42}-\ln\frac{\mbS(\sman)}{s} \right]
                                \left(\frac{\als(\sman)}{\pi}\right)^2
\nll 
        & &   + \frac{\mqQ(\sman)}{\smans} \Biggl\{C^A_{40} 
              + C^A_{41}       \frac{\als(\sman)}{\pi}
              + \left[C^A_{42}+C^{A,L}_{42}\ln\frac{\mqS(\sman)}{s}\right]
                                \left(\frac{\als(\sman)}{\pi}\right)^2
                                    \Biggr\}
\nll
        & &   -12\frac{\mqpQ(\sman)}{\smans}
                                \left(\frac{\als(\sman)}{\pi}\right)^2.    
\label{rafact}
\eqa
\eqns{ALQF2Z}{rafact} are strictly valid only in the inclusive
setup, i.e. when no cuts are applied in the final state. 
For this setup
the appearance of $\als(\sman)$ and $\alpha(\sman)$ is strictly proved,
see \cite{Kataev:1992dg} for the QED case and \cite{Chetyrkin:1994js3}
for the QCD case.

However, in the case of QCD FSR corrections to the hadronic total
cross-section,  
the feasibility of the inclusive formulae is justified not
only for rather loose cuts 
by the fact that hadrons are 
always inclusive with respect to gluon bremsstrahlung.

For QED corrections with loose cuts,
when we apply nevertheless formulae depending on cuts, it is reasonable
to use the running QED coupling at scale $\alpha(\sman)$ rather than
at scale $\alpha(0)$ in order to have a smooth transition to the case
of inclusive setup.

Moreover, it is understood that: 
\begin{itemize}
\item[$-$]
only $\fc$-and $\fb$-quark finite mass corrections are retained, 
i.e. $\mq=0$ for $\fq=\fu,\fd,\fs$. 
Therefore, 
these corrections are valid slightly above the
$\fb\barb$-threshold, say $\sqrt{\sman} \geq 13$ GeV, and below the
$\ft\bart$-threshold, say $\sqrt{\sman} < 350$ GeV;
\item[$-$]
$\mqp$ denotes {\em the other} mass in a quark doublet, i.e. it is $\mb$ if
$\fq=\fc$ and it is $\mc$ if $\fq=\fb$;
quark masses with the argument $(\sman)$ are $\MSB$ {\em running}
masses, while $M_q$ stands for {\em pole} masses.
The numerical coefficients in \eqn{rvfact} and \eqn{rafact} 
are listed here: 
\end{itemize}
\noindent
{{\em Massless non-singlet corrections}
\cite{Chetyrkin:1979bj,Dine:1979qh,Celmaster:1980xr,Gorishnii:1991hw}}:
\bqa
C_{02} &=& {\tt COEF02} = \frac{365}{24}-11\ztri
                         +\left[-\frac{11}{12}+\frac{2}{3}\ztri\right]\nf,
\\ \nll
C_{03} &=& {\tt COEF03} = \frac{87029}{288}-\frac{121}{8}\ztwo
                         -\frac{1103}{4}\ztri +\frac{275}{6}\zfiv
\nll
       & &+\left[-\frac{7847}{216}+\frac{11}{6}\ztwo
          +\frac{262}{9}\ztri-\frac{25}{9}\zfiv\right]\nf
\nll
       & &+\left[\frac{151}{162}-\frac{1}{18}\ztwo
          -\frac{19}{27}\ztri\right]\nfS;
\label{mass0}
\eqa
\noindent 
{{\em Quadratic massive corrections}~\cite{Chetyrkin:1994js3}}:
\bqa
C_{23}&=&{\tt COEFL3}=
-80+60\ztri+\left[\frac{32}{9}-\frac{8}{3}\ztri\right]\nf,
\\ \nll
C^V_{21}&=&{\tt COEFV1} = 12,
\\ \nll
C^V_{22}&=&{\tt COEFV2}= \frac{253}{2} - \frac{13}{3}\nf,
\\ \nll
C^V_{23}&=& {\tt COEFV3}= 
2522-\frac{855}{2}\ztwo+\frac{310}{3}\ztri-\frac{5225}{6}\zfiv
\nll 
        & &+\left[-\frac{4942}{27}+34\ztwo
           -\frac{394}{27}\ztri+\frac{1045}{27}\zfiv\right]\nf
            +\left[\frac{125}{54}-\frac{2}{3}\ztwo\right]\nfS,\qquad
\\ \nll
C^A_{20}&=& {\tt COEFA0} =-6,
\\ \nll
C^A_{21}&=& {\tt COEFA1} =-22,
\\ \nll
C^A_{22}&=& {\tt COEFA2} =-\frac{8221}{24}+57\ztwo+117\ztri
           +\left[\frac{151}{12}-2\ztwo-4\ztri\right]\nf,
\\ \nll
C^A_{23}&=& {\tt COEFA3} =-\frac{4544045}{864}+1340\ztwo+\frac{118915}{36}\ztri
                                -127\zfiv
\nll 
        & &+\left[\frac{71621}{162}-\frac{209}{2}\ztwo-216\ztri
                                +5\zfor+55\zfiv\right]\nf
\nll 
        & &+\left[-\frac{13171}{1944}+\frac{16}{9}\ztwo
                                +\frac{26}{9}\ztri\right]\nfS;
\label{mass2}
\eqa
\noindent 
{\em Quartic massive corrections}:
\bqa
C_{42}     &=& \frac{13}{3}-4\ztri,
\\ \nll
{\tt R4LC} &=& \frac{\mcQ(\sman)}{\smans}
               \lpar C_{42}-\ln \frac{\mcQ(\sman)}{s}\rpar 
               \frac{\alsS(\sman)}{\pi^2}\,,
\\ \nll 
C^V_{40}&=&-6,
\\ \nll 
C^V_{41}&=&-22,
\\ \nll
C^V_{42}&=&-\frac{3029}{12}+162\ztwo+112\ztri
           +\left[\frac{143}{18}-4\ztwo-\frac{8}{3}\ztri\right]\nf,
\\ \nll
{\tt RV40}&=& C^V_{40} + C^V_{41} \frac{\als(\sman)}{\pi}  
           +  C^V_{42} \frac{\alsS(\sman)}{\pi^2}\,,
\\ \nll
C^{V,L}_{42}&=&-\frac{11}{2}+\frac{1}{3}\nf,
\\ \nll
{\tt RV4L} &=& C^{V,L}_{42} \frac{\alsS(\sman)}{\pi^2}\,,
\\ \nll
C^A_{40}&=&6,
\\ \nll
C^A_{41}&=&10,
\\ \nll
C^A_{42}&=& \frac{3389}{12}-162\ztwo-220\ztri
           +\left[-\frac{41}{6}+4\ztwo+\frac{16}{3}\ztri\right]\nf,
\\ \nll
{\tt RA40} &=& C^A_{40}+C^A_{41}\frac{\als(\sman)}{\pi} + C^A_{42} 
              \frac{\alsS(\sman)}{\pi^2}\,,
\\ \nll
C^{A,L}_{42} &=& \frac{77}{2}-\frac{7}{3}\nf,
\\ \nll
{\tt RA4L}   &=& C^{A,L}_{42} \frac{\alsS(\sman)}{\pi^2}\,;
\label{mass4}
\eqa
\noindent
{\em Power suppressed $\ft$-mass correction}:
\bqa
C^t_2(x)&=&x\left(\frac{44}{675} - \frac{2}{135}\ln x \right);
\label{powsup}
\eqa
\noindent 
{\em Singlet axial corrections}:
\bqa
{\cal I}^{(2)}(x) &=&  - \frac{37}{12} + \ln x + \frac{7}{81}x
                       + \mbox{${\tt 0.0132}$}x^2,
\\ \nll   
{\cal I}^{(3)}(x) &=& - \frac{5075}{216} 
                      + \frac {23}{6}\ztwo + \ztri + \frac{67}{18}\ln x
                      + \frac{23}{12} \ln^2 x;
\eqa
\noindent 
{\em Singlet vector correction}:
\bqa
R^h_{\sss{V}}(\sman) &=& 
\left(\sum_f v_f\right)^2\left(-\mbox{${\tt 0.41317}$}\right)
               \left(\frac{\als(\sman)}{\pi}\right)^3;
\label{singlet}
\eqa 
\noindent 
{\em ABL corrections} for $A_{\sss{FB}}$ ~\cite{Arbuzov:1992pr}:
\bqa
R^{\fq}_{\sss{FB}}&=&f_1\frac{\mq}{\sqrt{\sman}}\,,
\\
f_1&=&\frac{16}{3}\frac{\als(\sman)}{\pi}\,.
\label{abl_corr}
\eqa 
\subsection{Running masses\label{runnmass}}
The running $\fc$-quark mass is calculated by function {\tt ZRMCMC}:
\bqa
\Mc &=& \mc(\McS) \Biggl\{1
   +\left[\frac{4}{3}+\ln\frac{\McS}{\mcS(\McS)}\right]\frac{\als(\McS)}{\pi}
\nll
&&+~\Biggl[K_c+\left(\frac{173}{24}-\frac{13}{36}\nf\right)
\ln\frac{\McS}{\mcS(\McS)}
+\left(\frac{15} {8} -\frac{1} {12}\nf\right)\ln^2\frac{\McS}{\mcS(\McS)}
\nll
&& +~\frac{4}{3}\Lambda \left(\frac{\ms(\McS)}{\mc(\McS)}\right)\Biggr]
\left(\frac{\als(\McS)}{\pi}\right)^2\Biggr\},
\label{mcrun1}
\eqa
with
\bqa
K_c&=& {\tt AKC} = \frac{2905}{288}+\frac{1}{3}\Biggl[7+2\ln(2)\Biggr]\ztwo
     -\frac{1}{6}\ztri
     -\frac{1}{3}\Biggl[\frac{71}{48}+\ztwo\Biggr]\nf,
\\
\Lambda\left(r\right)
&\approx&
 \frac{\pi^2}{8}r-0.597 r^2 + 0.230 r^3,
\eqa
which is solved numerically with respect to $\mc(\McS)$ at $\nf=4$,
followed by a RG-evolution
(in subroutine {\tt QCDCOF}) with two scales: 1) $\McS \to \MbS$,
2) $\MbS \to s$:
\bqa
\label{runmasc}
\mc(\sman) & = & {\tt AMQRUN}=\mc(\McS)
\Biggl[\frac{\als(\MbS)}{\als(\McS)}\Biggr]^{\gamma_{0}^{(4)}/\beta_{0}^{(4)}}
\Biggl\{1 + C_{1}(4)
\Biggl[\frac{\als(\MbS)}{\pi} - \frac{\als(\McS)}{\pi}\Biggr]
\\
&&  + \frac{1}{2} C^2_{1}(4)
\Biggl[\frac{\als(\MbS)}{\pi} - \frac{\als(\McS)}{\pi}\Biggr]^2
    + \frac{1}{2} C_{2}(4)
\Biggl[\Biggl(\frac{\als(\MbS)}{\pi}\Biggr)^2
     - \Biggl(\frac{\als(\McS)}{\pi}\Biggr)^2\Biggr]
\Biggr\}\qquad\qquad
\nll 
&&\times
\Biggl[\frac{\als(\sman)}{\als(\MbS)}\Biggr]^{\gamma_0^{(5)}/\beta_0^{(5)}}
\Biggl\{1 + C_1(5) 
\Biggl[\frac{\als(\sman)}{\pi} - \frac{\als(\MbS)}{\pi} \Biggr]
\nll
&&  + \frac{1}{2} C^2_{1}(5) 
\Biggl[\frac{\als(\sman)}{\pi} - \frac{\als(\MbS)}{\pi} \Biggr]^2
     + \frac{1}{2} C_2(5) 
\Biggl[\Biggl(\frac{\als(\sman)}  {\pi}\Biggr)^2
     - \Biggl(\frac{\als(\MbS)}{\pi}\Biggr)^2 \Biggr]
\Biggr\}.
\nonumber
\eqa
The running $\fb$-quark mass is calculated by an analogous chain with the first
step similar to~\eqn{mcrun1} with the obvious replacements $\fc\to\fb$,
$s \to s\;and\;c$ at $\nf=5$, followed by a `one step' RG-evolution:
$\MbS \to s$. 
We note that $\ms(\mu)$ and $\mc(\mu)$ {\em inside} $\Delta$
in turn evolve as appropriate\footnote{It should also be noted that
\eqn{mcrun1}) has a very bad perturbative convergence for the
case of $\fc$-quark, like $1+0.253+0.228$.This is due to the large numerical 
value
of $K_c$ and $\als(\McS)=0.375$. 
For the $\fb$-quark it looks a bit better:
$1+0.119+0.059$ due to a smaller value of $\als(\MbS)=0.225$ (we used 
$\als(\mzs)=0.1204$ in this evaluation).}.

In the equations above:
\bqa
C_{1}(\nf) & = & 
\displaystyle{ \frac{\gamma_{1}^{(\nf)}}{\beta_{0}^{(\nf)}} 
             - \frac{\beta_{1} ^{(\nf)}
               \gamma_{0}^{(\nf)}}{\left(\beta_{0}^{(\nf)}\right)^2}}\,,
\\
\nl
C_{2}(\nf) & = & 
\displaystyle{ \frac{\gamma_{2}^{(\nf)}}{\beta_{0}^{(\nf)}} 
             - \frac{\beta_{1} ^{(\nf)}
               \gamma_{1}^{(\nf)}}{\left(\beta_{0}^{(\nf)}\right)^2}  
             - \frac{\beta_{2} ^{(\nf)}
               \gamma_{0}^{(\nf)}}{\left(\beta_{0}^{(\nf)}\right)^2}  
             +
             \frac{\left(\beta_{1}^{(\nf)}\right)^2\gamma_{0}^{(\nf)}}
{\left(\beta_{0}^{(\nf)}\right)^3}}\,.  
\eqa
The Beta function coefficients are:
\bqa
\beta_{0}^{(\nf)} & = & {\tt BETA0} =
\displaystyle{ \frac{1}{4}  \left(11  -\frac{2}{3}   \nf \right) },
\\
\nl
\beta_{1}^{(\nf)} & = & {\tt BETA1} =
\displaystyle{ \frac{1}{16} \left(102 -\frac{38}{3}  \nf \right) },
\\
\nl
\beta_{2}^{(\nf)} & = &  {\tt BETA2} =
\displaystyle{ \frac{1}{64} \left(\frac{2857}{2}-\frac{5033}{18}\nf 
              +\frac{325}{54} \nfS \right) }.
\label{betas}
\eqa
The Gamma function coefficients are:
\bqa
\gamma_{0}^{(\nf)} & = & {\tt GAMA0} = 1,
\\
\nl
\gamma_{1}^{(\nf)} & = & {\tt GAMA1} =
\displaystyle{ \frac{1}{16} \left( \frac{202}{3} - \frac{20}{9}\nf
                                                          \right) },
\\
\nl
\gamma_{2}^{(\nf)} & = & {\tt GAMA2} = 
\displaystyle{ \frac{1}{64} \left\{ 1249 
             - \left[ \frac{2216}{27} + \frac{160}{3} \ztri \right] \nf
             - \frac{140}{81} \nfS \right\} }.
\label{gamas}
\eqa
The running $\als$ follows:
\bqa
\frac{\als\left(s,\LMSBn \right)}{\pi} & = &
\frac{1}{\beta_0 L}
 \left\{ 1 - \frac{\beta_1}{\beta^2_0 L}   \ln L
           + \frac{\beta^2_1}{\beta^4_0 L^2} 
\left[ \ln^2 L - \ln L - 1 + \frac{\beta_2\beta_0}{\beta^2_1} \right] \right\},
\\
\nl
L & = & \ln \frac{s}{\LMSBnS}\;.
\label{runals}
\eqa
Actually the code calculates $\LMSBv$
for an input value of $\bar{\als}(\mzs)$,
solving numerically the RG-equation:
\bqa
\frac{\als\left(\mzs,\LMSBv \right)}{\pi} 
-\frac{1}{\beta_0 L}
 \left\{ 1 - \frac{\beta_1}{\beta^2_0 L}   \ln L
           + \frac{\beta^2_1}{\beta^4_0 L^2} 
\left[ \ln^2 L - \ln L - 1 + \frac{\beta_2\beta_0}{\beta^2_1} \right]
\right\} \;=\;0.\quad
\eqa
Then $\LMSBf$ and $\LMSBt$ are calculated using the matching condition:
\bqa
\ln\left(\frac{\LMSBn}{\LMSBnml}\right)^2&=&\beta^{(\nf-1)}_0\Biggl\{
\left(\beta^{(\nf)}_0-\beta^{(\nf-1)}_0\right)L_{\sss{M}}
+\left(\frac{\beta^{(\nf)}_1}{\beta^{(\nf)}_0}-\frac{\beta^{(\nf-1)}_1}
{\beta^{(\nf-1)}_0}\right)\ln
L_{\sss{M}} 
\nll 
\nll
&&
-\frac{\beta^{(\nf-1)}_1}{\beta^{(\nf-1)}_0}\ln\frac{\beta^{(\nf)}_0}
{\beta^{(\nf-1)}_0}
+\frac{\beta^{(\nf)}_1}{\left(\beta^{(\nf)}_0\right)^2}
 \left(\frac{\beta^{(\nf)}_1}{\beta^{(\nf)}_0}-\frac{\beta^{(\nf-1)}_1}
{\beta^{(\nf-1)}_0}\right) 
 \frac{\ln L_{\sss{M}}}{L_{\sss{M}}}
\nll 
\nll
&&
+\frac{1}{\beta^{(\nf)}_0}\Biggl[\Biggl(\frac{\beta^{(\nf)
    }_1}{\beta^{(\nf)  }_0}\Biggl)^2 
-\Biggl(\frac{\beta^{(\nf-1)}_1}{\beta^{(\nf-1)}_0}\Biggl)^2 
-\frac{\beta^{(\nf)}_2}{\beta^{(\nf)  }_0} 
+\frac{\beta^{(\nf-1)}_2}{\beta^{(\nf-1)}_0} 
-\frac{7}{72}\Biggr]\frac{1}{L_{\sss{M}}}\Biggr\}, 
\nll
\eqa
where
\bqa
L_{\sss{M}} & = & \ln \frac{\MqS}{\LMSBnS}\;.
\eqa
The parameters $\LMSBn$ for different flavor numbers are used then
where appropriate 
in the \eqn{runals} in order to calculate the running $\als(\sman)$.

\newpage

\section{Subroutine {\tt EWCOUP} \label{ewcoup}}
\eqnzero
Subroutine {\tt EWCOUP(INTRF,INDF,S)} prepares coupling functions
dressed with EW and QCD corrections.
For the Born case they were introduced in \sect{i.born}.

In the argument list of subroutine {\tt EWCOUP(INTRF,INDF,S)} 
all parameters are {\tt INPUT}. 
For the default setting it is run for every $\sman$
and produces {\tt EW COUP}lings for requested values of 
{\tt INT}e{\tt RF}ace and {\tt INDF} of a process. 
If it is called
from {\tt ZUATSM}, the angular distribution branch, and if EW boxes are added
to EW form factors, it will be called for every $\tman$.
\subsection{Helicities and polarizations\label{helipol}}
First, we introduce the longitudinal polarizations of the electron
$(\lpoli{-})$ and positron $(\lpoli{+})$ and the helicities of the final
state fermions $\hpoli{\pm}$ in the following combinations for polarizations:
\bqa
\lpoli{1}&=&{\tt COMB1}=1-\lpoli{+} \lpoli{-}\,,
\\
\lpoli{2}&=&{\tt COMB2}=  \lpoli{+}-\lpoli{-}\,,
\label{polarizations}
\eqa
and for helicities, assuming that helicity may have only three
values $(0,-1,+1)$\footnote{The helicity values $(-1,+1)$ correspond
to eigenvalues and may be 
used for the determination of final state helicity asymmetries while
the value $(0)$ reflects helicity averaging.}:
\bqa
\begin{array}{ll}
\hpoli{+}\;\neq\;0\;{\tt .and.}\;\hpoli{-}\;\neq\;0&\left\{ 
\begin{array}{l}
\hpoli{1}={\tt HOMB1}={\ds{\frac{1}{4}(1-\hpoli{+} \hpoli{-})}},
\\ \\
\hpoli{2}={\tt HOMB2}={\ds{\frac{1}{4}(  \hpoli{+}-\hpoli{-})}},
\end{array}\right.
\\ \\
\hpoli{+}\;=\;0\;{\tt .and.}\;\hpoli{-}\;=\;0&\left\{ 
\begin{array}{l}
\hpoli{1}={\tt HOMB1}=1,
\\
\hpoli{2}={\tt HOMB2}=0,
\end{array}\right.
\\ \\
\hpoli{+}\;=\;0\;{\tt .and.}\;\hpoli{-}\;\neq\;0&\left\{ 
\begin{array}{l}
\hpoli{1}={\tt HOMB1}={\ds{\frac{1}{2}}},
\\ \\
\hpoli{2}={\tt HOMB2}={\ds{\frac{1}{2}(-\hpoli{-})}},
\end{array}\right.
\\ \\
\hpoli{+}\;\neq\;0\;{\tt .and.}\;\hpoli{-}\;=\;0&\left\{ 
\begin{array}{l}
\hpoli{1}={\tt HOMB1}={\ds{\frac{1}{2}}},
\\ \\
\hpoli{2}={\tt HOMB2}={\ds{\frac{1}{2}(+\hpoli{+})}}.
\end{array}\right.
\end{array}
\label{helicities}
\eqa
Then, if we neglect masses, \eqns{kgizsplit}{kgiz} for the coupling
functions (or alternatively {\em coupling factors}) change to:
\bqa
\begin{array}{rclcl}
K^{\ff}_{\sss{T}}(\ph) &\rightarrow&
K^{\ff}_{\sss{T}}(\ph,\lpoli{1,2},\hpoli{1,2}) &=&  
\lpoli{1}\hpoli{1}\qes\qfs\cf,
\\ \\
0&\rightarrow&
K^{\ff}_{\sss{FB}}(\ph,\lpoli{1,2},\hpoli{1,2})&=&
\lpoli{2} \hpoli{2}\qes\qfs\cf,
\\ \\
K^{\ff}_{\sss{T}}(I)   &\rightarrow&
K^{\ff}_{\sss{T}}(I,\lpoli{1,2},\hpoli{1,2})   &=&
2|\qe\qf|\bigl(\lpoli{1} v_e+\lpoli{2} a_e\bigr)
         \bigl(\hpoli{1} v_f+\hpoli{2} a_f\bigr)\cf,
\\ \\
K^{\ff}_{\sss{FB}}(I)  &\rightarrow&
K^{\ff}_{\sss{FB}}(I,\lpoli{1,2},\hpoli{1,2})  &=&
2|\qe\qf|\bigl(\lpoli{1} a_e+\lpoli{2} v_e\bigr)
         \bigl(\hpoli{1} a_f+\hpoli{2} v_f\bigr)\cf,
\\ \\
K^{\ff}_{\sss{T}}(Z)   &\rightarrow&
K^{\ff}_{\sss{T}}(Z,\lpoli{1,2},\hpoli{1,2})   &=&
\Bigl[\lpoli{1}(v_e^2 + a_e^2)+2\lpoli{2} v_ea_e\Bigr]
\Bigl[\hpoli{1}(v_f^2 + a_f^2)+2\hpoli{2} v_fa_f\Bigr]\cf,
\\ \\
K^{\ff}_{\sss{FB}}(Z) &\rightarrow& 
K^{\ff}_{\sss{FB}}(Z,\lpoli{1,2},\hpoli{1,2})  &=& 
\Bigl[2\lpoli{1}v_ea_e+\lpoli{2}\Bigl(v_e^2+a_e^2\Bigr)\Bigr]
\Bigl[2\hpoli{1}v_fa_f+\hpoli{2}\Bigl(v_f^2+a_f^2\Bigr)\Bigr]\cf.
\end{array}
\label{kgizl0}
\eqa
One may see that the vector- and axial-vector couplings $v_e,a_e$
and their squares $v_e^2,a_e^2$ merely have to be substituted by
``new'' couplings depending linearly on the polarization $\lambda$
(the same is true in an analogous manner for couplings and
polarization of the final-state).

If $\mfl\neq 0$, $a_f$ has to be
substituted in the above formulae by $a_f\beta_f$, with $\beta_f$
given by \eqn{mus}. 

Then, one has for $K^{\ff}_{\sss{T}}(I),\;K^{\ff}_{\sss{T}}(Z)$:
\bqa
K^f_{\sss{T}}(I,\lpoli{1,2},\hpoli{1,2})&\rightarrow&
K^f_{\sss{T}}(I,\lpoli{1,2},\hpoli{1,2},\beta_f(s))=
  2\big|\qe\qf\big|\big[\lpoli{1} v_e + \lpoli{2} a_e\big]
                   \big[\hpoli{1} v_f + \hpoli{2} a_f\beta_f(s)\big],
\nll \\
K^f_{\sss{T}}(Z,\lpoli{1,2},\hpoli{1,2})&\rightarrow&
K^f_{\sss{T}}(Z,\lpoli{1,2},\hpoli{1,2},\beta_f(s))=
\Bigl[\lpoli{1}(v^2_e+a^2_e) + 2\lpoli{2} v_e a_e\Bigr] 
\nll&& \times 
\Bigl[\hpoli{1}\Bigl(v^2_f+a^2_f{\beta}^2_f(s)\Bigr)
               + 2\hpoli{2} v_f a_f \beta_f(s)
\Bigr].
\eqa
There are further contributions
$K^{m}_{\sss{T}}(\ph)$, $K^{m}_{\sss{T}}(I)$ and $K^{m}_{\sss{T}}(Z)$ 
in the total cross-section $\sigma_{\sss{T}}$:
\bqa
K^{m}_{\sss{T}}(\ph,\lpoli{1,2},\hpoli{1,2})&=& 
          \lpoli{1}\bigl(2-\hpoli{1}\bigr)\qes\qfs, 
\\
K^{m}_{\sss{T}}(I,\lpoli{1,2},\hpoli{1,2})  &=&
 2\big|\qe\qf\big|\bigl(\lpoli{1} v_e+\lpoli{2} a_e\bigr)
                   \bigl(2-\hpoli{1}\bigr)v_f,
\\
K^{m}_{\sss{T}}(Z,\lpoli{1,2},\hpoli{1,2})  &=&
\Bigl[\lpoli{1}\bigl(v^2_e+a^2_e\bigr)+2\lpoli{2} v_e a_e\Bigr]
                   \bigl(2-\hpoli{1}\bigr)v^2_f.   
\label{kgizlmf}
\eqa
\subsection{Preparation of effective couplings for various interfaces
\label{effective_couplings}}
In the beginning {\tt EWCOUP} calculates the running electromagnetic coupling
$\alpha(\sman)$ by a call of {\tt XFOTF3} and the four complex-valued
effective couplings by a call of {\tt ROKANC}, see \fig{interfEWCOUP}:
\bqa
\rhoi{\fe\ff}&=&{\tt XRO},
\\
\Rva{\fe}    &=&{\tt XVEZ}=1-4|\qe|\Bigl(\kappai{\fe}\siws+\Imsi{\fe}\Bigr),
\\
\Rva{\ff}    &=&{\tt XVFZ}=1-4|\qf|\Bigl(\kappai{\ff}\siws+\Imsi{\ff}\Bigr),
\\
\Rva{\fe\ff} &=&{\tt XVEFZ}=-1+\Rva{\fe}+\Rva{\ff}+16|\qe||\qf|
\Bigl[\siwf\kappai{\fe\ff}
+\siws\Bigl(\kappai{\fe}\Imsi{\ff}+\kappai{\ff}\Imsi{\fe}\Bigr)
\Bigr].\quad
\label{XROVEFZ}
\eqa
In general, they are functions of two Mandelstam variables $(\sman,\tman)$.
These four complex-valued form factors  are generalizations of the
effective $\zb$ decay constants $\rZf$ and $\Rvaz{\ff}$ 
for the case of the scattering process $\fep\fem\to\ff\fbf$.  

At the $\zb$ resonance, some approximations and factorizations
are fulfilled:
\bqa
\label{appr_and_fact0}
\rhois{\fe\ff}(\mzs,\tman) &\approx&\rZdf{\fe}\rZdf{\ff}\,,
\\
\kappai{\fe\ff}(\mzs,\tman)&\approx&\kZdf{\fe}\kZdf{\ff}\,,
\\
\kappai{\fe}(\mzs,\tman)   &\approx&\kZdf{\fe}\,,
\\
\kappai{\ff}(\mzs,\tman)   &\approx&\kZdf{\ff}\,.
\label{appr_and_fact}
\eqa
We note that the $\tman$-dependence switches off at $\sman=\mzs$.

If the factorization property were fulfilled we would have
\bq
\Rva{\fe\ff}=\Rva{\fe}\Rva{\ff},
\eq
which, in turn, would greatly simplify all subsequent formulae. 
Although factorization holds with rather good precision at the $Z$
resonance, it 
deteriorates on the wings and the very high precision of \LEPI\ data demands
to refuse from an application of this approximation even at resonance.


Then {\tt EWCOUP} calculates $\sman$-dependent QCD corrections
by a call to {\tt QCDCOF}, providing the array {\tt QCDCOR(0:14)},
see \eqn{qcdcor_fst}. 
From {\tt QCDCOR(13)} one constructs three
vector singlet contributions
$R^{\sss{S},\{\ph,\sss{\zb}\ph,\sss{\zb}\}}_{\sss{V}}(\sman)$:
\bqa
R^{\sss{S},         \ph}_{\sss{V}}(\sman)&=&{\tt VSNGAA}=
\frac{1}{9}  R^{\sss{S}}_{\sss{V}}(\sman),
\\
R^{\sss{S},\sss{\zb}\ph}_{\sss{V}}(\sman)&=&{\tt VSNGZA}=
\lpar\frac{7}{3}-\frac{44}{9}\siws\rpar R^{\sss{S}}_{\sss{V}}(\sman),
\\
R^{\sss{S},   \sss{\zb}}_{\sss{V}}(\sman)&=&{\tt VSNGZZ}=
\lpar 1+\frac{4}{3}\siws\rpar R^{\sss{S}}_{\sss{V}}(\sman),
\eqa
which are shared `democratically' over five open quark channels:
$\fu,\fd,\fc,\fs,\fb$.

Depending on the flag {\tt FINR} the final state QCD$\otimes$QED 
correction factors $R^{\ff}_{\sss{V,A}}$ and $R^{\sss{S}}_{\sss{V}}$
are modified as follows:
\begin{itemize}
\item[]{\tt FINR}=--1: $R^{\ff}_{\sss{V,A}}=1$, 
$R^{\sss{S},\{\ph,\sss{\zb}\ph,\sss{\zb}\}}_{\sss{V}}=0$;
\item[]{\tt FINR}=0:  full expressions are used as given in \sect{qcdrun};
\item[]{\tt FINR}=1:  from $R^{\ff}_{\sss{V,A}}$ one subtracts the
inclusive $\ord{\alpha}$ QED correction 
$3\alpha\qfs/(4\pi)$ avoiding double counting in the case of a 
proper treatment of cuts with the aid of function {\tt FUNFIN},
see \subsect{IBORNA}; the factors $R^{\sss{S}}_{\sss{V}}$ remain unchanged.
\end{itemize}

Further, {\tt EWCOUP} provides non-factorized EW$\otimes$QCD 
corrections, governed by flag {\tt CZAK}.

This all is done for every of the ten values of {\tt INTRF} 
and of the twelf values of {\tt INDF} (see \fig{interfEWCOUP} for the 
meaning of {\tt INDF} and 
for the correspondence between interface name and its number {\tt INTERF}).

Then the calculations proceed in different streams for different interfaces
providing seven {\em coupling factors}
$K^{\ff}_{\sss{T,FB}}(\ph,I,\zb)$, \eqn{kgizl0},
and ${\bar{K}}^{\ff,m}_{\sss{T}}(\zb)$,
see \eqn{kgiz}.

We note that they bear also the process index $\ff$ = {\tt INDF}.
\subsection[Standard Model interfaces 
{\tt ZUTHSM}, {\tt ZUTPSM}, {\tt ZULRSM}, {\tt ZUATSM}. {\tt INTRF}=1]
{Standard Model interfaces 
{\tt ZUTHSM}, {\tt ZUTPSM}, {\tt ZULRSM}, {\tt ZUATSM}. 
\\
{\tt INTRF}=1 
\label{intrf1}
}
In the SM, the coupling factors take into account EW and QCD
RC by means of EWFF (\eqns{form1}{form4})
and mixed QEQ$\otimes$QCD FSR corrections by means of radiator factors
(\eqn{QCDradiators} and \sect{qcdrun}).
The Born formulae~\eqn{kgizl0} change to:
\bqa
K^{\ff}_{\sss{T}}(\ph,\lpoli{1,2},\hpoli{1,2})&=&{\tt VEFA}
=\lpoli{1} \hpoli{1}\qes\lrbr\qfs R^{\ff}_{\sss{V}}(\sman)
          +\frac{1}{5}R^{{\sss{S}},\ph}_{\sss{V}}(\sman)\rrbr\cf,
\label{gcouft1}
\\
K^{\ff}_{\sss{FB}}(\ph,\lpoli{1,2},\hpoli{1,2})&=&{\tt AEFA}
=\lpoli{2} \hpoli{2}\qes\qfs\cf,
\label{gcoufa1}
\\ 
\frac{1}{2}K^{\ff}_{\sss{T}}(I,\lpoli{1,2},\hpoli{1,2})&=&{\tt XVEFI}
=\rhoi{\fe\ff}\Biggl\{\lpoli{1}\hpoli{1}|\qe|
\lrbr|\qf|\Rva{\fe\ff}R^{\ff}_{\sss{V}}(\sman)
      +\Rva{\fe}\frac{1}{5}R^{\sss{S},\sss{\zb}\ph}_{\sss{V}}(\sman)\rrbr
\nll&&
+|\qe||\qf|\Bigl[\lpoli{2}\hpoli{1}\Rva{\ff}+\lpoli{1}\hpoli{2}\Rva{\fe}
                +\lpoli{1}\hpoli{2}\Bigr]
\Biggr\}\cf,
\label{icouft1}         
\\
\frac{1}{2}K^{\ff}_{\sss{FB}}(I,\lpoli{1,2},\hpoli{1,2})&=&{\tt XAEFI}
=\rhoi{\fe\ff}|\qe||\qf|
\Bigl\{\lpoli{1}\hpoli{1}R^{\fq}_{\sss{FB}}(\sman)
\nll &&
+\lpoli{2}\hpoli{1}\Rva{\fe}
+\lpoli{1}\hpoli{2}\Rva{\ff}+\lpoli{2}\hpoli{2}\Rva{\fe\ff}
\Bigr\}\cf,
\label{icoufa1}
\\
{\bar{K}}^{\ff,m}_{\sss{T}}(Z,\lpoli{1,2},\hpoli{1,2})&=&{\tt VEEZ}
=\big|\rhoi{\fe\ff}\big|^2\lpoli{1}\hpoli{1}\bigl(|\Rva{\fe}|^2+1\bigr)
R^{\ff}_{\sss{A}}(\sman)\cf,
\label{mcouft1}
\\
K^{\ff}_{\sss{T}}(Z,\lpoli{1,2},\hpoli{1,2})&=&{\tt VEFZ}=
\big|\rhoi{\fe\ff}\big|^2\Bigl[
\lpoli{1}\hpoli{1}V_{\sss{Z1}}+\lpoli{2}\hpoli{1}V_{\sss{Z2}}
\lpoli{1}\hpoli{2}A_{\sss{Z2}}+\lpoli{2}\hpoli{2}A_{\sss{Z1}}
\Bigr]\cf,
\label{zcouft1}
\\
K^{\ff}_{\sss{FB}}(Z,\lpoli{1,2},\hpoli{1,2})&=&{\tt AEFZ}=
\big|\rhoi{\fe\ff}\big|^2\Bigl[
\lpoli{1}\hpoli{1}A_{\sss{Z1}}+\lpoli{2}\hpoli{1}A_{\sss{Z2}}
\lpoli{1}\hpoli{2}V_{\sss{Z2}}+\lpoli{2}\hpoli{2}V_{\sss{Z1}}
\Bigr]\cf.\qquad\qquad
\label{zcoufa1}
\eqa
The vector and axial combinations are:
\bqa
V_{\sss{Z1}}&=&
\Bigl(\big|\Rva{\fe\ff}\big|^2+\big|\Rva{\ff}\big|^2\Bigr)
R^{\ff}_{\sss{V}}(\sman)
+\Bigl(\big|\Rva{\fe}\big|^2+1\Bigr)
\lrbr R^{\ff}_{\sss{A}}(\sman)+\frac{1}{5}
R^{\sss{S},\sss{\zb}}_{\sss{V}}(\sman)\rrbr,
\\
V_{\sss{Z2}}&=&2\Reb\Bigl[
\Rvac{\fe\ff}\Rva{\ff}R^{\ff}_{\sss{V}}(\sman)
            +\Rva{\fe}R^{\ff}_{\sss{A}}(\sman)\Bigr],
\\
A_{\sss{Z1}}&=&2\Reb\Bigl(\Rva{\fe}\Rvac{\ff}+\Rva{\fe\ff}\Bigr)
R^{\fq}_{\sss{FB}}(\sman)\,,
\\
A_{\sss{Z2}}&=&2\Reb\Bigl(\Rvac{\fe\ff}\Rva{\fe}+\Rva{\ff}\Bigr).
\eqa
For a proper account of non-factorized corrections one needs also 
vector factors without QCD corrections:
\bqa
V^{0}_{\sss{Z1}}&=&
 \big|\Rva{\fe\ff}\big|^2+\big|\Rva{\ff}\big|^2
+\big|\Rva{\fe}\big|^2+1,
\\
V^{0}_{\sss{Z2}}&=&2\Reb\Bigl(\Rvac{\fe\ff}\Rva{\ff}+\Rva{\fe}\Bigr),
\eqa
and a corresponding $K^{\ff,0}_{\sss{T}}$ 
(without helicities and polarizations)
made of $V^{0}_{\sss{Z1,2}}$ by an expression similar to \eqn{zcouft1}.

It is instructive to see that without QCD corrections and if 
the factorization properties \eqns{appr_and_fact0}{appr_and_fact} are
fulfilled 
the $K_{\sss{A}}$ factors reduce to familiar expressions \eqn{kgizl0} up to
trivial substitutions of couplings:
\ba
v_f&\to&\Rva{\ff},
\\
a_f&\to& 1,
\ea
and with corresponding prefactors made of $\rhoi{\fe\ff}$. 

The coupling factors are used in subroutines {\tt ZANCUT} and {\tt ZCUT} 
in order to calculate differential or integrated observables. This
is done in Subroutines {\tt COSCUT}, or {\tt SFAST} and 
{\tt SCUT} respectively,
which call {\tt BORN} for the calculation of the IBA cross-section 
$\sigma^0_{\sss{T}}$ and forward-backward difference $\sigma^0_{\sss{FB}}$.
These couplings are also used in those parts of the code which calculate
contributions from the initial-final interference, IFI.
\subsection{Interface {\tt ZUXSEC}. {\tt INTRF}=2 \label{intrf2}}
Here we discuss the parameter release scheme used in the interface 
{\tt ZUXSEC}.
It calculates only the cross-section, without initial or final state 
polarizations. 
This corresponds to the internal flag {\tt INTRF}=2 set in Subroutine 
{\tt EWCOUP} which 
provides the coupling functions for the subsequent use in {\tt BORN} 
within a scheme where one wants to have partial $\zb$ widths to be
released for a fit to experimental data.
Let us denote the partial width to be released by $\bgz{\ff}$.

Within this scheme $\ph$ exchange and $\zb\ph$ interference are
simply taken from the SM:
\bqa
K^{\ff}_{\sss{T}}(\ph)&=&{\tt VEFA}=\qes\lrbr\qfs R^{\ff}_{\sss{V}}(\sman)
             +\frac{1}{5}R^{{\sss{S}},\ph}_{\sss{V}}(\sman)\rrbr\cf,
\label{gcouft2}
\\
K^{\ff}_{\sss{FB}}(\ph)&=&{\tt AEFA}=0,
\label{gcoufa2}
\\ 
\frac{1}{2}K^{\ff}_{\sss{T}}(I)&=&{\tt XVEFI}
=\rhoi{\fe\ff}\Biggl\{|\qe|
\lrbr|\qf|\Rva{\fe\ff}R^{\ff}_{\sss{V}}(\sman)
      +\Rva{\fe}\frac{1}{5}R^{\sss{S},\sss{\zb}\ph}_{\sss{V}}(\sman)\rrbr
\Biggr\}\cf,         
\label{icouft2}
\\
\frac{1}{2}K^{\ff}_{\sss{FB}}(I)&=&{\tt XAEFI}
=\rhoi{\fe\ff}|\qe||\qf|R^{\fq}_{\sss{FB}}(\sman)\cf,
\label{icoufa2}
\\
K^{\ff}_{\sss{FB}}(Z)&=&{\tt AEFZ}=
\big|\rhoi{\fe\ff}\big|^2
2\Reb\Bigl(\Rva{\fe}\Rvac{\ff}+\Rva{\fe\ff}\Bigr)
R^{\fq}_{\sss{FB}}(\sman)\cf.\qquad
\label{zcoufa2}
\eqa
Therefore, they all should be understood as SM remnants.

The release of $\bgz{\ff}$ is realized through the $\zb$-couplings by the 
following set of equations:
\bqa
{\bar{K}}^{\ff,m}_{\sss{T}}(Z)={\tt VEEZ}&=&
\lrbr \frac{\bgz{\fe}}{A_{\sss{N}} c_{ee}}-\frac{\pgz{\fe}}{A_{\sss{N}} c_{ee}}
+\big|\rhoi{\fe\ff}\big|^2\bigl(|\Rva{\fe}|^2+1\bigr)
\rrbr R^{\ff}_{\sss{A}}(\sman)\cf,
\label{mcouft2}
\\
\label{zcouft2}
K^{\ff}_{\sss{T}}(Z)={\tt VEFZ}&=&
\frac{\bgz{\fe}}{A_{\sss{N}} c_{ee}}
\lrbr
\frac{\bgz{\ff}}{A_{\sss{N}}\beta_f}
+6\rhoi{\ff} r(\mfl) R^{\ff}_{\sss{A}}(\sman)\cf 
\rrbr
\frac{1}{c_2(\mfl)}
\\&-&
\frac{\pgz{\fe}}{A_{\sss{N}} c_{ee}}
\lrbr
\frac{\pgz{\ff}}{A_{\sss{N}}\beta_f}
+6\rhoi{\ff} r(\mfl) R^{\ff}_{\sss{A}}(\sman)\cf
\rrbr
\frac{1}{c_2(\mfl)}
\nll&+&
\big|\rhoi{\fe\ff}\big|^2
\lcbr
\Bigl(\big|\Rva{\fe\ff}\big|^2+\big|\Rva{\ff}\big|^2\Bigr)
R^{\ff}_{\sss{V}}(\sman)
+\big|\Rva{\fe}\big|^2+1
\lrbr R^{\ff}_{\sss{A}}(\sman)
+\frac{1}{5}R^{\sss{S},\sss{\zb}}_{\sss{V}}(\sman)
\rrbr
\rcbr\cf.
\nonumber
\eqa
The $\rhoi{\ff}$ is a short hand notation for the real part of the
$\zb$ decay form factor $\Reb\,\rZdf{\ff}$:
\ba
\rhoi{\ff} &=& \Reb\,\rZdf{\ff},
\ea 
see also the discussion in \subsect{intrf3} and \eqn{shorthand}. 
Here we also introduced the notations:
\bqa
r(\mfl) &=& \frac{\mfs}{\mzs}\,,
\\
c_2(\mfl)&=& 1+2r(\mfl),
\\
A_{\sss{N}} &=& \frac{\gf\mzc}{24\sqrt{2}\pi}\,,
\\
c_{ee}&=& 1+\frac{3}{4}\frac{\alpha(s)}{\pi}\qes\,,
\label{eq:c216}
\eqa
where $c_{ee}$ is a QED correction factor 
and $\pgz{\fe}$ and $\pgz{\ff}$  stand for partial $\zb$-boson decay widths
calculated within the SM by the {\tt DIZET} package.
Finally, again for a proper account of non-factorized corrections 
one needs a coupling function $K^{\ff,0}_{\sss{T}}(Z)$ without QCD corrections:
\bq
K^{\ff,0}_{\sss{T}}(Z)={\tt VEFZ0}.
\eq

Let us define the {\em SM-trajectory} by the set of equations
\bq
\bgz{\ff}=\pgz{\ff}.
\eq
We note that \eqns{mcouft2}{zcouft2} at the SM-trajectory
are identically equal to their SM analogs \eqns{mcouft1}{zcouft1}.
This is the reason for the invention of  the terminology 
``a Semi--Model-Independent Approach, SMIA'' for releases of such kind. 
Everything what makes 
coupling factors in \eqns{mcouft2}{zcouft2}
deviate from their {\em released} values will be called
{\em SM-remnants}.

The scheme which has been just described is realized for {\tt INTRF}=2 and
{\tt INDF}=0,9. For the calculation of the total hadronic cross-section 
special chains exist in subroutines {\tt EWCOUP} and {\tt BORN}. For these
chains, for the sake of CPU time saving the coupling functions are
stored in arrays: 
\bqa
K^{\ff={\tt J}}_{\sss{T}}(\ph) &=&{\tt AVEFA(J)},
\label{gcouft2s}
\\
K^{\ff={\tt J}}_{\sss{FB}}(\ph)&=&{\tt AAEFA(J)}=0,
\label{gcoufa2s}
\\ 
\frac{1}{2}K^{\ff={\tt J}}_{\sss{T}}(I) &=&{\tt XXVEFI(J)},
\label{icouft2s}
\\
\frac{1}{2}K^{\ff={\tt J}}_{\sss{FB}}(I)&=&{\tt XXAEFI(J)},
\label{icoufa2s}
\\
K^{\ff={\tt J}}_{\sss{FB}}(Z)&=&{\tt AAEFZ(J)},
\label{zcoufa2s}
\\
{\bar{K}}^{\ff={\tt J},m}_{\sss{T}}(Z) &=&{\tt AVEEZ(J)},
\label{mcouft2s}
\eqa
where $J$ counts the flavor index $\ff$ from $\fu$- to $\fb$-quarks.
The couplings are super-indexed with $\ff$.
It is worth mentioning that the EW form factors
were calculated by means of a call to {\tt ROKANC}
not five times, for every $\ff=\fu,\fd,\fc,\fs,\fd$, 
but only three times, for $\ff=\fu,\fd,\fb$. This also saves CPU time.

The coupling functions \eqns{gcouft2s}{mcouft2s}
are similar to \eqns{gcouft2}{mcouft2}, while $K^{\ff}_{\sss{T}}(Z)$
gets a special treatment:
\bqa
\label{zcouft2s}
K^{\ff={\tt J}}_{\sss{T}}(Z)&=&{\tt  AVEFZ(J)}=
\frac{\bgz{\had}}{\pgz{\had}}\Biggl\{
\frac{\bgz{\fe}}{A_{\sss{N}} c_{ee}}\frac{\pgz{\ff}}{A_{\sss{N}}}
-
\frac{\pgz{\fe}}{A_{\sss{N}} c_{ee}}\frac{\pgz{\ff}}{A_{\sss{N}}}
\\&+&
\big|\rhoi{\fe\ff}\big|^2
\lcbr
\Bigl(\big|\Rva{\fe\ff}\big|^2+\big|\Rva{\ff}\big|^2\Bigr)
R^{\ff}_{\sss{V}}(\sman)
+\big|\Rva{\fe}\big|^2+1
\lrbr R^{\ff}_{\sss{A}}(\sman)+\frac{1}{5}
R^{\sss{S},\sss{\zb}}_{\sss{V}}(\sman)
\rrbr
\rcbr\cf
\Biggr\},
\nonumber
\eqa
and also the corresponding coupling function $K^{\ff={\tt J},0}_{\sss{T}}(Z)$
without QCD corrections.

The arrays \eqns{gcouft2s}{zcouft2s} are needed not only for a use in
subroutine {\tt BORN}. 
They are also used in subroutine {\tt BOXINT}
as a part of the calculation of the IFI corrections to order $\ord{\alpha}$.
In order to save further CPU time, {\tt EWCOUP} prepares {\em
summed} coupling  
functions weighted with $\qf$, since $\sigma^{\rm{\sss{IFI}}}\propto\qe\qf$: 
\bq
K_{\sss{T}}(\ph)=\sum^{5}_{{\sss{J}}=1}\qf K^{\ff={\tt J}}_{\sss{T}}(\ph),
\label{summedcf}
\eq
etc. 
\subsection{Interfaces {\tt ZUXSA} and {\tt ZUTAU}. {\tt INTRF}=3
\label{intrf3}
}
Now we introduce a family of SMIA interfaces ({\tt ZUXSA, ZUTAU}),
which use the language of {\em effective couplings release}.
The effective couplings may then be fitted to experimental data.
Let us introduce short-hand notations for the real parts of on-resonance 
couplings calculated within the SM (see \eqnsc{varatiorez}{CDZRKZ}):
\ba
\rhoi{\ff}&=&\Reb\,\rZdf{\ff}\,,
\\
\rhopi{\ff}&=&(\rZdf{\ff})^{'}\,,
\\
\rva{\ff}  &=&\Reb\,\Rvaz{\ff}\,.
\label{shorthand}
\ea  
Let $\rhobi{\fe},\;\rhobi{\ff},\;\rvab{\fe},\;\rvab{\ff},\;\rab{\fe}$ 
and $\rab{\ff}$ denote constant, real valued couplings to be fitted
to the experimental data. 

In all MI subroutines using the language of effective couplings we have 
two modes:
\begin{itemize}
\item[] {\tt MODE}=0: the $\rvab{\ff}$ and $\rab{\ff}$ have the meaning of
vector and axial vector effective coupling constants in the decay
$\zb\to\ff\fbf$ and the  
$\rhobi{\ff}$ are set equal to one;
\item[] {\tt MODE}=1: the $\rvab{\ff}$ have the meaning of the ratio
of vector and axial vector couplings and instead of $\rabs{\ff}$ one
uses $\rhobi{\ff}$.
\end{itemize}

For {\tt MISC}=0 we use the ``barred'' quantities together with
the SM values $\rhoi{\fe}$ and $\rhoi{\ff}$ directly.
If instead one chooses {\tt MISC}=1, a scaling is performed first. 
For a discussion of this scaling and the definition of the
$\reni{\ff}$ factors see at the end of \subsect{POCOMM}.
We will use the abbreviation {\tt OSCAL} for such a scaling: 
\ba
\rhobi{\fe}&\to&\rhobi{\fe}/\reni{\fe}\,,
\\
\rhobi{\ff}&\to&\rhobi{\ff}/\reni{\ff}\,,
\\
\rvab{\fe} &\to&\rvab{\fe}/\sreni{\fe}\,,
\\
\rvab{\ff} &\to&\rvab{\ff}/\sreni{\ff}\,,
\\
\rab{\fe}  &\to&\rab{\fe}/\sreni{\fe}\,,    
\\
\rab{\ff}  &\to&\rab{\ff}/\sreni{\ff}\,,  
\\
\rhoi{\fe} &\to&\rhopi{\fe}/\reni{\fe}\,,
\\
\rhoi{\ff} &\to&\rhopi{\ff}/\reni{\ff}.
\label{oscal3}
\ea
For this case we
define the {\em SM trajectory} by another set of equalities,
which are {\tt MODE} dependent.
We give it for the example of {\tt MODE}=0:
\bqa
\rhobi{\fe}&=&\rhoi{\fe}\,,
\\
 \rhobi{\ff}&=&\rhoi{\ff},
\\
\rvab{\fe} &=&\rva{\fe}\,, 
\\
 \rvab{\ff} &=&\rva{\ff}.
\label{SM_trajectory}
\eqa
\subsubsection{$\zb$ resonance approximation and SM-remnants}
In the vicinity of the $\zb$ resonance the following approximations hold
with rather high accuracy:
\bqa
\rhoi{\fe\ff}&\approx&\sqrt{\rhoi{\fe}\rhoi{\ff}},
\\
\Rva{\fe}&\approx&\rva{\fe}\,,
\\
\Rva{\ff}&\approx&\rva{\ff}\,,
\\
\Rva{\fe\ff}&\approx&\rva{\fe}\rva{\ff}\,.
\label{interf3_approxim}
\eqa
This allows us to define a new set of couplings, which
at the {\em SM trajectory} exactly coincides with the SM analog:
\bqa
\label{interf3_release1}
\rhohi{\fe\ff}&=&\sqrt{\rhobi{\fe}\rhobi{\ff}}
                 -\sqrt{ \rhoi{\fe} \rhoi{\ff}}+\rhoi{\fe\ff}\,,
\\
\Rvah{\fe}&=&\rvab{\fe}-\rva{\fe}+\Rva{\fe}\,,
\\
\Rvah{\ff}&=&\rvab{\ff}-\rva{\ff}+\Rva{\fe}\,,
\\
\Rvah{\fe\ff}&=&\rvab{\fe}\rvab{\ff}-\rva{\fe}\rva{\ff}+\Rva{\fe\ff}\,.
\label{interf3_release}
\eqa
They are an example of a parameter release with {\em exact $\sman$-dependent
SM remnants} since at the SM trajectory:
\bq
\rhohi{\fe\ff}=\rhoi{\fe\ff}\,,
\eq
etc. 

We only mention that the quantities, accompanying the barred release
quantities in \eqns{interf3_release1}{interf3_release},
\bq
-\sqrt{ \rhoi{\fe} \rhoi{\ff}}+\rhoi{\fe\ff}\,,
\eq
etc., are another class of SM remnants compared to those appearing in
\eqns{mcouft2}{zcouft2}.

For the photonic coupling factors {\tt VEFI} and {\tt AEFI},
we use their SM values of \eqns{gcouft1}{gcoufa1}, 
because they have nothing to do with effective $\zb$ couplings. 
For the $\gamma Z$ interference and $\zb$ exchange coupling factors we
use expressions similar to those of the  SM, \eqns{icouft1}{zcoufa1},
but with two modifications: 
\begin{itemize}
\item[-] use $\rhohi{\fe\ff}$, $\Rvah{\ff}$, etc.
instead of $\rhoi{\fe\ff}$, $\Rva{\ff}$, etc.;
\item[-] proper restoration of axial couplings having in mind the two
options of {\tt MODE} described above. 
\end{itemize}
It is sufficient to show the modifications of $V,A$ and 
those of the $\gamma Z$ interference and of the ${\bar{K}}^{\ff,m}_{\sss{T}}$
coupling functions.  
The former are:
\bqa
V_{\sss{Z1}}&=&
\Bigl(\big|\Rvah{\fe\ff}\big|^2+\rabs{\fe}\big|\Rvah{\ff}\big|^2\Bigr)
R^{\ff}_{\sss{V}}(\sman)
+\Bigl(\big|\Rvah{\fe}\big|^2+\rabs{\fe}\Bigr)\rabs{\ff}
\lrbr R^{\ff}_{\sss{A}}(\sman)
+\frac{1}{5}R^{\sss{S},\sss{\zb}}_{\sss{V}}(\sman)\rrbr,
\\
V_{\sss{Z2}}&=&2\Reb\Bigl[
\rab{\fe}\Rvahc{\fe\ff}\Rva{\ff} R^{\ff}_{\sss{V}}(\sman)
+\Rvah{\fe}\rab{\fe}\rabs{\ff} R^{\ff}_{\sss{A}}(\sman)
\Bigr],
\\
A_{\sss{Z1}}&=&2\Reb\Bigl(\Rvah{\fe}\Rvahc{\ff}+\Rvah{\fe\ff}
\Bigr)\rab{\fe}\rab{\ff}
R^{\fq}_{\sss{FB}}(\sman)\,,
\\
A_{\sss{Z2}}&=&2\Reb\Bigl(\Rvah{\fe}\Rvahc{\fe\ff}
+\rabs{\fe}\Rvah{\ff}\Bigr)\rab{\ff},
\\
V^{0}_{\sss{Z1}}&=&
\big|\Rvah{\fe\ff}\big|^2+\rabs{\fe}\big|\Rvah{\ff}\big|^2
+\Bigl(\big|\Rvah{\fe}\big|^2+\rabs{\fe}\Bigr)\rabs{\ff},
\\
V^{0}_{\sss{Z2}}&=&2\Reb\Bigl(\rab{\fe}\Rvahc{\fe\ff}\Rva{\ff}
+\rab{\fe}\Rvah{\fe}\rabs{\ff}\Bigr),
\label{interf3_VAmod}
\eqa
and the latter:
\bqa
\label{icouf3}
\frac{1}{2}K^{\ff}_{\sss{T}}(I,\lpoli{1,2},\hpoli{1,2})&=&{\tt XVEFI}
=\rhohi{\fe\ff}\Biggl\{\lpoli{1}\hpoli{1}|\qe|
\lrbr|\qf|\Rvah{\fe\ff}R^{\ff}_{\sss{V}}(\sman)
      +\Rvah{\fe}\frac{1}{5}R^{\sss{S},\sss{\zb}\ph}_{\sss{V}}(\sman)\rrbr
\nll&&
+|\qe||\qf|\Bigl[\lpoli{2}\hpoli{1}\rab{\fe}\Rvah{\ff}
                +\lpoli{1}\hpoli{2}\Rvah{\fe}\rab{\ff}
                +\lpoli{1}\hpoli{2}\rab{\fe}\rab{\ff}\Bigr]
\Biggr\}\cf,         
\\
\frac{1}{2}K^{\ff}_{\sss{FB}}(I,\lpoli{1,2},\hpoli{1,2})&=&{\tt XAEFI}
=\rhohi{\fe\ff}|\qe||\qf|
\Bigl\{\lpoli{1}\hpoli{1}\rab{\fe}\rab{\ff}R^{\fq}_{\sss{FB}}(\sman)
\nll &&
+\lpoli{2}\hpoli{1}\Rvah{\fe}\rab{\ff}
+\lpoli{1}\hpoli{2}\rab{\fe}\Rvah{\ff}+\lpoli{2}\hpoli{2}\Rvah{\fe\ff}
\Bigr\}\cf,
\\
{\bar{K}}^{\ff,m}_{\sss{T}}(Z,\lpoli{1,2},\hpoli{1,2})&=&{\tt VEEZ}
=\big|\rhoi{\fe\ff}\big|^2\lpoli{1}\hpoli{1}
\bigl(|\Rvah{\fe}|^2+\rabs{\fe}\bigr)\rabs{\ff}
R^{\ff}_{\sss{A}}(\sman)\cf.
\label{interf3_cfmod}
\eqa
\subsection{Interface {\tt ZUXSA2}. {\tt INTRF}=4\label{intrf4}}
The interface {\tt ZUXSA2} is very similar to interface {\tt ZUXSA}.
It was invented only for a study of leptonic channels assuming lepton
universality 
and was designed ignoring polarizations and helicities. 
The formulae are particularly compact due to these limitations.
In {\tt ZUXSA2}, the quantities released for the fit are 
$\rhois{\fl},\rvabs{\fl},\rabs{\fl}$.
If {\tt MISC}=0 they are used directly.
For {\tt MISC}=1, the rescaling {\tt OSCAL} (introduced in
\subsect{intrf3}) is performed:
\bqa
\rhois{\fl}&\to&\rhois{\fl}/\renis{\fe}\,,
\\
\sqrt{\rvabs{\fl}}&\to&\sqrt{\rvabs{\fl}}/\sreni{\fe}\,,
\\
\sqrt{\rvabs{\fl}}&\to&\sqrt{\rvabs{\fl}}/\sreni{\fe}\,,
\\
\rabs{\fl}&\to&\rabs{\fl}/\reni{\fe}\,,
\\
\rvabs{\fl}&\to&\rvabs{\fl}/\reni{\fe}\,,
\\
\rhoi{\fe}&\to&\rhopi{\fe}/\reni{\fe}\,,
\\
\rhoi{\ff}&\to&\rhopi{\ff}/\reni{\ff}.
\label{oscal4}
\eqa
The parameter release with exact $\sman$-dependent SM remnants is
realized by: 
\bqa
\rhohi{\fe\ff}&=&\sqrt{\rhois{\fl}}
                  -\sqrt{\rhoi{\fe}\rhoi{\ff}}+\rhoi{\fe\ff}\,,
\\
\Rvah{\fe}&=&\sqrt{\rvabs{\fl}}-\rva{\fe}+\Rva{\fe}\,,
\\
\Rvah{\ff}&=&\sqrt{\rvabs{\fl}}-\rva{\ff}+\Rva{\ff}\,,
\\
\Rvah{\fe\ff}&=&\rvabs{\fl}-\rva{\fe}\rva{\ff}+\Rva{\fe\ff}\,.
\eqa
The coupling functions for this interface are written down without auxiliary
functions since they are compact:
\bqa
{\tt VEFA}&=&\qes\qfs,
\\
{\tt AEFA}&=&0,
\\
{\tt XVEFI}&=&\rhohi{\fe\ff}|\qe\qf|\Rvah{\fe\ff},       
\\
{\tt XAEFI}&=&\rhohi{\fe\ff}|\qe\qf|\rabs{\fl},
\\
{\tt VEEZ}&=&\bigl|\rhohi{\fe\ff}\bigr|^2
\Bigl(\big|\Rvah{\fe}\big|^2+\rabs{\fl}\Bigr)\rabs{\fl},
\\
{\tt VEFZ}&=&\bigl|\rhohi{\fe\ff}\bigr|^2
\Bigl[
\Bigl(\big|\Rvah{\fe\ff}\big|^2+\rabs{\fl}\big|\Rvah{\ff}\big|^2\Bigr)
+\bigl(\big|\Rvah{\fe}\big|^2+\rabs{\fl}\bigr)\rabs{\fl}
\Bigr],
\\
{\tt VEFZ0}&=&{\tt VEFZ},
\\
{\tt AEFZ}&=&\bigl|\rhohi{\fe\ff}\bigr|^2     
\rabs{\fl}2\Reb\Bigl(\Rvah{\fe}\Rvahc{\ff}+\Rvah{\fe\ff}\Bigr).
\eqa
\subsection{Interface {\tt ZUXAFB}. {\tt INTRF}=5\label{intrf5}}
Interface {\tt ZUXAFB} uses as parameters for release three combinations
of couplings: $(\vaeII)$, $(\vafII)$ and $\four$.
Like the previous one, it is also intended only for an application to
leptonic channels 
ignoring polarizations and helicities.
However, it doesn't make use of lepton universality. 
Moreover, it doesn't have {\tt MODE}=1; therefore,
$\rvab{\ff}$ should be treated as $\rvb{\ff}$.

If {\tt MISC}=0 these three parameters are used directly.
For {\tt MISC}=1 the scaling {\tt OSCAL} (introduced in
\subsect{intrf3}) looks as follows:
\bqa
(\four)&=&(\four)/\reni{\fe}/\reni{\ff},
\\
(\vaeII)&=&(\vaeII)/\reni{\fe}\,,
\\
(\vafII)&=&(\vafII)/\reni{\ff},
\\
\rhoi{\fe}&=&\rhopi{\fe}/\reni{\fe}\,,
\\
\rhoi{\ff}&=&\rhopi{\ff}/\reni{\ff}.
\eqa
Again, $\ph$-exchange and $\gamma Z$ interference couplings are taken
from the SM: 
\bqa
{\tt VEFA}&=&\qes\qfs,
\\
{\tt AEFA}&=&0,
\\
{\tt XVEFI}&=&\rhoi{\fe\ff}|\qe\qf|\Rva{\fe\ff},       
\\
{\tt XAEFI}&=&\rhoi{\fe\ff}|\qe\qf|,
\eqa
and the release is performed for the $\zb$-couplings only:
\bqa
{\tt VEEZ}&=&(\vaeII)-\rhoi{\fe}\bigl(\rva{\fe}^2+1\bigr)
+\bigl|\rhoi{\fe\ff}\bigr|^2\bigl(\big|\Rva{\fe}\big|^2+1\bigr),
\\
{\tt VEFZ}&=&(\vaeII)(\vafII)
           -\rhoi{\fe}\bigl(\rva{\fe}^2+1\bigr)
            \rhoi{\ff}\bigl(\rva{\ff}^2+1\bigr)
\nll &&    
+\bigl|\rhoi{\fe\ff}\bigr|^2
\bigl(\big|\Rva{\fe\ff}\big|^2+\big|\Rva{\ff}\big|^2+\big|\Rva{\fe}\big|^2+1
\bigr),  
\\
{\tt VEFZ0}&=&{\tt VEFZ},
\\
{\tt AEFZ}&=&4\bigl[(\four)-\rva{\fe}\rva{\ff}\rhoi{\fe}\rhoi{\ff}\bigr]
+2\bigl|\rhoi{\fe\ff}\bigr|^2
\Reb\bigl(\Rva{\fe}\Rvac{\ff}+\Rva{\fe\ff}\bigr).
\eqa
\subsection{Interface {\tt ZUALR}. {\tt INTRF}=6\label{intrf6}}
The interface {\tt ZUALR} is presently empty. 
It is foreseen for the analysis of $A_{\sss{LR}}$.

\section{Subroutine {\tt BORN}\label{IBORNA}}
\eqnzero
\subsection{Cross-sections and forward-backward asymmetry}
Subroutine {\tt BORN} delivers the two IBA cross-sections 
$\sigma^{\rm{\sss{IBA}}}_{\sss{T}}$ and
$\sigma^{\rm{\sss{IBA}}}_{\sss{FB}}$
for the production of leptons or quarks in the final-state (determined
by flag {\tt INDF})  
or only one $\sigma^{\rm{\sss{IBA}}}_{\sss{T}}$
for the total hadronic cross-section (for {\tt INDF}=10 and {\tt
INTRF}=2):  
\bqa
\label{IBAT}
\sigma^{\rm{\sss{IBA}}}_{\sss{T}}(R_1,R_2)=
{\tt SBORN}&=&\frac{1}{R_1}\beta_f\biggl\{S^{\ff}_{\rm{\sss{QED}}}c_2(\mfl)
\biggl[K^{\ff}_{\sss{T}}(\ph)|V_{pol}|^2
+\frac{1}{2}\Re e\Bigl(K^{\ff}_{\sss{T}}(I)\chi(R_1,R_2)V_{pol}^{*}\Bigr)
\biggr]
\nll&&
+\biggl[c_2(\mfl)\biggl(K^{\ff}_{\sss{T}}(Z)
+\Bigl(S^{\ff}_{\rm{\sss{QED}}}-1+S^{\ff}_{\rm{\sss{CKHSS}}}\Bigr)
K^{\ff,0}_{\sss{T}}(Z)\biggr)
\nll&&
+c_3(\mfl) 
{\bar{K}}^{\ff,m}_{\sss{T}}(Z)\biggr]
 \Re e\Bigl(\chi_{\sss{Z}}(R_1)\chi_{\sss{Z}}^*(R_2)\Bigr)\biggr\},
\\ \nll
\sigma^{\rm{\sss{IBA}}}_{\sss{FB}}(R_1,R_2)=
{\tt ABORN}&=&\frac{1}{R_1}\beta_f^2 A^{\ff}_{\rm{\sss{QED}}}
\biggl\{K^{\ff}_{\sss{FB}}(\ph)|V_{pol}|^2
+\frac{1}{2}\Re e\Bigl(K^{\ff}_{\sss{FB}}(I)\chi(R_1,R_2)V_{pol}^*\Bigr)
\nll&&
+K^{\ff}_{\sss{FB}}(Z)
\Re e \Bigl(\chi_{\sss{Z}}(R_1)\chi_{\sss{Z}}^*(R_2)\Bigr)\biggr\}.
\label{IBAFB}
\eqa
Now we explain all ingredients of \eqns{IBAT}{IBAFB}. 
It may be helpful to compare the IBA cross-section formulae with those
introduced in \sect{i.born} for the Born cross-sections and in
\sect{Zpropagator} for the $\zb$ propagator.
The mass-factors $\beta_f\equiv\beta_f(\sman), c_1(\mfl), c_2(\mfl)$
are defined in \eqns{thresh}{corf3}. 
They are basically used only for leptonic channels; for some exceptions see 
the descriptions of flags {\tt FINR} and {\tt POWR} in \appendx{zuflag}.

The $\ph$ and $\zb$ propagator ratios are:
\bqa
\chi(R_1,R_2)&=& \chi_{\sss{Z}}(R_1)+\chi_{\sss{Z}}(R_2),
\\ \nll
\label{chiz12}
\chi_{\sss{Z}}(\smanp) &=& \left\{
\begin{array}{l}
A_\kappa \frac{\ds{\smanp}}{\ds{\smanp-\mzs+\ib \mzl\gz}}
\quad\mbox{if}\quad{\tt GAMS=0} \\ \\
\kappa\frac{\ds{\smanp}}
{\ds{\smanp-\bar{M}^2_{\sss{Z}}+\ib\bar{M}_{\sss{Z}}\bar{\Gamma}_{\sss{Z}}}}
\quad\mbox{if}\quad{\tt GAMS=1},
\end{array}
\right.
\eqa
where
\bqa
\kappa&=& \frac{A_\kappa}{1+\ib\ds{{\gz}/{\mzl}}}\,,
\\
A_\kappa &=&\gf\frac{\mzs}{8\sqrt{2}\pi\alpha}\,,
\\
\bar{M}_{\sss{Z}} &=& \frac{\mzl}{c_\kappa}\,,
\\
\bar{\Gamma}_{\sss{Z}}&=&\frac{\gz}{c_\kappa}\,,
\\
c_\kappa &=& \sqrt{1+{\gzs}/{\mzs}}\,,
\\
R_1 &=& \frac{\smanp_1}{\sman}\,,
\\
R_2 &=& \frac{\smanp_2}{\sman}\,. 
\label{eq:c212}
\eqa
The default setting is {\tt GAMS}=1.
In this approach, the mass and width of the $\zb$ boson are considered 
to be $\mzl$ and $\gz$, although in the propagator the scaled parameters
are used.
This leads to the well-known phenomenon that fits to the  $\zb$
resonance line lead to different numerical rsults for mass and width
if in \eqn{chiz12} either the upper or lower definition is used (see
also \sect{Zpropagator} and references quoted therein).

The quantity $V_{pol}$ stands for the contributions to the running
$\alpha(\sman)$ 
from QED vacuum polarization by fermions, see \eqn{alpha_fer}. 
Here it is a complex valued quantity  
calculated by function {\tt XFOTF3}:
\bq
V_{pol}=\alpha(\sman)/\alpha=\frac{1}
{\ds{1-\frac{\alpha}{4\pi}{\tt (XFOTF3(IALEM,IALE2,IHVP,IQCD,1,DAL5H,-S))}}}\;.
\eq
The calculation of $V_{pol}$ depends on flags {\tt ALEM, ALE2} and {\tt CONV}.
Their interplay is thoroughly described in \appendx{zuflag} and shown
in \fig{interfEWCOUP} and \fig{subrBORN}.
Here we only explain why both {\tt EWCOUP} and {\tt BORN}
compute $\alpha(\sman)$. 
In regular use ({\tt CONV}=0,1),
{\tt EWCOUP} is supposed to deliver $\sman$ dependent quantities.
However, the results of {\tt BORN}, \eqns{IBAT}{IBAFB}, are in general 
convoluted and, as known, $\alpha(\sman)$ has to be convoluted too.
That's why {\tt BORN} calculates first of all $\alpha(\smanp)$.
We briefly mention a special option {\tt CONV}=2 when {\tt BORN} calls
{\tt EWCOUP} from inside and then not only $\alpha(\sman)$, but also 
all coupling functions supplied by {\tt EWCOUP} are convoluted.
It is a very CPU time consuming option. 
It was used once in \cite{Bardin:1999gt} 
to show that at \LEPI\ energies it is not necessary to convolute EWRC. 
This deserves however a dedicated study for \LEPII\ energies.

The factors $S^{\ff}_{\rm{\sss{QED}}}$ and
$A^{\ff}_{\rm{\sss{QED}}}$ are QED FSR correction factors which 
are calculated depending on flag {\tt FINR}.
If {\tt FINR}=--1, there are no QED-corrections, i.e.
\bq
S^{\ff}_{\rm{\sss{QED}}}=A^{\ff}_{\rm{\sss{QED}}}=1.
\eq
If {\tt FINR}=0, their inclusive values are used:
\bqa
S^{\ff}_{\rm{\sss{QED}}}&=& 1+\frac{3}{4}\frac{\alpha(\smanp)}{\pi}\qfs\,,
\\
A^{\ff}_{\rm{\sss{QED}}}&=&0.
\eqa
For {\tt FINR}=1, the $S^{\ff}_{\rm{\sss{QED}}}$ and
$A^{\ff}_{\rm{\sss{QED}}}$ are supplied by subroutine {\tt FUNFIN}.
In this case the returned values depend on the cuts that are used;
see the description of {\tt ZUCUTS} in \appendx{zucuts} and in 
\fig{zf_flow_main} and \fig{structZCUT}.

Finally, the  factors $S^{\ff}_{\rm{\sss{CKHSS}}}$ describe non-factorized
mixed QCD$\otimes$EW corrections of order $\ord{\alpha\als}$ 
\cite{Czarnecki:1996ei,Harlander:1998zb},
which are assumed to be equal to their on-resonance values, \eqn{ck_hhs}.

Another option in {\tt BORN} is the calculation of the total hadronic cross
section $\sigma_{\had}$. 
This is done in an extra chain for {\tt INDF}=10 and {\tt INTRF}=2.
The corresponding term {\tt SBORN}, proportional to $\sigma_{\had}$, 
is: 
\bqa
\label{IBAThad}
{\tt SBORN}&=&\sum\limits^{5}_{\sss{J}=1}
\frac{1}{R_1}\beta_f\biggl\{S^{\ff}_{\rm{\sss{QED}}}c_2(\mfl)
\biggl[K^{\ff=\sss{J}}_{\sss{T}}(\ph)|V_{pol}|^2
+\frac{1}{2}\Re e\Bigl(K^{\ff=\sss{J}}_{\sss{T}}(I)\chi(R_1,R_2)V_{pol}^{*}
\Bigr)
\biggr]
\nll&&
+\biggl[c_2(\mfl)\biggl(K^{\ff=\sss{J}}_{\sss{T}}(Z)
+\Bigl(S^{\ff}_{\rm{\sss{QED}}}-1+S^{\ff}_{\rm{\sss{CKHSS}}}\Bigr)
K^{\ff=\sss{J},0}_{\sss{T}}(Z)\biggr)
\nll&&
+c_3(\mfl) 
{\bar{K}}^{\ff=\sss{J},m}_{\sss{T}}(Z)\biggr]
 \Re e\Bigl(\chi_{\sss{Z}}(R_1)\chi_{\sss{Z}}^*(R_2)\Bigr)\biggr\}.
\eqa
Further, {\tt BORN} computes two additional objects:
\bqa
\label{IBATs}
{\tt SBORNS}&=&\frac{1}{R_1}\biggl[
K_{\sss{T}}(\ph)|V_{pol}|^2
+\frac{1}{2}\Re e\Bigl(K_{\sss{T}}(I)\chi(R_1,R_2)V_{pol}^{*}\Bigr)
+K_{\sss{T}}(Z)
 \Re e\Bigl(\chi_{\sss{Z}}(R_1)\chi_{\sss{Z}}^*(R_2)\Bigr)\biggr],
\nll
\\
{\tt ABORNS}&=&\frac{1}{R_1}
\biggl[K_{\sss{FB}}(\ph)|V_{pol}|^2
+\frac{1}{2}\Re e\Bigl(K^{\ff}_{\sss{FB}}(I)\chi(R_1,R_2)V_{pol}^*\Bigr)
+K^{\ff}_{\sss{FB}}(Z)
\Re e \Bigl(\chi_{\sss{Z}}(R_1)\chi_{\sss{Z}}^*(R_2)\Bigr)\biggr],
\nll
\label{IBAFBs}
\eqa
composed with the summed couplings given in \eqn{summedcf}. 
They are used at {\tt MISD}=1 in order to restore the SM remnants for
IFI completely.

The \eqns{IBAT}{IBAFB} and \eqns{IBAThad}{IBAFBs} represent the complete 
set of objects used for the subsequent convolution with ISR (see 
\chapt{ch-phot}) and for construction of realistic observables:
total and differential cross-sections and forward-backward asymmetries
with cuts\footnote{If at least one incoming and one outgoing fermion
are polarized, then the 
contribution to the forward-backward anti-symmetric Standard Model cross
section from pure photon exchange does {\em not} vanish as in \eqn{kgfb}.
This can be seen from formulae \eqns{helicities}{kgizl0}.}.
\subsection{More asymmetries}
In addition to the forward-backward 
asymmetry, several other asymmetries are interesting.
It is useful to define a generic `spin' asymmetry, $A(h)$:
\bq
A(h) =\frac{\sigma(h) - \sigma(-h)}{\sigma(h) + \sigma(-h)}\,,
\label{asym}
\eq
where $h$ can denote the polarization of an incoming fermion
or the helicity of an outgoing one.
 
Choosing $h$ to be $h_+=+1$ the helicity of a final-state $\tau^+$ and 
$\sigma(h)$ to be $\sigma_{\sss{T}}(h_+)$, $A(h)$ becomes the 
$\tau$~polarization,
$\lambda_\tau \equiv A^{\rm{pol}}$. Similarly, one can define:
$A_{\sss{F}}^{\rm{pol}}$,  $A_{\sss{B}}^{\rm{pol}}$ from (\ref{asym}) with
$\sigma(h) = \sigma_{\sss{A}}(h_+),\; A=F,B,FB$, respectively.
The subscript $F(B)$ is used to indicate that only data from
the forward (backward) hemisphere are in the measurement; the
corresponding theoretical relations are given by \eqns{sigmatn}{binsec}.
The forward-backward polarization asymmetry
$A_{\sss{FB}}^{\rm{pol}}$ may be defined as follows:
\bq
A_{\sss{FB}}^{\rm{pol}} =
\frac{\sigma_{\sss{FB}}(h) - \sigma_{\sss{FB}}(-h)}
{\sigma_{\sss{T}}}\,.
\label{lafb}
\eq
The $\tau$ polarization $\lambda_\tau \equiv A^{\rm{pol}}$ and
asymmetry \eqn{lafb} are constructed in interfaces {\tt ZUTPSM} and 
{\tt ZUTAU}, and left-right asymmetries \eqn{asym} -- in {\tt ZULRSM}.

\subsection{Bhabha scattering\label{bha}}
The present \zf\ package contains {\tt BHANG} version 4.67 (May 1991).  
We restrict ourselves to a description of effective Born
cross-sections although with the package {\tt BHANG} also photonic corrections
may be calculated. 
A sketch of the photonic corrections may be found in Section 2.4 of
\cite{Bardin:1992jc}.
Early comparisons of  the October 1990 version with other Fortran programs
indicated to
deviations at the per cent level \cite{SRiemann:1991,Beenakker:1991??}.
Since then, we have not performed a detailed study of the accuracy of the
numerical output.  

Concerning the effective Born cross-section, the situation is better.
For {\tt INDF}=11, the Bhabha cross-section is calculated.
Then, instead of subroutine {\tt ZCUT}, subroutine {\tt BHANG} is selected.
The first call is for subroutine {\tt BHAINI}.
Flag {\tt IPART}=0 selects the Bhabha cross-section and e.g. {\tt IPART}=1 the
$s$ channel terms only.
Depending on the pre-selected flag {\tt INTRF}, different interfaces are
chosen: 1 -- Standard Model, 2 -- partial widths, 3 -- couplings, 4 -- squared
couplings. 
This flag is active in subroutine {\tt BHACOU} where the actual couplings are
calculated. 
After {\tt BHAINI}, {\BHANG} calls subroutine {\tt BHATOT}.
After this call, the variables {\tt CROSB} (the
integrated effective Born cross-section)
and {\tt ASYB} (forward-backward asymmetry) are
determined in {\tt BHATOT}.
The angular integrations are done analytically.
In fact, for the determination of effective Born results {\tt BHATOT} calls
subroutine {\tt BHABBB} which returns expressions integrated from
$\cos\vartheta=0$ to a cut value.
Two calls suffice then to treat an arbitrary angular interval.
Alternatively, the calculation of QED corrections is performed with the
numerical integration subroutine {\tt SIMPS} and calls the differential
cross-section function {\tt BHAFUN(THET)}, with {\tt THET} the
scattering angle. 

If a polarization asymmetry is to be calculated, the (expert) user has to
select the beam polarizations {\tt ALAMP} and {\tt ALAMM} in subroutine {\tt
BHAINI}. 
When instead of {\tt BHATOT} the subroutine {\tt BHADIF} is called, a
differential 
cross-section is returned, also using function {\tt BHAFUN}.
Though, for this the interface has to be written by the user; so we
recommend such an option only to the expert user.

For the Standard Model interface, the following approximations were done
in order to simplify and speed up calculations.
The complex valued $s$ channel form factors $\alpha(s)$, $\rho(s)$,
$\kappa_e(s)$, 
$\kappa_{ee}(s)$ are calculated during the initialization phase when $s$ is
set; they depend through the weak box diagrams also on $t$ and $u$: 
\ba 
{\tt T}&=& \frac{\sman}{2}(1-\cos\vartheta)=-\tman,
\\
{\tt U}&=& \sman+\tman=-\uman.
\ea
For the Bhabha process, it is set arbitrarily
\ba
|\tman| = |\uman| = \frac{\sman}{2}
\ea
for the following call of subroutine {\tt ROKANC} (the exclusion is
the calculation of $ZZ$ and $WW$ box contributions when it is set {\tt
IBOX}=1). 
In the $t$ channel,  $\alpha(t)$ is 
calculated with function {\tt XFOTF1} while the other $t$ channel form
factors are set:  
\ba
\rho(t) = \kappa_e(t) = \kappa_{ee}(t) = 1.
\ea
At \LEPI\ this was a good approximation for years, since there the $\sman$ 
channel dominates, and the 
form factors are not too dependent on $\sman,\tman,\uman$.
When the $\sman$ channel looses its dominance and, additionally, the weak box
diagrams are less suppressed, one should give up this approximation.

The weak couplings are prepared in subroutine {\tt BHACOU}.
The effective Born cross-section functions returned are {\tt SBORN0, ABORN0}.
Further functions in use are {\tt BHABNS, BHABNT}; the former with pure $\sman$
channel terms, the latter with $\sman\tman$ interferences and the $\tman$ 
channel exchange contributions. 

The weak form factors are calculated with the package {\tt DIZET}.   
The formulae to be used for the cross-sections are described in
\cite{Bardin:1991xet}.
The numbers presented there have been produced with the Fortran
program 
{\tt BHASHA} \cite{Bardin:1990uu}\footnote{{\tt BHASHA} was derived
from a 1990 version of {\tt DIZET}.}. 

The form factors $\kappa_{e}, \kappa_{ee}, \rho_{ee}$ are calculated
in subroutine {\tt ROKANC}.
which depend on $\sman$,$\tman$,$\uman$ and the charges of the
initial and final state fermions (here minus one).
In the $\sman$ channel, one calls with {\tt QE}=--1, {\tt QF}=--1:
\ba
{\tt ROKANC}(\ldots,\uman,-\sman,\tman,{\tt QE},{\tt QF},{\tt XFZ},\ldots), 
\ea
and in the $t$ channel 
\ba
{\tt ROKANC}(\ldots,\sman,-\tman,\uman,{\tt QE},{\tt QF},{\tt XFZ},\ldots),
\ea
and sets:
\ba
\rho_{ee}&=&{\tt XFZ(1)},
\nll
v_e      &=&{\tt 1-4*SW2*XFZ(2)},
\nll
v_{ee}   &=&{\tt 1-8*SW2*XFZ(2)+16*SW2*SW2*XFZ(4)}.
\label{v8}
\ea
These corrections are switched on and off with flag {\tt WEAK}.
They introduce in the weak
corrections to the $s$ channel a dependence on the scattering angle.
In the $t$ channel, correspondingly,
 the weak corrections will depend not only on the scattering
angle, but also on $s$ -- if $WW$ and $ZZ$ boxes are not switched off.
The complex valued
$s$ channel corrections from the fermionic vacuum polarization are contained
in $F_{\sss{A}}(s)$,
\ba
F_{\sss{A}}(s)
&=& \frac{\alpha(-|s|)}{\alpha}\,,
\label{fas}
\ea
and in the $t$ channel:
\ba
F_{\sss{A}}(t)
&=& \frac{\alpha(|t|)}{\alpha}\,.
\label{fat}
\ea
This is described in Section \ref{qedrun}.

In what follows, we give the explicit expressions for the differential
cross-section in the Standard Model approach.
It is trivial then for the expert user to understand the other interfaces.

The improved Born approximation for the Bhabha differential cross-section
consists of the
sum of contributions with $t$ channel and $s$ channel exchange of the photon
and $Z$ 
boson and their interferences:
\ba
\frac{d \sigma^{SM}}{d\cos\vartheta}
=
\frac{\pi \alpha^2}{2 s} \sum_{A} \sum_{a} T_a(A)
& A=\gamma, \gamma Z, Z,
& a= s,st,t.
\label{v1}
\ea
We introduce the notation:
\ba
\lambda_1 &=& 1-\lambda_+ \lambda_-,
\\
\lambda_2 &=& \lambda_+ - \lambda_-,
\\
\lambda_3 &=& 1+\lambda_+ \lambda_-,
\label{lamda}
\ea
where {\tt ALAMP}=$\lambda_{+}$, {\tt ALAMM}=$\lambda_{-}$ are the degree of
longitudinal polarization  of the positron or electron beam.
 
The $s$ channel cross-section contributions are ($c=\cos\vartheta$):
\ba
 T_s(\gamma)
&=&
\left| F_{\sss{A}}(s)\right|^2
\left[ \lambda_1 \left( 1+c^2 \right) \right],
\\  
T_s(\gamma Z)
&=&
2 \Re e
\Bigl\{
\rho_{ee}(s) \chi(s)
F_{\sss{A}}^*(s)
\Bigl[
[ \lambda_1 v_{ee}(s) + \lambda_2 v_e(s) ] (1+c^2)
\nll 
&&
+~ [\lambda_1+\lambda_2 v_e(s) ] 2c
\Bigr] \Bigr\},
\\ 
T_s(Z)
&=&
\left|\rho_{ee}(s) \chi(s) \right|^2
\Biggl\{
\Bigl[
\lambda_1 \left( 1 + 2 |v_e(s)|^2 + |v_{ee}(s)|^2 \right)
\nll 
& & 
+~2 \lambda_2 \Re e \left( v_e(s) [1+v_{ee}(s)^*]\right) \Bigr] (1+c^2)
\nll
&&
+~2 \Re e
\left[
\lambda_1\left(|v_e(s)|^2 + v_{ee}(s)\right)+\lambda_2 v_e(s)
\left( 1+v_{ee}^*(s) \right)
\right]
2c
\Biggr\} .
\label{v2}
\ea

We now describe the $t$ channel contributions:
\ba
T_t(\gamma)
&=&
F_{\sss{A}}(t)^2
\left[
2 \lambda_1
\frac{(1+c)^2}{(1-c)^2}
+ 8 \lambda_3
\frac{1}{(1-c)^2}
\right],
\\  
T_t(\gamma Z)
&=&
2 \rho_{ee}(t) \chi(t) F_{\sss{A}}(t)
\nll
&& \times \left\{
2 \left[ \lambda_1 \left( 1+v_{ee}(t) \right) - \lambda_2 v_e(t) \right]
\frac{(1+c)^2}{(1-c)^2}
- 8 \lambda_3 \left(1-v_{ee}(t)\right)
\frac{1}{(1-c)^2}
\right\},
\nll
\\  
T_t(Z)
&=&
\left[\rho_{ee}(t) \chi(t) \right]^2
\Biggl\{
2\Biggl[
\lambda_1 \left( 1+4v_e(t)^2+2v_{ee}(t) +v_{ee}(t)^2\right)
\nll
&&+~4\lambda_2 v_e(t) \left(1+v_{ee}(t)\right) \Biggr]
\frac{(1+c)^2}{(1-c)^2}
+~8\lambda_3\left(1-v_{ee}(t) \right)^2
\frac{1}{(1-c)^2}
\Biggr\} .
\label{v5}
\ea
The following additional abbreviation is used:
\ba
\chi(t) &=& \frac{G_{\mu} \mzs}{\sqrt{2} 8\pi\alpha} \frac{t}{t+\mzs}\,.
\label{v6}
\ea
The form factors $F_{\sss{A}}(t), v_e(t), v_{ee}(t), \rho(t)$ are real valued.
 
Finally, we describe in this Section the contributions from the $\gamma
Z$ interference:
\ba
T_{st}(\gamma)
&=&
- 2 \Re e \left[ F_{\sss{A}}^*(s) F_{\sss{A}}(t) \right] \lambda_1
\frac{(1+c)^2}{(1-c)}\,,
\\  
T_{st}(\gamma Z)
&=&
- 2 \Re e
\left\{
\chi(t) \rho_{ee}(t) F_{\sss{A}}^*(s)
\left[ \lambda_1 \left( 1+v_{ee}(t) \right) - 2 \lambda_2 v_e(t) \right]
+ \left( t \leftrightarrow s \right)
\right\}
\frac{(1+c)^2}{(1-c)}\,,
\nll
\\  
T_{st}(Z)
&=&
-2 \Re e \left\{ \chi(s) \rho_{ee}(s) \chi(t) \rho_{ee}(t) \left[
\lambda_1 \left( [1+v_{ee}(s)]  [1+v_{ee}(t)]
\right. \right. \right.
\nll
& &
\left. \left. \left.
+ 4 v_e(s) v_e(t) \right)
- \lambda_2 \left( v_e(s) [1+v_{ee}(t)] + v_e(t) [1+v_{ee}(s)] \right)
\right] \right\}
\frac{(1+c)^2}{(1-c)} .
\nll
\label{v7}
\ea
 
It is straightforward to improve the weak corrections for the Bhabha 
channel.

In view of the recent update of the photonic corrections with
acollinearity cut for fermion pair production, we have to assume that
the photonic corrections for Bhabha scattering deserve also a complete 
(and quite involved) re-calculation.
This is planned to be done.

      
\chapter{Technical Description of the \zf\ package \label{ch-zfitter}}
\section{{\tt DIZET} User Guide\label{dizetug}}
\eqnzero
\subsection{Structure of {\tt DIZET}}
  The package {\tt DIZET} is a library that calculates
electroweak (EW), internal mixed QCD\-$\otimes$\-EW and QCD$\otimes$QED
final state radiative corrections for several processes in the
Standard Model. 
A first call of subroutine {\tt DIZET} returns
various pseudo-observables, the $\wb$-boson mass,
weak mixing angles, the $\zb$-boson width, the $\wb$-boson width and
other quantities.
After the first call to {\tt DIZET}, several subroutines of  {\tt DIZET}
might be used for the calculation of: 
\vspace*{-3mm}

\begin{itemize}
\item{} weak form factors and running $\alpha(\sman)$ for two-fermion
into two-fermion neutral current processes -- call subroutine {\tt
ROKANC}, see \subsect{rokanc};  
\vspace*{-2mm}

\item{} weak form factors for two-fermion into two-fermion charged
current processes -- call subroutine {\tt RHOCC};
\vspace*{-2mm}

\item{} weak form factors for $\nu_{e}\fe$ elastic and deep inelastic
NC and CC neutrino-nucleon scattering (earlier versions were
interfaced with the codes {\tt NUFITTER} \cite{Bardin:1984B} and {\tt
NUDIS2} \cite{Bardin:1986B}); 
\vspace*{-2mm}

\item{} weak form factors and running $\alpha(\sman)$ for $\fl-N$ NC
and CC deep inelastic scattering -- call from package {\tt
HECTOR}~\cite{Akhundov:1996} and~\cite{Arbuzov:1995id}). 
\end{itemize}
\vspace*{-3mm}

\subsection{Input and Output of {\tt DIZET}\label{iodizet}}
The {\tt DIZET} argument list contains
Input, Output, and Mixed (I/O) types of arguments:

\begin{small}
\centerline{
{\fbox{
{\tt 
CALL DIZET(NPAR,AMW,AMZ,AMT,AMH,DAL5H,ALQED,ALSTR,ALSTRT,ZPAR,PARTZ,PARTW)}}}
}
\end{small}
\subsubsection{Input and I/O parameters to be set by the user}
{\bf Input:}
\vspace{-.5mm}

\begin{description}
\item[] {\tt NPAR(1:21), INTEGER*4} vector of flags, see \subsect{dizetflags}
        and \subsect{zuflag}
\vspace*{-.5mm}

\item[] {\tt AMT} = $\mtl$ -- $\ft$-quark mass
\vspace*{-.5mm}

\item[] {\tt AMH} = $\mhl$ -- Higgs boson mass 
\vspace*{-.5mm}

\item[] {\tt ALSTR} = $\als(\mzs)$ -- strong coupling at $\sman = \mzs$
\end{description}
\newpage

{\bf I/O:}
\begin{description}
\item[] {\tt AMW} = $\mwl$, $\wb$ boson mass, input if {\tt NPAR(4)} = 2,3,
                  but is being calculated for {\tt NPAR(4)} = 1
\item[] {\tt AMZ} = $\mzl$, $\zb$ boson mass, input if {\tt NPAR(4)} = 
1,3,
                  but is being calculated for {\tt NPAR(4)} = 2 
\item[] {\tt DAL5H} = $\dalhv$, see \subsect{IPS1}
\end{description}
The $\mzs$, $\mws$, and $\dalhv$ cannot be assigned by a parameter 
statement (input/output variables).
\subsubsection{Output of the {\tt DIZET} package}
\begin{description}
\item[] {\tt ALQED~} = $\alpha(\mzs)$, calculated from $\dalhv$, see
description of flag {\tt ALEM} in \subsect{dizetflags}
\item[] {\tt ALSTRT} = $\als(\mts)$         
\item[] {\tt ZPAR(1)} = {\tt DR}=$\dr$, the loop corrections to the
muon decay constant $\gf$, see \sect{deltar} 
\item[] {\tt ZPAR(2)} = {\tt DRREM} = $\dr_{\rm{rem}}$, the remainder
contribution $\ord{\alpha}$ 
\item[] {\tt ZPAR(3)} = {\tt SW2} = $\siws$, squared of sine of the weak mixing
angle defined by weak boson masses   
\item[] {\tt ZPAR(4)} = {\tt GMUC} = $\gf$, muon decay constant, if
{\tt NPAR(4)} = 1,2, is set in {\tt CONST1} depending on flag
                                   {\tt NPAR(20)}, see \subsect{dizetflags} 
(It should be calculated if {\tt NPAR(4)}=3 from $\mzl,\mwl$, but 
then it will deviate from the experimental value.)   
\item[] {\tt ZPAR(5-14)} -- stores effective sines for all partial 
$\zb$-decay channels:
\vspace*{-1cm}

\[
\begin{array}{lcl}
         5 & - & \mbox{neutrino} \\
         6 & - & \mbox{electron} \\
         7 & - & \mbox{muon}     \\
         8 & - & \mbox{tau}      \\
         9 & - & \mbox{up}       \\
        10 & - & \mbox{down}     \\
        11 & - & \mbox{charm}    \\
        12 & - & \mbox{strange}  \\
        13 & - & \mbox{top (presently equal to up)}\\
        14 & - & \mbox{bottom}
\end{array}   
\]                    
\item[] {\tt ZPAR(15)} = {\tt ALPHST} $\equiv \als(\mzs)$  
\item[] {\tt ZPAR(16-30)} = {\tt QCDCOR(0-14)},
{\tt QCDCOR(I)} -- array of QCD correction factors for quark production 
processes and/or $\zb$ boson partial width (channel $i$) into quarks.
Enumeration as follows:
\bqa
\begin{array}{lcl}
\mbox{{\tt QCDCOR(0)} }&=&1\\
\mbox{{\tt QCDCOR(1)} }&=&R^{\fu}_{\sss{V}}(\mzs)\\
\mbox{{\tt QCDCOR(2)} }&=&R^{\fu}_{\sss{A}}(\mzs)\\
\mbox{{\tt QCDCOR(3)} }&=&R^{\fd}_{\sss{V}}(\mzs)\\
\mbox{{\tt QCDCOR(4)} }&=&R^{\fd}_{\sss{A}}(\mzs)\\
\mbox{{\tt QCDCOR(5)} }&=&R^{\fc}_{\sss{V}}(\mzs)\\
\mbox{{\tt QCDCOR(6)} }&=&R^{\fc}_{\sss{A}}(\mzs)\\
\mbox{{\tt QCDCOR(7)} }&=&R^{\fs}_{\sss{V}}(\mzs)\\
\mbox{{\tt QCDCOR(8)} }&=&R^{\fs}_{\sss{A}}(\mzs)\\
\mbox{{\tt QCDCOR(9)} }&=&R^{\fu}_{\sss{V}}(\mzs)\quad\mbox{foreseen for}\\
\mbox{{\tt QCDCOR(10)}}&=&R^{\fu}_{\sss{A}}(\mzs)\quad\ft\bart\mbox{-channel}\\
\mbox{{\tt QCDCOR(11)}}&=&R^{\fb}_{\sss{V}}(\mzs)\\
\mbox{{\tt QCDCOR(12)}}&=&R^{\fb}_{\sss{A}}(\mzs)\\
\mbox{{\tt QCDCOR(13)}}&=&R^{\sss{S}}_{\sss{V}} 
\mbox{--singlet vector correction, see \sect{qcdrun}} \\
\mbox{{\tt QCDCOR(14)}}&=&f_1,\quad\mbox{ABL--correction},
\mbox{\eqn{abl_corr}},~\cite{Arbuzov:1992pr}
\end{array}
\label{qcdcor_fst}
\eqa
\item[] {\tt PARTZ(I)} -- array of partial decay widths
of the $\zb$-boson (see definitions in \eqnsc{defzwidthl}{defzwidthq})
for the channels:  
\[
\begin{array}{ll}
{\tt I} = 0 & \mbox{neutrino}                 \\                      
{\tt I} = 1 & \mbox{electron}                 \\             
{\tt I} = 2 & \mbox{muon}                     \\                             
{\tt I} = 3 & \mbox{tau}                      \\                           
{\tt I} = 4 & \mbox{up}                       \\                             
{\tt I} = 5 & \mbox{down}                     \\             
{\tt I} = 6 & \mbox{charm}                    \\         
{\tt I} = 7 & \mbox{strange}                  \\
{\tt I} = 8 & \mbox{top (foreseen, not realized)}\\
{\tt I} = 9 & \mbox{bottom}                   \\
{\tt I} = 10& \mbox{hadrons}              \\
{\tt I} = 11& \mbox{total}  
\end{array}
\]
\item[] {\tt PARTW(I)} -- array of partial decay widths of the 
$\wb$-boson\footnote{The calculation of the $\wb$
width~\cite{Bardin:1986fi} 
follows the same principles as that of the
$\zb$ width and is realized in subroutine {\tt ZWRATE} of {\tt DIZET}.
Since the $\wb$ width is not that important for the description of
fermion pair production, we do not go into details.}
for the channels:
\[
\begin{array}{ll}
{\tt I} = 1 & \mbox{one leptonic } \\ 
{\tt I} = 2 & \mbox{one quarkonic} \\
{\tt I} = 3 & \mbox{total} 
\end{array}
\]
\end{description}

\newpage

\subsection{The flags used by {\tt DIZET} \label{dizetflags}}
Since the {\tt DIZET} package may be used as  stand-alone in order
to compute POs and since it is being used by other codes
(e.g. {\tt KORALZ}~\cite{Jadach:1994yv}, {\tt
BHAGENE3}~\cite{Field:1996}, {\tt
HECTOR}~\cite{Akhundov:1996,Arbuzov:1995id}), 
we present here a short description of all flags in  
{\tt DIZET}. 
The flag values 
must be filled in vector {\tt NPAR(1:21)} by the the user.
Most of these flags overlap with the flags set in user subroutine
{\tt ZUFLAG} called by  {\tt ZFITTER},
however, in the stand-alone mode {\tt ZUFLAG} need not be
called. 
In order to explcitely show the correspondence between the flag names
called with {\tt ZUFLAG} 
and the flags used inside {\tt DIZET} we will provide boxes:
\\
\centerline{$\fbox{\tt ZUFLAG('CHFLAG', IVALUE-inside DIZET)}$\,.}
\\
The description is given in the order of vector {\tt NPAR(1:21)}.
Flag values marked as {\bf presently not supported} are not recommended.
For instance, they may be chosen for backward compatibility with
respect to earlier versions of the code.

\begin{description}       
\item[{\tt NPAR(1)} = {\tt IHVP}] $\rightarrow\;\;$ 
\fbox{\tt ZUFLAG('VPOL',IHVP)} --
Handling of hadronic vacuum polarization:
 \begin{description}
  \item[{\tt IHVP} = 1]  (default) by the parameterization 
                       of~\cite{Eidelman:1995ny}
  \item[{\tt IHVP} = 2]  by effective quark masses
of~\cite{Jegerlehner:1991dq,Jegerlehner:1991ed}~~~{\bf presently not} 
  \item[{\tt IHVP} = 3]  by the parameterization 
                       of~\cite{Burkhardt:1989ky} $\hspace{19mm}$ 
                       {\bf supported}
 \end{description}
\end{description}
\begin{description}
\item[{\tt NPAR(2)} = {\tt IAMT4}] $ \rightarrow $ 
\fbox{\tt ZUFLAG('AMT4',IAMT4)} --
Re-summation of the leading $\ord{\gf\mts}$ electroweak corrections,
see \subsect{dr_beyond}:
 \begin{description}
  \item[{\tt IAMT4} = 0] no re-summation

  \item[{\tt IAMT4} = 1] with re-summation recipe of~\cite{Consoli:1989pc}~~
                      {\bf presently}    
  \item[{\tt IAMT4} = 2] with re-summation recipe of~\cite{Halzen:1991je}  
                      $\hspace{7mm}$ {\bf not}     
  \item[{\tt IAMT4} = 3] with re-summation recipe of~\cite{Fanchiotti:1991kc}~
                      $\hspace{1mm}$ {\bf supported}
  \item[{\tt IAMT4} = 4] (default) with two-loop sub-leading corrections
                       and re-summation recipe \\
  of~\cite{Degrassi:1994a0,Degrassi:1995ae,Degrassi:1995mc,%
Degrassi:1996mg,Degrassi:1996ZZ,Degrassi:1999jd}  
 \end{description}
\end{description}
\begin{description}
\item[{\tt NPAR(3)} = {\tt IQCD}] $ \rightarrow $ 
\fbox{\tt ZUFLAG('QCDC',IQCD)}
-- Handling of internal QCD corrections of order $\ord{\alpha\als}$,
see \subsect{mixed}:
\begin{description}
\item[{\tt IQCD} = 0]  no internal QCD corrections 
\item[{\tt IQCD} = 1]  by Taylor expansions (fast option)
of~\cite{Bardin:1989aa} 
\item[{\tt IQCD} = 2]  by exact formulae of~\cite{Bardin:1989aa}
\item[{\tt IQCD} = 3]  (default) by exact formulae of~\cite{Kniehl:1990yc}  
\end{description}
\end{description}
\begin{description}
\item[{\tt NPAR(4)} = {\tt IMOMS}]
-- Choice of two input/output parameters from the three parameters
$\big\{\gf,$ $\mzl,$ $\mwl\bigr\}$, see \subsect{IPS1}:    
 \begin{description}
  \item[{\tt IMOMS} = 1] (default) input $\gf,\mzl$ (output: $\mwl$),
   see~\eqn{wmass}   
  \item[{\tt IMOMS} = 2] input $\gf,\mwl$ (output: $\mzl$),
   see~\eqn{zmass}   
  \item[{\tt IMOMS} = 3] input $\mzl,\mwl$ (output: $\gf$), {\bf foreseen, not
  realized}, see~\eqn{Gmu} 
 \end{description}
\end{description}
\begin{description}
\item[{\tt NPAR(5)} = {\tt IMASS}]
-- Handling of hadronic vacuum polarization in $\dr$; for tests only:
 \begin{description}
  \item[{\tt IMASS} = 0] (default) uses a fit to data 
  \item[{\tt IMASS} = 1] uses effective quark masses, see \subsect{IPS1}
 \end{description}
\end{description}
\begin{description}
\item[{\tt NPAR(6)} = {\tt ISCRE}] $\rightarrow$ 
\fbox{\tt ZUFLAG('SCRE',ISCRE)}
-- Choice of the scale of the two-loop remainder terms of $\dr$
with the aid of a conversion factor $f$, see \eqn{conversionf} and 
discussion in the beginning of \subsect{options_tu}:
\begin{description}
\item[{\tt ISCRE} = 0] scale of the remainder terms is $\Ksc=1$ 
\item[{\tt ISCRE} = 1] scale of the remainder terms is $\Ksc=f^2$ 
\item[{\tt ISCRE} = 2] scale of the remainder terms is
$\Ksc={\ds{\frac{1}{f^2}}}$ 
  \end{description}
\end{description}
\begin{description}
\item[{\tt NPAR(7)} = {\tt IALEM}] $ \rightarrow $
\fbox{\tt ZUFLAG('ALEM',IALEM)}
-- Controls the usage of $\alpha(\mzs)$, see flowchart \fig{subrBORN}.
Inside {\tt DIZET}, however, its meaning is limited:
 \begin{description}
  \item[{\tt IALEM} = 0 or 2] $\dalhv$ must be supplied 
   by the user as input to the {\tt DIZET} package        
  \item[{\tt IALEM} = 1 or 3] $\dalhv$ is calculated by the program       
                        using a parameterization {\tt IHVP} 
 \end{description}
\end{description}
For details see the complete discussion about this flag in 
\sect{IBORNA} and \subsect{zuflag}.
\begin{description}
\item[{\tt NPAR(8)} = {\tt IMASK}]
-- Historical relict of earlier versions. {\bf Presently unused.}
\end{description}
\begin{description}
\item[{\tt NPAR(9)} = {\tt ISCAL}]  $ \rightarrow $ 
\fbox{\tt ZUFLAG('SCAL',ISCAL)}
-- Choice of the scale of $\als(\xi\mtl)$:
 \begin{description}
  \item[{\tt ISCAL} = 0] (default) exact {\tt AFMT} 
                       correction \cite{Avdeev:1994db}
  \item[{\tt ISCAL} = 1]\hspace{-2mm}{\bf ,2,3} options used
in~\cite{Kniehl:1995yr},  
        {\bf presently not supported}
  \item[{\tt ISCAL} = 4] Sirlin's scale $\xi= 0.248$ ~\cite{Sirlin:1995yr}
 \end{description}
\end{description}
\begin{description}
\item[{\tt NPAR(10)} =  {\tt IBARB}]  $ \rightarrow $
\fbox{\tt ZUFLAG('BARB',IBARB)}
-- Handling of leading $\ord{\gfs\mtq}$ corrections: 
 \begin{description}
  \item[{\tt IBARB} = 0] corrections are not included 
  \item[{\tt IBARB} = 1]  corrections are applied in the 
                    limiting case: Higgs mass negligible with
                    respect to the top mass, \cite{vanderBij:1987hy}
  \item[{\tt IBARB} = 2] (default) analytic results of 
\cite{Barbieri:1993ra} approximated by a polynomial 
\cite{Barbieri:1999bbo}
 \end{description}
These options are inactive for {\tt AMT4} = 4.
\end{description}
\begin{description}
\item[{\tt NPAR(11)} = {\tt IFTJR}] $ \rightarrow $
\fbox{\tt ZUFLAG('FTJR',IFTJR)} 
-- Treatment of $\ord{\gf\als\mts}$ {\tt FTJR} corrections
\cite{Fleischer:1992fq}, see \eqn{taubb} in \subsect{zbbff} 
 \begin{description}  
  \item[{\tt IFTJR} = 0]  without {\tt FTJR} corrections
  \item[{\tt IFTJR} = 1]\hspace{-2mm}{\bf ,2}  with~~~~~{\tt FTJR} corrections
 \end{description}
\end{description} 
Inside {\tt DIZET} its meaning is limited.
See complete discussion about this flag in \subsect{zuflag}.
\begin{description}
\item[{\tt NPAR(12)} = {\tt IFACR}] $ \rightarrow $
\fbox{\tt ZUFLAG('EXPR',IFACR)}
-- Realizes different expansions of $\dr$:
\bqa
\left.
\begin{array}{lr}
{\hspace{-2mm} {\tt  IFACR} = {\bf 0}}  \quad \mbox{\rm first row} \\ [1mm] 
{\hspace{-2mm} {\tt  IFACR} = {\bf 1}}  \quad \mbox{\rm second row}\\ [1mm]
{\hspace{-2mm} {\tt  IFACR} = {\bf 2}}  \quad \mbox{\rm third row }\nonumber
\end{array}
\right\} & \mbox{of~\eqn{dr_facr}\phantom{ and (1.74)}}
\nonumber
\eqa
\end{description}
\begin{description}
\item[{\tt NPAR(13)} = {\tt IFACT}] $ \rightarrow $
\fbox{\tt ZUFLAG('EXPF',IFACT)}
-- Realizes different expansions of $\rho$ and $\kappa$:
\bqa
\left.
\begin{array}{lr}
{\hspace{-2mm} {\tt  IFACT} = {\bf 0}}  \quad \mbox{\rm first row} \\ [1mm]
{\hspace{-2mm} {\tt  IFACT} = {\bf 1}}  \quad \mbox{\rm second row}\\ [1mm] 
{\hspace{-2mm} {\tt  IFACT} = {\bf 2}}  \quad \mbox{\rm third row }\nonumber
\end{array}
\right\} & \mbox{of~\eqnsc{drho_fact}{dkap_fact}}
\nonumber
\eqa
\end{description}
\begin{description}
\item[{\tt NPAR(14)} = {\tt IHIGS}] $ \rightarrow $
\fbox{\tt ZUFLAG('HIGS',IHIGS)} --
Switch on/off resummation of the leading Higgs contribution,
~see discussion around ~\eqn{drhiggs}:
 \begin{description}
  \item[{\tt IHIGS} = 0] leading Higgs contribution is not re-summed
  \item[{\tt IHIGS} = 1] leading Higgs contribution is re-summed 
 \end{description}
\end{description}
\begin{description}
\item[{\tt NPAR(15)} = {\tt IAFMT}]  $ \rightarrow $
\fbox{\tt ZUFLAG('AFMT',IAFMT)}
-- {\tt AFMT} corrections ~\cite{Avdeev:1994db}, see discussion after 
\eqn{deltarho_d} in \subsect{drho} and also \eqn{TBQCD0}: 
 \begin{description}  
  \item[{\tt IAFMT} = 0 ]  without {\tt AFMT} correction
  \item[{\tt IAFMT} = 1 ]  with~~~~~{\tt AFMT} correction (see also
description of flag {\tt SCAL})
 \end{description}
\end{description} 
\begin{description}
\item[{\tt NPAR(16)} = {\tt IEWLC}] -- Treatment of the remainder
terms of $\rho$ and $\kappa$
(used in {\tt ROKAPP} together with obsolete option {\tt AMT4} = 1-3):
 \begin{description}  
  \item[{\tt IEWLC} = 0] all remainders are set equal to zero
  \item[{\tt IEWLC} = 1] (default) standard treatment
 \end{description}
\end{description} 
\begin{description}
\item[{\tt NPAR(17)} = {\tt ICZAK}] $ \rightarrow $
\fbox{\tt ZUFLAG('CZAK',ICZAK)} --
{\tt CKHSS} non-factorized $\ord{\alpha\als}$ corrections, 
~\cite{Czarnecki:1996ei,Harlander:1998zb}, see \eqn{ck_hhs} in 
\subsect{notation} and \fig{subrBORN}:
  \begin{description} 
   \item[{\tt ICZAK} = 0] without {\tt CKHSS} corrections
   \item[{\tt ICZAK} = 1]\hspace{-2mm}{\bf ,2} with~~~{\tt CKHSS}
corrections 
  \end{description}
\end{description} 
Inside {\tt DIZET} its meaning is limited.
See complete discussion about this flag in \subsect{zuflag}.
\begin{description}
\item[{\tt NPAR(18)} = {\tt IHIG2}] $ \rightarrow $
\fbox{\tt ZUFLAG('HIG2',IHIG2)} -- Handling of the 
leading Higgs contribution,~\eqn{dr_h} 
and~\cite{vanderBij:1984bw,vanderBij:1984aj}:
 \begin{description}
  \item[{\tt IHIG2} = 0] without Higgs corrections 
  \item[{\tt IHIG2} = 1] with~~~~~Higgs corrections
 \end{description}
\end{description}
\begin{description}
\item[{\tt NPAR(19)} = {\tt IALE2}] $ \rightarrow $
\fbox{\tt ZUFLAG('ALE2',IALE2)} --
Treatment of leptonic corrections for $\Delta\alpha$, \fig{interfEWCOUP}:
 \begin{description}
  \item[{\tt IALE2} = 0]  for backward compatibility with versions up to 
                        v.5.12
  \item[{\tt IALE2} = 1]  with  one-loop corrections~\eqn{alpha_fer}      
  \item[{\tt IALE2} = 2]  with two-loop corrections~\cite{Kallen:1955ks}
  \item[{\tt IALE2} = 3]  with three-loop corrections~\cite{Steinhauser:1998rq}
 \end{description}
\end{description}

\begin{description}
\item[{\tt NPAR(20)} = {\tt IGFER}]  $ \rightarrow $
\fbox{\tt ZUFLAG('GFER',IGFER)} --
Handling of QED corrections to the Fermi constant:
 \begin{description}
  \item[{\tt IGFER} = 0] for backward compatibility with versions
                       up to v.5.12
  \item[{\tt IGFER} = 1] one-loop QED corrections for Fermi constant
                    \cite{Berman:1958,Kinoshita:1959,Kallen:1968}
  \item[{\tt IGFER} = 2] two-loop QED corrections for Fermi constant
                    \cite{vanRitbergen:1998yd,vanRitbergen:1998hn}
 \end{description}
\end{description}
\begin{description}
\item[{\tt NPAR(21)} = {\tt IDDZZ}] --  Used in {\tt ZWRATE} for
internal tests: 
 \begin{description}
  \item[{\tt IDDZZ} = 0] {\tt RQCDV(A)} are set to $0$
  \item[{\tt IDDZZ} = 1] (default) standard treatment of FSR QCD corrections
 \end{description}
\end{description}
\subsection{\label{xfotf3}Calculation of $\alpha(\sman)$. Function {\tt
XFOTF3}} 
the running QED coupling at scale $\sman$ is calculated with function
{\tt XFOTF3} as follows:
\bq
\alpha(\sman)=\frac{\alpha}
{\ds{1-
\frac{\alpha}{4\pi}{\tt DREAL(XFOTF3(IALEM,IALE2,IHVP,IQCD,1,DAL5H,-S))}}}\;.
\eq

\section{\zf\ User Guide\label{zfguide}}
\eqnzero
\zf\ is coded in {\tt FORTRAN 77} and it has been implemented
on different computers with different operating systems: IBM, IBM PC,
VMS, APOLLO, HP, SUN, and PC-linux. 
Double-precision variables have been used throughout the program in order to
obtain maximum accuracy, which is especially important for resonance physics.
The package consists of eleven {\tt FORTRAN} files with the number of
lines as shown below, giving a total of 23671 lines: 
\[
\begin{array}{lr}
{\tt zf514\_aux.f }\quad & \quad 1616 \\
{\tt zfbib6\_21.f }\quad & \quad 4790 \\
{\tt zfmai6\_21.f }\quad & \quad 20   \\
{\tt zfusr6\_21.f }\quad & \quad 3344 \\
{\tt acol6\_1.f   }\quad & \quad 3610 \\
{\tt bcqcdl5\_14.f}\quad & \quad 727  \\
{\tt bhang4\_67.f }\quad & \quad 2054 \\
{\tt bkqcdl5\_14.f}\quad & \quad 484  \\
{\tt dizet6\_21.f }\quad & \quad 5653 \\
{\tt m2tcor5\_11.f}\quad & \quad 848  \\
{\tt pairho.f     }\quad & \quad 525  
\end{array}
\]
Some flowcharts, which may help to understand the internal
organization of \zf\ are shown in 
\fig{zf_flow_main},
\fig{structZCUT},
\fig{structZANCUT},
\fig{interfEWCOUP},
\fig{subrBORN}. 
 
The following routines are normally called in the initialization phase
of programs using the \zf\ package. 
Normally they are called in the order listed below.
See, however, an example of different use as described in \subsect{fullPOs}.
 
\subsection{Subroutine {\tt ZUINIT}\label{zuinit}}
 
Subroutine {\tt ZUINIT} is used to initialize variables with
their default values.
This routine {\em must} be called before any other \zf\ routine.
 
\SUBR{CALL ZUINIT}

\vspace*{-5mm}
 
\subsection{Subroutine {\tt ZUFLAG}\label{zuflag}}
 
Subroutine {\tt ZUFLAG} is used to modify the default
values of flags which control various \zf\ options.
 
\SUBR{CALL ZUFLAG(CHFLAG,IVALUE)}
 
\noindent \underline{Input Arguments:}
\smallskip
\begin{description}
  \item[\tt CHFLAG] is the character identifier of a \zf\ flag.
                
  \item[\tt IVALUE] is the value of the flag. See \tbn{ta7} for
                        a list of the defaults. 
\end{description}

Sixteen flags that may be modified by {\tt ZUFLAG} overlap with flags
used by {\tt DIZET} package. 
They are described in \subsect{dizetug}.
Here we describe 21 flags: the remaining 18 flags and 3 overlapping ones,
whose meaning is broader than described in \subsect{dizetug}.

\noindent Possible combinations of {\tt CHFLAG} and {\tt IVALUE} 
for these 21 flags are listed below:\footnote{It is worth noting that
not for all flags 
the default value is necessarily the preferred value.
A typical example is flag {\tt FINR}, distinguishing two different treatments
of FSR, which are relevant in different experimental setups.}

\begin{description} 
  \item[\tt AFBC] --
  Controls the calculation of the forward
  backward asymmetry for interfaces {\tt ZUTHSM}, {\tt ZUXSA},
  {\tt ZUXSA2}, and {\tt ZUXAFB}:
  \begin{description}
    \item[{\tt IVALUE} = 0]
    asymmetry calculation is inhibited (can speed up the program
    if asymmetries are not desired)
    \item[{\tt IVALUE} = 1]
    (default) both cross-section and asymmetry calculations are done
  \end{description}
\end{description}
 
\begin{description} 
  \item[\tt AFMT] -- see {\tt NPAR(15)} in subsection \ref{dizetflags}
\end{description}

\begin{description} 
  \item[\tt ALEM] --
  Controls the treatment of the running QED coupling $\alpha(\sman)$:
  \begin{description}
  \item[{\tt IVALUE} = 0 or 2] $\dalhv$ must be supplied 
   by the user as input to the {\tt DIZET} package;
   using this input {\tt DIZET} calculates {\tt ALQED} = $\alpha(\mzs)$
  \item[{\tt IVALUE} = 1 or 3] $\dalhv$ and $\alpha(\mzs)$ are calculated 
   by the program using a parameterization {\tt IHVP}.
  \end{description}
\end{description}

   The scale of $\alpha($scale$)$ is governed in addition by the flag 
   {\tt CONV}, see description below and the flowchart \fig{subrBORN}.
   Values {\tt ALEM} = 0,1 are accessible only at {\tt CONV} = 0.  
   Then for {\tt ALEM} = 0,1 $\alpha(\mzs)$ 
   and for {\tt ALEM} = 2,3  $\alpha(\sman)$ are calculated.             
   Values {\tt ALEM} = 2,3 are accessible for {\tt CONV} = 0,1,2.  
   Then for {\tt CONV} = 0 $\alpha(\sman)$ 
   and for {\tt CONV} = 1,2 $\alpha(\smanp)$ are calculated. 
   Recommended values: {\tt ALEM} = 2,3.                    

\begin{description} 
  \item[\tt ALE2] -- see {\tt NPAR(19)} in subsection \ref{dizetflags}
\end{description}

\begin{description} 
  \item[\tt AMT4] -- see {\tt NPAR(2)} in subsection \ref{dizetflags}
\end{description}

\begin{description} 
  \item[\tt BARB] -- see {\tt NPAR(10)} in subsection \ref{dizetflags}
\end{description}

\begin{description} 
  \item[\tt BORN] --
  Controls calculation of QED and Born observables:
  \begin{description}
    \item[{\tt IVALUE} = 0]
    (default) QED convoluted observables
    \item[{\tt IVALUE} = 1]
    electroweak observables corrected by Improved Born Approximation
  \end{description}
\end{description}
 
\begin{description}
  \item[\tt BOXD] --
  Determines calculation of $ZZ$ and $WW$ box
  contributions,
  (see \sect{ew_boxes}):
  \begin{description}
    \item[{\tt IVALUE} = 0]
    no box contributions are calculated
    \item[{\tt IVALUE} = 1]
    (default) the boxes are calculated as additive separate contribution
    to the cross-section  
    \item[{\tt IVALUE} = 2]
    box contributions are added to all four form factors    
  \end{description}
\end{description}

\begin{description} 
  \item[\tt CONV] --
  Controls the energy scale of running $\alpha$ and EWRC,
  see \fig{subrBORN}:
  \begin{description} 
    \item[{\tt IVALUE} = 0]
    $\alpha(s)$
    \item[{\tt IVALUE} = 1]
    (default) $\alpha(\smanp)$ convoluted
    \item[{\tt IVALUE} = 2]
    both electroweak radiative correction and $\als$ are convoluted
  \end{description}
\end{description}

\begin{description} 
  \item[\tt CZAK] --
  Treatment of {\tt CKHSS} non-factorized corrections,
  \cite{Czarnecki:1996ei},~\cite{Harlander:1998zb},~see~\fig{subrBORN}:
  \begin{description} 
    \item[{\tt IVALUE} = 0] without {\tt CKHSS} corrections 
    \item[{\tt IVALUE} = 1] (default) with {\tt CKHSS} corrections everywhere
    \item[{\tt IVALUE} = 2] {\tt CKHSS} corrections are taken into account only 
                            in POs, the option is used for tests only
  \end{description}
\end{description}

\begin{description} 
  \item[\tt DIAG] --
  Selects type of diagrams taken into account:
  \begin{description} 
    \item[{\tt IVALUE} = --1]
    only $\zb$- exchange diagrams are taken into account 
    \item[{\tt IVALUE} = 0]
    $\zb$ and $\ph$ - exchange diagrams are taken into account 
    \item[{\tt IVALUE} = 1]
    (default) $\zb$ and $\ph$ exchange and $\zb\ph$ interference are 
    included
  \end{description}
\end{description}

\begin{description} 
  \item[\tt EXPF] -- see {\tt NPAR(13)} in subsection \ref{dizetflags}
\end{description}

\begin{description} 
  \item[\tt EXPR] -- see {\tt NPAR(12)} in subsection \ref{dizetflags}
\end{description}

\begin{description} 
  \item[\tt FINR] --
  Controls the calculation of final-state radiation,
  see \fig{interfEWCOUP}:
  \begin{description}
    \item[{\tt IVALUE} = --1]
    final-state QED and QCD correction are not applied;
    \item[{\tt IVALUE} = 0]
    by $\smanp$ cut,
    final-state QED correction is described with the factor
    $1 + 3 \alpha(\sman) / (4 \pi ) \qfs$
    \item[{\tt IVALUE} = 1]
    (default) $M^{2}_{ff}$ cut,
    includes complete treatment of final-state radiation
    with common soft-photon exponentiation
  \end{description}
\end{description}
 
\begin{description}
  \item[\tt FOT2] --
  Controls second-order leading log and next-to-leading 
  log QED corrections:
  \begin{description}
    \item[{\tt IVALUE} = --1]
    no initial state radiation QED convolution at all
    \item[{\tt IVALUE} = 0]
    complete $\alpha$ additive radiator
    \item[{\tt IVALUE} = 1]
    with logarithmic hard corrections
    \item[{\tt IVALUE} = 2]
    complete $\alpha^2$ additive radiator 
    \item[{\tt IVALUE} = 3] 
    (default) complete $\alpha^3$ additive radiator 
    \item[{\tt IVALUE} = 4] 
    optional $\alpha^3$ additive radiator for estimation of 
    theoretical errors: QED-E radiator, Eqs. (3.31) to (3.32)
\cite{Bardin:1989qr} 
    \item[{\tt IVALUE} = 5]  
    ``pragmatic'' LLA third order corrections in a factorized
form~\cite{Skrzypek:1992vk} 
  \end{description}
\end{description}

\begin{description}
  \item[\tt FSRS] --
  Final state radiation scale:
  \begin{description}
    \item[{\tt IVALUE} = 0] $\alpha(0)$, preferred for tight cuts
    \item[{\tt IVALUE} = 1] (default) $\alpha(s)$, preferred for loose cuts
  \end{description}
\end{description}

\begin{description} 
  \item[\tt FTJR] --
  Treatment of {\tt FTJR} corrections \cite{Fleischer:1992fq}, see
\fig{subrBORN}:  
  \begin{description} 
    \item[{\tt IVALUE} = 0] without {\tt FTJR}  corrections 
    \item[{\tt IVALUE} = 1] (default) with {\tt FTJR}  corrections everywhere
    \item[{\tt IVALUE} = 2] {\tt FTJR}  corrections are taken into account only 
                            in POs, the option is used for tests only
  \end{description}
\end{description}

\begin{description}
  \item[\tt GAMS] --
  Controls the $\sman$ dependence of ${\cal G}_Z$,
  the \Z-width function, see \sect{Zpropagator}:
  \begin{description}
    \item[{\tt IVALUE} = 0]
    forces ${\cal G}_Z$ to be constant
    \item[{\tt IVALUE} = 1]
    (default) allows ${\cal G}_Z$
    to vary as a function of $\sman$ \cite{Bardin:1988xt}.
  \end{description}
\end{description}

\begin{description} 
  \item[\tt GFER] -- see {\tt NPAR(20)} in subsection \ref{dizetflags}
\end{description}

\begin{description} 
  \item[\tt HIGS] -- see {\tt NPAR(14)} in subsection \ref{dizetflags}
\end{description}

\begin{description} 
  \item[\tt HIG2] -- see {\tt NPAR(18)} in subsection \ref{dizetflags}
\end{description}

\begin{description}
  \item[\tt INTF] --
  Determines if the ${\cal O} (\alpha)$ initial-final state
  QED interference (IFI) is calculated:
  \begin{description}
    \item[{\tt IVALUE} = 0]
    the interference term is ignored
    \item[{\tt IVALUE} = 1]
    (default) with IFI in the $\ord{\alpha}$
  \end{description}
\end{description}

\begin{description}
  \item[\tt IPFC] --
  Pair flavour content for the pair production corrections:
  \begin{description}
    \item[{\tt IVALUE} = 1]
       only electron pairs
    \item[{\tt IVALUE} = 2]       
            only muon pairs
    \item[{\tt IVALUE} = 3]
              only tau-lepton pairs
    \item[{\tt IVALUE} = 4]
              only hadron pairs
    \item[{\tt IVALUE} = 5]
              (default) all channels summed            
    \item[{\tt IVALUE} = 6]
              leptonic pairs (without hadrons)
  \end{description}
\end{description}

\begin{description}
  \item[\tt IPSC] --
 Pair production singlet-channel contributions (works with ISPP = 2): 
  \begin{description}
    \item[{\tt IVALUE} = 0]
         (default) only non-singlet pairs                              
    \item[{\tt IVALUE} = 1]
         LLA singlet pairs according to~\cite{Berends:1988ab}       
    \item[{\tt IVALUE} = 2]
        complete $O(\alpha^2)$ singlet pairs, {\it ibid}
    \item[{\tt IVALUE} = 3]
        singlet pairs up to order $(\alpha L)^3$, {\it ibid}
  \end{description}
\end{description}

\begin{description}
  \item[\tt IPTO] --
    Third (and higher) order pair production contributions
\cite{Arbuzov:1999uq} 
    (works with {\tt ISPP} = 2):              
  \begin{description}
     \item[{\tt IVALUE} = 0]  
 only $O(\alpha^2)$ contributions        
     \item[{\tt IVALUE} = 1] 
  $O(\alpha^3)$ pairs                    
    \item[{\tt IVALUE} = 2]  
  some "non-standard" $O(\alpha^3)$ LLA pairs added       
    \item[{\tt IVALUE} = 3]
(default) $O(\alpha^4)$ LLA electron pairs added                 
  \end{description}
\end{description}

\begin{description}
  \item[\tt ISPP] --
  Treatment of ISR pairs:
  \begin{description}
    \item[{\tt IVALUE} = --1]
    pairs are treated with a ``fudge'' factor as in versions up to v.5.14
    \item[{\tt IVALUE} = 0]
    without ISR pairs
    \item[{\tt IVALUE} = 1]
    with ISR pairs,~\cite{Kniehl:1988id} with a re-weighting, 
    see \sect{secondo}
   \item[{\tt IVALUE} = 2]
   (default) with ISR pairs according to~\cite{Arbuzov:1999uq}
   \item[{\tt IVALUE} = 3]
   with ISR pairs according to~\cite{Jadach:1992aa}
   \item[{\tt IVALUE} = 4]
   with ISR pairs~\cite{Jadach:1992aa} with extended treatment of 
   hadron pair production
  \end{description}
\end{description}
 
\begin{description}
  \item[\tt MISC] --
  Controls the treatment of scaling of $\rho$ in the Model Independent 
  approach, see discussion in \subsect{POCOMM}:
  \begin{description}
    \item[{\tt IVALUE} = 0]
    (default) non-scaled $\rho$'s are used, {\tt AROTFZ}-array
    \item[{\tt IVALUE} = 1]
    scaled $\rho$'s, absorbing imaginary parts, are used, {\tt ARROFZ}-array 
  \end{description}
\end{description}

\begin{description}
  \item[\tt MISD] --
  Controls the $\sman$ dependence of the Model Independent approach,
  see \sect{ewcoup}:
  \begin{description}
    \item[{\tt IVALUE} = 0]
    fixed $\sman = \mzs$ in EWRC, old treatment  
    \item[{\tt IVALUE} = 1]
    (default) ensures equal numbers from all interfaces and for all partial
    channelds but {\tt INDF=10} for all $\sqrt{\sman}$ and for {\tt INDF=10} 
    up to 100 GeV\footnote{The reason of that will be investigated while
    working on the \LEPII\ version of \zf.}
  \end{description}
\end{description}

\begin{description}
  \item[\tt PART] --
  Controls the calculation of various parts of Bhabha scattering:
  \begin{description}
    \item[{\tt IVALUE} = 0]
    (default) calculation of full Bhabha cross-section and asymmetry
    \item[{\tt IVALUE} = 1]
    only $\sman$ channel
    \item[{\tt IVALUE} = 2]
    only $\tman$ channel
    \item[{\tt IVALUE} = 3]
    only $\sman-\tman$ interference
  \end{description}
\end{description}

\begin{description}
  \item[\tt POWR] --
  Controls inclusion of final-state fermion masses
  in kinematical factors, see ~\fig{subrBORN}. 
  It acts differently for quarks and leptons. For leptons: 
  \begin{description}
    \item[{\tt IVALUE} = 0]
    final state lepton masses are set equal to zero
    \item[{\tt IVALUE} = 1]
    (default) final state lepton masses are retained
    in all kinematical factors
  \end{description}
  For quarks it is active only for {\tt FINR} = --1 and then: 
  \begin{description}
    \item[{\tt IVALUE} = 0]
    final state quark masses are set equal to zero
    \item[{\tt IVALUE} = 1]
    (default) final state quark masses are set to their running values
    (that is, for $\fc\barc$ and $\fb\barb$ channels)
    and retained in all kinematical factors
  \end{description}
\end{description}

\begin{description}
  \item[\tt PREC] --
  is an integer number which any precision governing any 
  numerical integration is divided by, increasing thereby the numerical   
  precision of computation:
  \begin{description}
    \item[{\tt IVALUE} = 10]
    (default)
    \item[{\tt IVALUE} = 1 -- 99]
    in some cases when some numerical instability 
    while running v.5.10 was registered, it was 
    sufficient to use {\tt PREC} = 3,     
    in some other cases (e.g. with $P_\tau$) only {\tt PREC} = 30 
    solved the instability         
  \end{description}
\end{description}

\begin{description}
  \item[\tt PRNT] --
  Controls {\tt ZUWEAK} printing:
  \begin{description}
    \item[{\tt IVALUE} = 0]
    (default) printing by subroutine {\tt ZUWEAK} is suppressed
    \item[{\tt IVALUE} = 1]
    each call to {\tt ZUWEAK} produces some output
  \end{description}
\end{description}
 
\begin{description} 
  \item[\tt QCDC] -- see {\tt NPAR(3)} in subsection \ref{dizetflags}
\end{description}

\begin{description} 
  \item[\tt SCAL] -- see {\tt NPAR(9)} in subsection \ref{dizetflags}
\end{description}

\begin{description} 
  \item[\tt SCRE] -- see {\tt NPAR(6)} in subsection \ref{dizetflags}
\end{description}

\begin{description} 
  \item[\tt VPOL] -- see {\tt NPAR(1)} in subsection \ref{dizetflags}
\end{description}

\begin{description}
  \item[\tt WEAK] --
  Determines if the weak loop calculations are to be performed
  \begin{description}
    \item[{\tt IVALUE} = 0]
    no weak loop corrections to the cross-sections are calculated
    and weak parameters are forced to their Born values,  i.e.
    $\rho_{ef} = \kappa_{e,f,ef} = 1$
    \item[{\tt IVALUE} = 1]
    (default) weak loop corrections to the cross-sections are
    calculated
  \end{description}
\end{description}

In \tbn{tab:xxxx} an overview over all flags used in
{\tt DIZET} and {\tt ZFITTER} is given. 
The internal flag names and their default values  are listed
as well as their corresponding positions in the vectors 
{\tt NPAR(1:21)} in {\tt DIZET} (or  {\tt NPARD(1:21)}
in {\tt ZFITTER}) and  {\tt NPAR(1:30)}
in {\tt ZFITTER}.

\subsection{Subroutine {\tt ZUWEAK}\label{zuweak}}
Subroutine {\tt ZUWEAK} is used to perform the weak sector calculations.
These are done internally with {\tt DIZET}, see~\sect{dizetug}.
The routine calculates a number of important electroweak parameters
which are stored in common blocks for later use (see ~\sect{fullPOs}).
If any \zf\ flag has to be modified this must be done before
calling {\tt ZUWEAK}.
 
\SUBR{CALL ZUWEAK(ZMASS,TMASS,HMASS,DAL5H,ALFAS)}

\BS
\noindent \underline{Input Arguments:}
\smallskip
\begin{description}
  \item[\tt ZMASS] is the \Z\  mass  $\mzl$  in GeV.
  \item[\tt TMASS] is the top quark mass  $\mtl$  in GeV, [10-400].
  \item[\tt HMASS] is the Higgs mass  $\mhl$  in GeV, [10-1000].
  \item[\tt DAL5H] is the value of $\dalhv$.
  \item[\tt ALFAS] is the value of the strong coupling constant
     $\als$  at $q^2 = \mzs $ 
(see factors {\tt QCDCOR} in \tbn{qcdcor_fst}).
\end{description}
 
Computing time may be saved by performing
weak sector calculations only once during the initialization of the \zf\
package.
This is possible since weak parameters are nearly
independent of $s$ near the \Z\ peak, \EG\ $\sim \ln s/\mzs$.
However, the incredible precision of \LEPI\ data forced us to give
up this option, see description of flag {\tt MISD}.
\subsection{Subroutine {\tt ZUCUTS}\label{zucuts}}
 
Subroutine {\tt ZUCUTS} is used to define kinematic and geometric cuts
for each fermion channel:
it selects
the appropriate QED calculational {\em chain}.
 
\SUBR{CALL  ZUCUTS(INDF,ICUT,ACOL,EMIN,S$\_$PR,ANG0,ANG1,SIPP)}

\BS
\noindent \underline{Input Arguments:}
\smallskip
\begin{description}
 \item[\tt INDF] is the fermion index (see \tbn{indf} and \fig{interfEWCOUP}).
 \item[\tt ICUT] controls the kinds of cuts ({\em chain}) to be used.
   \begin{itemize}
     \item[{\tt ICUT} = --1:] (default) 
       allows for an $\smanp$ cut (a cut on $M^2_{\ff\fbf}$, the
         fermion and antifermion   
       invariant mass); the fastest branch based on \cite{Bardin:1991det}
     \item[{\tt ICUT} = 0:] {\bf not recommended}; branch is known to
       contain bugs. It
       allows for a cut on the acollinearity {\tt ACOL} of the \FF\ pair,
       on the minimum energy {\tt EMIN} of both fermion and antifermion,
       and for a geometrical acceptance cut
\cite{Bilenkii:1989zg}\footnote{As was shown recently 
       \cite{Christova:1999cct,Christova:1998tc,Christova:1999gh}, the old
       results of \cite{Bilenkii:1989zg} 
       contained bugs which occasionally didn't show up in comparisons
       as e.g. in \cite{Bardin:1995a2}. 
       The option is retained for
       back-compatibility with older versions only.}
     \item[{\tt ICUT} = 1:] $\smanp$ or $M^2_{\ff\fbf}$ cuts and geometrical 
       acceptance cut, based on \cite{Bardin:1991fut}
     \item[{\tt ICUT} = 2:]  new branch, replaces  {\tt ICUT} = 0 for
        realistic cuts {\tt ACOL} and {\tt EMIN}, based
       on~\cite{Christova:1999cct} 
     \item[{\tt ICUT} = 3:]   
       the same  branch, using {\tt ACOL} cut and  {\tt EMIN} cut
       but also with possibility to impose an additional acceptance
       cut \cite{Christova:1999gh}
   \end{itemize}
  \item[\tt ACOL] is the maximum acollinearity angle $ \xi^{\max} $ of the
    \FF\ pair in degrees ({\tt ICUT} = 0,2,3).
  \item[\tt EMIN] is the minimum energy $ E^{\min}_f $ of the fermion and
    antifermion in GeV ({\tt ICUT} = 0,2,3).
  \item[\tt S\_PR] is the minimum allowed invariant \FF\ mass $M^2_{\ff\fbf}$ 
    in GeV ({\tt ICUT} = --1,1) or, with some approximations, 
    the minimum allowed invariant mass of the 
    propagator after ISR\footnote{The invariant mass of the propagator
is not an observable quantity
unless specific assumptions on ISR and FSR are made.}
(This is related to the maximum photon energy by
    \eqn{cuts2}, \eqn{spmin}. See also description of {\tt FINR} flag.).
  \item[\tt ANG0] (default = $0^{\circ}$) is the minimum polar angle
    $\vartheta$ in degrees of the final-state antifermion.
  \item[\tt ANG1] (default = $180^{\circ}$) is the maximum polar angle
    $\vartheta$ in degrees of the final-state antifermion.
  \item[\tt SIPP]  a cut parameter governing the calculation of
     corrections due to 
    initial pairs, actually it should be chosen equal to
$\smanp$.
\end{description}

\subsection{Subroutine {\tt ZUINFO}\label{zuinfo}}
 
Subroutine {\tt ZUINFO} prints the values of  \zf\ flags and cuts.
 
\SUBR{CALL ZUINFO(MODE)}
 
\BS
\noindent \underline{Input Argument:}
\smallskip
 
\begin{description}
  \item[\tt MODE] controls the printing of \zf\ flag and cut values.
\begin{itemize}
  \item[{\tt MODE} = 0:] Prints all flag values.
  \item[{\tt MODE} = 1:] Prints all cut values.
\end{itemize}
\end{description}
 
 
\section
{Interface Routines of \zf\ \label{interfaces} }
\setcounter{equation}{0}
 
The calculational chains of \zf\ can be combined with several interfaces.
These interfaces will be described here.
For the Standard Model branch the cross-section and asymmetry interface
is subroutine {\tt ZUTHSM}, while for the tau polarization it is
subroutine {\tt ZUTPSM}. 
Subroutine {\tt ZUATSM} allows to calculate
the differential cross-section, $d \sigma /d \cos \theta$, predicted by
the Standard Model. 
With subroutine {\tt ZULRSM} the left-right asymmetry
is determined within the Standard Model.
Subroutines {\tt ZUXSA},
{\tt ZUXSA2}, {\tt ZUXAFB}, and {\tt ZUTAU} are interfaces using
effective couplings. 
The interface for the partial widths is {\tt ZUXSEC}. 
The interface using an S-matrix inspired language is realized
with the  {\tt SMATASY} package
\cite{Leike:1991pq,Riemann:1992gv,Kirsch:1995cf1,Kirsch:1995cf}. 

Note that subroutine {\tt ZUWEAK} must be called prior to any of the
interfaces to be described below. As a consequence, flags used in this
subroutine can influence the calculation of cross-sections and
asymmetries in the interfaces described now.
 
All subroutines need the following \underline{input arguments}:
\begin{description}
  \item[\tt INDF] is the fermion index (see \tbn{indf}).
  \item[\tt SQRS] is the centre-of-mass energy  \RS  ~in GeV.
  \item[\tt ZMASS] is the \Z\ mass  $\mzl$  in GeV.
\end{description}

\subsection
{Subroutine {\tt ZUTHSM}
\label{zuthsm}}
 
Subroutine {\tt ZUTHSM} is used to calculate Standard Model
cross-sections and 
forward-back\-ward asymmetries as described in ~\sect{intrf1}.
 
\SUBR{CALL ZUTHSM(INDF,SQRS,ZMASS,TMASS,HMASS,DAL5H,ALFAS,XS*,AFB*)}

 \BS
\noindent \underline{Input Arguments:}
\smallskip
\begin{description}
  \item[\tt TMASS] is the top quark mass  $\mtl$  in GeV, [10-400].
  \item[\tt HMASS] is the Higgs mass  $\mhl$  in GeV, [10-1000].
  \item[\tt DAL5H] is the value of $\dalhv$.
  \item[\tt ALFAS] is the value of the strong coupling constant
     $\alpha_s$  at $q^2 = \mzs$ (see also flag {\tt  QCDC} and
      factors {\tt QCDCOR}).
\end{description}
 
\nn \underline{Output Arguments}\footnote{An asterisk (*) following an
argument in a calling sequence is used to denote an output argument.}:
\smallskip
\begin{description}
  \item[\tt XS] is the total cross-section $\sigma_{\sss T}$  in nb.
  \item[\tt AFB] is the forward-backward asymmetry \afb .
\end{description}

\nn \underline{Output Internal Flag}:
\smallskip
\begin{description}
  \item[\tt INTRF=1]
\end{description}

\subsection
{Subroutine {\tt ZUATSM}
\label{zuatsm}}
 
Subroutine {\tt ZUATSM} is used to calculate 
differential cross-sections, $d \sigma/ d\cos \theta$,
in the Standard Model as described in ~\sect{intrf1}.
 
\SUBR{CALL ZUATSM(INDF,SQRS,ZMASS,TMASS,HMASS,DAL5H,ALFAS,CSA*,DXS*)}

\BS
\noindent \underline{Input Arguments:}
\smallskip
\begin{description}
  \item[\tt TMASS] is the top quark mass  $\mtl$  in GeV, [10-400].
  \item[\tt HMASS] is the Higgs mass  $\mhl$  in GeV, [10-1000].
  \item[\tt ALQED] is the value of the running electromagnetic
                   coupling constant.
  \item[\tt ALFAS] is the value of the strong coupling constant
     $\alpha_s$  at $q^2 = \mzs $ (see factors {\tt QCDCOR}).
  \item[\tt CSA] is the cosine of the scattering angle.
\end{description}
 
\nn \underline{Output Arguments}:
\smallskip
\begin{description}
  \item[\tt DXS] is the theoretical differential cross-section.
\end{description}

\nn \underline{Output Internal Flag}:
\smallskip
\begin{description}
  \item[\tt INTRF=1]
\end{description} 
 
\subsection
{Subroutine {\tt ZUTPSM}
\label{zutpsm}}
 
Subroutine {\tt ZUTPSM} is used to calculate the 
tau
polarization and tau polarization asymmetry in the Standard Model 
as described in ~\sect{intrf1}.
 
\SUBR{CALL ZUTPSM(SQRS,ZMASS,TMASS,HMASS,DAL5H,ALFAS,TAUPOL*,TAUAFB*)}

\BS
\noindent \underline{Input Arguments:}
\smallskip
\begin{description}
  \item[\tt HMASS] is the Higgs mass  $\mhl$  in GeV, [10-1000].
  \item[\tt DAL5H] is the value of $\dalhv$.
  \item[\tt ALFAS] is the value of the strong coupling constant
     $\alpha_s$  at $ q^2 = \mzs $ (see factors {\tt QCDCOR}).
\end{description}
 
\nn \underline{Output Arguments}:
\smallskip
\begin{description}
  \item[\tt TAUPOL] is the tau polarization  $A_{\mathrm{pol}}$ 
  of (\eqn{asym}).
  \item[\tt TAUAFB] is the tau polarization forward-backward
  asymmetry
  $A^{\mathrm{pol}}_{\sss FB}$  as defined in (\eqn{lafb}).
\end{description}
 
\nn \underline{Output Internal Flag}:
\smallskip
\begin{description}
  \item[\tt INTRF=1]
\end{description} 
 
\subsection
{Subroutine {\tt ZULRSM}
\label{zutlrm}}
 
Subroutine {\tt ZULRSM} is used to calculate the 
left-right asymmetry in the Standard Model 
as described in ~\sect{intrf1}.
 
\SUBR{CALL ZULRSM(INDF,SQRS,ZMASS,TMASS,HMASS,DAL5H,ALFAS,POL,XSPL*,XSMI*)}
          
\BS
\noindent \underline{Input Arguments:}
\smallskip
\begin{description}
  \item[\tt TMASS] is the top quark mass  $\mtl$ in GeV, [40-300].
  \item[\tt HMASS] is the Higgs mass  $\mhl$ in GeV, [10-1000].
  \item[\tt DAL5H] is the value of $\dalhv$.
  \item[\tt ALFAS] is the value of the strong coupling constant
     $\alpha_s$ at $q^2 = \mzs $ (see also flag {\tt ALST}).
  \item[\tt POL] is the degree of longitudinal polarization of
                 electrons.
\end{description}
 
\nn \underline{Output Arguments}:
\smallskip
\begin{description}
  \item[\tt XSPL] is the cross-section for {\tt POL} $>$ 0
  \item[\tt XSMI] is the cross-section for {\tt POL} $<$ 0
\end{description}
 
\nn \underline{Output Internal Flag}:
\smallskip
\begin{description}
  \item[\tt INTRF=1]
\end{description} 

\subsection{Subroutine {\tt ZUXSA}\label{zuxsa}}
 
Subroutine {\tt ZUXSA} is used to calculate cross-section and
forward-backward asymmetry as described in ~\sect{intrf3} as functions 
of \RS, $\mzl$, $\gz$.
 
\SUBR{CALL ZUXSA(INDF,SQRS,ZMASS,GAMZ0,MODE,GVE,XE,GVF,XF,XS*,AFB*)}

\BS
\noindent \underline{Input Arguments:}
\smallskip
\begin{description}
  \item[\tt GAMZ0] is the total \Z\ width  $\gz$ in GeV.
  \item[\tt MODE] determines which weak couplings are used:
  \begin{itemize}
  \item[{\tt MODE} = 0:] {\tt XE} ({\tt XF}) are effective
    axial-vector couplings
     $\rab{\fe,\ff}$ for electrons (final-state fermions).
    \item[{\tt MODE} = 1:] {\tt XE} ({\tt XF}) are the effective weak
    neutral-current 
    amplitude normalizations  $\rhobi{\fe,\ff}$ for electrons (final-state
    fermions).
  \end{itemize}
  \item[\tt GVE] is the effective vector coupling for electrons
   $\rvab{\fe}$.
  \item[\tt XE] is  the effective axial-vector coupling  $\rab{\fe}$ or
    weak neutral-current amplitude normalization  $\rhobi{\fe}$
    for electrons (see {\tt MODE}).
  \item[\tt GVF] is the effective vector coupling for the final-state
    fermions  $\rvab{\ff}$.
  \item[\tt XF] is  the effective axial-vector coupling  $\rab{\ff}$ or
    the weak neutral-current amplitude normalization  $\rhobi{\ff}$
    for the final-state fermions (see {\tt MODE}).
\end{description}
 
\noindent \underline{Output Arguments:}
\smallskip
\begin{description}
  \item[\tt XS] is the cross-section $ \sigma_{\sss T}$ in nb.
  \item[\tt AFB] is the forward-backward asymmetry  \afb.
\end{description}
 
\nn \underline{Output Internal Flag}:
\smallskip
\begin{description}
  \item[\tt INTRF=3]
\end{description} 

\subsection{Subroutine {\tt ZUXSA2}\label{zuxsa2}}

Subroutine {\tt ZUXSA2} is used to calculate lepton cross-section and
forward-backward asymmetry as functions of \RS, $\mzl$, $\gz$, and of the weak
couplings {\em assuming lepton universality} as introduced in \sect{intrf4}.
This routine is similar to {\tt ZUXSA} except that the couplings
are squared.
 
\SUBR{CALL ZUXSA2(INDF,SQRS,ZMASS,GAMZ0,MODE,GV2,X2,XS*,AFB*)}
 
\BS
\noindent \underline{Input Arguments:}
\smallskip
\begin{description}
  \item[\tt GAMZ0] is the total \Z\ width $\gz$ in GeV.
  \item[\tt MODE] determines which weak couplings are used:
  \begin{itemize}
    \item[{\tt MODE} = 0:] {\tt X2} is the square of the effective
    axial-vector coupling $\rab{\fl}$ for leptons.
    \item[{\tt MODE} = 1:] {\tt X2} is the square of the effective
neutral-current 
      amplitude normalization $\rhobi{\fl}$ for leptons.
  \end{itemize}
  \item[\tt GV2] is the square of the effective vector coupling
    $\rvab{\fl}$ for leptons.
  \item[\tt X2] is the square of the effective axial-vector coupling
    $\rab{\fl}$ or neutral-current amplitude normalization
    $\rhobi{\fl}$ for leptons (see {\tt MODE}).
\end{description}
 
\noindent \underline{Output Arguments:}
\smallskip
\begin{description}
  \item[\tt XS] is the cross-section $\sigma_{\sss T}$ in nb.
  \item[\tt AFB] is the forward-backward asymmetry \afb.
\end{description}
 
\nn \underline{Output Internal Flag}:
\smallskip
\begin{description}
  \item[\tt INTRF=4]
\end{description} 
 
\subsection{Subroutine {\tt ZUTAU}\label{zutau}}
 
Subroutine {\tt ZUTAU}  is used to calculate the $\tau^+$ polarization
as a function of \RS, $\mzl$, $\gz$, and the weak couplings,
 see \sect{intrf3}.
 
\SUBR{CALL ZUTAU(SQRS,ZMASS,GAMZ0,MODE,GVE,XE,GVF,XF,TAUPOL*,TAUAFB*)}
 
\BS
\noindent \underline{Input Arguments:}
\smallskip
\begin{description}
  \item[\tt GAMZ0] is the total \Z\ width $\gz$ in GeV.
  \item[\tt MODE] determines which weak couplings are used:
  \begin{itemize}
    \item[{\tt MODE} = 0:] {\tt XE} ({\tt XF}) is the effective
    axial-vector coupling $\rab{\fe,\ff}$  for electrons (final-state fermions).
    \item[{\tt MODE} = 1:] {\tt XE} ({\tt XF}) is the effective weak
neutral-current 
    amplitude normalization $\rhobi{\fe,\ff}$  for electrons (final-state
    fermions).
  \end{itemize}
  \item[\tt GVE] is the effective vector coupling for electrons
    $\rvab{\fe}$ .
  \item[\tt XE] is  the effective axial-vector coupling $\rab{\fe}$  or
    weak neutral-current amplitude normalization $\rhobi{\fe}$ 
    for electrons (see {\tt MODE}).
  \item[\tt GVF] is the effective vector coupling for the final-state
   fermions $\rvab{\ff}$ .
  \item[\tt XF] is  the effective axial-vector coupling $\rab{\ff}$  or
    weak neutral-current amplitude normalization $\rhobi{\ff}$ 
    for the final-state fermions (see {\tt MODE}).
\end{description}
 
\noindent \underline{Output Arguments:}
\smallskip
\begin{description}
  \item[\tt TAUPOL] is the $\tau$-lepton polarization $\lambda_{\tau}$
defined in (\eqn{asym}).
  \item[\tt TAUAFB] is the forward-backward asymmetry for polarized $\tau$-leptons $A^{\mathrm{pol}}_{\sss FB}$  as defined in (\eqn{lafb}).
\end{description}

\nn \underline{Output Internal Flag}:
\smallskip
\begin{description}
  \item[\tt INTRF=3]
\end{description} 

\subsection{Subroutine {\tt ZUXSEC}\label{zuxsec}}
 
Subroutine {\tt ZUXSEC} is  used to calculate the cross
section as a function of \RS, $\mzl$, $\gz$, $\Gamma_e$ and $\Gamma_f$
as described in \sect{intrf2}.
 
\SUBR{CALL ZUXSEC(INDF,SQRS,ZMASS,GAMZ0,GAMEE,GAMFF,XS*)}
 
\BS
\noindent \underline{Input Arguments:}
\smallskip
\begin{description}
  \item[\tt GAMZ0] is the total \Z\ width $\gz$  in GeV.
  \item[\tt GAMEE] is the partial \Z\ decay width $\Gamma_e$  in GeV.
  \item[\tt GAMFF] is the partial \Z\ decay width $\Gamma_f$  in GeV;
  if {\tt INDF}=10, {\tt GAMFF}=$\Gamma_{h}$.
\end{description}
 
\nn \underline{Output Internal Flag}:
\smallskip
\begin{description}
  \item[\tt INTRF=2]
\end{description} 

\subsection{Subroutine {\tt ZUXAFB}\label{zulr}}

Subroutine {\tt ZUXAFB} is  used to calculate the cross
section as a function of \RS, $\mzl$, $\gz$, $\Gamma_e$ and $\Gamma_f$
as described in ~\sect{intrf5}.
 
\SUBR{CALL ZUXAFB(INDF,SQRS,ZMASS,GAMZ0,PFOUR,PVAE2,PVAF2,XS*,AFB*)}
 
\BS
\noindent \underline{Input Arguments:}
\smallskip
\begin{description}
  \item[\tt GAMZ0] is the total \Z\ width $\gz$  in GeV.
  \item[\tt PFOUR] is the product of vector and axial-vector couplings
                   $\four$.
  \item[\tt PVAE2] is $\vaeII$.
  \item[\tt PVAF2] is $\vafII$.
\end{description}
 
\noindent \underline{Output Argument:}
\smallskip
\begin{description}
  \item[\tt XS] is the cross-section $\sigma_{\sss T}$ in nb.
  \item[\tt AFB] is the forward-backward asymmetry \afb.
\end{description}
 
\nn \underline{Output Internal Flag}:
\smallskip
\begin{description}
  \item[\tt INTRF=5]
\end{description} 

\subsection
{Subroutine {\tt ZUALR}
\label{zualr}}

Subroutine {\tt ZUALR} is reserved for the fit of $A_{\sss LR}$,
see \sect{intrf6}.
     
\SUBR{CALL ZUALR(SQRS,ZMASS,GAMZ0,MODE,GVE,XE,GVF,XF,TAUPOL,TAUAFB)}

\BS
\nn \underline{Output Internal Flag}:
\smallskip
\begin{description}
  \item[\tt INTRF=6]
\end{description}

\section{\zf\ Common Blocks\label{zupars}}
\eqnzero
Only a few \zf\ common blocks are of potential interest.
to the user. They are documented here.
\vspace*{-1mm} 

\begin{verbatim}
      COMMON /ZUPARS/QDF,QCDCOR(0:14),ALPHST,SIN2TW,S2TEFF(0:11),WIDTHS(0:11)
\end{verbatim}
\vspace*{-1mm}

The common block {\tt /ZUPARS/} contains some \zf\ parameters:

\begin{description}
  \item[\tt QDF] is the final-state radiation factor 
  $3\alpha/(4\pi)$ 
  \item[\tt QCDCOR] is an array of QCD corrections,
  defined in~\eqn{qcdcor_fst}
  \item[\tt ALPHST] is $\als(\mzs)$.
  \item[\tt SIN2TW] is $\siws$, Sirlin's weak mixing angle as defined
  in~\eqn{defsw2}.
  \item[\tt S2TEFF] are the values of $\kappa_f\stws$ at $\sman=\mzs$ 
  for each fermion channel $f$ (see \tbn{indf}). 
They are some auxiliary
  quantities, which do not coincide with the {\em effective weak mixing angles}
  defined by \eqn{sw2eff} and may be of interest for experts only. 
  Note that {\tt S2TEFF(10:11)} are set to zero.
  \item[\tt WIDTHS] are the partial decay widths \eqns{defzwidthl}{defzwidthq}
  of the $\zb$ for all fermion channels enumerated as in \tbn{indf}.
  {\tt WIDTHS(11)} is the total $\zb$ width.
\end{description}
 
\begin{verbatim}
      COMMON/ZFCHMS/ALLCH(0:11),ALLMS(0:11)
\end{verbatim}
 
The common block {\tt ZFCHMS} contains the charges {\tt ALLCH} 
and masses {\tt ALLMS} of the fermions. See \tbn{indf} for enumeration.

Note that {\tt ALLCH(10)} = 1 for technical reasons and {\tt ALLCH(11)}
is undefined.
Also, {\tt ALLMS(10)} = 0 and {\tt ALLMS(11)} again is undefined.
 
We would also like to mention that the 
variables {\tt FAA}, {\tt FZA}, {\tt FZZ}, 
which are introduced as {\tt DATA} in subroutine {\tt EWCOUP}
allow us to switch on/off the $\ph\ph$, $\ph\zb$, $\zb\zb$ parts
of the cross-sections, respectively.

 
\chapter{{\tt ZFITTER} availability\label{avail}}

The Fortran files composing the \zf\ package, this description, and
also related information may be found at the following locations: 

\bigskip

{\tt /afs/cern.ch/user/b/bardindy/public/}

\bigskip

{\tt http://www.ifh.de/$\sim$riemann/Zfitter/zf.html}

\bigskip

\noindent
E-mail addresses of the authors of this description:

\[
\begin{array}{ll}
\mbox{D. Bardin}       &  \mbox{bardin@nusun.jinr.ru}  \\
                       &  \mbox{Dmitri.Bardin@cern.ch} \\
\mbox{P. Christova}    &  \mbox{penchris@nusun.jinr.ru}\\
                       &  \mbox{penka@ifh.de}          \\
\mbox{M. Jack}         &  \mbox{jack@ifh.de}           \\
\mbox{L. Kalinovskaya} &  \mbox{kalinov@nusun.jinr.ru} \\
                       &  \mbox{Lidia.Kalinovskaia@cern.ch} \\
\mbox{A. Olshevsky}    &  \mbox{Alexander.Olchevski@cern.ch}\\
\mbox{S. Riemann}      &  \mbox{Sabine.Riemann@ifh.de}      \\
\mbox{T. Riemann}      &  \mbox{Tord.Riemann@ifh.de}        \\
\end{array}
\]

\clearpage
\addcontentsline{toc}{section}{Acknowledgements}
\section*{Acknowledgments}
We would like to thank the colleagues who joined us for longer or shorter
periods:
A.~Chizhov,
O.~Fedorenko,
M.~Sachwitz,
A.~Sazonov,
Yu.~Sedykh,
I.~Sheer,
and
L.~Vertogradov.
Without their contributions the {\tt ZFITTER} package would not be what it is.

We acknowledge the contributions 
to the first description of the package, ref.~\cite{Bardin:1992jc},
from 
A.~Chizhov, 
O.~Fedorenko, 
S.~Ganguli, 
A.~Gurtu, 
M.~Lokajicek, 
K.~Mazumdar, 
G.~Mitselmakher, 
J.~Ridky, 
M.~Sachwitz, 
A.~Sazonov,
D.~Schaile, 
Yu.~Sedykh, 
I.~Sheer, 
and
L.~Vertogradov.

In the process of package development we profited heavily from 
the fruitful discussions with
G.~Altarelli,
S.~Arkadova,
W.~Hollik,
E.~Kuraev,
A.~Leike,
L.~Maiani, 
L.~Okun and 
D.~Shirkov.  

The implementation of higher order corrections into the package was supported
by 
A.~Arbuzov,
R.~Barbieri,
F.~Berends,
K.~Chetyrkin,
G.~Degrassi,
P.~Gambino,
F.~Je\-ger\-leh\-ner,
R.~Harlander,
A.~Kataev,
H.~K\"uhn,
S.~Larin,
M.~Steinhauser,
and
O.~Tarasov.

Without the dedicated comparisons with other programs performed in close 
collaboration with   
A.~Borrelli,
W.~Hollik,
S.~Jadach,
L.~Maiani,
M.~Martinez,
G.~Passarino,
B.~Pietrzyk,
A.~Rozanov,
F.~Teubert,
and
Z.~Was 
the package would not be as reliable as it is.

Special thanks goes to the users of \zf, who were constantly 
providing us with critical commments and ideas for improvements:
M.~Gr\"unewald,
T.~Kawamoto,
C.~Pauss,
and
G.~Quast. 

We would like to acknowledge also the 
contributions 
of: 
D.~Bourilkov,
I.~Boyko,
R.~Cla\-re,
P.~Clarke,
G.~Duckeck, 
F.~Filthaut,
D.~Haidt,
J.~Holt,
J.~Kellogg,
M.~Kobel,
K.~Mazumdar,
J.~Mnich,
K.~M\"onig,
P.~Renton,
D.~Schaile,
M.~Winter,
and of many others for useful discussions and suggestions, which 
were constantly stimulating the \zf\ package development. 

The authors are grateful to the JINR and LNP directorates, in particular: 
V.~Kadyshevsky, 
A.~Sissakian, 
N.~Rusakovitch,
 and A.~Kurilin 
for their constant interest and support.

D.B., P.C., and L.K. are very much indebted to DESY Zeuthen
for financial support and hospitality extended to them during several 
stays in 1994--1998 when a notable part of the stays was 
devoted to this project.
They are indebted for the generous help of DESY Zeuthen, especially to
Paul S{\"o}ding, for strengthening their computing equipments.

D.B. is indebted to the Theory Division of CERN for financial support 
and hospitality extended to him during several short stays in 1996--1999 
when part of the time was devoted to this project, in particular for the
stay in May--June 1999, when it was completed.

D.B. would like to express special thanks to Giampiero Passarino
for the numerous selfless comparisons of \zf\ and {\tt TOPAZ0} results 
during several workshops on \LEPI\ and \LEPII\ physics and in collaboration 
with the LEP Electroweak Working Group.

        
\appendix                
\def\theequation{\Alph{chapter}.\arabic{section}.\arabic{equation}}
\def\thefigure{\Alph{chapter}.\arabic{figure}}
\chapter{One-Loop Core of {\tt ZFITTER}\label{app-ew}}
\section{Introduction to One-Loop EWRC\label{zbb_introduction}}
\eqnzero
The aim of this Appendix is to give a condensed presentation 
of the one-loop core of {\tt ZFITTER}.
It outlines the way from Feynman diagrams to the amplitude form
factors; the latter are the
main building blocks of the code.

This Appendix consists of three Sections.  
\sect{auxil} introduces the Passarino-Veltman functions
\cite{Passarino:1979jh} used in \zf,
\sect{building-blocks} contains the building blocks for the
construction of the one-loop electroweak {\em amplitudes}, and 
\sect{amplitudes} contains the {\bf amplitudes} themselves.
A complete presentation of the OMS renormalization scheme,
on which the electroweak core of {\tt ZFITTER} is based upon, was
presented recently in \cite{BardinPassarino:1999}. 
This Appendix is not a plane extraction from there, but contains 
some minimum of formulae needed to understand the amplitude form factors
for the process $\fep\fem\to\ff\barf$. 
\section{Passarino--Veltman Functions\label{auxil}}
\eqnzero
Here we present scalar integrals over Feynman parameters which one meets 
in the calculation of ${\cal O}(\alpha)$ electroweak radiative corrections,
EWRC. 
We use the Passarino-Veltman functions
$\saff{0},\;\sbff{0},\;\scff{0}$ and 
$\sdff{0}$. 
The defining integrals and the answers in the dimensional
regularization are given.
\subsubsection{One-point function\label{opf}}
Defining integral:
\bq
\ib\pi^2\aff{0}{\Mind{}} = \tHs^{4-n}\int
\frac{d^n\imom}{\imoms+\Minds{}-\ib\ep}\,.
\label{a0def}
\eq
Here and below $\tHs$ is the so-called t'Hooft scale, an arbitrary quantity
with mass dimensionality which is introduced in order to make 
the dimensionality of the integral of \eqn{a0def} independent of the 
variation of space-time dimension $n$.
Its answer is
\bq
\aff{0}{\Mind{}}=\Minds{}\lpar-\dlt+\ln\frac{\Minds{}}{\tHss}-1\rpar.
\label{a0answer}
\eq
\subsubsection{Two-point functions\label{tpf}}
Defining integral:
\bqa
\ib\pi^2\bff{0}{\pmoms}{\Mind{1}}{\Mind{2}}
&=&
\tHs^{4-n}\int\frac{d^n\imom}{d_0 d_1},
\\
d_0&=&\imoms+\Minds{1}-\ib\ep,
\\
d_1 &=&(\imom+\pmom)^2+\Minds{2}-\ib\ep.
\label{b0def}
\eqa
The most general answer reads
\bqa
\bff{0}{\pmoms}{\Mind{1}}{\Mind{2}} &=&
 \dlt-\ln\frac{\Mind{1}\Mind{2}}{\tHss}
+\frac{\Lambda}{\pmoms}
 \ln\frac{-\pmoms-\ib\epsilon+\Minds{1}+\Minds{2}-\Lambda}{2\Mind{1}\Mind{2}}
\nll &&
 +~\frac{\Minds{1}-\Minds{2}}{2\pmoms}\ln\frac{\Minds{1}}{\Minds{2}}+2,
\label{b0answer-old}
\eqa
where $\Lambda^2=\lkall{\pmoms}{\Minds{1}}{\Minds{2}}$ is the K\"allen 
$\lambda$-function:
\ba
\lkall{x}{y}{z}=x^2+y^2+z^2-2xy-2xz-2yz.
\ea
Several particular cases are needed. 
The case of equal masses:
\ba
\bff{0}{\pmoms}{\Mind{}}{\Mind{}} = 
\dlt - \ln\frac{\Minds{}}{\tHss} + 2 - \beta\ln\frac{\beta+1}{\beta-1}\,,
\label{b0mm}
\ea
with
\ba
\beta^2=1+4\frac{\Minds{}}{\pmoms-\ib\ep}.
\ea
Cases with some vanishing arguments:
\bqa
\bff{0}{0}{\Mind{1}}{\Mind{2}}&=&
\frac{\aff{0}{\Mind{2}}-\aff{0}{\Mind{1}}}{\Minds{1}-\Minds{2}}\,,
\\
\bff{0}{0}{0}{\Mind{}}&=& -\frac{\aff{0}{\Mind{}}}{\Minds{}}\,,
\\
\bff{0}{\pmoms}{0}{0} &=&\dlt-\ln\frac{\pmoms -\ib\epsilon}{\tHss}+2.
\eqa
It is convenient to split $\saff{0}$- and $\sbff{0}$-functions into two
parts: (i) one containing the ultraviolet (UV) singularity and the $\tHs$-dependent 
part, i.e. the {\em pole} part, and (ii)  the {\em finite} rest.  
For $\saff{0}$ this is already achieved in \eqn{a0def},
while for $\sbff{0}$ we define in addition a $\sfbff{0}$ function:
\bq
\bff{0}{\pmoms}{\Mind{1}}{\Mind{2}}= 
\dlt-\ln\frac{\mws}{\tHss} + \fbff{0}{\pmoms}{\Mind{1}}{\Mind{2}},
\label{b0m1m2-old}
\eq
with an artificial scale $\mwl$.

Higher-rank $\sbff{}$-functions, $\sbff{1}$ and $\sbff{21}$,
are also needed. 
We simply list for them the integral representations
with separated UV-poles:
\bqa
\bff{1}{\pmoms}{\Mind{1}}{\Mind{2}}
&=&-\frac{1}{2}\dlt + \intfx{\xvar}\xvar
\ln\frac{\pmoms x(1-x)+\Mind{1}(1-x)+\Mind{2} x}{\tHss}
\nl &=&
\frac{1}{2\pmoms}
  \biggl[\aff{0}{\Mind{1}}-\aff{0}{\Mind{2}}
 +\Bigl(\Delta\Minds{12}-\pmoms\Bigr)
  \bff{0}{\pmoms}{\Mind{1}}{\Mind{2}}\biggr],
\label{b1m1m2}
\eqa
and 
\bqa
\bff{21}{\pmoms}{\Mind{1}}{\Mind{2}} 
&=&\frac{1}{3}\dlt - \intfx{\xvar}\xvars
\ln\frac{\pmoms x(1-x)+\Mind{1}(1-x)+\Mind{2} x}{\tHss}
\nl &=&
 \frac{3\lpar \Minds{1}+ \Minds{2} \rpar + \pmoms}{18 \pmoms }
+\frac{\Delta\Minds{12}- \pmoms}{3\pmomq}\aff{0}{\Mind{1}}
-\frac{\Delta\Minds{12}-2\pmoms}{3\pmomq}\aff{0}{\Mind{2}}
\nll &&
+\frac{\lkall{-\pmoms}{\Minds{1}}{\Minds{2}}-3\pmoms\Minds{1}}{3\pmomq} 
\bff{0}{\pmoms}{\Mind{1}}{\Mind{2}},
\nll
\mbox{with}\qquad\qquad
\Delta\Minds{12}&=&\Minds{1}-\Minds{2}.
\label{bffdifM}
\eqa
They enter in the combination 
\bqa
\bff{f}{\pmoms}{\Mind{1}}{\Mind{2}}
&=&
2\lrbr\bff{21}{\pmoms}{\Mind{1}}{\Mind{2}}+\bff{1}{\pmoms}{\Mind{1}}{\Mind{2}}
 \rrbr
\nll &=& -\frac{1}{3\epsb}
+2\int_0^1 dx x(1-x)\ln\frac{\pmoms x(1-x)+\Mind{1}(1-x)+\Mind{2} x}{\tHss}\,,
\qquad\quad
\label{Bf_function-old}
\eqa
which is often met in {\em fermionic} parts of self-energy functions
(see \sect{building-blocks}). 
For the case of equal masses 
$(\Mind{1}=\Mind{2})$,
the finite part of \eqn{Bf_function} may be cast in the following compact 
form:
\bq
\pmoms\fbff{f}{\pmoms}{\Mind{}}{\Mind{}} = 
 \frac{\pmoms}{9} 
+\frac{2\Minds{}}{3}\ln\frac{\Minds{}}{\mws}
+\frac{1}{3}\lpar 2\Minds{}-\pmoms \rpar \fbff{0}{\pmoms}{\Mind{}}{\Mind{}},
\eq
where
\bq
\fbff{0}{\pmoms}{\Mind{}}{\Mind{}} =
 -\ln\frac{\Minds{}}{\mws} + 2 - \beta \ln \frac{\beta+1}{\beta-1}\,.
\label{BfF_function}
\eq
The counter-terms involve the derivative of $\sfbff{0}$, 
which in the case of equal masses is very compact:
\ba
\fbff{0p}{\pmoms}{\Mind{}}{\Mind{}} 
&\equiv& 
\frac{\pd\;\fbff{0}{\pmoms}{\Mind{}}{\Mind{}}}{\pd\;\pmoms}
\nl
&=&-
\frac{1}{\pmoms}+\frac{2\Minds{}}{\pmomq}\frac{1}{\beta}
\ln\frac{\beta+1}{\beta-1}\,.
\ea
Finally, we use a subtracted $\sbff{0}$ function defined by
\bq
\Delta\bff{ }{\pmoms}{\Mind{1}}{\Mind{2}}=
      \bff{0}{\pmoms}{\Mind{1}}{\Mind{1}}
     -\bff{0}{0}{\Mind{1}}{\Mind{2}}.
\label{deltab}
\eq
\subsubsection{Three- and four-point functions\label{tfpf}}
The above presentation of scalar $\saff{0}$ and $\sbff{0}$ functions
is rather general. 
In what follows,
only those $\scff{0}$ and $\sdff{0}$ functions are described, which
are used in \zf. 
We will use a truncated argument list for these functions and 
omit masses of external particles, which are ignored in the 
calculations.
There are only two generic scalar three-point integrals which one meets in our
calculations.
The first one arises in $V\ff\fbf$ vertices if external 
fermion masses are neglected and if the internal boson is not massless 
(i.e. not a photon).
It is:
\bqa
\label{threepoint}
C_0(\pmoms;\Mind{1},\Mind{2},\Mind{1}) &=&
 \int^1_0 y dy \int^1_0 dx 
            \frac{1}{\Minds{1} (1-y) + \Minds{2} y + \pmoms y^2 x (1 - x )}
\nll &=&
\frac{1}{\pmoms}\Biggl[\Litwo\lpar\frac{x_0  }{x_0-x_1}\rpar
                   -\Litwo\lpar\frac{x_0-1}{x_0-x_1}\rpar
                   +\Litwo\lpar\frac{x_0  }{x_0-x_2}\rpar
\nll && \hspace{4mm}
                   -\Litwo\lpar\frac{x_0-1}{x_0-x_2}\rpar
                   -\Litwo\lpar\frac{x_0  }{x_0-x_3}\rpar
                   +\Litwo\lpar\frac{x_0-1}{x_0-x_3}\rpar\Biggr],\qquad
\eqa
with
\bqa
x_0 = \frac{\Minds{1}-\Minds{2}}{\pmoms}\,,
\\
x_{1,2} = \frac{1}{2}\lpar1 \mp \sqrt{1+\frac{4 \Minds{2}}{\pmoms}}\rpar,
\\
x_3 = \frac{\Minds{1}}{\Minds{1}-\Minds{2}}\,.
\label{threexses}
\eqa
The generic four-point integral is:
\bqa
\label{fourpoint}
\dff{0}{\pmoms}{I}{\Mind{1}}{0}{\Mind{1}}{\Mind{2}}
&=&
\int^1_0 dz \int^1_0 ydy \int^1_0 dx
\nl
&&\hspace{-.65cm}\times
\frac{1}{\left[\Minds{1} (1-y) + \Minds{2} y + I(1-y)(1-z) 
                 + \pmoms z^2 y^2 x (1-x)\right]^2}
\nll 
&=&\frac{1}{\pmoms(I+\Minds{1})\sqrt{d_4}}
\\ &&\times
\sum_{ {i}= 1,2 }
\Biggl[
       \Litwo\lpar\frac{{\bar x}_i  }{{\bar x}_i-x_1}\rpar
      -\Litwo\lpar\frac{{\bar x}_i-1}{{\bar x}_i-x_1}\rpar
\nll &&
      +\Litwo\lpar\frac{{\bar x}_i  }{{\bar x}_i-x_2}\rpar
      -\Litwo\lpar\frac{{\bar x}_i-1}{{\bar x}_i-x_2}\rpar 
      -\Litwo\lpar\frac{{\bar x}_i  }{{\bar x}_i-x_3}\rpar
\nll &&
      +\Litwo\lpar\frac{{\bar x}_i-1}{{\bar x}_i-x_3}\rpar 
      +\Litwo\lpar\frac{{\bar x}_i  }{{\bar x}_i-x_4}\rpar
      -\Litwo\lpar\frac{{\bar x}_i-1}{{\bar x}_i-x_4}\rpar 
\nonumber
\Biggr],
\eqa 
where the new roots are:
\bqa
{\bar x}_{1,2} 
&=&\frac{x_4}{2}\lpar1 \mp \sqrt{d_4}\rpar,
\\
d_4 &=& 1+\frac{4 \Minds{2} I(I+\Minds{1}-\Minds{2})}{\pmoms(I+\Minds{1})^2}\,,
\\
x_4 &=& \frac{I+\Minds{1}}{I+\Minds{1}-\Minds{2}}\,,\quad
\label{fourxses}
\eqa
and the $x_{1,2,3}$ are defined in \eqn{threexses}. 

 The analytic continuation
\bqa
\pmoms \to -\sman,
\eqa
is well-defined if one understands all masses appearing in 
\eqns{threepoint}{fourpoint} with infinitesimal imaginary parts
\bq
\Minds{i}\to\Minds{i}-\ib\ep.
\eq
The second 
generic scalar three-point integral
is met in the infrared divergent $V\ff\fbf$
vertex with a virtual photon and with external momenta on-shell.
Here we must keep all arguments in the $\scff{0}$ function:
\bqa
\cff{0}{-\sman}{\mfl}{0}{\mfl}
&\equiv&
\scff{0}\lpar -\mfs,-\mfs,-\sman;\mfl,0,\mfl\rpar
\nl
&=&
\frac{\tHs^{4-n}}{\ib\pi^2}\intmomi{n}{\imom}
\frac{1}{\imoms\lrbr\lpar\imom+\pmomi{1}\rpar^2+\mfs\rrbr
               \lrbr\lpar\imom-\pmomi{2}\rpar^2+\mfs\rrbr}\,.
\label{IRDC01}
\eqa
We limit ourselves to the case $\sman\gg\mfs$, where we have:
\bq
\sman\cff{0}{-\sman}{\mfl}{0}{\mfl} 
\approx
-\lpar\frac{1}{\epsh}+\ln\frac{\mfl}{\tHss}\rpar
\ln\frac{\sman}{\mfs}-\frac{1}{2}\ln^2\frac{\sman}{\mfs}+\frac{2}{3}\pi^2.
\label{IRDC02}
\eq
\section{ Building Blocks in the OMS Approach\label{building-blocks}}
\eqnzero
In this Section we describe various {\it building blocks} used to construct
the one-loop amplitudes of the processes $\zb\to\ff\fbf$ and
$\fep\fem\to\ff\fbf$ in terms of the  
$\saff{0},\;\sbff{0},\;\scff{0}$ and $\sdff{0}$ functions introduced
in the previous Section. 
They are presented in the order of increasing complexity:
self-energies, vertices, and boxes. 
The {\em on-mass-shell (OMS)} renormalization scheme is used in the
unitary gauge \cite{Bardin:1980fet,Bardin:1982svt}. 
\subsection{Bosonic self-energies \label{bosonicse}}
\subsubsection{$\zb,\ph$ bosonic self-energies and $\zb$--$\ph$ transition
\label{bsetran}}
In the unitary gauge there are only five diagrams which contribute to the
{\em total} $\zb$ and $\ph$ bosonic self-energies and to the $\zb$--$\ph$ 
transition. They are shown in \fig{ssssez}.

With $S_{\sss{\zb\zb}}$, $S_{\zg}$ and $S_{\ph\ph}$ standing for the sum of all
diagrams, depicted by a grey circle in \fig{ssssez}, we define the three 
corresponding self-energy functions $\Sigma$:
\ba
S_{\sss{\zb\zb}}&=&(2\pi)^4\ib\frac{g^2}{16\pi^2\cows}\Sigma_{\sss{\zb\zb}},
\\
S_{\zg}   &=&(2\pi)^4\ib\frac{g^2\siw}{16\pi^2\cow}\Sigma_{\zg},
\\
S_{\ph\ph}   &=&(2\pi)^4\ib\frac{g^2\siws}{16\pi^2}\Sigma_{\ph\ph}.
\label{sefunct}
\ea
All {\bf bosonic} self-energies and transitions may be naturally split into
{\em bosonic} and {\em fermionic} components.

\begin{itemize}
\item {Bosonic components of $\zb,\ph$ self-energies and
        $\zb$--$\ph$ transitions}, diagrams \fig{ssssez}(2-5), are:
\end{itemize}
\bqa
\Sigma^{\bos}_{\sss{\zb\zb}}(\sman)&=&
\mws
\Biggl\{
\cowf\lpar -4-\frac{17}{3}\frac{1}{\Rw }
             +\frac{4}{3 }\frac{1}{\Rws}
             +\frac{1}{12}\frac{1}{\Rwc}
     \rpar
\bff{0}{-\sman}{\mwl}{\mwl}
\nll  &&
+\frac{1}{12} \Biggl[
\lpar\frac{1}{\cowf}-\frac{2}{\cows}\rw +\rw^2 \rpar \Rw
    -\frac{10}{\cows} + 2\rw - \frac{1}{\Rw}
 \Biggr]
\bff{0}{-\sman}{\mhl}{\mzl}
\nll &&
+\cows \lpar - 4+\frac{4}{3}\frac{1}{\Rw } 
                +\frac{1}{6}\frac{1}{\Rws}\rpar
     \frac{\aff{0}{\mwl}}{\mzs}
\nll &&
+\frac{1}{12}\lpar\frac{\mzs-\mhs}{\sman}+1\rpar
 \frac{\Bigl[ \aff{0}{\mzl} - \aff{0}{\mhl} \Bigr]}{\mzs}
 -\frac{1}{12} \frac{\aff{0}{\mhl}}{\mzs}
\nll &&
-\Biggl[
\frac{1}{6\cows}+4\cowf+\frac{1}{6}\rw
-\lpar\frac{1}{18}+\frac{4}{3}\cowf\rpar \frac{1}{\Rw} 
+\frac{1}{9} \cowf
\lpar 5 - \frac{1}{2}\frac{1}{\Rw} \rpar \frac{1}{\Rws}
 \Biggr]
 \Biggr\},\qquad\quad
\label{sig_zz}
 \\ \nll
\Sigma^{\bos}_{\ph\ph}(\sman) &=& -\sman \Pgg^{\bos}(\sman),
\label{sig_gg} 
\\ \nll
\Sigma^{\bos}_{\zg}(\sman) &=&
          -\sman \Pzg^{\bos}\lpar\sman \rpar
   = -\cows\sman \Pgg^{\bos}\lpar\sman \rpar. 
\label{sig_zy}
\eqa

Here and below the following abbreviations are used:
\ba
\cows&=&\frac{\mws}{\mzs},
\\
\siws&=&1-\cows,
\\
\rw&=&\frac{\mhs}{\mws},
\\
\rz&=&\frac{\mhs}{\mzs},
\\
\Rw&=&\frac{\mws}{\sman},
\\
\Rz&=&\frac{\mzs}{\sman}.
\label{abbrev-old}
\ea

\vspace*{-5.5mm}

\begin{figure}[tbhp]
\[
\begin{array}{ccccccc}
\begin{picture}(75,20)(0,8)
\SetScale{2.}
  \Photon(0,5)(13.5,5){1.5}{5}
  \Photon(25,5)(37,5){1.5}{5}
\SetScale{1.}
   \GCirc(37.5,10){12.5}{0.5}
  \Text(0,18)[bl]{$\zb,\gamma$}
  \Text(12.5, 2)[tc]{$\mu$}
  \Text(75,18)[br]{$\zb,\gamma$}
  \Text(62.5, 2)[tc]{$\nu$}
\end{picture}
&=&
\begin{picture}(75,20)(0,8)
  \Photon(0,10)(25,10){3}{5}
  \ArrowArcn(37.5,10)(12.5,0,180)
  \ArrowArcn(37.5,10)(12.5,180,0)
   \Vertex(25,10){2.5}
   \Vertex(50,10){2.5}
  \Photon(50,10)(75,10){3}{5}
  \Text(37.5,-8)[tc]{$\fbu,\fbd$}
  \Text(37.5,30)[bc]{$u,d$}
  \Text(0,-8)[lt]{$(1)$}
\end{picture}
&+&
\begin{picture}(75,20)(0,8)
  \Photon(0,10)(25,10){3}{5}
  \PhotonArc(37.5,10)(12.5,0,180){3}{7}
  \PhotonArc(37.5,10)(12.5,180,0){3}{7}
    \Vertex(25,10){2.5}
    \Vertex(50,10){2.5}
  \Photon(50,10)(75,10){3}{5}
    \ArrowLine(37.6,22.2)(38.3,24.2)
    \ArrowLine(37.5,-2.2)(36.7,-4)
  \Text(37.5,-8)[tc]{$\wbm$}
  \Text(37.5,30)[bc]{$\wbp$}
  \Text(0,-8)[lt]{$(2)$}
\end{picture}
&+&
\begin{picture}(75,20)(0,8)
  \Photon(0,10)(25,10){3}{5}
  \PhotonArc(37.5,10)(12.5,0,180){3}{7}
  \DashCArc(37.5,10)(12.5,180,0){3.}
    \Vertex(25,10){2.5}
    \Vertex(50,10){2.5}
  \Photon(50,10)(75,10){3}{5}
    \Text(37.5,-8)[tc]{$\hb$}
    \Text(37.5,30)[bc]{$\zb$}
    \Text(0,-8)[lt]{$(3)$}
\end{picture}
\nll \nll \nll \nll \nll \nll
&+&
\begin{picture}(75,20)(0,8)
  \Photon(0,10)(75,10){3}{15}
  \PhotonArc(37,28)(12.5,17,197){3}{8}
  \PhotonArc(37,28)(12.5,197,17){3}{8}
     \Vertex(37,13.){2.5}
  \Text(36.26,50)[bc]{$\wb$}
  \Text(0,48)[lb]{$(4)$}
\end{picture}
&+&
\begin{picture}(75,20)(0,8)
  \Photon(0,10)(75,10){3}{15}
  \DashCArc(37,26)(12.5,0,180){3}
  \DashCArc(37,26)(12.5,180,0){3}
    \Vertex(37,13){2.5}
  \Text(36.26,50)[bc]{$\hb$}
  \Text(0,48)[lb]{$(5)$}       
\end{picture}
\end{array}
\]
\vspace*{-7.5mm}
\caption[Photon and $\zb$-boson self-energies and the $\zb$--$\ph$ transition]
{\it 
Photon and $\zb$-boson self-energies and the $\zb$--$\ph$ transition
\label{ssssez}}
\end{figure}
\vspace*{-3mm}

In the unitary gauge both $\Sigma^{\bos}_{\ph\ph}$ and 
$\Sigma^{\bos}_{\zg}$ may be expressed in terms of one function,
$\Pgg^{\bos}$, the bosonic component of photonic vacuum polarization:
\bqa
\Pgg^{\bos}(\sman)&=&
\Rw \Biggl\{ \Biggl[ 4+\frac{17}{3}\frac{1}{\Rw}
    -\frac{4}{3}\frac{1}{\Rws}-\frac{1}{12} \frac{1}{\Rwc}
              \Biggr] 
\bff{0}{-\sman}{\mwl}{\mwl}
\nll &&
+\lpar 4 - \frac{4}{3}\frac{1}{\Rw}
         - \frac{1}{6}\frac{1}{\Rws}\rpar 
 \lrbr \frac{\aff{0}{\mwl}}{\mws} + 1 \rrbr
+\frac{1}{18}\frac{1}{\Rws} \lpar \frac{1}{\Rw}-13 \rpar
\Biggr\}.\qquad
\label{pigg}
\eqa

Since only finite parts will contribute to resulting expressions for
physical amplitudes which should be free from ultraviolet poles 
(after renormalization, or adding counter-terms, see below),
it is convenient to split every divergent function into singular and 
finite parts:
\bqa
\Pgg^{\bos}(\sman) & = &
    \lpar 
   - 7+\frac{7}{6 } \frac{1}{\Rw }     
   + \frac{1}{12} \frac{1}{\Rws} 
         \rpar \pole
        + \Pgg^{{\bos},F}(\sman),
\label{pigg_pf}
\\   
\label{pigg_pf1}
\Pgg^{{\bos},F}(\sman)&=& \lpar
        4\Rw +\frac{17}{3} 
        - \frac{4}{3 }\frac{1}{\Rw }
        - \frac{1}{12}\frac{1}{\Rws}
                 \rpar 
        \fbff{0}{-\sman}{\wml}{\wml}
        + \frac{1}{18}\frac{1}{\Rw}
   \lpar  \frac{1}{\Rw} - 13\rpar.
\nonumber
\\
\eqa
With the $\zb$ boson self-energy, $\Sigma_{\sss{\zb\zb}}$, one constructs
a useful ratio:
\bqa
\Dz{}{\sman}&=&\frac{1}{\cows}
            \frac{\Sigma^{}_{\sss{\zb\zb}}\lpar s\rpar
                 -\Sigma^{}_{\sss{\zb\zb}}\lpar \mzs\rpar}{\mzs-\sman}\,,
\label{usefulratio}
\eqa
which also has bosonic and fermionic parts, which in turn may be 
subdivided into pole and finite parts. 
The bosonic component is:
\bqa
\Dz{\bos}{\sman}&=&
-\Biggl[
\frac{1}{6\cows}+\frac{7}{6}-7\cows
 +\lpar\frac{1}{12\cows}+\frac{7}{6}\rpar\frac{1}{\Rz}
 +\frac{1}{12\cows}\frac{1}{\Rzs}
 \Biggr]
\lpar\dlt-\ln \frac{\mws}{\tHss}\rpar
\nll &&
+\Dz{{\bos},F}{\sman},
\\
\Dz{{\bos},F}{\sman}&=&
 - \Biggl[ 
  4\cowf+\lpar\frac{1}{12\cows}+\frac{4}{3}\rpar\frac{1}{\Rz}
  +\frac{1}{12\cows}\frac{1}{\Rzs}
  \Biggr]
\fbff{0}{-\sman}{\mwl}{\mwl}
\nll &&
+\lpar\frac{1}{12\cows}+\frac{4}{3}-\frac{17}{3}\cows-4\cowf\rpar
\nll &&\times
 \frac{s\fbff{0}{-\sman}{\mwl}{\mwl}-\mzs\fbff{0}{-\mzs}{\mwl}{\mwl}}{\mzs-s}
\nll &&
+\frac{1}{\cows}\lpar 1-\frac{1}{3}\rhz+\frac{1}{12}\rhzs\rpar
\frac{\mzs\Bigl[\fbff{0}{-\sman}{\mhl}{\mzl}-\fbff{0}{-\mzs}{\mhl}{\mzl}\Bigr]}
{\mzs-s}
\nll &&
 -\frac{1}{12\cows}\lpar 1-\rhz\rpar^2 \Rz 
\Bigl[
-\fbff{0}{-\sman}{\mhl}{\mzl}+\ln\cows+1
\Bigr]
\nll &&
-\frac{1}{12\cows}
\Biggl[
\fbff{0}{-\sman}{\mhl}{\mzl}-\rhz\lpar 1+\rhz\rpar\frac{1}{\Rz}\ln\rhz
\Biggr]
\nll &&
 -\frac{1}{18}
\Biggl[
 \frac{1}{\cows}
+\lpar\frac{1}{\cows}-13\rpar \lpar 1 + \frac{1}{\Rz} \rpar
+\frac{1}{\cows}\frac{1}{\Rzs}
\Biggr].        
\label{Dz_bF}
\eqa
\vspace{2mm}

\begin{itemize}
 \item{Fermionic components of the $\zb$ and $\ph$ bosonic self-energies and
of the $\zb$--$\ph$ transition.}
\end{itemize}
The fermionic components are represented as sums over all fermions of
the theory, $\sum_f$. They, of course, depend on vector and axial couplings
of fermions to the $\zb$ boson and photon,
$v_f$, and $a_f$, and $Q_f$, correspondingly, 
and color factor $c_f$ and fermion mass $\mfl$.
The couplings defined here deviate from \eqn{kg1} to \eqn{kg3}:
\ba
v_f&=&I^{(3)}_f-2Q_f\siws,
\\
a_f&=&I^{(3)}_f,
\ea
with weak isospin $I^{(3)}_f$, and
\ba
Q_f&=&-1\quad\mbox{for leptons},\quad+\frac{2}{3}\quad\mbox{for up-},
\quad-\frac{1}{3}\quad\mbox{for down-quarks},
\\
c_f&=&1\quad\mbox{for leptons},\quad 3\quad\mbox{for quarks}.
\ea

The three main self-energy functions are 
\bqa
\Sigma^{\fer}_{\sss{\zb\zb}}(\sman) &=& 
\sum_f c_f \Biggl[
-\lpar v^2_f+a^2_f \rpar\sman  
\bff{f}{-\sman}{\mfl}{\mfl}
-2a^2_f\mf\bff{0}{-\sman}{\mfl}{\mfl}
           \Biggr],
\\
\Sigma^{\fer}_{\ph\ph}(\sman)   &=& -\sman \Pgg^{\fer}(\sman),
\\
\Sigma^{\fer}_{\zg}(\sman)&=& -\sman \Pzg^{\fer}\lpar\sman\rpar. 
\label{3_fer_se}
\eqa
However, $\Pgg^{\fer}$ and $\Pzg^{\fer}$ are now different owing to
different couplings, but still proportional to one function $\sbff{f}$:
\bqa
\Pgg^{\fer}(\sman) &=& 4 \sum_f c_f  Q^2_f  \bff{f}{-\sman}{\mfl}{\mfl},
\label{pigg_fer-old}
\\
\Pzg^{\fer}(\sman) &=&
 \sum_f c_f \lpar |Q_f|-4\stws Q^2_f \rpar \bff{f}{-\sman}{\mfl}{\mfl}.
\label{pizg_fer}
\eqa
As usual, we split them into singular and finite parts:
\bqa
\label{fermionicsplit}
\Pzg^{\fer}(\sman) &=&
  -  \frac{1}{3} 
     \Biggl(\frac{1}{2}\Nf-4\siws\asums{\ff}\cf\qfs\Biggr)
     \lpar\dlt-\ln\frac{\mws}{\tHss} \rpar
  +  \Pzg^{{\fer},F}(\sman),
\\
\Sigma^{\fer}_{\sss{\zb\zb}}(\sman)&=&
\Biggl[-\frac{1}{2}\asums{\ff}\cf\mfs + \frac{\sman}{3}
\Biggl(\frac{1}{2}\Nf-\siws\Nf+4\siwf\asums{\ff}\cf\qfs\Biggr)\Biggr]
\lpar \dlt - \ln\frac{\mws}{\tHss} \rpar
 + \Sigma^{{\fer},F}_{\sss{\zb\zb}}(\sman).
\nonumber
\eqa
In \eqn{fermionicsplit}, $\Nf=24$ is the total number of fermions in the SM.
We do not show explicit expressions for finite parts, marked with superscript 
$F$ because these might be trivially derived from 
\eqns{3_fer_se}{pizg_fer} by replacing complete 
expressions for $\sbff{f}$ and $\sbff{0}$ with their finite parts
$\sfbff{f}$ and $\sfbff{0}$, correspondingly.
 
\subsubsection{$\wb$ boson self-energy}
Now we consider the $\wb$ boson self-energy which in the unitary gauge
is described by seven diagrams, shown if \fig{sssewb}.

Firstly, we present an explicit expression for its bosonic component:
\bqa
\Sigma^{\bos}_{_{\wb\wb}}(\sman)&=&
\mws
\Biggl\{
\Biggl[
\lpar
\frac{1}{12\cowf}+\frac{2}{3}\frac{1}{\cows}
      -\frac{3}{2}+\frac{2}{3}\cows+\frac{1}{12}\cowf \rpar\Rw 
\nll && 
+\frac{2}{3}
  \lpar\frac{1}{\cows}-4-4\cows+\cowf \rpar
- \lpar\frac{3}{2}+\frac{8}{3}\cows+\frac{3}{2}\cowf \rpar
\frac{1}{\Rw}
\nll &&
+\frac{2}{3}
\cows\lpar1+\cows \rpar
\frac{1}{\Rws}
+\frac{1}{12} \cowf \frac{1}{\Rwc}
\Biggr]
\bff{0}{-\sman}{\mzl}{\wml}
\nll &&
-\frac{\siws}{6}
\lpar -5\Rw+17+17\frac{1}{\Rw}-5\frac{1}{\Rws} \rpar
\bff{0}{-\sman}{0}{\wml}
\nll &&
-\frac{1}{12} 
\Biggl[ -\lpar 1-\rw\rpar^2\Rw-10+2\rw-\frac{1}{\Rw} \Biggr]
\bff{0}{-\sman}{\mhl}{\wml}
\nll &&
-\frac{1}{12} \Biggl\{
 \lpar\frac{1}{\cows} - 2 
    +\cows-\cowf-\frac{1}{\Rw} \rpar\Rw - 24 + 2 \cows - \cowf
\nll &&
+\left[ 10-\frac{\cows}{\Rw}
 \lpar 1+\cows-\frac{\cows}{\Rw} \rpar \right]  
\frac{\aff{0}{\wml}}{\mws}
\nll &&
+\Biggl[
\lpar\frac{1}{\cows} + 9 - 9 \cows - \cowf  \rpar\Rw 
+ 1 + 14 \cows + 9 \cowf
\nll &&
+  \cows\Rw \biggl[ \lpar 1 - 9 \cows \rpar 
-  \cows\Rw \biggr] \Biggr]
   \frac{\aff{0}{\mzl}}{\mws}                           
-  \lpar \Rw + \frac{1}{\rw } + 2 \rpar
\frac{\aff{0}{\mhl}}{\mws}
\Biggr\}
\nll &&
-\frac{1}{6}\lpar\frac{1}{\cows} + 22 + \cows + \cowf +\rw  \rpar
+\frac{1}{9}  \Biggl[
 \lpar 6 + 3 \cows + \frac{7}{2}\cowf \rpar \Rw
\nll && 
+ \lpar 1 + \frac{3}{2}\cows + \frac{5}{2} \cowf \rpar 
                              - \frac{\cowf}{2} \frac{1}{\Rwc}
              \Biggr] \Biggr\}.
\label{sig_ww}
\eqa
Secondly, we give its fermionic component:
\bqa
\label{sig_wwf}
\Sigma^{\fer}_{_{\wb\wb}}(\sman)&=&
    -\sman\asum{f}{d}{} c_f\bff{f}{-\sman}{\mfpl}{\mfl} 
         +\asum{f}{d}{} c_f\minds{\ff}\bff{1}{-\sman}{\mfpl}{\mfl},
\eqa
where summation extends on all {\em doublets} of the SM.

\begin{figure}[thbp]
\[
 \begin{array}{cccccccc}
\begin{picture}(75,20)(0,8)
\SetScale{2.}
  \Photon(0,5)(13.5,5){1.5}{5}
  \Photon(25,5)(37,5){1.5}{5}
\SetScale{1.}
  \GCirc(37.5,10){12.5}{0.5}
  \ArrowLine(10,9.6)(10.55,11.6)
  \ArrowLine(65,9.6)(65.55,11.6)
\Text(12.5,18)[bc]{$\wbp$}
\Text(12.5, 2)[tc]{$\mu$}
\Text(62.5,18)[bc]{$\wbm$}
\Text(62.5, 2)[tc]{$\nu$}
\end{picture}
&=&
\begin{picture}(75,20)(0,8)
  \Photon(0,10)(25,10){3}{5}
  \ArrowArcn(37.5,10)(12.5,0,180)
  \ArrowArcn(37.5,10)(12.5,180,0)
  \Vertex(25,10){2.5}
  \Vertex(50,10){2.5}
  \Photon(50,10)(75,10){3}{5}
  \ArrowLine(10,9.6)(10.55,11.6)
  \ArrowLine(65,9.6)(65.55,11.6)
\Text(37.5,-8)[tc]{$\fbd$}
\Text(37.5,30)[bc]{$u$}
\Text(0,-8)[lt]{$(1)$}
\end{picture}
&&  && \nll \nll \nll \nll 
&+&
\begin{picture}(75,20)(0,8)
  \Photon(0,10)(25,10){3}{5}
  \PhotonArc(37.5,10)(12.5,0,180){3}{7}
  \PhotonArc(37.5,10)(12.5,180,0){3}{7}
  \Vertex(25,10){2.5}
  \Vertex(50,10){2.5}
 \Photon(50,10)(75,10){3}{5}
 \ArrowLine(10,9.6)(10.55,11.6)
 \ArrowLine(65,9.6)(65.55,11.6)
 \ArrowLine(37.6,22.2)(38.3,24.2)
\Text(37.5,-8)[tc]{$\zb$}
\Text(37.5,30)[bc]{$\wbp$}
\Text(0,38)[lt]{$(2)$}
\end{picture}
&+&
\begin{picture}(75,20)(0,8)
  \Photon(0,10)(25,10){3}{5}
  \PhotonArc(37.5,10)(12.5,0,180){3}{7}
  \PhotonArc(37.5,10)(12.5,180,0){3}{13}
  \Vertex(25,10){2.5}
  \Vertex(50,10){2.5}
 \Photon(50,10)(75,10){3}{5}
 \ArrowLine(10,9.6)(10.55,11.6)
 \ArrowLine(65,9.6)(65.55,11.6)
 \ArrowLine(37.6,22.2)(38.3,24.2)
\Text(37.5,-8)[tc]{$\gamma$}
\Text(37.5,30)[bc]{$\wbp$}
\Text(0,38)[lt]{$(3)$}
\end{picture} 
&+&
\begin{picture}(75,20)(0,8)
  \Photon(0,10)(25,10){3}{5}
  \PhotonArc(37.5,10)(12.5,0,180){3}{7}
  \DashCArc(37.5,10)(12.5,180,0){3.}
  \Vertex(25,10){2.5}
  \Vertex(50,10){2.5}
 \Photon(50,10)(75,10){3}{5}
 \ArrowLine(10,9.6)(10.55,11.6)
 \ArrowLine(65,9.6)(65.55,11.6)
 \ArrowLine(37.6,22.2)(38.3,24.2)
\Text(37.5,-8)[tc]{$\hb$}
\Text(37.5,30)[bc]{$\wbp$}
\Text(0,38)[lt]{$(4)$}
\end{picture} 
\nll \nll \nll \nll \nll 
    &+&
\begin{picture}(75,20)(0,8)
  \Photon(0,10)(75,10){3}{15}
  \PhotonArc(37,28)(12.5,17,197){3}{8}
  \PhotonArc(37,28)(12.5,197,17){3}{8}
 \Vertex(37,13.){2.5}
 \ArrowLine(10,9.6)(10.55,11.6)
 \ArrowLine(65,9.6)(65.55,11.6)
\Text(36.26,50)[bc]{$\wb$}
\Text(0,48)[lb]{$(5)$}
\end{picture}
&+&
\begin{picture}(75,20)(0,8)
  \Photon(0,10)(75,10){3}{15}
  \PhotonArc(37,28)(12.5,17,197){3}{8}
  \PhotonArc(37,28)(12.5,197,17){3}{8}
 \Vertex(37,13.){2.5}
 \ArrowLine(10,9.6)(10.55,11.6)
 \ArrowLine(65,9.6)(65.55,11.6)
\Text(36.26,50)[bc]{$\zb$}
\Text(0,48)[lb]{$(6)$}
\end{picture}
&+&
\begin{picture}(75,20)(0,8)
  \Photon(0,10)(75,10){3}{15}
  \PhotonArc(37,28)(12.5,3,183){3}{13}
  \PhotonArc(37,28)(12.5,183,3){3}{13}
 \Vertex(37,13.){2.5}
 \ArrowLine(10,9.6)(10.55,11.6)
 \ArrowLine(65,9.6)(65.55,11.6)
\Text(36.26,50)[bc]{$\gamma$}
\Text(0,48)[lb]{$(7)$}
\end{picture}
\end{array} 
\]
\caption[The $\wb$-boson self-energy]
{\it
The $\wb$-boson self-energy
\label{sssewb}}
\end{figure}

\subsubsection{Bosonic self-energies and counter-terms\label{bosonic_ct}}
Bosonic self-energies and transitions enter one-loop amplitudes either
directly via functions $\Dz{}{\sman}$, \eqn{usefulratio}, or by means 
of bosonic counter-terms, which are made of self-energy functions at
zero argument, owing to {\em electric charge renormalization} or  
at $\pmoms=-\Minds{}$; that is, on-a-mass-shell, owing to {\em 
on-mass-shell renormalization}, (OMS scheme).

\newpage

\begin{itemize}
\item Electric charge renormalization.
\end{itemize}

The electric charge renormalization introduces the quantity $\Pgg(0)$
with bosonic and fermionic components:
\bqa
\lpar z_{\ph}-1 \rpar &=&  \Pgg(0),\nll        
\Pgg^{\bos}(0) &=& 7 \lpar \dlt-\Lnw \rpar + \frac{2}{3}\,,  
\nll
\Pgg^{\fer}(0) &=&-\frac{4}{3}\Trqf\lpar\dlt-\Lnw \rpar+\zaOfer.
\label{elcharen}
\eqa
\begin{itemize}
\item $\zb$ boson wave-function renormalization.
\end{itemize}

In the treatment of $\zb$ decay one deals with the renormalization of
an external $\zb$ boson line, which involves the on-shell derivative:
\ba
\lpar z_{_{\zb}} - 1 \rpar 
&=& 
\frac{1}{\ctws}
              \frac{\pd{\;} \Sigma^{}_{\sss{\zb\zb}}(\pmoms)} 
             {\pd{\;} \pmoms } \Bigg|_{\pmoms= - \mzs}
\\
&\equiv&
\frac{1}{\ctws}\Sigma^{'}_{\sss{\zb\zb}}(\mzs),                          
\ea 
The bosonic components are:
\bqa
\frac{1}{\cows}
\biggl[\Sigma^{'}_{\sss{\zb\zb}}(\mzs) \biggr]^{\bos} 
&=&
                 \left[-\frac{1}{ 3 \ctws}
                 -\frac{7}{3}+7 \ctws\right] \lpar\dlt-\Lnw \rpar 
+\frac{1}{\cows}
\biggl[\Sigma^{'}_{\sss{\zb\zb}}(\mzs)\biggr]^{{\bos},F}, 
\nll \\
\frac{1}{\cows}
\biggl[\Sigma^{'}_{\sss{\zb\zb}}(\mzs)\biggr]^{{\bos},F}
&=& 
-\lpar\frac{1}{4 \cows}+\frac{8}{3}-\frac{17}{3} \cows \rpar
                 \fbff{0}{-\mzs}{\mwl}{\mwl} 
\nll &&
-\lpar\frac{1}{12\cows}+\frac{4}{3\cows}
            -\frac{17}{3}-4\cows \rpar \mws
             \fbff{0p}{-\mzs}{\mwl}{\mwl}
\nll &&
-\frac{1}{6}\rw\lpar 1-\frac{1}{2}\rz \rpar \fbff{0}{-\mzs}{\mhl}{\mhl}
\nll &&
-\lpar 1-\frac{1}{3}\rz+\frac{1}{12}\rzs \rpar\frac{\mzs}{\cows}
\fbff{0p}{-\mzs}{\mhl}{\mhl}
\nll &&
  - \frac{1-\rz}{12\cows}\lpar \ln \cows + \rz\ln\rw \rpar
\nll &&
  - \frac{11}{36\cows} + \frac{13}{9} 
  + \frac{1}{6}\rw\lpar 1-\frac{1}{2}\rz\rpar.
\label{derzbosonic}
\eqa
The fermionic components are:
\bqa 
\frac{1}{\cows}
\biggl[\Sigma^{'}_{\sss{\zb\zb}}(\mzs)\biggr]^{\fer}                          
&=&  \left[\frac{1}{6} \lpar\frac{1}{\ctws}-2 \rpar N_f
                - \frac{4}{3} \frac{\stwf}{\ctws} \Trqf \right]
         \lpar \dlt - \Lnw \rpar 
\nll &&
+\frac{1}{\cows}
\Biggl[\Sigma^{'}_{\sss{\zb\zb}}(\mzs)\Biggr]^{{\fer},F}, 
\\
\frac{1}{\cows}
\biggl[\Sigma^{'}_{\sss{\zb\zb}}(\mzs)\biggr]^{{\fer},F} &=&
  \frac{1}{3 \cows} \sum_f c_f \lpar v^2_f+a^2_f \rpar
  \Biggl[
  \ln\frac{\mf}{\mws} + 2 \frac{\mf}{\mzs} - \frac{2}{3}
\nll &&
+ \lpar 1 - 2\frac{\mfs}{\mzs} + 4 \frac{\mindf{\ff}}{\mzq} \rpar   
  \frac{1}{\beta} \ln\frac{1+\beta}{1-\beta}
  \Biggr] 
\nll &&
 -\frac{1}{2} \sum_f c_f \frac{\mf}{\mws}
   \lpar 1 + 2\frac{\mf}{\mzs} \frac{1}{\beta} 
           \ln\frac{1+\beta}{1-\beta}  \rpar.
\label{derzfermionic}
\eqa                          
\begin{itemize}
\item{Renormalization constant due to $\zb$--$\ph$ mixing:}
\end{itemize}
\bq
\frac{1}{\mzs}\Sigma_{\zg}(\mzs)=\frac{1}{\mzs}\Sigma^{\bos}_{\zg}(\mzs)
                                +\frac{1}{\mzs}\Sigma^{\fer}_{\zg}(\mzs).
\eq
Its bosonic and fermionic components are
\bqa 
\frac{1}{\mzs}\Sigma^{\bos}_{\zg}(\mzs) &=&     
 \lpar \frac{1}{12\ctws} + \frac{7}{6} - 7 \ctws \rpar
 \lpar \dlt - \Lnw\rpar  - \Pzg^{{\bos},F}(\mzs),
\\
\frac{1}{\mzs}\Sigma^{\fer}_{\zg}(\mzs) &=&     
 -\frac{1}{2}
 \lpar\frac{8}{3}\stws\Trqf - \frac{1}{3} N_f \rpar
 \lpar \dlt - \Lnw \rpar    - \Pzg^{{\fer},F}(\mzs).\quad
\eqa
\begin{itemize}
\item $\rho$-parameter.
\end{itemize}

Finally, two self-energy functions enter Veltman's parameter $\Delta\rho$,
a gauge-invariant combination of self-energies, which naturally appears 
in one-loop calculations:
\bqa
\Delta \rho &=&
\frac{1}{\mws}
\lrbr \Sigma_{_{WW}}(\mws) - \Sigma_{_{ZZ}}(\mzs) \rrbr,
\label{rhodef-old}
\eqa
with individual components where we explicitly show the pole parts: 
\bqa
\Delta \rho^{\bos}
&=& \lpar-\frac{1}{ 6\ctws} - \frac{41}{6} + 7 \ctws \rpar
      \lpar \dlt - \Lnw \rpar + \Delta \rho^{{\bos},F},
\\
\Delta \rho^{\fer} &=&
   \Biggl[ \frac{1}{6} N_f-\frac{1}{6}
          \lpar2-\frac{1}{\ctws}\rpar N_f
   -\frac{4}{3} \frac{\stws}{\ctws} \Trqf \Biggr]
         \lpar \dlt- \Lnw \rpar + \Delta \rho^{{\fer},F}.\qquad
\eqa
The finite parts are not shown since they 
are trivially derived from the defining equation 
\eqn{rhodef} replacing the total self-energies with their finite parts.
Having all building blocks made of bosonic self-energies at our disposal,
we continue with {\bf fermionic} self-energy diagrams.

\subsection{Fermionic self-energies}
\subsubsection{Fermionic self-energy diagrams}
In the unitary gauge there are only four diagrams contributing to the
total self-energy function of a fermion, see \fig{fsesm}.

\begin{figure}[th]
\[
\begin{array}{ccccccc}
\begin{picture}(75,20)(0,7.5)
\SetScale{2.}
  \ArrowLine(0,5)(13.5,5)
  \ArrowLine(25,5)(37.5,5)
\SetScale{1.}
\Text(12.5,14)[bc]{$\ff$}
\Text(62.5,14)[bc]{$\ff$}
\GCirc(37.5,10){12.5}{0.5}
\end{picture}
&=&
\begin{picture}(75,20)(0,7.5)
\Text(12.5,13)[bc]{$\ff$}
\Text(62.5,13)[bc]{$\ff$}
\Text(37.5,26)[bc]{$\ff$}
\Text(37.5,-8)[tc]{$\gamma$}
\Text(0,-7)[lt]{$(1)$}
  \ArrowLine(0,10)(25,10)
  \ArrowArcn(37.5,10)(12.5,180,0)
  \PhotonArc(37.5,10)(12.5,180,0){3}{15}
  \Vertex(25,10){2.5}
  \Vertex(50,10){2.5}
  \ArrowLine(50,10)(75,10)
\end{picture}
&+&
\begin{picture}(75,20)(0,7.5)
\Text(12.5,13)[bc]{$\ff$}
\Text(62.5,13)[bc]{$\ff$}
\Text(37.5,26)[bc]{$\ff$}
\Text(37.5,-8)[tc]{$\zb$}
\Text(0,-7)[lt]{$(2)$}
  \ArrowLine(0,10)(25,10)
  \ArrowArcn(37.5,10)(12.5,180,0)
  \PhotonArc(37.5,10)(12.5,180,0){3}{9}
  \Vertex(25,10){2.5}
  \Vertex(50,10){2.5}
  \ArrowLine(50,10)(75,10)
\end{picture}
\nll \nll \nll \nll
&+&
\begin{picture}(75,20)(0,7.5)
\Text(12.5,13)[bc]{$\ff$}
\Text(62.5,13)[bc]{$\ff$}
\Text(37.5,26)[bc]{$\ffp$}
\Text(37.5,-8)[tc]{$\wb$}
\Text(0,-7)[lt]{$(3)$}
  \ArrowLine(0,10)(25,10)
  \ArrowArcn(37.5,10)(12.5,180,0)
  \PhotonArc(37.5,10)(12.5,180,0){3}{7}
  \Vertex(25,10){2.5}
  \Vertex(50,10){2.5}
  \ArrowLine(50,10)(75,10)
\end{picture}
&+&
\begin{picture}(75,20)(0,7.5)
\Text(12.5,13)[bc]{$\ff$}
\Text(62.5,13)[bc]{$\ff$}
\Text(37.5,26)[bc]{$\ff$}
\Text(37.5,-8)[tc]{$\hb$}
\Text(0,-7)[lt]{$(4)$}
  \ArrowLine(0,10)(25,10)
  \ArrowArcn(37.5,10)(12.5,180,0)
  \DashCArc(37.5,10)(12.5,180,0){3}
  \Vertex(25,10){2.5}
  \Vertex(50,10){2.5}
  \ArrowLine(50,10)(75,10)
\end{picture}
\end{array}
\]
\vspace*{3mm}
\caption[Fermion self-energies]
{\it
Fermion self-energies 
\label{fsesm}}
\end{figure}
\vspace*{3mm}

In \fig{fsesm} and in what follows, 
$\ffp$ denotes the weak isospin partner of fermion $\ff$.

Each self-energy diagram contains a $B$-boson line and will be
denoted as $\Sigma_{_B}({\sla{\pmom}})$,
\bqa
\Sigma_{_B}(\sla{\pmom}) = \lpar 2 \pi \rpar^4 
\ib \frac{g^2}{16\,\pi^2} A{_{_B}},
\eqa
with
\ba
\sla{\pmom}=\pmomi{\alpha} \gamma_{\alpha}.
\ea

The first three $A{_{_{B}}}$-functions in the unitary gauge read 
\bqa
\label{ferseung}
 A{_{\ph}}\lpar{-\sman};{\mfl},{0}\rpar
 &=& \stws\,\qfs 
\Biggl\{
\ib\sla{\pmom} 
\Biggl[ 2 \bff{1}{-\sman}{\mfl}{0} + 1 \Biggr] 
 -2\mf 
\Biggl[ 2 \bff{0}{-\sman}{\mfl}{0} - 1 \Biggr]
\Biggr\},
\nl
\\
A_{_{Z}} \lpar{-\sman};{\mzl},{\mfl}\rpar
&=& -\frac{1}{4\ctws}
\Biggl\{
\ib\sla{\pmom}\lpar\vpa{\ff}{(2)}+2\vc{\ff}\ac{\ff}\gfd \rpar
\nll &&
\times \lrbr \frac{-\sman + \mfs}{\mzs}
\bff{1}{-\sman}{\mzl}{\mfl}
+A\lpar{-\sman};{\mzl},{\mfl} \rpar
\rrbr
\nll &&
+\mf \vma{\ff}{(2)} 
\Biggl[3\bff{0}{-\sman}{\mzl}{\mfl}
       +\frac{\aff{0}{\mfl}}{\mzs}-2\Biggr]
\Biggr\},
\\ \nll
A_{_{W}} \lpar{-\sman};{\wml},{\mfl}\rpar
&=& -\frac{1}{4} i \sla{\pmom} \lpar 1+\gfd \rpar
\Biggl\{
\frac{-\sman+\minds{f^{'}}}{\mws} 
\bff{1}{-\sman}{\wml}{\mfl}
+ A\lpar {-\sman};\wml,{\mfl} \rpar
\Biggr\},
\nonumber
\eqa
while the fourth function, $A_{_{H}}$, vanishes when masses of external 
fermions are ignored\footnote{Beware
of two different definitions of $\vc{\ff}$ and $\ac{\ff}$ 
used in this description. 
We use here
the definitions 
\eqn{ratio}.}

In \eqn{ferseung} one uses an auxiliary function
\bqa
A\lpar -\sman;\Mind{},\mind{}\rpar =
 2\bff{1}{-\sman}{\Mind{}}{\mind{}}
 +\bff{0}{-\sman}{\Mind{}}{\mind{}}
 +\frac{\aff{0}{\mind{}}}{\Minds{}}-1,
\eqa       
and new short hand notations
\bqa
\vpa{\ff}{(2)}&=&\vcs{\ff}+\acs{\ff},
\\
\vma{\ff}{(2)}&=&\vcs{\ff}-\acs{\ff}.
\eqa
\subsubsection{Fermionic renormalization constants}
As can be seen from the previous subsection, in the Standard Model 
the contribution of any fermionic self-energy diagram
with a virtual boson $B$, \fig{fsesmc}(a),
can be parameterized with three coefficients $a_1,\;a_3$ and $a_4$
($a_2$ in the term $a_{2}\gfd$ is always equal to zero in the SM): 

\vspace*{-2mm}

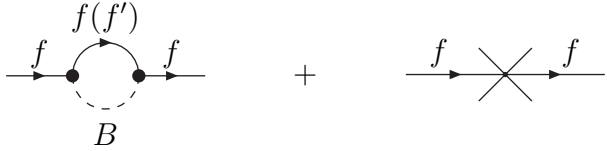
\begin{figure}[thbp]
\[
\begin{array}{ccc}
\begin{picture}(75,20)(0,7.5)
 \Text(12.5,13)[bc]{$\ff      $}
 \Text(62.5,13)[bc]{$\ff      $} 
 \Text(37.5,26)[bc]{$\ff(\ffp)$}
 \Text(37.5,-8)[tc]{$    B    $}
  \ArrowLine(0,10)(25,10)
  \ArrowArcn(37.5,10)(12.5,180,0)
  \DashCArc(37.5,10)(12.5,180,0){3}
  \Vertex(25,10){2.5}
  \Vertex(50,10){2.5}
  \ArrowLine(50,10)(75,10)
\end{picture}
\qquad
&+&
\qquad
\begin{picture}(75,20)(0,7)
  \Text(12.5,13)[bc]{$\ff$}
  \Text(62.5,13)[bc]{$\ff$}
  \ArrowLine(0,10)(37.5,10)
  \ArrowLine(37.5,10)(75,10)
  \Line(27.5,0)(47.5,20)
  \Line(27.5,20)(47.5,0)
  \Vertex(37.5,10){1}
\SetScale{1.}
\end{picture}
\end{array}
\]
\caption[Generic self-energy and counter-term diagrams]
{\it
Generic self-energy (a) and counter-term (b) diagrams
\label{fsesmc}}
\end{figure}
\vspace*{-7.5mm}

\bqa
\Sigma_f\lpar \ib \sla{\pmom} \rpar = 
        \lpar 2\pi \rpar^4 \ib
        \Bigl[ a_1+a_3 \ib \sla{\pmom} + a_4 \ib \sla{\pmom}\gfd \Bigr].
\eqa
Kinetic and mass terms of a counter-term Lagrangian may be depicted as
shown in \fig{fsesmc}(b).

From the requirement that the sum of pairs of these diagrams vanishes
on the  
fermion mass shell (a basic renormalization requirement of the OMS-scheme), 
one may fix all relevant counter-terms 
(mass and field renormalization constants) in the fermionic sector of the SM.
In particular, we derive the following relations 
for the two renormalization constants that we need:
\bqa
\bigg| \sqrt{z_{_L}} \bigg|^2 - I &=& 
            a_3 - 2 \minds{} a^{'}_{3} + 2\mind{} a^{'}_{1} + a_4,
\\
\bigg| \sqrt{z_{_R}} \bigg|^2 - I &=& 
            a_3 - 2 \minds{} a^{'}_3  - a_4,
\eqa
with derivatives 
\ba
a^{'}_i = \spd a_i/\spd \pmoms|_{\pmoms = - \minds{}}.
\ea
Calculating derivatives straightforwardly and substituting the $a_i$'s, one
obtains explicit expressions for $\sqrt{z_{_{L,R}}}$. 
It is convenient
to distinguish the electromagnetic components,
\bqa 
\lpar \sqrt{z_{_{L}}} - I \rpar^{em}_f &=& 
          \lpar \sqrt{z_{_{R}}} - I \rpar^{em}_f
        = \stws Q^2_f \lpar
        - \frac{1}{2\epsb} + \frac{1}{\epsh}
        + \frac{3}{2} \ln \frac{\mf}{\tHss} - 2 \rpar,
\label{wfren_em}
\eqa
and weak components,
\bqa
\bigg|\sqrt{z_{_R}}\bigg|^2 - I &=& 
                  \frac{3}{8} \frac{1}{\cows} \vma{f}{2},
\\
\bigg|\sqrt{z_{_L}}\bigg|^2 - I &=& 
          \frac{3}{4} \rt
          \lpar \frac{1}{\epsb} -\ln\frac{\mws}{\tHss} \rpar 
         +\frac{3}{8\cows} \vpa{f}{2} + \frac{1}{2} w^F_{_W},
\label{wfren_ew}
\eqa
where we introduced the short hand notation:
\bq
\rt=\frac{\mts}{\mws}\,,
\eq
and $w^F_{_{W}}$ is a finite part:
\bqa
w^F_{_{W}} &=& 
       \frac{5}{4}\rt + 3 \lpar 1-\frac{1}{2} \rt \rpar
       \frac{1}{(1-1/\rt)^2} \Lnrt-\frac{3}{2}\frac{1}{\rt - 1}\,.
\eqa
In \eqn{wfren_em} for the first time there appeared 
an infrared pole, ${1/\epsh}$, which is explicitly distinguished
from the ultraviolet pole, ${1/\epsb}$;
the infrared pole originates from a derivative. 
In the dimensional regularization scheme one may identify
\bq
\frac{1}{\epsh}=-\frac{1}{\epsb}\;.
\eq
However, it is reasonable to distinguish them since the poles cancel
separately. 

\subsection{The $\zb\ff\fbf$  and $\ph\ff\fbf$ vertices}
The {\em total} $\ph(\zb)\ff\fbf$ vertex 
depicted by a grey circle in \fig{zavert} consists of seven individual 
vertices, see \fig{zavert}.

 Diagrams \fig{zavert}(5)-(7) do not contribute for massless external fermions.
 For the sum of all vertices in the unitary gauge and  for 
$\ph\to\ff\fbf$ and $\zb\to\ff\fbf$ transitions, we use the standard
normalization
\bq
\ib\pi^2=\tpfi\frac{1}{16\pi^2}\,,
\eq
and define
\bqa
\Vverti{\mu}{\ph}{\sman} &=&
\tpfi\frac{1}{16\pi^2}\Gverti{\mu}{ }{\sman},
\\
\Vverti{\mu}{\zb}{\sman} &=&
\tpfi\frac{1}{16\pi^2}\Zverti{\mu}{ }{\sman},
\eqa
while the individual vertices \fig{zavert}(1-4) we denote as follows
\bqa
\Gverti{\mu}{}{\sman}&=&
 \Gverti{\mu}{\rm{QED}}{\sman}+\Gverti{\mu}{\zb  }{\sman}
+\Gverti{\mu}{\wb a}{\sman}   +\Gverti{\mu}{\wb n}{\sman},
\\
\Zverti{\mu}{}{\sman}&=&
 \Zverti{\mu}{\rm{QED}}{\sman}+\Zverti{\mu}{\zb  }{\sman}
+\Zverti{\mu}{\wb a}{\sman}   +\Zverti{\mu}{\wb n}{\sman}.
\eqa

All vertices in the unitary gauge have the following structures:
\bqa
\Gverti{\mu}{\rm{QED}}{\sman}&=&\ib\gbc\qfc\siwc\gadu{\mu} 
\vvertil{}{\ph}{\sman},
\label{G_QED}
\\[1mm]
\Gverti{\mu}{\zb }{\sman}&=&\ib\gbc\qf\frac{\siw}{4\cows}\gadu{\mu} 
\Bigl[
\vma{\ff}{2}+2\vc{\ff}\ac{\ff}\gap
\Bigr]
\vvertil{}{\sss{\zb}}{\sman},
\label{G_z}
\\ 
\Gverti{\mu}{\wb a}{\sman}&=&\ib\gbc\qfp\frac{\siw}{4}\gadu{\mu}\gap 
\vverti{}{\wb a}{\sman},
\label{G_Wa}
\\ 
\Gverti{\mu}{\wb n}{\sman}&=&\ib\gbc\frac{\siw}{2}
\lpar -\tcif\rpar\gadu{\mu}\gap 
\vverti{}{\wb n}{\sman},
\label{G_Wn}
\\ 
\Zverti{\mu}{\rm{QED}}{\sman}&=&
   \ib\gbc\qfs\frac{\siws}{2\cow}\gadu{\mu} 
\Bigl[
\vma{\ff}{}+\ac{\ff}\gap
\Bigr]
\vvertil{}{\ph}{\sman},
\label{Z_QED}
\\ 
\Zverti{\mu}{\zb  }{\sman}&=&\ib\gbc
\frac{1}{8{c^3_{_W}}}
\gadu{\mu} 
\Bigl[
\vma{\ff}{3}+\lpar 3\vcs{\ff}+\acs{\ff}\rpar\ac{\ff}\gap
\Bigr] 
\vvertil{}{\sss{\zb} }{\sman},
\label{Z_z}
\\
\Zverti{\mu}{\wb a}{\sman}&=&\ib\gbc\frac{1}{4\cow}\gadu{\mu}\gap 
\lrbr\frac{\vpa{\ffp}{}}{2}\vverti{}{\wb a}{\sman}+\ac{\ffp} 
\averti{}{\wb a}{\sman}\rrbr,
\label{Z_Wa}
\\ 
\Zverti{\mu}{\wb n}{\sman}&=&\ib\gbc\frac{\cow}{2}
\lpar-\tcif\rpar\gadu{\mu}\gap 
\vverti{}{\wb n}{\sman},
\label{Z_Wn}
\eqa
with the scalar form factors presented in the next subsection.
We recall that $\ffp$ stands for the isospin partner of a fermion $\ff$,
moreover:

\begin{figure}[bthp]
\vspace{-10mm}
\[
\begin{array}{ccccccccc}
\begin{picture}(65,88)(0,41)
  \GCirc(37.5,44){12.5}{0.5}
\SetScale{2.}
    \Photon(0,22)(12.5,22){1.5}{5}
    \ArrowLine(31.25,44)(22,27)
    \ArrowLine(21.75,16.5)(31.25,0)
\SetScale{1.}
  \Text(0,47.5)[bl]{$Z,\ph$}
  \Text(45,75)[lb]{$\bar{f}$}
  \Text(45,13)[lt]{$f$}
\end{picture}
&=&
\begin{picture}(55,88)(0,41)
  \Photon(0,44)(25,44){3}{5}
  \Vertex(25,44){2.5}
  \PhotonArc(0,44)(44,-28,28){3}{15}
  \Vertex(37.5,22){2.5}
  \Vertex(37.5,66){2.5}
  \ArrowLine(50,88)(37.5,66)
  \ArrowLine(37.5,66)(25,44)
  \ArrowLine(25,44)(37.5,22)
  \ArrowLine(37.5,22)(50,0)
  \Text(33,75)[lb]{$\bar{f}$}
  \Text(21,53)[lb]{$\bar{f}$}
  \Text(49,44)[lc]{$A$}
  \Text(21,35)[lt]{$f$}
  \Text(33,13)[lt]{$f$}
  \Text(1,1)[lb]{(1)}
\end{picture}
&+&
\begin{picture}(55,88)(0,41)
  \Photon(0,44)(25,44){3}{5}
  \Vertex(25,44){2.5}
  \PhotonArc(0,44)(44,-28,28){3}{9}
  \Vertex(37.5,22){2.5}
  \Vertex(37.5,66){2.5}
  \ArrowLine(50,88)(37.5,66)
  \ArrowLine(37.5,66)(25,44)
  \ArrowLine(25,44)(37.5,22)
  \ArrowLine(37.5,22)(50,0)
  \Text(33,75)[lb]{$\bar{f}$}
  \Text(21,53)[lb]{$\bar{f}$}
  \Text(49,44)[lc]{$Z$}
  \Text(21,35)[lt]{$f$}
  \Text(33,13)[lt]{$f$}
  \Text(1,1)[lb]{(2)}
\end{picture}
&+&
\begin{picture}(55,88)(0,41)
  \Photon(0,44)(25,44){3}{5}
  \Vertex(25,44){2.5}
  \PhotonArc(0,44)(44,-28,28){3}{9}
  \Vertex(37.5,22){2.5}
  \Vertex(37.5,66){2.5}
  \ArrowLine(50,88)(37.5,66)
  \ArrowLine(37.5,66)(25,44)
  \ArrowLine(25,44)(37.5,22)
  \ArrowLine(37.5,22)(50,0)
  \Text(33,75)[lb]{$\bar{f}$}
  \Text(21,53)[lb]{$\bar{f'}$}
  \Text(47.5,44)[lc]{$W$}
  \Text(21,35)[lt]{$f'$}
  \Text(33,13)[lt]{$f$}
  \Text(1,1)[lb]{(3)}
\end{picture}
&+&
\begin{picture}(55,88)(0,41)
  \Photon(0,44)(25,44){3}{5}
  \Vertex(25,44){2.5}
  \ArrowArcn(0,44)(44,28,-28)
  \Vertex(37.5,22){2.5}
  \Vertex(37.5,66){2.5}
  \ArrowLine(50,88)(37.5,66)
  \ArrowLine(37.5,22)(50,0)
  \Photon(25,44)(37.5,66){3}{5}
  \Photon(37.5,22)(25,44){3}{5}
  \Text(33,75)[lb]{$\bar{f}$}
  \Text(15,53)[lb]{$W$}
  \Text(48,44)[lc]{$f'$}
  \Text(15,35)[lt]{$W$}
  \Text(33,13)[lt]{$f$}
  \Text(1,1)[lb]{(4)}
\end{picture}
\vspace{1cm}   
\\
 &+& 
\begin{picture}(55,88)(0,41)
  \Photon(0,44)(25,44){3}{5}
  \Vertex(25,44){2.5}
  \DashCArc(0,44)(44,-28,28){3}
  \Vertex(37.5,22){2.5}
  \Vertex(37.5,66){2.5}
  \ArrowLine(50,88)(37.5,66)
  \ArrowLine(37.5,66)(25,44)
  \ArrowLine(25,44)(37.5,22)
  \ArrowLine(37.5,22)(50,0)
  \Text(33,75)[lb]{$\bar{f}$}
  \Text(21,53)[lb]{$\bar{f}$}
  \Text(47.5,44)[lc]{$H$}
  \Text(21,35)[lt]{$f$}
  \Text(33,13)[lt]{$f$}
  \Text(1,1)[lb]{(5)}
\end{picture}
&+&
\begin{picture}(55,88)(0,41)
  \Photon(0,44)(25,44){3}{5}
  \Vertex(25,44){2.5}
  \ArrowArcn(0,44)(44,28,-28)
  \Vertex(37.5,22){2.5}
  \Vertex(37.5,66){2.5}
  \ArrowLine(50,88)(37.5,66)
  \ArrowLine(37.5,22)(50,0)
  \Photon(25,44)(37.5,66){3}{5}
  \DashLine(37.5,22)(25,44){3}
  \Text(33,75)[lb]{$\bar{f}$}
  \Text(-1,51)[lb]{$(Z)$}
  \Text(17,53)[lb]{$Z$}
  \Text(48,44)[lc]{$f$}
  \Text(17,35)[lt]{$H$}
  \Text(33,13)[lt]{$f$}
  \Text(1,1)[lb]{(6)}
\end{picture}
&+&
\begin{picture}(55,88)(0,41)
  \Photon(0,44)(25,44){3}{5}
  \Vertex(25,44){2.5}
  \ArrowArcn(0,44)(44,28,-28)
  \Vertex(37.5,22){2.5}
  \Vertex(37.5,66){2.5}
  \ArrowLine(50,88)(37.5,66)
  \ArrowLine(37.5,22)(50,0)
  \DashLine(25,44)(37.5,66){3}
  \Photon(37.5,22)(25,44){3}{5}
  \Text(33,75)[lb]{$\bar{f}$}
  \Text(-1,51)[lb]{$(Z)$}
  \Text(17,53)[lb]{$H$}
  \Text(48,44)[lc]{$f$}
  \Text(17,35)[lt]{$Z$}
  \Text(33,13)[lt]{$f$}
  \Text(1,1)[lb]{(7)}
\end{picture}
&&
\end{array}
\]
\vspace{10mm}
\caption[The $\zb\ff\fbf$ and $\ph\ff\fbf$~vertices]
{\it 
The $\zb\ff\fbf$ and $\ph\ff\fbf$~vertices. The symbol $(\zb)$ in
some graphs indicates  
that they contribute only to the $\zb$ vertex\label{zavert}}
\end{figure}

\ba
\delta_f &=& v_f-a_f =  -2\qd\stws,
\\
\sigma_f &=& v_f+a_f.
\ea
\subsubsection{Scalar form factors for the $\ph\ff\fbf$ 
and $\zb\ff\fbf$ vertices}
In QED diagrams, \eqnsc{G_QED}{Z_QED}, a common QED scalar form factor enters
\bqa
\vvertil{}{\ph}{\sman} &=&
-2\lpar-\sman+2\mfs\rpar\cff{0}{-\sman}{\mfl}{0}{\mfl}
\nll &&
+~ \bff{0}{\Trmoms}{\mfl}{\mfl}-4\bff{ff}{\Trmoms}{\mfl}{\mfl}-2,
\label{QEDscff}
\eqa
with a subtracted $\sbff{}$-function,
\bqa
\bff{ff}{\Trmoms}{\mfl}{\mfl}&=&\bff{0}{-\sman}{\mfl}{\mfl}
                               -\bff{0}{-\mfs}{\mfl}{0}.
\eqa
The $\cff{0}{-\sman}{\mfl}{0}{\mfl}$ function contains an infrared divergence
(IRD), see \eqns{IRDC01}{IRDC02}.
Every scalar form factor is presented as a sum of its pole and 
finite parts:
\bqa
\vvertil{ }{\ph}{\sman} &=& \pole + \vvertil{F}{\ph}{\sman},
\\
\vvertil{F}{\ph}{\sman} &=& - 4\ln\frac{\mfs}{\mws}- 3\ln\Rw 
                           + 2\sman\cff{0}{-\sman}{\dml}{0}{\dml}. 
\eqa
The diagrams with the virtual $\zb$ boson, \eqnsc{G_z}{Z_z}, are described
by another common scalar form factor:
\bqa
\vvertil{}{\sss{\zb}}{\sman}  
    &=&  2s(1+R_{_Z})^2\cff{0}{-\sman}{0}{\zml}{0}
      +\lpar 2 \Rz - 3 \rpar \Delta\bff{}{-\sman}{0}{\zml} - 2,\quad
\eqa
with $\Delta\bff{}{-\sman}{0}{\zml}$ defined in \eqn{deltab}.
Here we have no ultraviolet pole term since the  poles from the two 
$\sbff{}$-functions in $\Delta\sbff{}$ cancel, and 
for the finite part we use a notation omitting the superscript $F$:
\bqa
\vvertil{}{\sss{\zb}}{\sman}   &=&
       - 5 +\lpar 2 \Rz - 3 \rpar \ln\Rz + 2\Rz
       + 2\sman\lpar 1 + \Rz\rpar^2\cff{0}{-\sman}{0}{\zml}{0}.
\eqa

Next, one has {\em abelian} diagrams with virtual $\wb$ bosons,
\eqnsc{G_Wa}{Z_Wa}. Again, there is a common scalar function:
\bqa
\vverti{}{\wb a}{\sman} &=&
\Biggl[( 2 +  \rt) ( 1 - \rt)^2 \Rw
       + \lpar 2 - \rt \rpar^2 + \frac{2}{\Rw}
\Biggr]\mws\cff{0}{-\sman}{\uml}{\wml}{\uml}
\nll &&
        -~\frac{1}{2} \rt \bff{0}{-\sman}{\uml}{\uml}
        -\biggl[\lpar 2 + \rt \rpar
                \lpar 1 - \rt \rpar \Rw - 3 + \rt 
         \biggr]\Delta\bff{}{-\sman}{\uml}{\wml}
\nll && 
      +~ \frac{\aff{0}{\uml}}{\mws} - \frac{1}{2}\rt - 2,
\eqa
which is subdivided into pole and finite parts:
\bqa
\vverti{}{\wb a}{\sman} &=&
   - \frac{3}{2} \rt \pole
   + \vverti{F}{\wb a}{\sman},
\\
   \vverti{F}{\wb a}{\sman}
&=&
\Biggl[( 2 +  \rt) ( 1 - \rt)^2 \Rw
       + \lpar 2-\rt\rpar^2 + \frac{2}{\Rw}
\Biggr]\mws\cff{0}{-\sman}{\uml}{\wml}{\uml}
\nll &&       
         + \lrbr 3- \frac{1}{2}\rt +
   \lpar 2 - \rt - \rts \rpar\Rw \rrbr
\fbff{0}{-\sman}{\uml}{\uml}
\\ &&
 +\biggl[ \lpar \rt + 2 \rpar\rt\Rw 
          + \lpar 2+\frac{2}{\rt-1}\rpar\rt -2\biggr] \ln \rt        
          + 1 - \frac{5}{2}\rt -\lpar \rts + \rt - 2 \rpar \Rw.
\nonumber
\eqa
However, the abelian diagram, \fig{zavert}(3), with incoming $\zb$ boson line,
\eqn{Z_Wa}, has an additional contribution: 
\bqa
\averti{}{\wb a}{\sman} 
&=& \rt \Biggl\{
        \left[ -(1-\rt)^2 \Rw - 2 \right]
        \mws\cff{0}{-\sman}{\uml}{\wml}{\uml}
\nll &&
     -~ \frac{1}{2}\bff{0}{-\sman}{\uml}{\uml}
+ (1-\rt)\Rw\Delta\bff{ }{-\sman}{\uml}{\wml}
        + \frac{1}{2}\Biggr\}.
\eqa
Its pole and finite parts of which vanish in the massless approximation, 
$\rt\to 0$:
\bqa
\averti{}{\wb a}{\sman} &=& - \frac{1}{2} \rt \pole 
+\averti{F}{\wb a}{\sman},
\\
\averti{F}{\wb a}{\sman}&=&
\rt 
\Biggl\{
        -  \Bigl[\lpar 1-\rt \rpar^2\Rw + 2 \Bigr]
            \mws \cff{0}{-\sman}{\uml}{\wml}{\uml} 
\nll &&
+\lrbr\lpar \rt-1\rpar\Rw+\frac{1}{2}\rpar\fbff{0}{-\sman}{\uml}{\uml} 
-\lrbr\rt\lpar\Lnrt-1\rpar+1\rrbr\Rw  + \frac{1}{2}
\Biggr\}.\qquad\quad
\eqa

Finally, there is a common scalar function entering
\eqnsc{G_Wa}{Z_Wn}, which corresponds to a {\em non-abelian} diagram
with two virtual $\wb$ bosons, \fig{zavert}(4):
\bqa
\vverti{}{\wb n}{\sman} &=&      
\Biggl[
     (2+\rt) (1-\rt)^2 \Rw
     +4-\frac{5}{2} \rt + 2 \rts - \frac{1}{2} \rtc
\nll &&
     +~\rt \lpar 2-\frac{1}{2} \rt \rpar\frac{1}{\Rw} 
\Biggr] \mws \cff{0}{-\sman}{\wml}{\uml}{\wml}
\nll &&
   -~\Biggl[\frac{2}{3}-\frac{1}{2} \rt
   +\lpar  \frac{3}{2}-\frac{1}{4} \rt \rpar \frac{1}{\Rw}
   +       \frac{1}{12 \Rws} \Biggr]\bff{0}{-\sman}{\wml}{\wml}
\nll &&
   +~\left[ (2 + \rt)(1-\rt)\Rw + 3 - \frac{3}{2} \rt
              + \frac{1}{2} \rts \right] 
\Delta\bff{}{-\sman}{\wml}{\uml}
\nll &&
  -~\frac{1}{3}\lpar 2 - \frac{1}{2 \Rw} \rpar
   \frac{\aff{0}{\wml}}{\mws}
  -\frac{\aff{0}{\uml}}{\mws}
\nll &&
     -~\frac{2}{3}-\frac{1}{2} \rt
     +\lpar\frac{4}{9}+\frac{1}{4} \rt \rpar \frac{1}{\Rw}
     -\frac{1}{ 18\Rws}
\nl
 &=&
  \left[ 
\frac{3}{2} \rt
 + \lpar \frac{4}{3}  
  -\frac{1}{4}\rt\rpar\frac{1}{\Rw} 
  +\frac{1}{12}\frac{1}{\Rws}
 \right]     
   \pole          
  +\vverti{F}{\wb n}{\sman} ,
\ea
with
\ba
\vverti{F}{\wb n}{\sman} 
&=&
\Biggl[
     (2+\rt) (1-\rt)^2 \Rw
     +4-\frac{5}{2} \rt + 2 \rts - \frac{1}{2} \rtc
     +\rt \lpar 2-\frac{1}{2} \rt \rpar\frac{1}{\Rw} 
\Biggr] \mws 
\nll &&
\times\cff{0}{-\sman}{\wml}{\uml}{\wml} 
\nll &&
-~ \Biggl[
\lpar 2- \rt- \rts \rpar\Rw
+\frac{7}{3} -\rt\lpar 1 + \frac{\rt}{2} \rpar
-\lpar \frac{3}{2} -\frac{\rt}{4}\rpar\frac{1}{\Rw}
-\frac{1}{12}\frac{1}{\Rws}\Biggr]
\nll &&
\times \fbff{0}{-\sman}{\wml}{\wml}
\nll &&
      - ~\Biggl[  
\lpar 2 + \rt \rpar \Rw + 2 - \frac{2}{\rt-1}+\frac{1}{2}\rt\Biggr]
     \rt \Lnrt        
\nll &&
+~\rt\Rw-3+2\rt-\frac{1}{2}\rts +
     \lpar \frac{11}{18} + \frac{\rt}{4} \rpar \frac{1}{\Rw}
     -\frac{1}{18}\frac{1}{\Rws}\,.
\eqa

\subsubsection{Construction of $\vvertil{\Vvert}{\gZQ}{\sman}$ 
and $\vvertil{\Vvert}{\gZL}{\sman}$: vertices and counter-terms}  
Having in mind applications to $\zb$ resonance physics, in particular
calculations of one-loop EWRC for the $\zb$ decay, it is useful to construct
off-shell vertices: the sum of all vertex diagrams, \fig{zavert}, and 
fermionic counter-terms, \eqnsc{wfren_em}{wfren_ew}. 
We will denote this sum by a black circle, see \fig{zavert1}.

The corresponding vertex functions may be parameterized by four scalar
form factors $F_{\gZL}\lpar\sman\rpar$ and $F_{\gZQ}\lpar\sman\rpar$:
\bqa
\label{fourformdefg}
V^{\ph\ff\fbf}_{\mu}\lpar\sman\rpar
 &=& \lpar 2\pi\rpar^4\ib\frac{\ib g^3\siw}{16\pi^2}\gamma_\mu
\lrbr\frac{1}{2}I^{(3)}_f
     F_{\gL}\lpar\sman\rpar\lpar 1+\gfd \rpar 
+\qd F_{\gQ}\lpar\sman\rpar\rrbr,
\\
V^{\zb\ff\fbf}_{\mu}\lpar\sman\rpar
 &=& \lpar 2\pi\rpar^4\ib\frac{\ib g^3}{16\pi^2 2\cow}\gamma_\mu
\lrbr I^{(3)}_f
 F_{\ZL}\lpar\sman\rpar \lpar 1+\gfd \rpar - 2\qd\stws
 F_{\ZQ}\lpar\sman\rpar \rrbr.
\label{fourformdefz}
\eqa
We note that the Born approximation would correspond to the replacements
\ba
\frac{g^2}{16\pi^2}F_{\gQ}&\to& 1,
\\
\frac{g^2}{16\pi^2}F_{\ZL}&\to& 1,
\\
\frac{g^2}{16\pi^2}F_{\ZQ}&\to& 1.
\label{formalreplacements}
\ea

This property justifies
the normalization of these terms. 
Since the term 
with $F_{\gL}$ has no Born analog, its normalization
is chosen in a convenient way.

We will understand that the 
second diagram of
\fig{zavert1} stands only for those counter-terms that originate from
fermionic self-energy diagrams; i.e. those of \eqns{wfren_em}{wfren_ew}.
They will be referred to as {\em fermionic} counter-terms. The rest will be 
called {\em bosonic} counter-terms. 
To distinguish the two types of counter-terms, we will use 
different drawings 
in \fig{zavert1} and \fig{zavert3}
below.

The four contributions from the fermionic counter-terms 
to the four scalar form factors above are:

\begin{figure}[th]
\vspace*{-20mm}
\[
\begin{array}{cccccccc}
& \begin{picture}(125,86)(0,40)
  \Vertex(100,43){12.5}
\SetScale{2.}
  \Photon(13,22)(50,22){1.5}{15}
  \ArrowLine(62.5,43)(50,22.5)
  \ArrowLine(50,21.5)(62.5,0)
  \Text(108,74)[lb]{$\fbf$}
  \Text(62.5,50)[bc]{$(\ph,\zb)$}
  \Text(108,12)[lt]{$\ff$}
\end{picture}
&=&
\begin{picture}(125,86)(0,40)
  \Photon(25,43)(100,43){3}{15}
  \GCirc(100,43){12.5}{0.5}
  \ArrowLine(125,86)(107,53)
  \ArrowLine(106,31)(125,0)
  \Text(108,74)[lb]{$\fbf$}
  \Text(62.5,50)[bc]{$(\ph,\zb)$}
  \Text(108,12)[lt]{$\ff$}
\end{picture}
&+&
\begin{picture}(125,86)(0,40)
  \Photon(25,43)(100,43){3}{15}
\SetScale{2.0}
  \Line(45.25,16.75)(54.75,26.25)
  \Line(45.25,26.25)(54.75,16.75)
\SetScale{1.0}
  \ArrowLine(125,86)(100,43)
  \Vertex(100,43){2.5}
  \ArrowLine(100,43)(125,0)
  \Text(108,74)[lb]{$\fbf$}
  \Text(62.5,50)[bc]{$(\ph,\zb)$}
  \Text(108,12)[lt]{$\ff$}
\end{picture}
\end{array}
\]
\vspace{10mm}
\caption[Off-shell $\zb\ff\fbf$ and $\ph\ff\fbf$ vertices]
{\it
Off-shell $\zb\ff\fbf$ and $\ph\ff\fbf$ vertices
\label{zavert1}}
\end{figure}

\bqa
F^{ct}_{\gQ} &=& 2\lpar\sqrt{z_{_R}}-I\rpar^{em} 
= \qds\stws\lpar -\frac{1}{\epsb}+\frac{2}{\epsh}+3\ln\frac{\mws}{\tHss}\rpar
+F^{ct,F}_{\gQ},
\\
F^{ct,F}_{\gQ}&=&
 \qds \stws \lpar 3\Lnfw - 4\rpar
+\frac{3}{8}\frac{1}{\ctws}\vmads,
\\ \nll
F^{ct}_{\gL}  &=& \lpar \sqrt{z_{_L}} - I \rpar
      - \lpar \sqrt{z_{_R}} - I \rpar 
      = \frac{3}{8} \rt \pole + F^{ct,F}_{\gL},
\\
F^{ct,F}_{\gL}&=&
        \frac{3}{4} \frac{\ad\vd}{\ctws}+\frac{1}{4} \wwUf,
\\ \nll
F^{ct}_{\ZQ} &=& \vmad \lpar \sqrt{z_{_R}} - I \rpar^{em}
=\qdc\stwf\lpar\frac{1}{\epsb}-\frac{2}{\epsh}-3\ln\frac{\mws}{\tHss}\rpar
+F^{ct,F}_{\ZQ},
\\
F^{ct,F}_{\ZQ}&=&
      - \frac{3}{8} \qd \frac{\stws}{\ctws} \vmads
      - \qdc \stwf \lpar 3 \Lnfw  - 4 \rpar,   
\\ \nll
F^{ct}_{\ZL}  &=& \sigma^2_f \lpar \sqrt{z_{_L}} - I \rpar
      - \delta^2_f  \lpar \sqrt{z_{_R}} - I \rpar  
\nll
      &=&
      \frac{3}{2} \qds \ad \stws 
      \lpar\frac{1}{\epsh}+\ln\frac{\mws}{\tHss}\rpar
     + \frac{3}{16} \rt \vpad  \pole
      + F^{ct,F}_{\ZL},
\\
F^{ct,F}_{\ZL}&=&
           \frac{3}{2} \qds \ad \stws
        \lpar \Lnfw - \frac{4}{3} \rpar
      + \frac{3}{16}  \frac{1}{\ctws}\ad \tvpad
      + \frac{1}{8} \vpad \wwUf.
\eqa

\begin{itemize}
\item{Off-shell vertices.}
\end{itemize}
Now we construct the four off-shell vertices, which are
derived from the next four equations, two for $\ph\ff\fbf$ vertices:
\bqa
\vvertil{\Vvert}{\gQ}{\sman}
&=&\frac{1}{\qf\siw}
\Bigl\{\Gverti{\mu}{}{\sman}[\gadu{\mu}\ib\gbc]    
+\qf\siw\fverti{ct}{\gQ}\Bigr\},
\\ 
\vvertil{\Vvert}{\gL}{\sman}&=&\frac{2}{\tcif\siw}
\Bigl\{ \Gverti{\mu}{}{\sman}[\gadu{\mu}\gap\ib\gbc]
+\qf\siw\fverti{ct}{\gL}\Bigr\};
\eqa
and two for $\zb\ff\fbf$ vertices:
\bqa
\vverti{\Vvert}{\ZQ}{\sman}&=&
\frac{2\cow}{\lpar -2\qf\siws\rpar}
\Biggl\{
\Zverti{\mu}{}{\sman}[\gadu{\mu}\ib\gbc]
+\frac{1}{\cow}\fverti{ct}{_{\ZQ}}
\Biggr\},
\\
\vverti{\Vvert}{\ZL}{\sman}&=&\frac{2\cow}{\tcif}
\Biggl\{\Zverti{\mu}{}{\sman}[\gadu{\mu}\gap\ib\gbc]
+\frac{1}{\cow}\fverti{ct}{_{\ZL}}
\Biggr\}.
\eqa
The factors $1/(\siw\qf)$ and $2/(\siw \tcif)$ 
for $\ph\ff\fbf$ vertices, and the 
factors $2\cow/\lpar -2\qf\siws\rpar$ and $2\cow/\tcif$ 
for $\zb\ff\fbf$ are due to the form factor definitions 
\eqns{fourformdefg}{fourformdefz}.

The four off-shell vertices with separated ultraviolet poles read
\bqa
F^V_{\gQ}(\sman)&=&
F^V_{\ZQ}(\sman) = 2\qds\stws\lpar\frac{1}{\epsh}+\ln\frac{\mws}{\tHss}\rpar
+ F^{V,F}_{\gQ}(\sman), 
\\
F^{V,F}_{\gQ}(\sman)&=&
          \qds \stws \left[ 3\Lnfw  + 
          \vvertil{F}{\ph}{\sman} 
           - 4 \right] 
         +\frac{\vmads}{4\ctws} 
          \left[ 
          \vvertil{F}{\sss{Z}}{\sman} 
        + \frac{3}{2} \right], 
\\ \nll
F^V_{\gL}(\sman) &=&
         - \frac{1}{4}\frac{\rt}{\Rw}\pole + F^{V,F}_{\gL}(\sman),
\\
F^{V,F}_{\gL}(\sman) &=&
         \frac{1}{\Rw} \lpar\frac{4}{3}
        +\frac{1}{12}\frac{1}{\Rw} \rpar
        +\frac{\qd}{\ctws}\vd 
            \left[
          \vvertil{F}{\sss{Z}}{\sman} 
        +\frac{3}{2}\right]  
        +\frac{1}{2}\frac{\qd}{\tcif} \wwUf
        -\qum\vverti{F}{\wb a}{\sman}-\vverti{F}{\wb n}{\sman},
\nll \\
F^V_{\ZQ}(\sman) &=&
2\qds\stws\lpar\frac{1}{\epsh}+\ln\frac{\mws}{\tHss}\rpar
+ F^{V,F}_{\ZQ}(\sman), 
\\
F^{V,F}_{\ZQ}(\sman) &=&
  \qds \stws \left[ 3\Lnfw+\vvertil{F}{\ph}{\sman}-4\right]
             +\frac{\vmads}{4\ctws} 
        \left[\vvertil{F}{\sss{Z}}{\sman}+\frac{3}{2}\right],
\\ \nll
F^V_{\ZL}(\sman) &=&
      2\qds\stws\lpar\frac{1}{\epsh}+\ln\frac{\mws}{\tHss}\rpar
      +\Biggl[\frac{\ctws}{\Rw} 
       \lpar\frac{1}{4}\rt+\frac{4}{3}+\frac{1}{12}\frac{1}{\Rw} \rpar
      + \frac{1}{4}\rt\Biggr]\pole
\nll &&
      + F^{V,F}_{\ZL}(\sman),
\\
F^{V,F}_{\ZL}(\sman) &=&
      \qds \stws \left[3 \Lnfw  
      + \vvertil{F}{\ph}{\sman} 
      - 4\right]  
      + \frac{1}{4} \frac{1}{\ctws} \tvpad
        \left[ 
        \vvertil{F}{\sss{Z}}{\sman} 
      + \frac{3}{2}\right] 
\nll &&   
      - \ctws \vverti{F}{\wb n}{\sman} 
      + \frac{1}{4\tcif}\left[2\au\averti{F}{Wa}{\sman}   
      + \vpad \wwUf+\vpau 
        \vverti{F}{\wb a}{\sman}          
        \right].
\label{offshellmassive}
\eqa
The expressions for 
these vertices are valid in case of a heavy virtual fermion (e.g. the top-quark).
They may be used to construct the particular case when the virtual fermion 
is massless. 
It will be used to calculate one-loop EWRC for the initial electron
vertex.

\begin{itemize}
\item Construction of the electron vertex.
\end{itemize}  
 The four individual off-shell electron vertices are derived from the general
case \eqn{offshellmassive} by setting:  
\ba
\minds{\ffp} = \minds{\nu_e} &=& 0, 
\\
\qf &=& \qe,
\\
Q_{\ffp} = Q_{\nu_e} &=& 0, 
\\
\sewti{_{\wb}}{F(0)} &=& \frac{3}{2},
\\
\averti{F}{\wb a}{\sman}&=&0.
\ea
They are
\bqa
F^{Ve}_{\gQ}(\sman)  &=& F^{Ve}_{\ZQ}(\sman) =
2\qes \stws\lpar\frac{1}{\epsh}+\ln\frac{\mws}{\tHss}\rpar
+ F^{Ve,F}_{\gQ}(\sman),
\\
F^{Ve,F}_{\gQ}(\sman)&=& 
        \qes \stws \left[ 3 \Lnew  
      + \vvertil{F}{\ph}{\sman} - 4 \right]
      + \frac{1}{4}\frac{\vmaes}{\ctws} 
        \lrbr 
       \vvertil{}{\sss{Z}}{\sman}
      + \frac{3}{2} \rrbr, 
\\ \nll
F^{Ve}_{\gL}(\sman)  &=&
      \frac{1}{3\Rw}\lpar 4 + \frac{1}{4}\frac{1}{\Rw} \rpar \pole
     + F^{Ve,F}_{\gL}(\sman), 
\\
F^{Ve,F}_{\gL}(\sman)&=& \frac{\qe}{\ctws} \ve
      \lrbr
      \vvertil{}{\sss{Z}}{\sman}
     -\frac{3}{2} \rrbr 
     +\frac{3}{4} \frac{\qe}{\tcie} 
     -\big|Q_{\nu_{e}} \big|
      \vverti{F(0)}{\wb a}{\sman} - \vverti{F(0)}{\wb n}{\sman},
\\ \nll
F^{Ve}_{\ZQ}(\sman)  &=& 
2\qes\stws\lpar\frac{1}{\epsh}+\ln\frac{\mws}{\tHss}\rpar
+ F^{Ve,F}_{\ZQ}(\sman),
\\
F^{Ve,F}_{\ZQ}(\sman)&=&
      \qes \stws 
      \lrbr 3 \ln \frac{\mes}{\mws} + \vvertil{F}{\ph}{\sman}  - 4\rrbr
     +\frac{\vmaes}{4\ctws}  
      \left[
      \vvertil{}{\sss{Z}}{\sman}
    + \frac{3}{2} \right], 
\\ \nll
F^{Ve}_{\ZL}(\sman)  &=&
2\qes\stws\lpar\frac{1}{\epsh}+\ln\frac{\mws}{\tHss}\rpar
+\frac{1}{3}\frac{\ctws}{\Rw}
\lpar 4 + \frac{1}{4} \frac{\ctws}{\Rw}\rpar\pole
+ F^{Ve,F}_{\ZL}(\sman),
\nll \\
F^{Ve,F}_{\ZL}(\sman)&=&
      \qes \stws \left[ 3 \ln \frac{\mes}{\mws} 
    + \vvertil{F}{\ph}{\sman} - 4 \right]
\nll &&
    + \frac{1}{4} \frac{\tvpae}{\ctws} 
      \lrbr \vvertil{}{\sss{Z}}{\sman}
    + \frac{3}{2}\rrbr
    - \ctws \vverti{F(0)}{\wb n}{\sman} + \frac{1}{4\tcie}
      \lrbr \frac{3}{2} {\vpa{e}{}} 
    + \vpa{\nu_e}{}\vverti{F(0)}{\wb a}{\sman}\rrbr.\quad
\nl
\label{offshellmassless}
\eqa
Having constructed all these vertices, we may move to the last building blocks:
electroweak boxes.
\subsection{The $\wb\wb$ box}
Here we discuss only the $\wb\wb$ box, see \fig{WWboxes}.
Only one diagram contributes for a given channel. The contribution of each 
diagram may be parameterized by only one scalar function:

\bq
WW^{d(c)}_{box}=(2\pi)^4\ib\frac{g^4}{16\pi^2}
\gadi{\mu}\gap\otimes\gadi{\mu}\gap
\boxc{{_{\wb\wb}}}{d(c)}\lpar\sman,\tman\rpar,
\eq
where $\sman$, $\tman$, and $\uman$ are the usual Mandelstamm
variables
and superscripts $d$ and $c$ denote the direct and crossed box diagrams.
In this Appendix, the Mandelstam variables are defined such that they 
satisfy the identity:
\bq
\sman+\tman+\uman=0.
\eq
\begin{itemize}
\item Direct box.
\end{itemize}
Only the direct box is the source of $m_t$-dependent terms.
The general answer for the first diagram of \fig{WWboxes}, valid for the case 
of a heavy virtual fermion, e.g. for {\em up=top}, reads:
\bqa
\boxc{{_{\wb\wb}}}{d}\lpar\sman,\tman\rpar &=&
\Biggl\{-t\lpar 1+\frac{t^2}{u^2}\rpar
        -4\frac{\mws t^2}{u^2}+2\frac{\mwq}{u} 
         \lpar 1+2\frac{\sman}{u}\rpar                           
\nll &&
+\um  \lrbr 2+3\frac{\sman}{u}+2\frac{s^2}{u^2}
    -2\frac{\mws}{u}\lpar 1+2\frac{\sman}{u}\rpar\rrbr
     +\frac{\umf\sman}{u^2} \Biggr\}                         
\dff{0}{-\sman}{-\tman}{\wml}{0}{\wml}{\uml}                
\nll &&
     -\Biggl\{2+2\frac{\sman}{u}+\frac{s^2}{u^2}-2\frac{\mws\sman}{u^2}
-\frac{1}{2}\rt\lrbr4-\Rw\lpar 1+2\frac{s^2}{u^2}\rpar\rrbr 
\nll &&
     +\frac{1}{2}\rts\lpar1-2\Rw\rpar-\frac{1}{2}\rtc\Rw\Biggr\}
      \cff{0}{-\sman}{\wml}{\uml}{\wml}                      
\nll &&
-\Biggl(2+2\frac{\sman}{u}+\frac{s^2}{u^2}-2\frac{\mws\sman}{u^2}
       +\frac{\um s}{u^2}\Biggr)C_0\lpar-s;\wml,0,\wml\rpar  
\nll &&
+\Biggl(2+3\frac{\sman}{u}+\frac{s^2}{u^2}
       +2\frac{\mws\tman}{u^2}-\frac{\mt t}{u^2}\Biggr)          
\Bigl[\cff{0}{-\sman}{0}{\wml}{\uml}
           +\cff{0}{-\sman}{\uml}{\wml}{0}\Bigr]             
\nll &&
-\frac{1}{\sman}
\Biggl[ 2\frac{\sman}{u}+\frac{5}{3}\frac{1}{\Rw}+\frac{1}{12}\frac{1}{\Rws}
        +\frac{1}{4}\rt \lpar 2-\frac{1}{\Rw}\rpar
        -\frac{1}{2}\rts\Biggr]                   
\nll &&
\times   \bff{0}{-\sman}{\wml}{\wml} +\frac{2}{u}\bff{0}{-\sman}{\uml}{0}
        +\frac{1}{2}\frac{\rt}{\sman}\frac{\aff{0}{\uml}}{\mws}  \nll &&
-\frac{1}{3\mws}\lrbr
 \frac{1}{2}\lpar 1+3\rt\Rw \rpar \frac{\aff{0}{\wml}}{\mws}
 - 1-\frac{3}{4}\rt+\frac{1}{6}\frac{1}{\Rw} \rrbr.
\label{wwdirect}
\eqa

The expression for the $\wb\wb$ box diagram with extracted pole is:
\bqa
\boxc{{_{\wb\wb}}}{d}\lpar\sman,\tman\rpar&=&
        \frac{1}{\mws}
        \lpar -\frac{3}{2}+\frac{1}{4}\rt-\frac{1}{12 \Rw}
        \rpar \lpar\dlt-\ln \frac{\mws}{\tHss}\rpar
+\boxc{{_{\wb\wb}}}{d,F}\lpar\sman,\tman\rpar,
\\
 \boxc{{_{\wb\wb}}}{d,F}\lpar\sman,\tman\rpar&=&     
 \Biggl\{ -t\lpar 1 + \frac{t^2}{u^2}\rpar
      -4\frac{\mws t^2}{u^2}
      +2\frac{\mwq}{u}\lpar 1+2\frac{\sman}{u}\rpar     
\\ &&
+\um\Biggl[2+3\frac{\sman}{u}+2 \frac{s^2}{u^2} 
      -2\frac{\mws}{u}\lpar 1+2\frac{\sman}{u}\rpar\Biggr]
      +\frac{\umf\sman}{u^2}\Biggr\}                   
\dff{0}{-\sman}{-\tman}{\mwl}{0}{\mwl}{\uml}                 
\nll &&
-\Biggl\{2+2\frac{\sman}{u}+\frac{s^2}{u^2}-2\frac{\mws s}{u^2}
      -\frac{1}{2} \rt\Biggl[4-\Rw \lpar 1
     + 2\frac{s^2}{u^2} \rpar \Biggr]                   
\nll &&
+\frac{1}{2} \rts \lpar 1 - 2 \Rw \rpar +\frac{1}{2}\rtc \Rw
             \Biggr\}\cff{0}{-\sman}{\mwl}{\uml}{\mwl}  
\nll &&
-\lpar 2+2\frac{\sman}{u}+\frac{s^2}{u^2}
            -2\frac{\mws s}{u^2}+\frac{\um s}{u^2}\rpar 
            \cff{0}{-\sman}{\mwl}{0}{\mwl}              
\nll &&
+\lpar 2+3\frac{\sman}{u}+\frac{s^2}{u^2}
           +2\frac{\mws t}{u^2}-\frac{\um t}{u^2}\rpar  
\Bigl[\cff{0}{-\tman}{0}{\mwl}{\uml}
            +\cff{0}{-\tman}{\uml}{\mwl}{0} \Bigr]      
\nll &&          
- \frac{1}{\sman}\Biggl[
          \frac{2s }{u}+\frac{5}{3}\frac{1}{\Rw}+\frac{1}{12\Rws}
         +\frac{\rt}{4}\lpar 2-\frac{1}{\Rw}\rpar -\frac{\rts}{2 }
          \Biggr] \fbff{0}{-\sman}{\mwl}{\mwl}          
\nll &&
+\frac{2}{u}\fbff{0}{-t}{\uml}{0}
         -\frac{\rts}{2 s}\lpar 1-\ln\rt \rpar
         +\frac{1}{6\mws}\lpar 3\rt\Rw+3+\frac{3}{2}\rt
         -\frac{1}{3\Rw}\rpar.\qquad
\nonumber
\eqa
We note that $\scff{0}$ and $\sdff{0}$ are both infrared finite.  

\begin{itemize}
\item Crossed box.
\end{itemize}
The second diagram of \fig{WWboxes}
may be computed 
below the $\ft\bar{t}$ production threshold 
ignoring fermion masses. 
Its answer is very compact:
\bqa
\label{box_cross}
\boxc{{_{\wb\wb}}}{c}\lpar\sman,\uman\rpar&=&
   -2u^2 \dff{0}{-\sman}{-\uman}{\wml}{0}{\wml}{0}
   +4\cff{0}{-\sman}{\mwl}{0}{\mwl}              
\\
&& +\frac{1}{3\mws} \lrbr
   \frac{1}{4}\lpar 20+\frac{1}{\Rw}\rpar
   \bff{0}{-\sman}{\wml}{\wml}
  +\frac{1}{2}\frac{\aff{0}{\wml}}{\mws}   
  -1+\frac{1}{6\Rw}\rrbr.
\nonumber
\eqa 
And the same expression with extracted pole is
\bqa
\boxc{{_{\wb\wb}}}{c}\lpar\sman,\uman\rpar&=&
            \frac{1}{6\mws}\lpar 9+\frac{1}{2\Rw} \rpar \pole
           +\boxc{{_{\wb\wb}}}{c,F}\lpar\sman,\uman\rpar,
\\
\boxc{{_{\wb\wb}}}{c,F}\lpar\sman,\uman\rpar&=&
   -2u^2 \dff{0}{-\sman}{-\uman}{\wml}{0}{\wml}{0}
   +4\cff{0}{-\sman}{\mwl}{0}{\mwl}              
\nll
&& +\frac{1}{3\mws} \lrbr
   \frac{1}{4}\lpar 20+\frac{1}{\Rw}\rpar
   \fbff{0}{-\sman}{\wml}{\wml}
    -\frac{3}{2}+\frac{1}{6\Rw}\rrbr.
\label{wwcrossed}
\eqa 
It is instructive to compare the pole parts of \eqn{wwdirect} in the massless
approximation with that of \eqn{wwcrossed}. 
They agree modulo an overall
sign;
 that should be the case if one remembers that the Born amplitude
has  different sign for {\em down} and {\em up} final state fermions.

\begin{figure}[th]
\vspace*{-10mm}
\[
\begin{array}{ccc}
\begin{picture}(100,70)(0,33)
  \ArrowLine(0,10)(25,10)
  \ArrowLine(25,10)(25,60)
  \ArrowLine(25,60)(0,60)
  \Photon(25,10)(75,10){3}{10}
  \Photon(25,60)(75,60){3}{10}
  \Vertex(25,10){2.5}
  \Vertex(25,60){2.5}
  \Vertex(75,10){2.5}
  \Vertex(75,60){2.5}
  \ArrowLine(100,60)(75,60)
  \ArrowLine(75,60)(75,10)
  \ArrowLine(75,10)(100,10)
  \Text(12.5,70)[tc]{$\fbe$}
  \Text(50,72.5)[tc]{$\wb$}
  \Text(87.5,72)[tc]{$\fbd$}
  \Text(12.5,35)[lc]{$\fnue$}
  \Text(85,35)[rc]{$u$}
  \Text(12.5,0)[cb]{$\fe$}
  \Text(50,-2.5)[bc]{$\wb$}
  \Text(87.5,-2)[cb]{$\fd$}
\end{picture}
\qquad
 &&
\qquad
\begin{picture}(100,70)(0,33)
  \ArrowLine(0,10)(25,10)
  \ArrowLine(25,10)(25,60)
  \ArrowLine(25,60)(0,60)
  \Photon(25,10)(75,60){3}{14}
  \Photon(25,60)(75,10){3}{14}
  \Vertex(25,10){2.5}
  \Vertex(25,60){2.5}
  \Vertex(75,10){2.5}
  \Vertex(75,60){2.5}
  \ArrowLine(100,60)(75,60)
  \ArrowLine(75,60)(75,10)
  \ArrowLine(75,10)(100,10)
  \Text(12.5,70)[tc]{$\fbe$}
  \Text(50,70)[tc]{$\wb$}
  \Text(87.5,70)[tc]{$\fbu$}
  \Text(12.5,35)[lc]{$\fnue$}
  \Text(85,35)[rc]{$\fd$}
  \Text(12.5,0)[cb]{$\fe$}
  \Text(50,0)[bc]{$\wb$}
  \Text(87.5,0)[cb]{$u$}
\end{picture}
\end{array}
\]
\vspace{5mm}
\caption[The $\wb\wb$ boxes]
{\it
The $\wb\wb$ boxes
\label{WWboxes}}
\end{figure}

Finally, for the $\fd\bar{\fd}$ and $\fs\bar{\fs}$ channels one needs
the direct $\wb\wb$ box in top-less approximation:  $\mt\to 0$. 
It can trivially be derived from \eqn{wwdirect} and does not need to 
be additionally presented.

The contributions from the two $\zb\zb$ boxes are presented in
\subsect{ew_boxes}. 

\section{Amplitudes\label{amplitudes}}
\eqnzero
\subsection{Born amplitudes}
We begin with the Born amplitude for the process $\fe\fbe\to\ff\fbf$
described by the two Feynman diagrams with $\ph$ and $\zb$ exchange,
\fig{zavert2}:

\bqa
A^{\sss{B}}_{\ph} &=&e\qe\,e\qf
                  \gadu{\mu} \otimes \gadu{\mu} \frac{-\ib}{Q^2}
\nl
               &=& -\ib\,e^2\frac{\qe\qf}{Q^2}\gadu{\mu}\otimes\gadu{\mu}
\nl
               &=& -\ib\,4\pi\alpha(0)\frac{\qe\qf}{Q^2}
                  \gadu{\mu} \otimes \gadu{\mu}\,,
\\ 
A^{\sss{B}}_{\sss{\zb}} &=&
       \frac{e}{2\stwl\ctwl}\,\frac{e}{ 2\stwl\ctwl}  
                         \gadu{\mu}
       \lrbr \tcie \lpar 1+\gfd\rpar -2\qe \stws \rrbr 
                 \otimes \gadu{\mu}  
       \lrbr \tcif \lpar 1+\gfd\rpar -2\qf \stws \rrbr 
       \frac{-\ib}{Q^2+\mzs}                                      
\nll &=& -                   
\ib e^2\frac{ \tcie \tcif}{ 4 \stws \ctws (Q^2 +\mzs )}
       \gadu{\mu}\lrbr 1+\gfd-4|\qe|\stws\rrbr
       \otimes \gadu{\mu}\lrbr 1+\gfd-4|\qf|\stws\rrbr 
\nll &=& -                    
\ib e^2\frac{ \tcie \tcif}{4\stws\ctws(Q^2 +\mzs)}
\biggl[
\gadu{\mu}\bigl(1+\gfd\bigr)\otimes\gadu{\mu}\bigl(1+\gfd\bigr) 
-4|\qe|\stws\gadu{\mu}\otimes\gadu{\mu}\bigl(1+\gfd\bigr) 
\nll\nll && \hspace{38mm}
-4|\qf|\stws\gadu{\mu}\bigl(1+\gfd\bigr)\otimes\gadu{\mu} 
+16|\qe||\qf|\stwf\gadu{\mu}\otimes\gadu{\mu} 
\biggr].
\label{amplborn}
\eqa

The last representations for both amplitudes are identity transformations 
for the Born case. 
They are useful for the subsequent discussion of 
one-loop amplitudes showing explicitly five structures to which the 
complete amplitude may be reduced: one in the $\ph$-exchange amplitude and
four in the $\zb$-exchange amplitude. 
The latter may be called 
$LL$, $QL$, $LQ$, and $QQ$ structures, correspondingly;
see last \eqn{amplborn}.

In the following we will use the propagator factor:  
\bqa
\chi_{\sss{Z}}(\sman)&=&\frac{1}{16\siws\cows}\frac{\sman}
{\ds{\sman - \mzs + \ib\frac{\gz}{\mzl}\sman}}\,.
\label{propagators}
\eqa
In terms of these factors the Born amplitudes are 
\bqa
A^{\sss{B}}_{\ph} &=& \ib\,4\pi\alpha(0)\frac{\qe\qf}{\sman}
                  \gadu{\mu}\otimes\gadu{\mu}\,,
\\
A^{\sss{B}}_{\sss{\zb}}&=&
\ib\,e^2\,4\tcie\tcif\frac{\chi_{\sss{Z}}(\sman)}{\sman}
\biggl[
\gadu{\mu}\bigl(1+\gfd\bigr)\otimes\gadu{\mu}\bigl(1+\gfd\bigr) 
-4|\qe|\stws\gadu{\mu}\otimes\gadu{\mu}\bigl(1+\gfd\bigr) 
\nll\nll && \hspace{30mm}
-4|\qf|\stws\gadu{\mu}\bigl(1+\gfd\bigr)\otimes\gadu{\mu} 
+16|\qe||\qf|\stwf\gadu{\mu}\otimes\gadu{\mu} 
\biggr].\qquad\qquad
\label{amplborn1}
\eqa

In this Section the presentation develops in two parallel streams:
one for the $\zb\to\ff\fbf$ decay, and another one for the process 
scattering $\fe\fbe\to(\zb,\ph)\to\ff\fbf$. 

Here we discussed the Born amplitude only for the scattering process.
For the decay it was already given in the second \eqn{fourformdefz}
followed by the formal replacement \eqn{formalreplacements}.
\subsection{Towards one-loop amplitudes\label{towards_all}}
\subsubsection{The decay $\zb\to\ff\fbf$\label{towards_decay}}
There are two basic ingredients contributing to the one-loop amplitude of 
the $\zb$-decay. 
\begin{itemize}
\item[1.] Vertex correction for the $\zb$-decay.
\end{itemize}

\noindent
We recall diagram \fig{zavert1}, the second \eqn{fourformdef}, which 
can be considered as the definition of the one-loop amplitude, 
and the two corresponding off-shell functions $F^V_{\ZQ}(\sman)$ and
$F^{V,F}_{\ZL}(\sman)$ defined in \eqn{offshellmassive}.
For the description of the $\zb$-decay one should consider their on-mass-shell
limit. 
For these two functions we introduce shortened notations omitting 
the subscript $\zb$ and the argument list (they are constants 
for the decay):
\bqa
\label{fvq}
F^V_{\sss{Q}}&\equiv&F^V_{\ZQ}(\mzs)=
2\qds\stws\lpar\frac{1}{\epsh}+\ln\frac{\mws}{\tHss}\rpar
+F^{V,F}_{\ZQ}(\mzs),      
\\
\label{fvl}
F^V_{\sss{L}}&\equiv&F^V_{\ZL}(\mzs)=
2\qds\stws\lpar\frac{1}{\epsh}+\ln\frac{\mws}{\tHss}\rpar
+\Biggl(
\frac{1}{2}\rt+\frac{4}{3}+\frac{1}{12}\frac{1}{\Rw} 
 \Biggr)\pole 
\nll &&
+ ~F^{V,F}_{\ZL}(\mzs).
\eqa

\begin{itemize}
\item[2.] Bosonic counter-terms.
\end{itemize}
Contributions to the decay amplitude from counter-terms originate from 
all relevant bosonic self-energy diagrams.
They may be depicted 
the vertex in \fig{zavert3}.

The corresponding contributions are derived from the counter-term Lagrangian 
and may be expressed in terms of quantities introduced 
in \subsect{bosonic_ct}. 
We will call them {\em bosonic counter-terms}:
\bqa
\label{fqct}
F_{\sss{Q}}^{ct} &=&
       \frac{1}{2} \left[\left( z_{\sss{\zb}} - 1 \right)
     - \left( z_{\ph} - 1 \right)
     - \frac{1}{\stws} \delrho{} \right]
     + \frac{ \Sigma_{\zg}(\mzs)}{\mzs},  
\\
\label{flct}
F_{\sss{L}}^{ct} &=& 
       \frac{1}{2}\left[
       \left(z_{\sss{\zb}} - 1\right) - \left(z_{\ph} - 1 \right)
      +\frac{\ctws-\stws}{\stws}\delrho{} \right].
\eqa
Now we jump to the scattering process.

\begin{figure}[thbp]
\vspace*{-3mm}
\[
\begin{picture}(125,86)(0,0)
  \Photon(25,43)(100,43){3}{15}
\SetScale{2.0}
    \Line(45,16.5)(46.5,18)
    \Line(47.5,19)(52.5,24)
    \Line(53.5,25)(55,26.5)
    \Line(45,26.5)(46.5,25)
    \Line(47.5,24)(52.5,19)
    \Line(53.5,18)(55,16.5)
\SetScale{1.0}
  \ArrowLine(125,86)(100,43)
  \Vertex(100,43){2.5}
  \ArrowLine(100,43)(125,0)
 \Text(108,74)[lb]{$\fbf$}
 \Text(62.5,50)[bc]{$\zb$}
 \Text(108,12)[lt]{$\ff$}
\end{picture}
\]
\vspace{-9mm}
\caption[Bosonic counter-terms for $\zb\to\ff\fbf$]
{\it
Bosonic counter-terms for $\zb\to\ff\fbf$
\label{zavert3}}
\end{figure}
\vspace*{-7mm}

\subsubsection{The process $\fep\fem\to(\ph,\zb)\to\ff\fbf$
\label{towards_process}}
For the scattering process there are many more ingredients than for the decay.
Here we distinguish $\ph$- and $\zb$-exchanges; initial and final state vertex
corrections; contributions from bosonic self-energies, counter-terms,
and boxes. 

\begin{itemize}
\item The process $\fe\fbe\to(\ph)\to\ff\fbf$; final fermion vertex.
\end{itemize}

We begin with the final state vertex correction shown in \fig{zavert4}(b).
Its amplitude is 
 $A_{\ph} =$ Born electron vertex $\otimes$ dressed final
state $\ph$-vertex $\times$ $\ph$-propagator:
\bqa
A_{\ph} &=& e\qe\,e\gadu{\mu}\otimes\gadu{\mu} 
        \Biggl[\frac{1}{2}\tcif\lpar 1+\gfd\rpar\vvertil{V}{\gL}{\sman} 
        +\qf \vvertil{V}{\gQ}{\sman}
        \Biggr]\frac{-\ib}{Q^2}\,. 
\eqa

\begin{itemize}
\item Process $\fe\fbe\to(\ph)\to\ff\fbf$; initial electron vertex.
\end{itemize}

Similarly, there exists the initial state electron vertex correction shown 
in \fig{zavert4}(a).

We do not give its separate contribution but rather show both vertex 
corrections originating from the two $\ph$-exchange diagrams:
\bqa
A_{\ph} &=&
      (-\ib e^2)\Biggl\{\gadu{\mu}\lrbr
       \frac{1}{2}\tcie\lpar 1+\gfd \rpar\vvertil{V(0)}{\gL}{\sman} 
      +\qe\vvertil{V(0)}{\gQ}{\sman}
                                  \rrbr\otimes\gadu{\mu}\qf 
\nll && 
   +~\gadu{\mu}\qe\otimes\gadu{\mu}\lrbr
       \frac{1}{2}\tcif\lpar 1 +\gfd \rpar\vvertil{V}{\gL}{\sman} 
      +\qf\vvertil{V}{\gQ}{\sman}
                                  \rrbr\Biggr\}\frac{1}{Q^2}\,.       
\eqa 

\begin{itemize}
\item Process $\fe\fbe\to(\zb)\to\ff\fbf$; final fermion vertex.
\end{itemize}
Analogously, one constructs the final state vertex for the
$\zb$-exchange diagram, \fig{zavert6}(b).

\noindent
Its amplitude is the product $A_{\sss{\zb}} =$ Born electron vertex
$\otimes$ dressed final 
state $\zb$-vertex $\times$ $\zb$-propagator:
\bqa 
A_{\sss{\zb}} &=& \frac{e}{2\stwl\ctwl}\,\frac{e}{2\stwl\ctwl}  
\\ &&
\times \gadu{\mu}
 \lrbr \tcie \lpar 1+\gfd\rpar -2\qe \stws \rrbr 
\otimes \gadu{\mu}  
 \lrbr \tcif \lpar 1+\gfd\rpar 
  \vvertil{V}{\ZL}{\sman}-2\qf\stws \vvertil{V}{\ZQ}{\sman}\rrbr
\frac{-i}{Q^2+\mzs}\,.
\nonumber
\eqa

\begin{itemize}
\item Process $\fe\fbe\to(\zb)\to\ff\fbf$; initial electron vertex.
\end{itemize}

\begin{figure}[ht]
\vspace{-22mm}
\[
\begin{array}{ccc}
\begin{picture}(125,86)(0,40)
   \Photon(25,43)(100,43){3}{15}
   \Vertex(25,43){12.5}
   \ArrowLine(0,0)(25,43)
   \ArrowLine(25,43)(0,86)
       \ArrowLine(125,86)(100,43)
       \Vertex(100,43){2.5}
       \ArrowLine(100,43)(125,0)
\Text(14,74)[lb]{$\fbe$}
\Text(108,74)[lb]{$\fbf$}
\Text(62.5,50)[bc]{$(\ph)$}
\Text(14,12)[lt]{$\fe$}
\Text(108,12)[lt]{$\ff$}
\end{picture}
\qquad
&&
\qquad
\begin{picture}(125,86)(0,40)
  \Photon(25,43)(100,43){3}{15}
  \Vertex(100,43){12.5}
  \ArrowLine(125,86)(100,43)
  \ArrowLine(100,43)(125,0)
    \ArrowLine(0,0)(25,43)
    \Vertex(25,43){2.5}
    \ArrowLine(25,43)(0,86)
\Text(14,74)[lb]{$\fbe$}
\Text(108,74)[lb]{$\fbf$}
\Text(62.5,50)[bc]{$(\ph)$}
\Text(14,12)[lt]{$\fe$}
\Text(108,12)[lt]{$\ff$}
\end{picture}
\end{array}
\]
\vspace{4mm}
\caption[Electron and final fermion vertices in $\fe\fbe\to(\ph)\to\ff\fbf$]
{\it
Electron (a) and final fermion (b) vertices in $\fe\fbe\to(\ph)\to\ff\fbf$
\label{zavert4}}
\end{figure}
\vspace*{-2mm}

The initial state vertex for the $\zb$-exchange diagram,
\fig{zavert6}(a), is combined with the final state fermion vertex into:
\bqa 
A_{\sss{\zb}} &=&  \frac{-\ib e^2}{4\stws \ctws} 
\Biggl\{
                         \gadu{\mu}
 \lrbr \tcie \lpar 1+\gfd\rpar  \vvertil{V(0)}{\ZL}{\sman} 
                          - 2\qe \stws  \vvertil{V(0)}{\ZQ}{\sman}\rrbr 
                 \otimes \gadu{\mu}  
 \lrbr \tcif \lpar 1+\gfd\rpar   -2\qf \stws \rrbr  
\nll &&
+                        \gadu{\mu}
 \lrbr \tcie \lpar 1+\gfd\rpar  - 2\qe \stws \rrbr 
                 \otimes \gadu{\mu}  
 \lrbr \tcif \lpar 1+\gfd\rpar  \vvertil{V}{\ZL}{\sman}
  -2\qf \stws \vvertil{V}{\ZQ}{\sman}\rrbr  
\Biggr\}
\frac{1}{Q^2+\mzs}\,.
\nll
\eqa

\begin{itemize}
\item{All vertex corrections.}
\end{itemize}

It is instructive to combine all four vertex corrections together, into
a common expression where we restored omitted normalization factor 
${\ds\gspi}$:  
\bqa
\label{zOLA}
{\cal A}^{\sss{OLA}}_{\sss{\zb}}&=&
-\ib \gspi \frac{e^2 \tcie \tcif }{4 \stws \ctws (Q^2+\mzs) }     
\\ &&
\times \Biggl\{\gadu{\mu} {\lpar 1+\gfd \rpar} 
       \otimes \gadu{\mu} {\lpar 1+\gfd \rpar} 
       \lrbr  \vvertil{V(0)}{\ZL}{\sman}+\vvertil{V}{\ZL}{\sman} \rrbr 
\nll &&
-4|\qe|\stws \gadu{\mu}\otimes \gadu{\mu} \lpar 1+\gfd \rpar  
       \lrbr \vvertil{V(0)}{\ZQ}{\sman}+ \vvertil{V}{\ZL}{\sman}+
       \frac{(Q^2 +\mzs) 4\stws\ctws |\qe| }{Q^2(-4|\qe| \stws)}
       \vvertil{V}{\gL}{\sman}  \rrbr                                    
\nll &&
-4|\qf|\stws \gadu{\mu} {\lpar 1+\gfd \rpar } \otimes\gadu{\mu} 
       \lrbr \vvertil{V(0)}{\ZL}{\sman}+ \vvertil{V}{\ZQ}{\sman}+
       \frac{(Q^2 +\mzs) 4\stws\ctws |\qf| }{Q^2(-4|\qf| \stws)}
       \vvertil{V(0)}{\gL}{\sman}  \rrbr\nll &&
+16 |\qe|| \qf| \stwf \gadu{\mu} \otimes \gadu{\mu}                    
\nll &&
\times \Biggl[ \vvertil{V(0)}{\sss{LQ}}{\sman}                            
             + \vvertil{V}{\ZQ}{\sman}
             + \frac{(Q^2 +\mzs) 4\stws\ctws |\qe| |\qf|}
       {(16|\qe||\qf| \stwf) Q^2}\lpar \vvertil{V(0)}{\gQ}{\sman}
       +\vvertil{V}{\gQ}{\sman}\rpar \Biggr] \Biggr\}. 
\nonumber
\eqa
It has the following Born-like structure:
\bqa
{\cal A}^{\sss{OLA}}_{\sss{\zb}}&=&
\ib\gspi\,e^2\,4\tcie\tcif\frac{\chi_{\sss{Z}}(\sman)}{\sman}
\nll &&
\times \Biggl\{\gadu{\mu} {\lpar 1+\gfd \rpar } \otimes
       \gadu{\mu} {\lpar 1+\gfd \rpar } \vvertil{}{\sss{LL}}{\sman,\tman}     
-4 |\qe | \stws \gadu{\mu}     
      \otimes \gadu{\mu} {\lpar 1+\gfd \rpar}\vvertil{}{\sss{QL}}{\sman,\tman} 
\nll &&
-4 |\qf | \stws
\gadu{\mu}{\lpar 1+\gfd\rpar}\otimes\gadu{\mu}\vvertil{}{\sss{LQ}}{\sman,\tman}
+16|\qe\qf|\stwf\gadu{\mu}\otimes\gadu{\mu}
\vvertil{}{\sss{QQ}}{\sman,\tman}\Biggr\}.\qquad\quad
\label{structures-old}
\eqa

\begin{figure}[th]
\vspace{-23mm}
\[
\begin{array}{ccc}
\begin{picture}(125,86)(0,40)
    \Photon(25,43)(100,43){3}{15}
    \Vertex(25,43){12.5}
    \ArrowLine(0,0)(25,43)
    \ArrowLine(25,43)(0,86)
      \ArrowLine(125,86)(100,43)
      \Vertex(100,43){2.5}
      \ArrowLine(100,43)(125,0)
\Text(14,74)[lb]{$\fbe$}
\Text(108,74)[lb]{$\fbf$}
\Text(62.5,50)[bc]{$(\zb)$}
\Text(14,12)[lt]{$\fe$}
\Text(108,12)[lt]{$\ff$}
\end{picture}
\qquad
&&
\qquad
\begin{picture}(125,86)(0,40)
 \Photon(25,43)(100,43){3}{15}
 \Vertex(100,43){12.5}
 \ArrowLine(125,86)(100,43)
 \ArrowLine(100,43)(125,0)
   \ArrowLine(0,0)(25,43)
   \Vertex(25,43){2.5}
   \ArrowLine(25,43)(0,86)
\Text(14,74)[lb]{$\fbe$}
\Text(108,74)[lb]{$\fbf$}
\Text(62.5,50)[bc]{$(\zb)$}
\Text(14,12)[lt]{$\fe$}
\Text(108,12)[lt]{$\ff$}
\end{picture}
\end{array}
\]
\vspace{4mm}
\caption[Electron and final fermion vertices in $\fe\fbe\to(\zb)\to\ff\fbf$]
{\it
Electron (a) and final fermion (b) vertices in $\fe\fbe\to(\zb)\to\ff\fbf$
\label{zavert6}}
\end{figure}
\vspace{-3mm}

If one compares with the last \eqn{amplborn}, it is seen that the Born-factors
`$1$' in front of the $LL$ etc. structures 
are replaced by four scalar form factors $\vvertil{}{\sss{LL}}{\sman,\tman}$,
$\vvertil{}{\sss{QL}}{\sman,\tman}$, $\vvertil{}{\sss{LQ}}{\sman,\tman}$ and 
$\vvertil{}{\sss{QQ}}{\sman,\tman}$.
We allow for a $\tman$-dependence having in mind the contribution 
of EW boxes, see \subsect{constff}. 

\begin{itemize}
\item Bosonic self-energies and bosonic counter-terms.
\end{itemize}
The contributions to form factors from bosonic self-energy diagrams and 
counter-terms originating from bosonic self-energy diagrams
come from four classes of diagrams; their sum is depicted by a black
circle in \fig{zavert8}.

The contribution of these diagrams to the four scalar form factors is derived
straightforwardly:
\bqa
\vvertil{ct}{\sss{LL}}{\sman} &=& 
      \Dz{}{\sman}-\stws\lpar\zcont{\ph}{}-1 \rpar
     +\frac{\cows-\siws}{\siws} \delrho{},                    
\\
\vvertil{ct}{\QL(\LQ)}{\sman}&=& 
      \Dz{}{\sman}-\Pzg^{}(\sman)
     -\stws \lpar\zcont{\ph}{}-1\rpar-\delrho{}, 
\\
\vvertil{ct,\bos}{\sss{QQ}}{\sman}&=&
      \Dz{\bos}{\sman}-2\Pzg^{\bos}(\sman)
     +\cows\lpar1-\Rz\rpar\Bigl[\Pgg^{\bos}\lpar s\rpar 
     -\lpar\zcont{\ph}{\bos}-1\rpar\Bigr]                  
\nll &&
-\stws \lpar\zcont{\ph}{\bos}-1\rpar
     -\frac{1}{\siws} \delrho{\bos},                          
\\
\vvertil{ct,\fer}{\sss{QQ}}{\sman}&=&
      \Dz{\fer}{\sman}-2\Pzg^{\fer}(\sman)
     -\stws \lpar\zcont{\ph}{\fer}-1\rpar
     -\frac{1}{\siws} \delrho{\fer}.
\label{zct}
\eqa
We note that the term $\cows\lpar1-\Rz\rpar\bigl[\Pgg^{\fer}\lpar\sman\rpar 
-\bigl(\zcont{\ph}{\fer}-1\bigr)\bigr]$ is conventionally extracted from 
$\vvertil{ct,\fer}{\sss{QQ}}{\sman}$. 
Recalling electron charge renormalization,
\eqn{elcharen}, one may easily verify that this term gives rise to 
a `dressed' $\ph$-exchange amplitude:
\bq
A^{\sss{OLA}}_{\ph} = \ib\frac{4\pi\qe\qf}{\sman}
\alpha(\sman) \gadu{\mu} \otimes \gadu{\mu}\,,
\label{Born_modulo-old}
\eq
which is identical to the Born amplitude \eqn{amplborn} 
modulo the replacement of $\alpha(0)$ by the
running electromagnetic coupling $\alpha(\sman)$:
\bq
\alpha(\sman)=\frac{\alpha}
{\ds{1-\frac{\alpha}{4\pi}\Bigl[\Pgg^{\fer}(\sman)-\Pgg^{\fer}(0)\Bigr]}}\,.
\label{alpha_fer-old}
\eq

\begin{figure}[thbp]
\vspace{-17mm}
\[
\begin{array}{ccc}
\begin{picture}(125,86)(0,40)
    \Photon(25,43)(50,43){3}{5}
    \Vertex(62.5,43){12.5}
    \Photon(75,43)(100,43){3}{5}
  \ArrowLine(125,86)(100,43)
  \Vertex(100,43){2.5}
  \ArrowLine(100,43)(125,0)
    \ArrowLine(0,0)(25,43)
    \Vertex(25,43){2.5}
    \ArrowLine(25,43)(0,86)
\Text(14,74)[lb]{$\fbe$}
\Text(108,74)[lb]{$\fbf$}
\Text(37.5,50)[bc]{$(\ph,\zb)$}
\Text(87.5,50)[bc]{$(\ph,\zb)$}
\Text(14,12)[lt]{$\fe$}
\Text(108,12)[lt]{$\ff$}
\end{picture}
\qquad
 &=&
\nll  \nll 
\begin{picture}(125,86)(0,40)
    \Photon(25,43)(50,43){3}{5}
    \GCirc(62.5,43){12.5}{0.5}
    \Photon(75,43)(100,43){3}{5}
  \ArrowLine(125,86)(100,43)
  \Vertex(100,43){2.5}
  \ArrowLine(100,43)(125,0)
    \ArrowLine(0,0)(25,43)
    \Vertex(25,43){2.5}
    \ArrowLine(25,43)(0,86)
\Text(14,74)[lb]{$\fbe$}
\Text(108,74)[lb]{$\fbf$}
\Text(37.5,50)[bc]{$(\ph,\zb)$}
\Text(87.5,50)[bc]{$(\ph,\zb)$}
\Text(14,12)[lt]{$\fe$}
\Text(108,12)[lt]{$\ff$}
\end{picture}
\qquad
&+&
\qquad
\begin{picture}(125,86)(0,40)
     \Photon(25,43)(100,43){3}{15}
\SetScale{2.0}
     \Line(26.5,16.75)(28,18.25)
     \Line(29,19.25)(34,24.25)
     \Line(35,25.25)(36.5,26.75)
      \Line(26.5,26.75)(28,25.25)
      \Line(29,24.25)(34,19.25)
      \Line(35,18.25)(36.5,16.75)
\SetScale{1.0}
     \ArrowLine(125,86)(100,43)
     \Vertex(100,43){2.5}
     \ArrowLine(100,43)(125,0)
  \ArrowLine(0,0)(25,43)
  \Vertex(25,43){2.5}
  \ArrowLine(25,43)(0,86)
\Text(14,74)[lb]{$\fbe$}
\Text(108,74)[lb]{$\fbf$}
\Text(37.5,50)[bc]{$(\ph,\zb)$}
\Text(87.5,50)[bc]{$(\ph,\zb)$}
\Text(14,12)[lt]{$\fe$}
\Text(108,12)[lt]{$\ff$}
\end{picture} \nll \nll 
\begin{picture}(125,86)(0,40)
  \Photon(25,43)(100,43){3}{15}
\SetScale{2.0}
    \Line(45,16.5)(46.5,18)
    \Line(47.5,19)(52.5,24)
    \Line(53.5,25)(55,26.5)
    \Line(45,26.5)(46.5,25)
    \Line(47.5,24)(52.5,19)
    \Line(53.5,18)(55,16.5)
\SetScale{1.0}
  \ArrowLine(125,86)(100,43)
  \Vertex(100,43){2.5}
  \ArrowLine(100,43)(125,0)
    \ArrowLine(0,0)(25,43)
    \Vertex(25,43){2.5}
    \ArrowLine(25,43)(0,86)
\Text(14,74)[lb]{$\fbe$}
\Text(108,74)[lb]{$\fbf$}
\Text(62.5,50)[bc]{$(\ph,\zb)$}
\Text(14,12)[lt]{$\fe$}
\Text(108,12)[lt]{$\ff$}
\end{picture}
\qquad
&+&
\qquad
\begin{picture}(125,86)(0,40)
  \Photon(25,43)(100,43){3}{15}
\SetScale{2.0}
    \Line(7.5,16.5)(9,18)
    \Line(10,19)(15,24)
    \Line(16,25)(17.5,26.5)
    \Line(7.5,26.5)(9,25)
    \Line(10,24)(15,19)
    \Line(16,18)(17.5,16.5)
\SetScale{1.0}
    \Vertex(25,43){2.5}
    \ArrowLine(0,0)(25,43)
    \ArrowLine(25,43)(0,86)
 \ArrowLine(125,86)(100,43)
 \Vertex(100,43){2.5}
 \ArrowLine(100,43)(125,0)
\Text(14,74)[lb]{$\fbe$}
\Text(108,74)[lb]{$\fbf$}
\Text(62.5,50)[bc]{$(\ph,\zb)$}
\Text(14,12)[lt]{$\fe$}
\Text(108,12)[lt]{$\ff$}
\end{picture}
\end{array}
\]
\vspace{10mm}
\caption[Bosonic self-energies
and bosonic counter-terms for $\fe\fbe\to(\zb,\ph)\to\ff\fbf$]
{\it
Bosonic self-energies
and bosonic counter-terms for $\fe\fbe\to(\zb,\ph)\to\ff\fbf$
\label{zavert8}} 
\end{figure}

\clearpage

\subsection{The form factors
$\deltar$, $F_{\sss{L}}, F_{\sss{Q}}$, and 
$F^{}_{\sss{LL}},
F^{}_{\sss{QL}},
F^{}_{\sss{LQ}},
F^{}_{\sss{QQ}}
$
\label{constff}
}

\subsubsection{Preliminaries} 
\begin{itemize}
\item Unified form factors.
\end{itemize}
If one looks at \eqnsc{offshellmassive}{offshellmassless} one may note that
the finite parts of vertex functions, together with certain terms 
originating from fermionic counter-terms are forming several well-defined
couples or larger families of terms.
We will call them 
{\em unified form factors} and will denote them by a corresponding
calligraphic symbol. 
In what follows we list all the unified form factors which are met. 

The first family consists of various QED contributions:
\bqa
\cvetril{f}{\ph }{\sman}&=&
\qfs\siws\Biggl[2\lpar\frac{1}{\epsh}+\ln\frac{\mws}{\tHss}\rpar
+3\ln\frac{\mfs}{\mws}+\vvertil{f,F}{\ph}{\sman}-4\Biggr].
\eqa

The second is a simple couple associated with diagrams with a virtual
$\zb$ boson:
\bqa
\cvetril{}{\sss{\zb}}{\sman} &=& \vvertil{F}{\sss{\zb}}{\sman}+\frac{3}{2}  
\nll &=& 2 \lpar  \Rz+1 \rpar^2\sman\cff{0}{-\sman}{0}{\mzl}{0}
    -\lpar 2\Rz+3 \rpar \ln\Rz - 2\Rz -\frac{7}{2}\,.\qquad
\label{cal_Fz}
\eqa

Now consider abelian and non-abelian diagrams with virtual $\wb$ bosons:
\bqa 
\cvetri{ }{\sss{\wb}}{\sman} =-\ctws \vverti{F}{\wb n }{\sman} 
  +\frac{1}{2} \left[ \vpada \wwUf 
  -\vpadpa \vverti{F}{\wb a}{\sman}-\averti{F}{\wb a}{\sman}\right],
\eqa
where 
\ba
\vpada = |\vpad| = 1 - 2|Q_f|\stws,
\ea
and
\ba
\vpada + \vpadpa = 2\ctws.
\ea
It may be naturally rewritten in terms of three form factors:
\bqa
\cvetri{ }{\sss{\wb}}{\sman} = \ctws \cvetri{}{\wb n}{\sman}  
  -\frac{1}{2} \vpadpa \cvetri{ }{\wb a }{\sman}  
  -\frac{1}{2} \avetri{ }{\wb a }{\sman},  
\eqa
where
\bqa
\cvetri{ }{\wb a }{\sman}&=&  \vverti{F}{\wb a}{\sman}  + \wwUf, 
\\
\cvetri{ }{\wb n }{\sman}&=& -\vverti{F}{\wb n }{\sman} + \wwUf, 
\\
\avetri{ }{\wb a }{\sman}&=& -\averti{F}{\wb a }{\sman}.
\label{cal_all_Fw}
\eqa
These three unified form factors, originating from diagrams with virtual  
$\wb$ bosons as well as the contribution of the direct $\wb\wb$ box diagram
$\boxc{\sss{\wb\wb}}{d}(\sman,\tman)$, contain the $\mt$-dependent terms. 
They
may be naturally split into two terms:
(i) their massless limit $(\mt=0)$; 
(ii) top quark addition, vanishing when $\mt\to 0$.
We denote massless parts with superscript $(0)$ and the additions with
superscript $(\ft)$: 
\bqa
\cvetri{}{}{\sman}&=&\cvetri{0}{}{\sman} + \cvetri{t}{}{\sman},
\\
 \boxc{\sss{\wb\wb}}{d}(\sman,\tman)&=&
 \boxc{\sss{\wb\wb}}{d,0}(\sman,\tman)
+\boxc{\sss{\wb\wb}}{d,\ft}(\sman,\tman).
\label{boxadditions-t}
\eqa
Below we present both parts of the three $\wb$ unified form factors.
  
\begin{itemize}
\item Top-less limit.
\end{itemize}
The massless limit is compact:
\bqa
\cvetri{0}{\wb a}{\sman} &=&
   2\lpar \Rw + 1 \rpar^2\sman\cff{0}{-\sman}{0}{\mwl}{0}
   -\lpar 2 \Rw + 3 \rpar \ln(-\Rw) -2\Rw -\frac{7}{2}\,,\qquad\quad
\\
\avetri{0}{\wb n}{\sman} &=& 0,                       
\\
\cvetri{0}{\wb n}{\sman} &=& 
  -2\lpar \Rw + 2 \rpar \mws\cff{0}{-\sman}{\mwl}{0}{\mwl}    
\nll &&
   -\lpar 2\Rw +\frac{7}{3}-\frac{3}{2 \Rw}
   -\frac{1}{12 \Rws}\rpar \fbff{0}{-\sman}{\mwl}{\mwl}    
\nll &&
   +2\Rw+\frac{9}{2}+\frac{11}{18\Rw} - \frac{1}{18\Rws}\,.
\label{calfwtopless}
\eqa

\begin{itemize}
\item Top quark additions.
\end{itemize}
They are straightforwardly derived from the definitions:
\bqa
\cvetri{t}{\wb a}{\sman}&=&
\cvetri{ }{\wb a}{\sman}-\cvetri{0}{\wb a}{\sman}        
\\
\avetri{t}{\wb a}{\sman}&=&
\avetri{ }{\wb a}{\sman}-\avetri{0}{\wb a }{\sman}       
\\
\cvetri{t}{\wb n}{\sman}&=&
\cvetri{ }{\wb n}{\sman} - \cvetri{0}{\wb n }{\sman}.
\label{calfwtopadd}
\eqa
Now we give the list of the additions:
\bqa
\cvetri{t}{\wb a }{\sman}  &=&
2(\Rw+1)^2\sman\Bigl[\cff{0}{-\sman}{\uml}{\wml}{\uml}
                    -\cff{0}{-\sman}{0}{\wml}{0}\Bigr]
\nll &&
+(2\Rw+3)\left[-\fbff{0}{-\sman}{\uml}{\uml}+\ln\Rw+2\right]
\nll &&
-\rt\Biggl\{\lpar 3\Rw+2-\rt-\rts\Rw \rpar 
                 \mws \cff{0}{-\sman}{\uml}{\wml}{\uml}          
\nll &&
+\lpar \Rw+\frac{1}{2}+\rt\Rw \rpar        
       \Bigl[1 - \fbff{0}{-\sman}{\uml}{\uml}\Bigr]   
\nll &&
-\left( 2 \Rw+\frac{1}{2}-\frac{2}{\rt-1}
   +\frac{3}{2}\frac{1} {(\rt-1)^2}+\rt\Rw\right) \Lnrt
   +\frac{3}{2}\frac{1}{\rt-1}+\frac{3}{4}\Biggr\},
\label{calFwat}
\\
\avetri{t}{\wb a }{\sman}  &=&
 -\rt \Biggl\{ \Bigr[ \Rw+2-\rt (2-\rt) \Rw \Bigr]
  \mws \cff{0}{-\sman}{\uml}{\wml}{\uml}                           
\nll &&
-\lpar \frac{1}{2}-\Rw+\rt \Rw \rpar
  \Bigl[\fbff{0}{-\sman}{\uml}{\uml} + 1\Bigr]+\rt\Rw \Lnrt\Biggr\},
\\
\label{calFwnt}
\cvetri{t}{\wb n }{\sman}  &=&
-2 (\Rw+2) \mws \Bigl[
\cff{0}{-\sman}{\wml}{\uml}{\wml}-\cff{0}{-\sman}{\wml}{0}{\wml}\Bigr]
\\ &&
+\rt \Biggl\{\Biggl[ 3 \Rw+\frac{5}{2}
  -\frac{2}{\Rw}-\rt\left( 2-\frac{1}{2\Rw} \right)   
\nll &&
+\rt^2 \left(\frac{1}{2}-\Rw \right)  \Biggr]
    \mws \cff{0}{-\sman}{\wml}{\uml}{\wml}                            
\nll &&
+\left[\Rw+1-\frac{1}{ 4 \Rw}
  -\rt \left( \frac{1}{2}-\Rw\right)\right]
 \Bigl[\fbff{0}{-\sman}{\wml}{\wml}-1\Bigr]  
\nll &&
+\Biggl[2 \Rw+\frac{1}{2}
  -\frac{2}{\rt-1}+\frac{3}{2}\frac{1}{(\rt-1)^2}
  -\rt \left(\frac{1}{2}-\Rw \right) \Biggr] \Lnrt
  +\frac{3}{2}\frac{1}{\rt-1}+\frac{1}{4}\Biggr].\qquad
\nonumber
\eqa
\subsubsection{The decay $\zb\to\ff\fbf$}
The two decay form factors with restored normalization factors read: 
\bqa
F_{\sss{Q}}&=&\gspi\lpar F^V_{\sss{Q}}+F_{\sss{Q}}^{ct}\rpar,
\label{decay_ffQ}
\\
F_{\sss{L}}&=&\gspi\lpar F^V_{\sss{L}}+F_{\sss{L}}^{ct}\rpar,
\label{decay_ffL}
\eqa
where $F^V_{\sss{Q}}$ and  $F^V_{\sss{L}}$ are defined in \eqn{fvq}
and \eqn{fvl} and $F_{\sss{Q}}^{ct}, F_{\sss{L}}^{ct}$ in \eqn{fqct}
and \eqn{flct}. 
And the final expressions look quite compact:
\bqa
F_{\sss{Q}} &=& \gspi \Biggl[
      \qd^2 \stws \cvetril{}{\ph}{\mzs}
    + \frac{\sigma^2_f}{4\ctws} \cvetril{}{\sss{\zb}}{\mzs}
    - \Pzg(\mzs)    
\nll &&
    - \frac{\stws}{2} \Pgg^{F}(0) + \frac{1}{2}  \zmz
    - \frac{1}{ 2\stws} \delrho{F} \Biggr],
\\
F_{\sss{L}} &=&
      \gspi \Biggl[
      \qd^2 \stws    \cvetril{}{\ph}{\mzs}
     +\frac{3\vd^2+\ad^2}{4\ctws}\cvetril{}{\sss{\zb}}{\sman} \nll &&
     +\cvetril{}{\sss{\wb}}{\mzs}
     -\frac{\stws}{2}\Pgg^{F}(0) 
     +\frac{1}{2}\zmz
     -\frac{\ctws-\stws}{2\stws}\delrho{F} \Biggr].
\label{decay_ff}
\eqa
where
\ba
\Pgg^{F}(0) &=& \Pgg^{\fer,F}(0) + \Pgg^{\bos}(0), 
\\
\Pgg^{\bos}(0) &=& \frac{2}{3}\;. 
\label{se_ggF}
\ea
\subsubsection{The process $\fe\fbe\to(\ph,\zb)\to\ff\fbf$}
For the scattering process we have to collect contributions
from all  
off-shell vertices from self-energies and counter-terms, \eqn{zct},
and additionally from all boxes. 
However, conventionally the $\zb\ph$ and $\ph\ph$ boxes
are shifted to the QED-corrections.
So, in the electroweak form factors we are
left with the purely $L\otimes L\;$ $\wb\wb$ box and with the direct 
and crossed $\zb\zb$
boxes.
The latter two contribute to all four form factors
and are free of ultraviolet
and infrared divergences;
they are {\em not} covered by the presentation below.
From equations \eqnsc{zOLA}{zct} and adding only the $\wb\wb$ box, one
derives:  
\bqa
\vvertil{}{\sss{LL}}{\sman,\tman}&=&
\vvertil{\Vvert(0)}{\ZL}{\sman}+\vvertil{\Vvert}{\ZL}{\sman}
+\vvertil{ct}{\sss{LL}}{\sman}
-\cows\lpar \Rz - 1\rpar\sman\boxc{{\sss{\wb\wb}}}{d}(\sman,\tman),
\label{F_LL}
\\
\vvertil{}{\sss{QL}}{\sman,\tman}&=&
\vvertil{\Vvert(0)}{\ZQ}{\sman}+\vvertil{\Vvert}{\ZL}{\sman}
+\cows\lpar \Rz - 1\rpar\vvertil{\Vvert}{\gL}{\sman}
+\vvertil{ct}{\sss{QL}}{\sman},
\label{F_QL}
\\
\vvertil{}{\sss{LQ}}{\sman,\tman}&=&
\vvertil{\Vvert(0)}{\ZL}{\sman}+\vvertil{\Vvert}{\ZQ}{\sman}
+\cows\lpar \Rz - 1\rpar\vvertil{\Vvert(0)}{\gL}{\sman}
+\vvertil{ct}{\sss{LQ}}{\sman},
\label{F_LQ}
\\
\vvertil{}{\sss{QQ}}{\sman,\tman}&=&
\vvertil{\Vvert(0)}{\ZQ}{\sman}+\vvertil{\Vvert}{\ZQ}{\sman}
-\frac{\cows}{\siws}\lpar \Rz - 1 \rpar
\Bigl[\vvertil{\Vvert(0)}{\gQ}{\sman}
     +\vvertil{\Vvert}{\gQ}{\sman}\Bigr]
     +\vvertil{ct}{\sss{QQ}}{\sman},\qquad\quad
\label{F_QQ}
\eqa
where
\bq
\vvertil{ct}{\sss{AB}}{\sman}=
\vvertil{ct,\bos}{\sss{AB}}{\sman}+\vvertil{ct,\fer}{\sss{AB}}{\sman}.
\eq
The final expressions for the basic quantities
$\vvertil{}{\sss{LL}}{\sman,\tman}$, $\vvertil{}{\sss{QL}}{\sman,\tman}$,  
$\vvertil{}{\sss{LQ}}{\sman,\tman}$ and $\vvertil{}{\sss{QQ}}{\sman,\tman}$, 
are free of ultraviolet poles.
But, as one may see from an inspection of the partial contributions, 
for instance from \eqn{calfwtopless}, there are terms rising with $\sman$,
\bq
\biggl[\frac{1}{\Rw},\frac{1}{\Rws}\biggr]\otimes
\biggl[1,\fbff{0}{-\sman}{\wml}{\wml}\biggr],
\label{nuni}
\eq
violating thereby unitarity.
All these terms, which are called {\em non-unitary} terms, cancel
identically in the sum. 
We introduce {\em hatted} quantities which are 
derived from the corresponding non-hatted objects by simply neglecting 
terms of the four kinds given in \eqn{nuni}. 
The answers, now free of ultraviolet 
poles and of non-unitary terms are very compact:
\bqa
\vvertil{}{\sss{LL}}{\sman,\tman}&=& 
  \gspi 
   \Biggl\{     
   \cvetril{e}{\ph }{\sman}+\cvetril{f}{\ph }{\sman}
  -\siws \Pgg^{F}(0) 
  +\frac{\cows-\siws}{\siws} \delrho{F}+\Dz{F}{\sman}         
\nll &&+\frac{1}{4\cows}
   \lpar 3\vcs{\fe} +\acs{\fe}+3\vcs{\ff}+\acs{\ff}\rpar
   \cvetril{}{\sss{\zb}}{\sman} 
  +\hvetri{(0)}{\sss{\wb} }{\sman} +\hvetri{}{\sss{\wb}}{\sman}   
\nll &&
-\cows\lpar \Rz - 1 \rpar s\hboxc{\sss{\wb\wb}}{d}{\sman,\tman}
  +\frac{5}{3}\fbff{0}{-\sman}{\mwl}{\mwl}-\frac{1}{2}        
\nll &&
-\frac{\rt}{4}\Bigl[\fbff{0}{-\sman}{\mwl}{\mwl}+1\Bigr]
\Biggr\},
\\
\vvertil{}{\sss{QL}}{\sman,\tman}&=&
  \gspi \Biggl\{     
   \cvetril{e}{\ph}{\sman}+\cvetril{f}{\ph}{\sman}
  -\siws \Pgg^{F}(0) - \delrho{F}                       
  +\Dz{F}{\sman} + \Pzg^{F}(\sman)                                
\nll &&
+\frac{1}{4\cows}\lpar\vma{\fe}{2}+3\vcs{\ff}+\acs{\ff}\rpar 
   \cvetril{}{\sss{\zb}}{\sman}+\hvetri{}{\sss{\wb}}{\sman}             
\nll &&
+\cows\lpar \Rz-1 \rpar 
   \Biggl[\frac{|\qf|}{2\cows}
   \lpar 1-4|\qf|\siws\rpar\cvetril{}{\sss{\zb}}{\sman} 
  -|\qfp|\cvetri{}{\wb a}{\sman}+\hvetri{}{\wb n}{\sman} \Biggr]  
\nll &&
+\frac{3}{2}\fbff{0}{-\sman}{\mwl}{\mwl}-\frac{11}{18}-\frac{\rt}{4}
   \Bigl[\fbff{0}{-\sman}{\mwl}{\mwl}+1\Bigr] \Biggr\},
\\
\vvertil{}{\sss{LQ}}{\sman,\tman}&=&
  \gspi \Biggl\{     
   \cvetril{e}{\ph}{\sman}+\cvetril{f}{\ph}{\sman}
  -\siws\Pgg^{F}(0)-\delrho{F}+\Dz{F}{\sman}-\Pzg^{F}(\sman)       
\nll &&
+\frac{1}{4\cows}\lpar 3\vcs{\fe}+\acs{\fe}+\vma{\ff}{2}\rpar
   \cvetril{}{\sss{\zb}}{\sman}+\hvetri{(0)}{\sss{\wb}}{\sman}     
\nll &&
+\cows\lpar\Rz-1\rpar
   \Biggl[\frac{|\qe|}{2\cows}\lpar 1-4|\qe|\siws\rpar
   \cvetril{}{\sss{\zb}}{\sman} + \hvetri{(0)}{\wb n}{\sman}\Biggr] 
\nll &&
+\frac{3}{2}\fbff{0}{-\sman}{\mwl}{\mwl}-\frac{11}{18}\Biggr\},
\\ 
\vvertil{}{\sss{QQ}}{\sman,\tman}&=&
  \gspi \Biggl\{     
   \Bigl[\cvetril{e}{\ph}{\sman}+\cvetril{\ff}{\ph}{\sman}\Bigr] 
   \Biggl[1+\frac{\cows}{\siws}\lpar 1
          -\Rz \rpar\Biggr]-\siws \Pgg^{F}(0)              
\nll &&
-\frac{1}{\siws}\delrho{F}+\Dz{F}{\sman}-2\Pzg^{F}(\sman)
-\cows\lpar\Rz-1\rpar
    \lrbr{\hat\Pgg}^{{\bos},F}(\sman)-\frac{2}{3}\rrbr                      
\nll &&
-\frac{(\Rw-1)}{4\cows\siws} 
   \Bigl(\vma{\fe}{2} + \vma{\ff}{2}\Bigr)\cvetril{}{\sss{\zb}}{\sman}
  +\frac{4}{3}\fbff{0}{-\sman}{\mwl}{\mwl}-\frac{13}{18}
\Biggr\},
\label{process_ff}
\eqa
where we use the definitions
\bqa
\Dz{F}{\sman}  &=& \hDz{\bos}{\sman}+\bDz{\fer}{\sman},              
\\
\Pzg^{F}(\sman)&=& {\hat\Pzg}^{{\bos},F}(\sman)+\Pzg^{\fer,F}(\sman),
\eqa
and as usually decompose:
\bq
\delrho{F} = \delrho{{\bos},F}+\delrho{{\fer},F}.
\eq
We complete this Subsection by presenting the massless limit of  the
$\wb\wb$ box contribution:
\bqa
\hboxc{\sss{\wb\wb}}{d,0}{\sman,\tman}&=&
   \Biggl[-t\lpar 1+\frac{t^2}{u^2}\rpar
   -4\frac{\mws t^2}{u^2}+2\frac{\mwq}{u}
\lpar 1+2\frac{\sman}{u}\rpar\Biggr]\dff{0}{-\sman}{-\tman}{\mwl}{0}{\mwl}{0}
\nll &&
-2\lpar 2+2\frac{\sman}{u}+\frac{s^2}{u^2}
-2\frac{\mws s}{u^2}\rpar\cff{0}{-\sman}{\mwl}{0}{\mwl}    
\nll &&
+2\lpar 2+3\frac{\sman}{u}+\frac{s^2}{u^2}+2\frac{\mws t}{u^2}\rpar
 \cff{0}{-\tman}{0}{\mwl}{0}                                 
\nll &&
-\frac{2}{u}\Bigl[\fbff{0}{-\sman}{\mwl}{\mwl}-\fbff{0}{-\tman}{0}{0}\Bigr],
\label{box_ww_d0}
\eqa
and of its addition with the $\mtl$ dependence:
\bqa
\hboxc{\sss{\wb\wb}}{d,t}{\sman,\tman}
&=&\hboxc{\sss{\wb\wb}}{d}{\sman,\tman}-\hboxc{\sss{\wb\wb}}{d,0}{\sman,\tman}
\nll
&=& \Biggl[-\tman\lpar 1+\frac{t^2}{u^2}\rpar-4\frac{\mws t^2}{u^2}
+2\frac{\mwq}{\uman}\lpar 1+2\frac{\sman}{u}\rpar\Biggr]                 
\nll &&
\times\Bigl[ \dff{0}{-\sman}{-\tman}{\mwl}{0}{\mwl}{\uml}
            -\dff{0}{-\sman}{-\tman}{\mwl}{0}{\mwl}{0}\Bigr]                 
\nll &&
-\lpar 2+2\frac{\sman}{u}+\frac{s^2}{u^2}-2\frac{\mws s}{u^2}\rpar   
\Bigl[
\cff{0}{-\sman}{\mwl}{\uml}{\mwl} - \cff{0}{-\sman}{\mwl}{0}{\mwl}\Bigr]     
\nll &&
+\lpar 2+3\frac{\sman}{u}+\frac{s^2}{u^2}+2\frac{\mws t}{u^2}\rpar   
\nll &&
\times\Bigl[\cff{0}{-\tman}{0}{\mwl}{\uml}
           +\cff{0}{-\tman}{\uml}{\mwl}{0}
          -2\cff{0}{-\tman}{0}{\mwl}{0}\Bigr]                         
\nll &&
+\frac{2}{u}\Bigl[\fbff{0}{-\tman}{\uml}{0}-\fbff{0}{-\tman}{0}{0}\Bigr] 
\nll &&
+\rt\Biggl\{\Biggl[2+3\frac{\sman}{u}+2\frac{s^2}{u^2}-2\frac{\mws}{u} 
  \lpar 1+2\frac{\sman}{u}\rpar+\frac{\um s}{u^2}\Biggr]             
\nll &&
\times\mws\dff{0}{-\sman}{-\tman}{\mwl}{0}{\mwl}{\uml}     
\nll &&
-\frac{1}{2}\Biggl[\Rw\lpar 1+2\frac{s^2}{u^2}\rpar-4
     +\rt \lpar 1-2\Rw \rpar+\frac{\rts}{\Rw}\Biggr]             
\nll &&
\times\cff{0}{-\sman}{\mwl}{\uml}{\mwl}-\frac{\mws s}{u^2} 
      \cff{0}{-\sman}{\mwl}{0}{\mwl}                                        
\nll &&
-\frac{\mws t}{u^2} 
\Bigl[\cff{0}{-\tman}{0}{\mwl}{\uml}
     +\cff{0}{-\tman}{\uml}{\mwl}{0}\Bigr]                                  
\nll &&
+\frac{1}{2s}\lrbr\lpar 1-\rt \rpar 
     \bigl[1-\fbff{0}{-\sman}{\mwl}{\mwl}\Bigr]+\rt\ln\rt\rrbr\Biggr\}.
\label{box_ww_dt}
\eqa
\subsubsection{One-loop expression for $\dr$}\label{oneloop_dr}
In order to construct final expressions for $\rho$'s and $\kappa$'s
we will need the well-known quantity $\deltar$.
It's derivation involves the full machinery of calculation of one-loop
corrections for $\flm$-decay that is similar to those presented for the 
decay and the scattering process here.
However, the calculation is lengthy and we
limit ourselves to the presentation of the resulting expression
\cite{Bardin:1980fet,Bardin:1982svt}.
It is free of any singularities and reads as follows: 
\bqa
\deltar&=&\gspi\Biggl\{ 
         -\frac{2}{3}-\Pgg^{\fer,F}(0)+\frac{\ctws}{\stws}
          \delrho{F} + \frac{1}{\stws} \Biggl[
          \delrho{F}_{\sss{\wb}}
         +\frac{11}{2}-\frac{5}{8} \ctws (1+\ctws)
         +\frac{9}{4} \frac{\ctws}{\stws} \ln \ctws \Biggr]
               \Biggr\},
\nll
\label{delta_r-old}
\eqa
where a new self-energy combination appears:
\bq
\Delta\rho^{F}_{_{W}}=\frac{1}{\mws}
\lrbr\Sigma^{F}_{\sss{WW}}(0) 
    -\Sigma^{F}_{\sss{WW}}(\mws)\rrbr.
\label{delta_rhoFW_old}
\eq
It is useful to present the {\em bosonic} components of the two
$\rho$-factors entering \eqn{delta_r}:
\bqa
\Delta\rho^{{\bos},F}_{_{W}}&=&
  \lpar \frac{1}{12\ctwf} + \frac{4}{3\ctws}
 -\frac{17}{3}-4\ctws\rpar\fbff{0}{-\mws}{\zml}{\wml}  
\nll &&
 +\lpar 1-\frac{1}{3}\rw+\frac{1}{12}\rws \rpar\fbff{0}{-\mws}{\hml}{\wml} 
\nll &&
 + \lpar \frac{3}{4(1-\rw)}+\frac{1}{4}-\frac{1}{12}\rw\rpar
   \rw\ln\rw+\lpar\frac{1}{12 \ctwf}+\frac{17}{12 \ctws} 
 - \frac{3}{\stws}+\frac{1}{4}\rpar \ln \ctws          
\nll &&
 + \frac{1}{12 \ctwf}+\frac{11}{8 \ctws}+\frac{139}{36}
 - \frac{177}{24}\ctws+\frac{5}{8}\ctwf-\frac{1}{12}\rw
   \lpar\frac{7}{2}-\rw \rpar,           
\label{bosonic_delrhoW}
\eqa
and
\bqa
\Delta\rho^{{\bos},F}
&=& \lpar\frac{1}{12 \ctws} + \frac{4}{3}  
        - \frac{17}{3}\ctws - 4 \ctwf \rpar
\nl&&\times    
\lrbr\fbff{0}{-\mzs}{\wml}{\wml} 
      - \frac{1}{\ctws}\fbff{0}{-\mws}{\mzl}{\wml} \rrbr
\nll  &&
 - \lpar 1 - \frac{1}{3}\rw +\frac{1}{12} \rws \rpar
            \fbff{0}{-\mws}{\hml}{\wml} 
\nll &&
 + \lpar 1 - \frac{1}{3}\rz+\frac{1}{12} \rzs \rpar
            \frac{1}{\ctws}\fbff{0}{-\mzs}{\hml}{\wml}
\nll &&
 + \frac{1}{12} \stws \rws \lpar\ln \rw - 1 \rpar
- \lpar\frac{1}{12 \ctwf} + \frac{1}{2 \ctws}
         -2 + \frac{1}{12} \rw \rpar \ln\ctws
\nll &&
-\frac{1}{12\ctwf} - \frac{19}{36\ctws} 
   - \frac{133}{18} + 8 \ctws.
\label{bosonic_delrho}
\eqa
The expressions for their {\em fermionic} components are derived 
straightforwardly from the defining equations.
\subsection{The form factors $\rho^{\ff}_{\sss{\zb}}, 
\kappa^{\ff}_{\sss{\zb}}$, and
$\rho_{ef}, \kappa_{ef}, \kappa_{e}, \kappa_{f}$
\label{constphokap}}
In this Subsection we complete the construction of one-loop corrected 
amplitudes for the decay and the scattering process. 
First of all we observe that the final expressions for the one-loop
form factors
\eqn{decay_ff} (for the decay) and \eqn{process_ff} (for the
scattering process) 
contain common factors:
 $\cvetril{f}{\ph}{\mzs}$ and 
$\cvetril{e}{\ph}{\sman}+\cvetril{f}{\ph}{\sman}$.
These factors
have to be excluded from the electroweak form factors and shifted
to the QED part of the corrections (together with $\ph\ph$ and $\ph\zb$ boxes)
in order to be eventually combined with the bremsstrahlung
contributions thus forming the infrared free complete QED correction.

Secondly, we change the normalization $\alpha(0)\to\gf$, thereby making profit 
of the precisely known quantity $\gf$.
\subsubsection{Expression for $\rho^{\ff}_{\sss{\zb}}$ and 
$\kappa^{\ff}_{\sss{\zb}}$ for the $\zb$-decay
}
The sum of the one-loop $\zb$-decay amplitude and the corresponding Born
amplitude is being rewritten as:
\bq
V^{\zb\ff\fbf}_{\mu}\lpar\mzs\rpar = 
\lpar 2\pi\rpar^4\ib\,\ib\sqrt{\sqrt{2}\gf\mzs} 
\sqrt{\rho^{\ff}_{\sss{\zb}}}\tcif\gamma_\mu
\lrbr\lpar 1+\gfd \rpar - 4|\qd|\stws\kappa^{\ff}_{\sss{\zb}}\rrbr.
\label{decayrhokappadef-old}
\eq
The new electroweak effective couplings 
$\rho^{\ff}_{\sss{\zb}}$ and $\kappa^{\ff}_{\sss{\zb}}$
are related to the one-loop decay form factors of \eqn{decay_ff}:
\bqa
\rho^{\ff}_{\sss{\zb}}  &=& 1 +2F_{\sss{L}} - \stws \deltar,  
\\
\kappa^{\ff}_{\sss{\zb}}&=& 1 + F_{\sss{Q}} - F_{\sss{L}}.
\eqa
One easily derives the final
expressions for $\rho^{\ff}_{\sss{\zb}}$ and $\kappa^{\ff}_{\sss{\zb}}$:
\bqa
\rho^{\ff}_{\sss{\zb}}  &=& 1+\gspi\lrbr 
   \zmz -\delrho{F}_{\sss{\zb}}
  -\frac{11}{2}+\frac{5}{8} \ctws(1+\ctws)
  -\frac{9}{4}\frac{\ctws}{\stws} \ln \ctws + 2 u_f\rrbr,\qquad
\label{rho_Z_f}
\\
\kappa^{\ff}_{\sss{\zb}}&=& 1+\gspi\lrbr
              - \frac{\ctws}{\stws} \delrho{F}
              + \Pzg^{F}(\mzs)
              + \frac{\vmads}{4\ctws}\cvetril{}{\sss{\zb}}{\mzs} - u_f\rrbr,
\label{kappa_Z_f}
\eqa
where two new quantities are used:
\bqa 
 \delrho{F}_{\sss{\zb}}&=& \frac{1}{\mws} 
\lrbr \Sigma^{F}_{\sss{\wb\wb}}(0)
     -\Sigma^{F}_{\sss{\zb\zb}}(\mzs) 
\rrbr,
\label{delrhozb}
\\
u_f&=&\frac{3 \vd^2+\ad^2 }{4\ctws} 
      \cvetril{}{\sss{\zb}}{\mzs} +\cvetri{}{\sss{\wb}}{\mzs}.
\label{uf}
\eqa 
The first combination is a 'variation' of Veltman's $\rho$-parameter,
which has the same asymptotic behaviour as $\delrho{F}$ and, as usually,
may be presented as the sum of bosonic and fermionic components:
\bq  
\delrho{F}_{\sss{\zb}} =
\delrho{{\bos},F}_{\sss{\zb}}+\delrho{\fer,F}_{\sss{\zb}}. 
\eq
\subsubsection{Construction of $\rho_{ef}$ and $\kappa_{ij}$ for the
scattering process
\label{constructionrhokappafortheprocess}}
Similarly to \eqn{decayrhokappadef}, we define the net sum of Born and 
one-loop amplitudes for the scattering process 
in terms of four effective form factors $\rho_{ef}(\sman,\tman)$ and 
$\kappa_{ij}(\sman,\tman)$:
\bqa
{\cal A}^{\sss{OLA}}_{\sss{\zb}}(\sman,\tman)&=&\ib
        \sqrt{2} G_\mu \tcie \tcif \mzs  
        \chi_{\sss{Z}}(\sman) \rho_{ef}(\sman,\tman)
        \biggl\{
        \gadu{\mu}{\lpar 1+\gfd \rpar }
        \otimes \gadu{\mu} { \lpar 1+\gfd \rpar}    
\nll &&
-4 |\qe | \stws \kappa_e(\sman,\tman)
        \gadu{\mu} \otimes \gadu{\mu}{\lpar1+\gfd\rpar}
       -4 |\qf | \stws \kappa_f(\sman,\tman)
        \gadu{\mu} {\lpar 1+\gfd \rpar } 
        \otimes \gadu{\mu}                              
\nll &&
+16 |\qe \qf| \stwf \kappa_{e,f}(\sman,\tman)
        \gadu{\mu} \otimes \gadu{\mu} \biggr\}.
\label{processrhokappa-old}
\eqa
The form factors are simply related to the one-loop form factors
\eqn{process_ff}: 
\bqa
\rho_{ef}  &=& 1+\vvertil{}{\sss{LL}}{\sman,\tman}-\siws\Delta r,       
\\
\kappa_{e} &=& 1+\vvertil{}{\sss{QL}}{\sman,\tman}
                -\vvertil{}{\sss{LL}}{\sman,\tman},  
\\
\kappa_{f} &=& 1+\vvertil{}{\sss{LQ}}{\sman,\tman}
                -\vvertil{}{\sss{LL}}{\sman,\tman},  
\\
\kappa_{ef}&=& 1+\vvertil{}{\sss{QQ}}{\sman,\tman}
                -\vvertil{}{\sss{LL}}{\sman,\tman}.
\eqa
After some trivial algebra one derives the final expressions:
\bqa
\label{rhoef}
\rho_{ef}&=&
  1+ \gspi \Biggl\{
  -\delrho{F}_{\sss{\zb}}+\Dz{F}{\sman}
  +\frac{5}{3}\fbff{0}{-\sman}{\mwl}{\mwl}
  -\frac{9}{4}\frac{\cows}{\siws}\ln\cows
  -6
\nll &&
+\frac{5}{8}\cows\lpar 1+\cows\rpar       
  +\frac{1}{4\cows}
  \lpar 3\vcs{\fe}+\acs{\fe}+3\vcs{\ff}+\acs{\ff}\rpar
  \cvetril{}{\sss{\zb}}{\sman}
  +\hvetri{0}{\sss{\wb}}{\sman}+\hvetri{}{\sss{\wb}}{\sman}  
\nll &&
  -\frac{\rt}{4}\Bigl[\fbff{0}{-\sman}{\mwl}{\mwl}+1\Bigr]
  -\cows\lpar\Rz-1\rpar\sman\hboxc{\sss{\wb\wb}}{d}{\sman,\tman} 
\Biggr\},
\\ 
\label{kappae}
\kappa_{e}&=&1+\gspi\Biggl\{
-\frac{\cows}{\siws} \delrho{F}-\Pzg^{F}(\sman)
-\frac{1}{6}\fbff{0}{-\sman}{\mwl}{\mwl}-\frac{1}{9}         
-\frac{\vc{\fe}\vpa{\fe}{}}{2\cows}\cvetril{}{\sss{\zb}}{\sman}
\nll &&
              - \hvetri{0}{\sss{\wb}}{\sman}               
+\lpar \Rz-1 \rpar \Biggl[
  \frac{|\qf|}{2}\lpar 1-4|\qf|\siws\rpar
\cvetril{}{\sss{\zb}}{\sman}
+\cows\Bigl[\hvetri{}{\wb n}{\sman}
\nll &&
-|\qfp|\cvetri{}{\wb a}{\sman}
    +\sman\hboxc{\sss{\wb\wb}}{d}{\sman,\tman}\Bigr]\Biggr]\Biggr\},
\\ 
\label{kappaf}
\kappa_{f}&=&1+\gspi\Biggl\{ 
-\frac{\cows}{\siws}\delrho{F}-\Pzg^{F}(\sman)
-\frac{1}{6}\fbff{0}{-\sman}{\mwl}{\mwl}-\frac{1}{9}      
-\frac{\vc{\ff}\vpa{\ff}{}}{2\cows} 
      \cvetril{}{\sss{\zb}}{\sman}
\nll &&
-\hvetri{}{\sss{\wb}}{\sman}
+\lpar \Rz-1 \rpar\Biggl[
  \frac{|\qe|}{2}\lpar 1-4|\qe|\siws\rpar
\cvetril{}{\sss{\zb}}{\sman}    
+\cows\Bigl[\hvetri{0}{\wb n}{\sman}
\nll &&
 -|\qep|\cvetri{}{\wb a}{\sman}
 +\sman\hboxc{\sss{\wb\wb}}{d}{\sman,\tman}\Bigr]\Biggr]          
-\frac{\rt}{4}\Bigl[\fbff{0}{-\sman}{\mwl}{\mwl}+1\Bigr]
\Biggr\},
\\ 
\kappa_{ef}&=& 
 1+ \gspi \Biggl\{ 
  -2\frac{\cows}{\siws} \delrho{F}-2\Pzg^{F}(\sman) 
  -\frac{1}{3}\fbff{0}{-\sman}{\mwl}{\mwl}-\frac{2}{9}    
\nll &&
-\frac{1}{4\cows}\Biggl[ \frac{\vma{\fe}{2}+\vma{\ff}{2}}{\siws}
  \lpar \Rw-1 \rpar+3\vcs{\fe}+\acs{\fe}+3\vcs{\ff}+\acs{\ff}
  \Biggr] \cvetril{}{\sss{\zb}}{\sman}                    
\nll &&
-\hvetri{0}{\sss{\wb}}{\sman}-\hvetri{}{\sss{\wb}}{\sman}
-\frac{\rt}{4}\Bigr[\fbff{0}{-\sman}{\mwl}{\mwl}+1\Bigr]
\nll &&
+\cows\lpar \Rz-1 \rpar 
\lrbr\frac{2}{3}-{\hat\Pgg}^{{\bos},F}(\sman) 
   +\sman\hboxc{\sss{\wb\wb}}{d}{\sman,\tman}\rrbr\Biggr\}.\qquad\qquad
\label{processrhokappas}
\eqa
In \eqn{processrhokappas} we used the definitions:
\bqa
 \hvetri{ }{\sss{\wb}}{\sman}&=& 
  \ctws\hvetri{}{\wb n}{\sman}  
 -\frac{1}{2} \vpadpa \cvetri{ }{\wb a }{\sman}  
 -\frac{1}{2} \avetri{ }{\wb a }{\sman},  
\\
\hvetri{0}{\sss{\wb}}{\sman}&=& \cows\hvetri{0}{\wb n}{\sman}
                             -\frac{1}{2} \cvetri{0}{\wb a}{\sman},
\\[1mm]
\Dz{F}{\sman}&=&\hDz{\bos}{\sman}+\bDz{\fer}{\sman},
\\[2.5mm]
\Pzg^{F}(\sman)&=&{\hat \Pzg}^{{\bos},F}(\sman)+\Pzg^{{\fer},F}(\sman).
\label{forprocessrhokappadef}
\eqa
These effective form factors are used in {\tt ZFITTER} for the calculation
of the one-loop EWRC.

\section{Relations Between Earlier and Actual Notations \label{bridges}}
Here we provide some useful relations between notations used in 
this description and notations of Section 4.4 of \cite{Bardin:1997xq}.
\subsection{N-point functions}
Relations between definitions of two-point functions:
\bqa
\fbff{0}{-\sman}{\mwl}{\mwl}-2&=&\frac{1}{2\sman}L_{_{\wb\wb}}(\sman)
=\lpar\smans-4\mws\sman\rpar{\cal F}(-\sman,\mws,\mws),
\\
\fbff{0}{-\sman}{\mhl}{\mzl}&=&
   \ln\cows - \frac{1}{2}\ln\rhz + 2
   -\frac{1}{2}\Rz\lpar\rhz-1\rpar\ln\rhz
   +\frac{1}{2s}L_{_{\zb\hb}}(\sman).\qquad\qquad
\eqa
The definitions of ${\cal F}(-\sman,\mws,\mws)$, 
${\cal F}_3(\sman,\mws)$, and ${\cal F}_4(\sman,I_1,\mws)$ below
were introduced in~\cite{Bardin:1989dit}.

Relations between definitions of three- and four-point functions:
\bqa
\begin{array}{lclclcl} 
C_0(-\sman;\uml,\wml,\uml)        &=&{\tt XS3T},   & &    \\
C_0(-\sman;0,\wml,0)              &=&{\tt XS3T0},  & &    \\ 
C_0(-\sman;\wml,\uml,\wml)        &=&{\tt XS3W},   & &    \\ 
C_0(-\sman;\wml,0,\wml)           &=&{\tt XS3W0}   &=&
    -{\ds{\frac{1}{s}}}{\cal F}_3(\sman,\mws),            \\ 
C_0(-\sman;\uml,\wml,0)           &=&{\tt XJ3WT},  & &    \\
C_0(-\sman;0,\wml,0)              &=&{\tt XJ3W0},  & &    \\ 
D_0(-\sman,\tman;\wml,0,\wml,\uml)&=&{\tt XS4WT},  & &    \\
D_0(-\sman,\tman;\wml,0,\wml,0)   &=&{\tt XS4W0}   &=&
     {\ds{\frac{1}{{s}^2}}}{\cal F}_4(\sman,-\uman,\mws).
\end{array}
\eqa
They are supplied by calls to different subroutines:
\[
\begin{array}{ll} 
&\SUBR{CALL S3WANA(AMT2,AMW2,-S,XJ0W,XS3W,XS3W0)} \\[-2mm]
&\SUBR{CALL S3WANA(AMW2,AMT2,-S,XJ0T,XS3T,XS3T0)} \\[-2mm]
&\SUBR{CALL J3WANA( 0D0,AMW2,AI12,XJ3W0)}         \\[-2mm]       
&\SUBR{CALL J3WANA(AMT2,AMW2,AI12,XJ3WT)}         \\[-2mm]      
&\SUBR{CALL S4WANA(AMT2,AMW2,-S,AI12,XS4WT)}      \\[-2mm]     
&\SUBR{CALL S4WANA( 0D0,AMW2,-S,AI12,XS4W0)}    
\end{array}
\]
\vspace*{-5mm}

Here are the relations between the definitions of the self-energy
contributions $\ph$ and $\zb$ (see \eqns{pigg_pf}{Dz_bF}):
\bqa
A(\sman) &=& {\hat \Pgg}^{{\bos},F}(\sman),
\\
D_{\sss{Z},f}(\sman) &=&\Dz{{\fer},F}{\sman},
\\
D_{\sss{Z},b}(\sman) &=&\hDz{\bos}{\sman},
\label{ssss5} 
\eqa
and the $\ph\zb$-mixing functions (see \eqns{pizg_fer}{fermionicsplit}):
\bq
M_f(\sman) = {\tt XAMM1F} = -\Pzg^{{\fer},F}(\sman).
\label{ssss8}
\eq

Relations between derivatives of the $\zb$ self-energy function (see
\eqns{derzbosonic}{derzfermionic}) are:
\bqa
Z^F_{b}(\mzs) &=& {\tt XZFM1} = \frac{1}{\cows}
\biggl[\Sigma^{'}_{\sss{\zb\zb}}(\mzs) \biggr]^{{\bos},F},
\nll
Z^F_{f}(\mzs) &=& {\tt XZFM1F}= \frac{1}{\cows}
\biggl[\Sigma^{'}_{\sss{\zb\zb}}(\mzs) \biggr]^{{\fer},F}.
\label{zfzft}
\eqa
Those for the bosonic parts of self-energy counter-terms (see
\eqn{sig_ww}) are:
\bqa
\label{Ol_W0}
W_{b}(0)   &=&{\tt W0}=\frac{1}{\mws}\Sigma^{\bos}_{_{\wb\wb}}(0),
\\
\label{Ol_W1}
W_{b}(\mws)&=&{\tt XWM1}=\frac{1}{\mws}\Sigma^{\bos}_{_{\wb\wb}}(\mws),
\\
\label{Ol_Z1}
Z_{b}({\mzs})&=&{\tt XZM1}=\frac{1}{\cows\mzs}\Sigma^{\bos}_{_{\wb\wb}}(\mzs),
\eqa
and for the fermionic parts  (see \eqnsc{fermionicsplit}{sig_wwf}):
\bqa
Z_{f}(\mzs)&=&{\tt XZM1F} = \frac{1}{\cows\mzs}
\Sigma^{{\fer},F}_{\sss{\zb\zb}}(\mzs),
\\
W_{f}(\mws)&=&{\tt XWM1F} = \frac{1}{\mws} 
\Sigma^{{\fer},F}_{_{\wb\wb}}(\mws).
\eqa
\subsection{Vertex functions}  
For the vertex functions (see \eqnsc{cal_Fz}{cal_all_Fw})
the relations between different notations are:
\ba
V_{1_Z}(\sman) &=& \cvetri{ }{\zb  }{\sman},
\\
V_{1_W}(\sman) &=& \cvetri{ }{\wb a}{\sman},
\\
V_{2_W}(\sman) &=& \hvetri{0}{\wb n}{\sman}.
\label{ssss4} 
\ea

 The vertex functions in the limit of vanishing fermion masses
are calculated in subroutine {\tt VERTZW}: 
\ba
{\tt V1ZZ} &=& \cvetri{ }{\zb   }{\mzs},   
\\
{\tt V1ZW} &=& \cvetri{0}{\wb a }{\mzs},
\\
{\tt V2ZWW}&=& \cvetri{0}{\wb n }{\mzs}.
\ea

\chapter{Subroutine {\tt ZFTEST} and Test Sample Output\label{ch-zftest}}

\section{Subroutine {\tt ZFTEST}\label{zftest} }
\eqnzero
The \zf\ distribution package includes subroutine {\tt ZFTEST}, which
serves essentially for three purposes:
\begin{itemize}
  \item[1.] It is an example of how to use \zf.
  \item[2.] It is an internal consistency check of the different \zf\
   branches.
  \item[3.] It allows one to check that \zf\ has been properly installed on
   the machine.
\end{itemize}
The routine creates a Table of cross-sections and asymmetries as
functions of \RS\ near the \Z\ peak, below, and above.
 
An example of how to run {\tt ZFTEST} is contained in the file {\tt
zfmai6$\_$21.f}, 
which contains some useful information:
\begin{verbatim}
*
* MAIN to call ZFTEST
*
* to compile and link:
*
* f77 -K zfmai6_21.f zfitr6_21.f dizet6_21.f acol6_1.f m2tcor5_11.f pairho.f
* bcqcdl5_14.f bkqcdl5_14.f bhang4_67.f zf514_aux.f -o zfitr6_21.exe
*
* Warning! option -K is mandatory on some WS's while on the others 
* leads to "illegal option -- K" while linking
*
* Starting 6.06, ZFITTER became Linux-compatible.
*
* In Dubna it is used under "Welcome to Linux 2.0.34"
* and compiled with the command suggested by G.Quast
*
* f77 *.f  -c -O3 -DCPU=596 -fno-automatic
*
      CALL ZFTEST(1)
      END
\end{verbatim}
After compiling, linking, and running of the whole package of eleven files
the results presented in \subsect{results} should be reproduced.
\subsection{Subroutine {\tt ZFTEST}\label{szftest}}
\begin{small}
\begin{verbatim}
      SUBROUTINE ZFTEST(IMISC)                                             *
*     ========== =============                                             *
****************************************************************************
*                                                                          *
*     SUBR. ZFTEST                                                         *
*                                                                          *
*     Example program to demonstrate the use of the ZFITTER package.       *
*                                                                          *
****************************************************************************
*
      IMPLICIT REAL*8(A-H,O-Z)
      COMPLEX*16 XALLCH,XFOTF
*
      DIMENSION XS(0:11,6),AFB(0:11,5),TAUPOL(2),TAUAFB(2),ALRI(0:11,2)
*
* constants
*
      PARAMETER(ALFAI=137.0359895D0,ALFA=1.D0/ALFAI,CONS=1.D0)
      PARAMETER(ZMASS=91.1867D0,TMASS=173.8D0,HMASS=100.D0)
      PARAMETER(ALFAS=.119D0)
      PARAMETER(RSMN=87.D0,DRS=1.D0,NRS=9)
      PARAMETER(ANG0=35D0,ANG1=145D0)
      PARAMETER(QE=-1.D0,AE=-.5D0,QU= 2.D0/3.D0,AU= .5D0,
     +                            QD=-1.D0/3.D0,AD=-.5D0)
*
* ZFITTER common blocks
*
      COMMON /ZUPARS/QDF,QCDCOR(0:14),ALPHST,SIN2TW,S2TEFF(0:11),
     & WIDTHS(0:11)
*
      COMMON /CDZRKZ/ARROFZ(0:10),ARKAFZ(0:10),ARVEFZ(0:10),ARSEFZ(0:10)
     &              ,AROTFZ(0:10),AIROFZ(0:10),AIKAFZ(0:10),AIVEFZ(0:10)
      COMMON /CDZAUX/PARTZA(0:10),PARTZI(0:10),RENFAC(0:10),SRENFC(0:10)
*
      COMMON /EWFORM/ XALLCH(5,4),XFOTF
*
*-----------------------------------------------------------------------
*
* initialize
*
      CALL ZUINIT
*
* set ZFITTER flags and print flag values
*
      CALL ZUFLAG('PRNT',1)
      CALL ZUFLAG('MISC',IMISC)
      CALL ZUFLAG('ISPP',1)
      ICUTC=-1
* 
* do weak sector calculations
*
      DAL5H=2.8039808929734D-02
*
      CALL ZUWEAK(ZMASS,TMASS,HMASS,DAL5H,ALFAS)
*
* define cuts for fermion channels and print cut values
*
      CALL ZUCUTS(11,0,15.D0,10.D0,0.D0,ANG0,ANG1,0.D0)
      CALL ZUINFO(0)
*
* make table of cross-sections and asymmetries
*
      PI   = DACOS(-1.D0)
*
* DO loop over S
*
      DO I = 1,8
*
* array of RS=SQRT(S)
*
        IF(I.EQ.1) RS=35D0
        IF(I.EQ.2) RS=65D0
        IF(I.EQ.3) RS=ZMASS-2D0
        IF(I.EQ.4) RS=ZMASS
        IF(I.EQ.5) RS=ZMASS+2D0
        IF(I.EQ.6) RS=100D0
        IF(I.EQ.7) RS=140D0
        IF(I.EQ.8) RS=175D0
*
* Changing of default, for instance:
* 1) NO PAIRS at PETRA and TRISTAN  
*
        IF(I.LT.3.) CALL ZUFLAG('ISPP',0)
        IF(I.GE.3.) CALL ZUFLAG('ISPP',1)
*
* The OUTPUT is adjusted for hp WS at Zeuthen
*
      IF(I.EQ.5.OR.I.EQ.8) THEN
        PRINT *
        PRINT *
        PRINT *
          ELSE
        PRINT *
      ENDIF
* table header
        PRINT *,' SQRT(S) = ',REAL(RS)
        PRINT *
        PRINT 1000
 1000   FORMAT(1X,'     '
     +  ,'<-------------- Cross-section -------------->'
     +  ,'  <------- Asymmetry ------->','  <--Tau_Pol-->',
     +   '  <----A_LR--->')
        PRINT 1001
 1001   FORMAT(1X,'INDF   ZUTHSM   ZUXSEC    ZUXSA   ZUXSA2   ZUXAFB'
     +,'   ZUTHSM  ZUXSA ZUXSA2 ZUXAFB ZUTPSM  ZUTAU  ZULRSM  ZUALR')
*
* loop over fermion indicies
*
        DO INDF = 0,11
         S=RS**2
         IF(INDF.NE.11) 
     +   CALL ZUCUTS(INDF,ICUTC,0D0,0D0,1D-2*S,0D0,180D0,.25D0*S)
* standard model interf. (INTRF=1)
         CALL ZUTHSM(INDF,RS,ZMASS,TMASS,HMASS,DAL5H,ALFAS,
     +     XS(INDF,1),AFB(INDF,1))
         IF(INDF.EQ.3) CALL ZUTPSM(RS,ZMASS,TMASS,HMASS,DAL5H,ALFAS,
     +     TAUPOL(1),TAUAFB(1))
* cross-section interf. (INTRF=2)
           GAME = WIDTHS(   1)/1000.
           GAMF = GAME          
           GAMZ = WIDTHS(11)/1000.
           IF(INDF.NE.11) GAMF = WIDTHS(INDF)/1000.          
           CALL ZUXSEC(INDF,RS,ZMASS,GAMZ,GAME,GAMF,XS(INDF,2))
* cross-section & forward-backward asymmetry interf. (INTRF=3)
         IF(INDF.NE.0 .AND. INDF.NE.10) THEN
           IF(IMISC.EQ.0) THEN
             ROEE= AROTFZ(1)
           ELSE
             ROEE= ARROFZ(1)
           ENDIF
          IF(IMISC.EQ.0) THEN
            GAE= SQRT(AROTFZ(1))/2
          ELSE
            GAE= SQRT(ARROFZ(1))/2
          ENDIF
          GVE  = ARVEFZ(1)*GAE
*
           IF(INDF.LT.11) THEN
             IF(IMISC.EQ.0) THEN
               ROFI= AROTFZ(INDF)
               GAF = SQRT(AROTFZ(INDF))/2
             ELSE
               ROFI= ARROFZ(INDF)
               GAF = SQRT(ARROFZ(INDF))/2
             ENDIF
             GVF = ARVEFZ(INDF)*GAF
           ELSEIF(INDF.EQ.11) THEN
             GAF = GAE
             GVF = ARVEFZ(1)*GAF
           ENDIF
           CALL ZUXSA(INDF,RS,ZMASS,GAMZ,0,GVE,GAE,GVF,GAF,
CB: MODE=1 CALL ZUXSA(INDF,RS,ZMASS,GAMZ,1,GVE/GAE,ROEE,GVF/GAF,ROFI,
     +     XS(INDF,3),AFB(INDF,3))
         ENDIF
* tau polarization interf. (INTRF=3)
         IF(INDF.EQ.3) CALL ZUTAU(RS,ZMASS,GAMZ,0,GVE,GAE,GVF,GAF,
     +     TAUPOL(2),TAUAFB(2))
* left-right polarization Asymmetry (IBRA=4)
         IF((INDF.GE.4.AND.INDF.LE.7).OR.INDF.EQ.9) THEN
          POL=.63D0
          CALL
     &    ZULRSM(INDF,RS,ZMASS,TMASS,HMASS,DAL5H,ALFAS,POL,XSPL,XSMI)
          ALRI(INDF,1)=(XSMI-XSPL)/(XSMI+XSPL)/POL
         ENDIF
         IF(INDF.EQ.10) THEN
          POL=.65D0
          XSMINS=0D0
          XSPLUS=0D0
          DO 2004 IALR=4,9
           IF(IALR.EQ.8) GO TO 2004
           CALL
     &     ZULRSM(IALR,RS,ZMASS,TMASS,HMASS,DAL5H,ALFAS,POL,XSPL,XSMI)
           XSMINS=XSMINS+XSMI
           XSPLUS=XSPLUS+XSPL
2004      CONTINUE
          ALRI(INDF,1)=(XSMINS-XSPLUS)/(XSMINS+XSPLUS)/POL
         ENDIF
*
* cross-section & forward-backward asymmetry interf. for gv**2 and ga**2(IBRA=4)
*
         IF(INDF.GE.1 .AND. INDF.LE.3 .OR. INDF.EQ.11) THEN
           IF(INDF.LT.11) THEN
             IF(IMISC.EQ.0) THEN
               ROFI= AROTFZ(INDF)
               GAF = SQRT(AROTFZ(INDF))/2
             ELSE
               ROFI= ARROFZ(INDF)
               GAF = SQRT(ARROFZ(INDF))/2
             ENDIF
             GVF = ARVEFZ(INDF)*GAF
           ELSEIF(INDF.EQ.11) THEN
             IF(IMISC.EQ.0) THEN
               ROFI= AROTFZ(1)
               GAF = SQRT(AROTFZ(1))/2
             ELSE
               ROFI= ARROFZ(1)
               GAF = SQRT(ARROFZ(1))/2
             ENDIF
             GVF = ARVEFZ(1)*GAF
           ENDIF
           GVF2= GVF**2
           GAF2= GAF**2
           CALL ZUXSA2(INDF,RS,ZMASS,GAMZ,0,GVF2,GAF2,
CB: MODE=1 CALL ZUXSA2(INDF,RS,ZMASS,GAMZ,1,GVF2/GAF2,ROEE**2,
     +     XS(INDF,4),AFB(INDF,4))
*
* cross-section & forward-backward asymmetry interf. for gve*gae*gvf*gaf, 
* gve**2+gae**2 and gvf**2+gaf**2 (IBRA=6)
*
           PFOUR=GVE*GVF*GAE*GAF
           PVAE2=GVE**2+GAE**2
           PVAF2=GVF**2+GAF**2
           IF(INDF.NE.11) THEN
            CALL ZUXAFB(INDF,RS,ZMASS,GAMZ,PFOUR,PVAE2,PVAF2,
     +      XS(INDF,5),AFB(INDF,5))
           ENDIF
         ENDIF
* S-matrix interf. (ISMA=1, via SMATASY)
* results
         IF(INDF.EQ.0) THEN
           PRINT 9000,INDF,(XS(INDF,J),J=1,2)
         ELSEIF(INDF.EQ.1) THEN
           PRINT 9010,INDF,(XS(INDF,J),J=1,5),AFB(INDF,1),
     +      (AFB(INDF,J),J=3,5)
         ELSEIF(INDF.EQ.2) THEN
           PRINT 9010,INDF,(XS(INDF,J),J=1,5),AFB(INDF,1),
     +      (AFB(INDF,J),J=3,5)
         ELSEIF(INDF.EQ.3) THEN
           PRINT 9015,INDF,(XS(INDF,J),J=1,5),AFB(INDF,1),
     +      (AFB(INDF,J),J=3,5),(TAUPOL(J),J=1,2)
         ELSEIF(INDF.EQ.10) THEN
           PRINT 9025,INDF,(XS(INDF,J),J=1,2),(ALRI(INDF,J),J=1,2)
         ELSEIF(INDF.EQ.11) THEN
           PRINT 9011,INDF,(XS(INDF,J),J=1,4),AFB(INDF,1),
     +      (AFB(INDF,J),J=3,4)
         ELSE
           PRINT 9020,INDF,(XS(INDF,J),J=1,3),AFB(INDF,1),AFB(INDF,3)
     +                  ,(ALRI(INDF,J),J=1,2)
          ENDIF
        ENDDO
        PRINT *
      ENDDO
      RETURN
 9000 FORMAT(1X,I4, 2F9.5)
 9010 FORMAT(1X,I4, 5F9.5, 2X,4F7.4)
 9011 FORMAT(1X,I4, 4F9.5,11X,4F7.4)
 9015 FORMAT(1X,I4, 5F9.5, 2X,6F7.4)
 9020 FORMAT(1X,I4, 3F9.5,20X,2F7.4,28X,2F7.4)
 9025 FORMAT(1X,I4, 2F9.5,71X,2F7.4)
*                                                             END ZFTEST
      END
\end{verbatim}
\end{small}
\subsection{{\tt ZFTEST} Results\label{results}}
Here we include the standard test-outputs, 
produced by a call to {\tt ZFTEST(0/1)} for two values of flag {\tt MISD}.

The first one corresponds to {\tt MISD=0}.
The numbers from
different interfaces agree only at $\sman=\mzs$.

The second output, produced with {\tt MISD=1}, exhibits a beautiful
agreement of numbers from different interfaces in a wide energy range,
$\sqrt{\sman}$ from $35$ to $175$ GeV (with the only exclusion for 
{\tt INDF=10} for $\sqrt{\sman}> 100$ GeV).
It is remarkable that due to 
a non-straightforward coding the CPU time increase with $\sman$-dependent 
remnants is only about $20\%$ compared to the use of {\tt MISD=0}.

\begin{figure}[t] 
\begin{center} 
\vspace*{-4.5cm}
\hspace*{-7.5cm}
\mbox{\epsfig{file=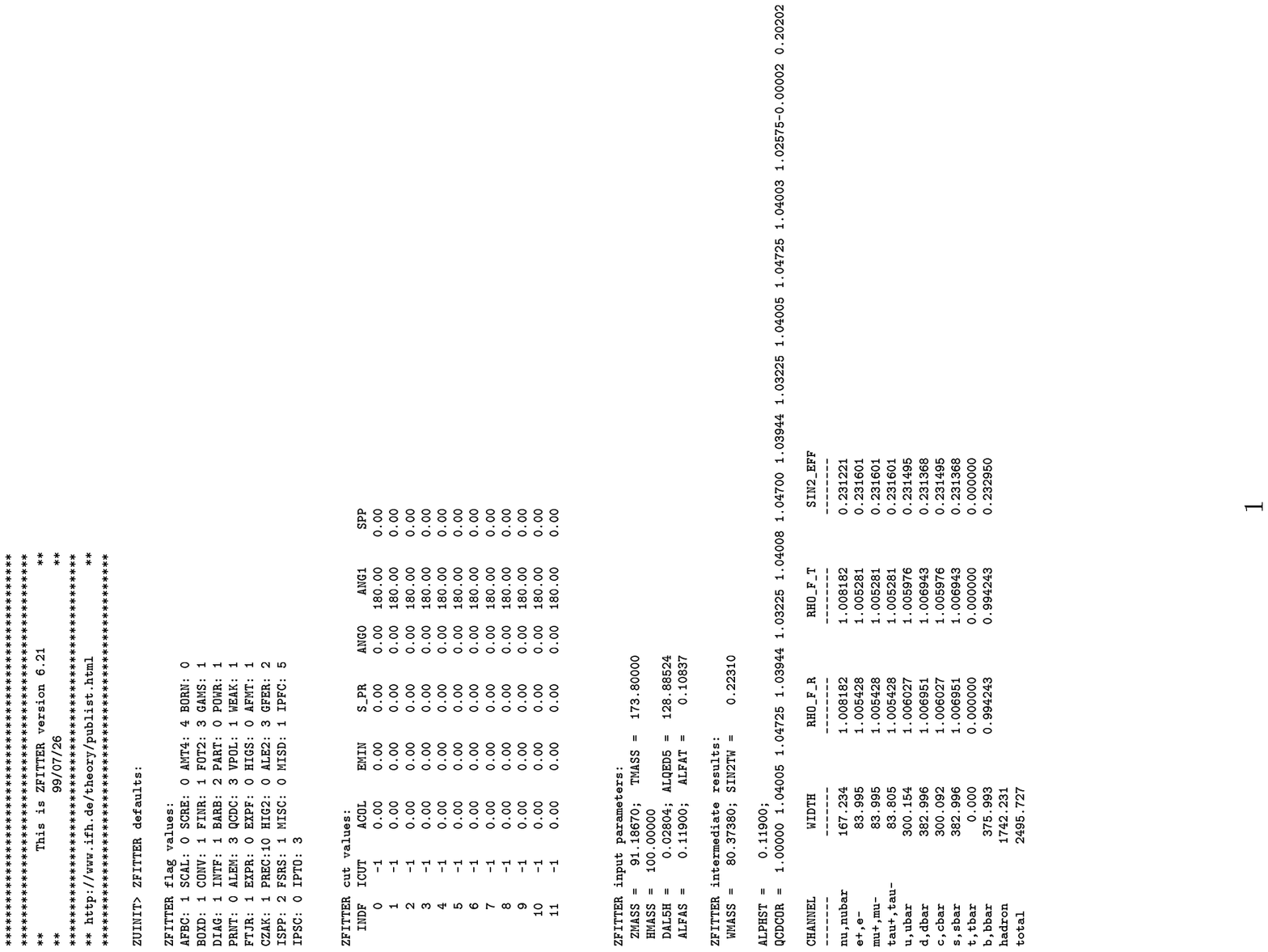
          ,height=35cm  
          ,width=35cm   
           ,angle=90   
      }}
\end{center}
\end{figure}

\begin{figure}[t] 
\begin{center} 
\vspace*{-4.5cm}
\hspace*{-7.8cm}
\mbox{\epsfig{file=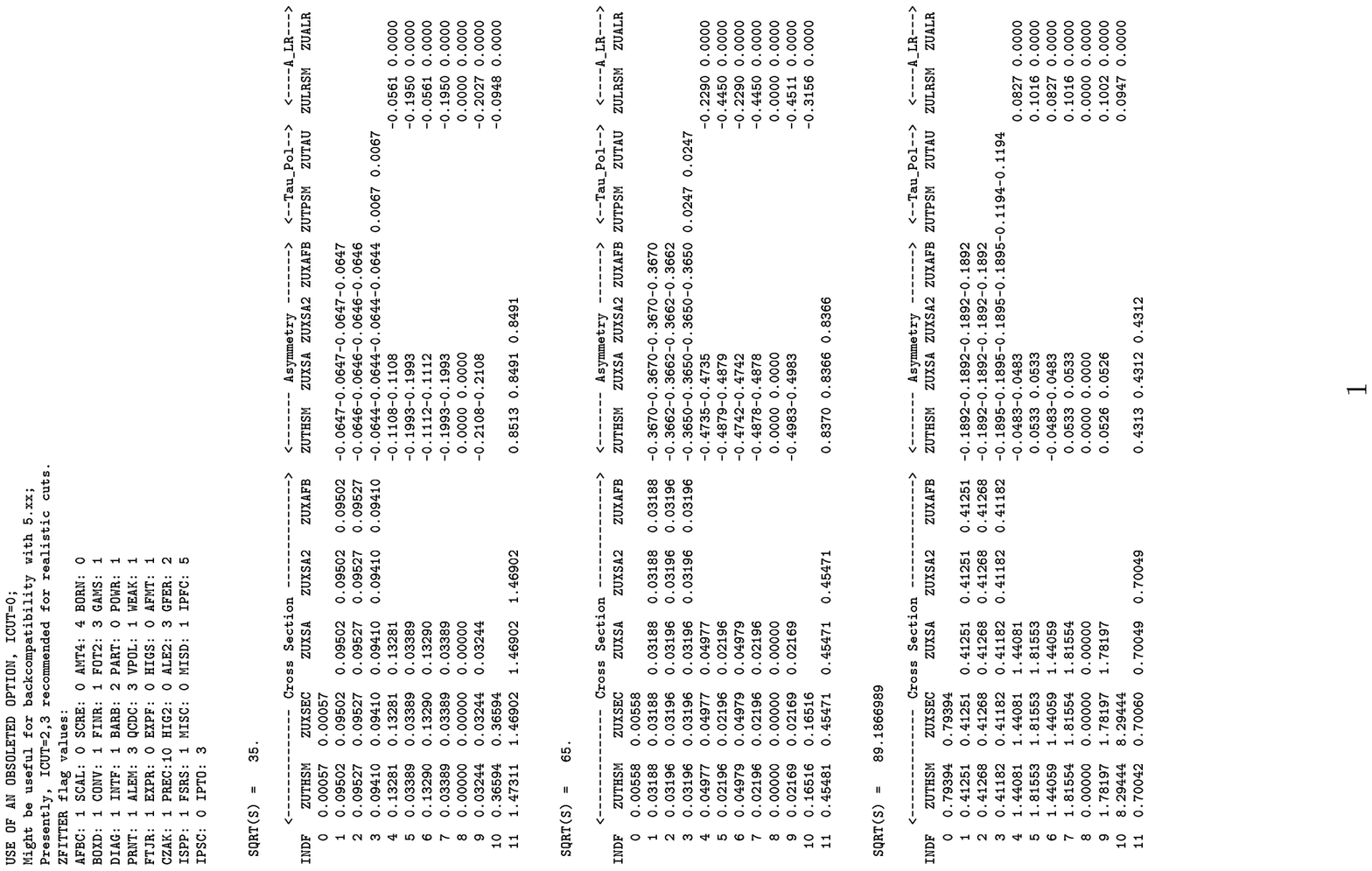
          ,height=35cm  
          ,width=35cm   
           ,angle=90   
      }}
\end{center}
\end{figure}

\begin{figure}[t] 
\begin{center} 
\vspace*{-4.5cm}
\hspace*{-7.5cm}
\mbox{\epsfig{file=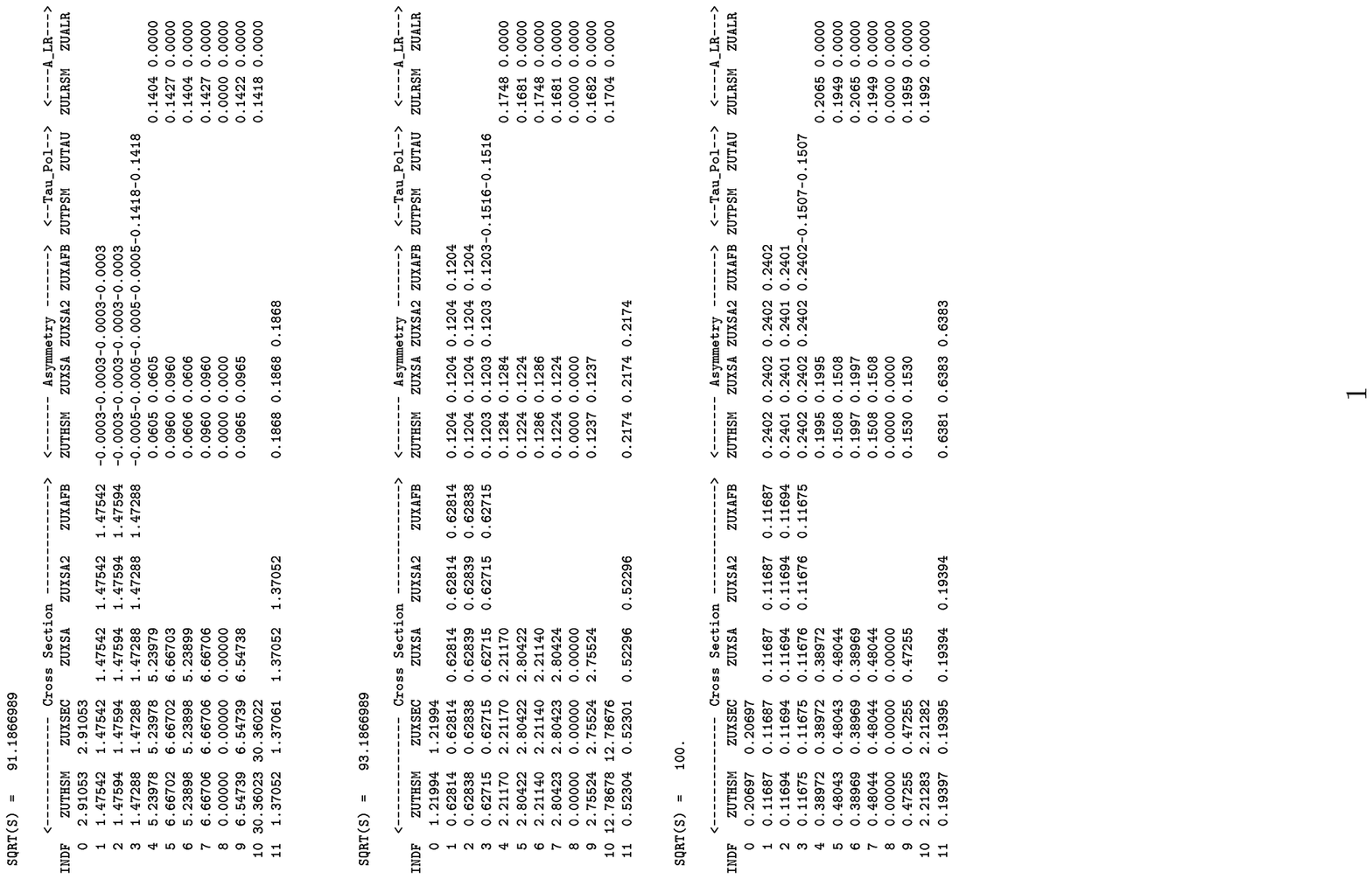
          ,height=35cm  
          ,width=35cm   
           ,angle=90   
      }}
\end{center}
\end{figure}

\begin{figure}[t] 
\begin{center} 
\vspace*{-4.5cm}
\hspace*{-7.5cm}
\mbox{\epsfig{file=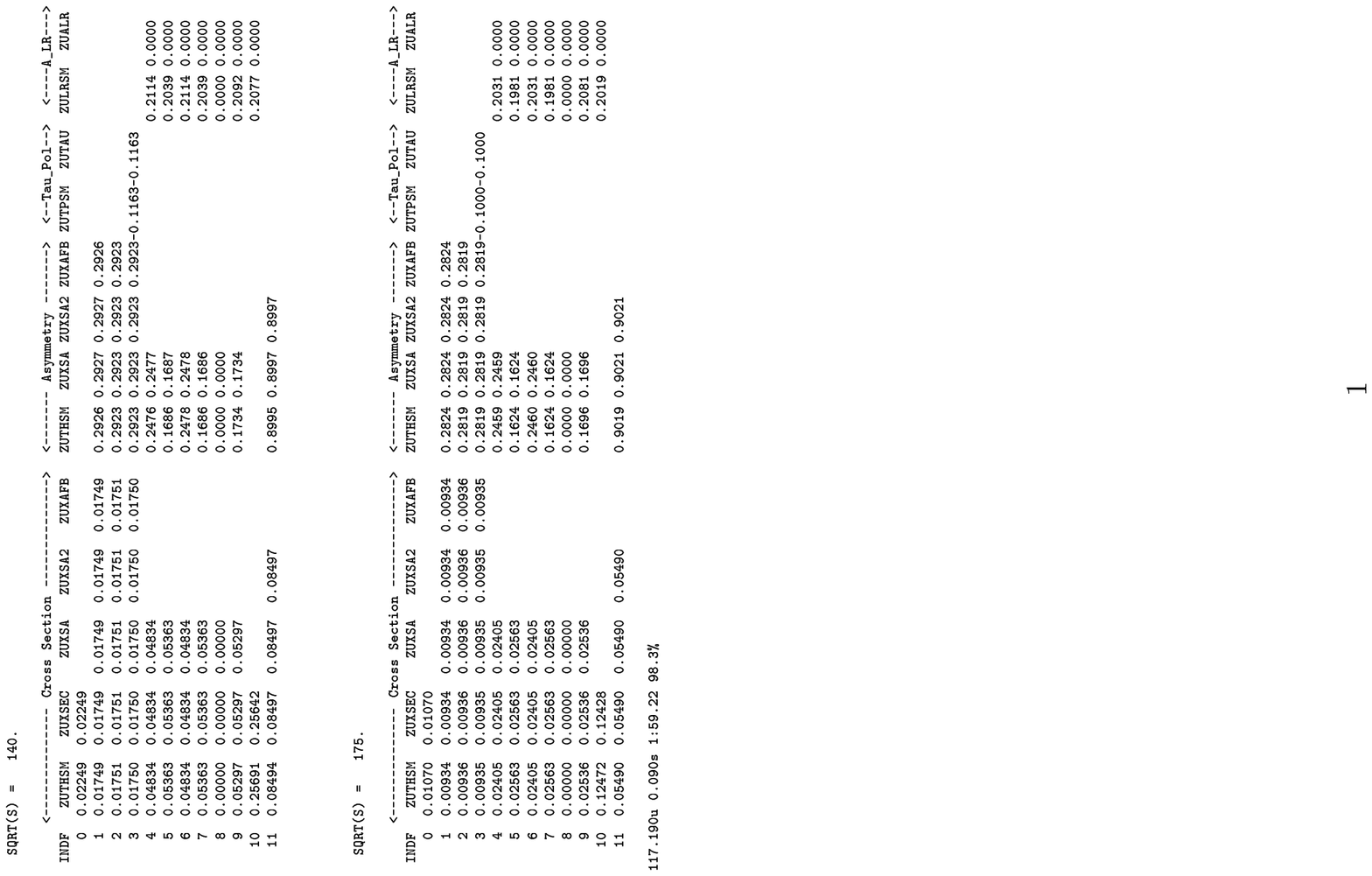
          ,height=35cm  
          ,width=35cm   
           ,angle=90   
      }}
\end{center}
\end{figure}

\clearpage
\def\href#1#2{#2}

\addcontentsline{toc}{chapter}{References}
\markboth{REFERENCES}{REFERENCES}
{\small 
\begingroup\begin{thebibliography}{100}

\bibitem{Bardin:1992jc1}
D.~Bardin, M.~Bilenky, A.~Chizhov, P.~Christova, O.~Fedorenko, M.~Jack,
  L.~Kalinovskaya, A.~Olshevsky, S.~Riemann, T.~Riemann, M.~Sachwitz,
  A.~Sazonov, Y.~Sedykh, I.~Sheer, and L.~Vertogradov, Fortran program {{\tt
  ZFITTER}}; \\ obtainable from {\tt afs/cern.ch/user/b/bardindy/public} \\ and
  from 
 {\tt http://www.ifh.de/$\sim$riemann/Zfitter/}.

\bibitem{Bardin:1989tq1}
A.~Akhundov, D.~Bardin, M.~Bilenky, P.~Christova, L.~Kalinovskaya, S.~Riemann,
  T.~Riemann, M.~Sachwitz, and H.~Vogt, Fortran program {{\tt DIZET}}.

\bibitem{Bardin:1980fet}
D.~Bardin, P.~Christova, and O.~Fedorenko, ``On the lowest order electroweak
  corrections to spin-$\frac{1}{2}$ fermion scattering {(I)}. {T}he one loop
  diagrammar'', {\em Nucl. Phys.} {\bf B175} (1980) 435.

\bibitem{Bardin:1982svt}
D.~Bardin, P.~Christova, and O.~Fedorenko, ``On the lowest order electroweak
  corrections to spin-$\frac{1}{2}$ fermion scattering {(II)}. {T}he one-loop
  amplitudes'', {\em Nucl. Phys.} {\bf B197} (1982) 1.

\bibitem{Bardin:1989tqt}
D.~Bardin, M.~Bilenky, P.~Christova, T.~Riemann, M.~Sachwitz, and H.~Vogt,
  ``{\tt DIZET}: A program package for the calculation of electroweak one loop
  corrections for the process $e^+ e^- \to f^+ f^-$ around the ${Z}^0$ peak'',
  {\em Comput. Phys. Commun.} {\bf 59} (1989) 303.

\bibitem{Bardin:1992jc}
D.~Bardin, M.~Bilenky, A.~Chizhov, O.~Fedorenko, S.~Ganguli, A.~Gurtu,
  M.~Lokajicek, G.~Mitselmakher, A.~Olshevsky, J.~Ridky, S.~Riemann, 
  T.~Riemann,
  M.~Sachwitz, A.~Sazonov, D.~Schaile, Y.~Sedykh, and L.~Vertogradov,
  ``{\tt ZFITTER} v.4.5: An analytical program for fermion pair production in
  $e^+e^-$ annihilation'', preprint CERN-TH. 6443/92 (1992),
  \href{http://xxx.lanl.gov/abs/hep-ph/9412201}{{\tt hep-ph/9412201}}.

\bibitem{zfitter:v6.21}
D.~Bardin, P.~Christova, M.~Jack, L.~Kalinovskaya, A.~Olshevski, S.~Riemann,
  and T.~Riemann, Fortran program {{\tt ZFITTER}} v.6.21 (26 July 1999).

\bibitem{Bardin:1989dit}
D.~Bardin, M.~S. Bilenky, G.~Mitselmakher, T.~Riemann, and M.~Sachwitz, ``A
  realistic approach to the standard ${Z}$ peak'', {\em Z. Phys.} {\bf C44}
  (1989) 493.

\bibitem{Bardin:1991fut}
D.~Bardin, M.~Bilenky, A.~Chizhov, A.~Sazonov, O.~Fedorenko, T.~Riemann, and
  M.~Sachwitz, ``Analytic approach to the complete set of {QED} corrections to
  fermion pair production in $e^+ e^-$ annihilation'', {\em Nucl. Phys.} {\bf
  B351} (1991) 1--48.

\bibitem{Bardin:1991det}
D.~Bardin, M.~Bilenky, A.~Sazonov, Y.~Sedykh, T.~Riemann, and M.~Sachwitz,
  ``{QED} corrections with partial angular integration to fermion pair
  production in $e^+ e^-$ annihilation'', {\em Phys. Lett.} {\bf B255} (1991)
  290--296.

\bibitem{Christova:1999cct}
P.~C. Christova, M.~Jack, and T.~Riemann, ``Hard photon emission in $e^+ e^-
  \to {\bar f} f$ with realistic cuts'', {\em Phys. Lett.} {\bf B456} (1999)
  264.

\bibitem{Akhundov:1986fct}
A.~Akhundov, D.~Bardin, and T.~Riemann, ``Electroweak one loop corrections to
  the decay of the neutral vector boson'', {\em Nucl. Phys.} {\bf B276} (1986)
  1.

\bibitem{Bardin:1989aa}
D.~Bardin and A.~Chizhov, ``On the ${O}(\alpha_{em}\alpha_s$) corrections to
  electroweak observables'', in {\em Proc. Int. Topical Seminar on Physics of
  $e^+e^-$ Interactions at {LEP} energies, 15-16 Nov 1988, JINR Dubna, USSR,
  {\rm JINR preprint E2-89-525 (1989)}} (D.~Bardin {\em et~al.}, eds.),
  pp.~42--48.

\bibitem{Bardin:1995a2}
D.~Bardin, G.~Passarino, and {W. Hollik (eds.)}, ``Reports of the working group
  on precision calculations for the {$Z$} resonance'', report CERN 95--03
  (1995).

\bibitem{Bardin:1989cwt}
D.~Bardin, M.~Bilenky, A.~Chizhov, A.~Sazonov, Y.~Sedykh, T.~Riemann, and
  M.~Sachwitz, ``The convolution integral for the forward - backward asymmetry
  in $e^+ e^-$ annihilation'', {\em Phys. Lett.} {\bf B229} (1989) 405.

\bibitem{Bardin:1991xet}
D.~Bardin, W.~Hollik, and T.~Riemann, ``Bhabha scattering with higher order
  weak loop corrections'', {\em Z. Phys.} {\bf C49} (1991) 485--490.

\bibitem{Kirsch:1995cf1}
M.~{Gr\"unewald}, S.~Kirsch, and T.~Riemann, Fortran program {{\tt SMATASY}}
  v.6.10 (27 May 1999), \\ obtainable from {\tt
  /afs/cern.ch/user/g/gruenew/public/smatasy/smata6$\_$10.fortran} \\ or from
  {\tt http://l3www.cern.ch/homepages/gruenew/welcome.html}.

\bibitem{Kirsch:1995cf}
S.~Kirsch and T.~Riemann, {\em Comp. Phys. Commun.} {\bf 88} (1995) 89--108.

\bibitem{Leike:1991pq}
A.~Leike, T.~Riemann, and J.~Rose, {\em Phys. Lett.} {\bf B273} (1991)
  513--518.

\bibitem{Riemann:1992gv}
T.~Riemann, {\em Phys. Lett.} {\bf B293} (1992) 451--456.

\bibitem{SRiemann:1997aa}
S.~Riemann, Fortran program {\tt ZEFIT}, obtainable from
   {\tt http://www.ifh.de/$\sim$riemanns/}.

\bibitem{Leike:1992uf}
A.~Leike, S.~Riemann, and T.~Riemann, {\em Phys. Lett.} {\bf B291} (1992)
  187--194.

\bibitem{Silin:19xy}
I.~N. Silin, Fortran subroutine {{\tt SIMPS}}.

\bibitem{Sedykh:19xy}
Y.~Sedykh, Fortran subroutine {{\tt FDSIMP}}.

\bibitem{Matsuura:1987}
T.~Matsuura, Fortran subroutines {{\tt TRILOG}} and {{\tt S12}}.

\bibitem{Jegerlehner:1995ZZ}
F.~Jegerlehner, Fortran function {\tt hadr5} (Feb 1995).

\bibitem{Eidelman:1995ny}
S.~Eidelman and F.~Jegerlehner, {\em Z. Phys.} {\bf C67} (1995) 585--602.

\bibitem{Burkhardt:1989ZZ}
H.~Burkhardt, Fortran function {\tt HADRQQ} (1989).

\bibitem{Burkhardt:1989ky}
H.~Burkhardt, F.~Jegerlehner, G.~Penso, and C.~Verzegnassi, {\em Z. Phys.} {\bf
  C43} (1989) 497.

\bibitem{Degrassi:1996ZZ}
G.~Degrassi, Fortran program {\tt m2tcor} (Oct 1996).

\bibitem{Degrassi:1994a0}
G.~Degrassi, S.~Fanchiotti, F.~Feruglio, B.~Gambino, and A.~Vicini, ``Two-loop
  electroweak top corrections: are they under control?'', in {\em Reports of
  the Working Group on Precision Calculations for the {$Z$} Resonance, {\rm
  report CERN 95--03 (1995)}} (D.~Bardin, W.~Hollik, and G.~Passarino, eds.),
  pp.~163--174.

\bibitem{Degrassi:1995ae}
G.~Degrassi, S.~Fanchiotti, F.~Feruglio, P.~Gambino, and A.~Vicini, ``Two loop
  corrections to the heavy top limit in the $\rho$ parameter'', New York Univ.
  preprint NYU-TH-01-04-95 (1995).

\bibitem{Degrassi:1995mc}
G.~Degrassi, F.~Feruglio, A.~Vicini, S.~Fanchiotti, and P.~Gambino, ``Two loop
  corrections for electroweak processes'', in {\em Proc. of the 30$^{th}$
  Rencontre de Moriond, Les Arcs, France, March 1995} (J.~{Tran Than Van},
  ed.), vol.~1, pp.~77--86, Editions Frontieres, 1995.

\bibitem{Degrassi:1996mg}
G.~Degrassi, P.~Gambino, and A.~Vicini, {\em Phys. Lett.} {\bf B383} (1996)
  219--226.

\bibitem{Degrassi:1997ps}
G.~Degrassi, P.~Gambino, and A.~Sirlin, {\em Phys. Lett.} {\bf B394} (1997)
  188--194.

\bibitem{Degrassi:1999jd}
G.~Degrassi and P.~Gambino, ``Two loop heavy top corrections to the ${Z}^0$
  boson partial widths'', Padua Univ. preprint DFPD-99-TH-19 (1999),
  \href{http://xxx.lanl.gov/abs/hep-ph/9905472}{{\tt hep-ph/9905472}}.

\bibitem{Arbuzov:1999uq}
A.~Arbuzov, ``Light pair corrections to $e^+e^-$ annihilation at LEP/SLC'',
  \href{http://xxx.lanl.gov/abs/hep-ph/9907500}{{\tt hep-ph/9907500}}.

\bibitem{Kniehl:1990yc}
B.~A. Kniehl, {\em Nucl. Phys.} {\bf B347} (1990) 86--104.




\bibitem{BardinPassarino:1999}
D. Bardin and G. Passarino,
``The Standard Model in the Making''
(Clarendon Press, Oxford, 1999).

\bibitem{Jack:1999af}
M.~Jack,
``QED radiative corrections to  $e^+ e^-\to {\bar f} f$ with realistic
cuts at LEP energies and beyond'',  
DESY 99-166 (1999), 
talk given at {\em 14th International Workshop on High Energy Physics and
Quantum Field Theory}, Moscow, Russia, 27 May - 2 Jun 1999, 
{\tt hep-ph/9911296}.


\bibitem{Christova:2000zu}
P.~Christova, M.~Jack, S.~Riemann and T.~Riemann,
``Radiative corrections to $e^+ e^-\to {\bar f} f$'', 
 LC-TH-2000-008 (2000), {\tt hep-ph/0002054};
to appear in: R. Heuer, F. Richard and P. Zerwas (eds.), ``Physics
Studies for a Future Linear Colider'', Proc. of {\em 2nd Joint ECFA/DESY
Study on Physics and Detectors for a Linear Electron-Positron
Collider}, held at Orsay, Lund, Frascati, Oxford, and Obernai from
April 1998 until Oct 1999, DESY report 123F (in preparation).


\bibitem{Bardin:1988xt}
D.~Bardin, A.~Leike, T.~Riemann, and M.~Sachwitz, {\em Phys. Lett.} {\bf B206}
  (1988) 539--542.

\bibitem{Christova:1999gh}
P.~Christova, M.~Jack, S.~Riemann, and T.~Riemann, ``{Predictions of {\tt
  {ZFITTER}} v.6 for fermion-pair production with acollinearity cut}'', DESY
  preprint 99-037 (1999), \href{http://xxx.lanl.gov/abs/hep-ph/9908289}{{\tt
  hep-ph/9908289}}.

\bibitem{Riemann:1989b}
T.~Riemann and Z.~Was, {\em Mod. Phys. Lett.} {\bf A4} (1989) 2487.

\bibitem{Bilenkii:1989zg}
M.~Bilenky and A.~Sazonov, ``{QED} corrections at {$Z^0$} pole with realistic
  kinematical cuts'', Dubna preprint JINR-E2-89-792 (1989).

\bibitem{Riemann:1988gy}
T.~Riemann, M.~Sachwitz, and D.~Bardin, ``The {Z} boson line shape at {LEP}'',
  in {\em Proc. of the 11th Warsaw Symp. on Elementary Particle Physics: New
  Theories in Physics, Kazimierz, Poland, May 23-27, 1988} (Z.~Ajduk,
  S.~Pokorski, and A.~Trautman, eds.), pp.~238--246, World Scientific,
  Singapore, 1989, Zeuthen preprint PHE 88-11 (1988), scanned image in KEK
  library, entry 8901077.

\bibitem{Bardin:1987hv}
D.~Bardin, O.~Fedorenko, and T.~Riemann, ``The electromagnetic ${O}(\alpha^3)$
  contributions to $e^+ e^-$ annihilation into fermions in the electroweak
  theory. {T}otal cross-section $\sigma_{T}$ and integrated asymmetry
  ${A}_{FB}$'', Dubna preprint JINR E2-87-663 (1987).

\bibitem{Bardin:1988ze}
D.~Bardin, M.~Bilenky, O.~Fedorenko, and T.~Riemann, ``The electromagnetic
  {$O(\alpha^3)$} contributions to $e^+ e^-$ annihilation into fermions in the
  electroweak theory. {T}otal cross-section {$\sigma_T$} and integrated
  asymmetry {$A_{FB}$}'', Dubna preprint JINR E2-88-324 (1988), scanned image
  in KEK library, entry 8808103.

\bibitem{Christova:2000x1}
P.~Christova, M.~Jack, and T.~Riemann, in preparation.

\bibitem{Passarino:1982zp}
G.~Passarino, {\em Nucl. Phys.} {\bf B204} (1982) 237--266.

\bibitem{Byckling:1973}
E.~Byckling and K.~Kajantie, {``Particle Kinematics''} (Wiley, London, 1973).

\bibitem{Bonneau:1971mkx}
G.~Bonneau and F.~Martin, {\em Nucl. Phys.} {\bf B27} (1971) 381.

\bibitem{Kuraev:1985}
E.~A. Kuraev and V.~S. Fadin, {\em Yad. Phiz. (in Russian)} {\bf 41} (1985)
  733--742.

\bibitem{Bardin:1989qr}
D.~Bardin {\em et~al.}, ``Z line shape'', in {\em Proc. of Workshop on Z
  Physics at LEP, Geneva, Switzerland, Feb 20-21 and May 8-9, 1989, {\rm report
  CERN 89--08 (1989)}} (G.~Altarelli, R.~Kleiss, and C.~Verzegnassi, eds.),
  vol.~1, pp.~89--128.

\bibitem{Skrzypek:1992vk}
M.~Skrzypek, {\em Acta Phys. Polon.} {\bf B23} (1992) 135.

\bibitem{Kniehl:1988id}
B.~A. Kniehl, M.~Krawczyk, J.~H. {K\"uhn}, and R.~G. Stuart, {\em Phys. Lett.}
  {\bf 209B} (1988) 337.

\bibitem{Bardin:1999gt}
D.~Bardin, M.~Gr{\"u}newald, and G.~Passarino, ``Precision calculation project
  report'', {\tt hep-ph/9902452}.

\bibitem{Jadach:1992aa}
S.~Jadach, M.~Skrzypek, and M.~Martinez, {\em Phys. Lett.} {\bf B280} (1992)
  129--136.

\bibitem{Berends:1988ab}
F.~A. Berends, G.~Burgers, and W.~L. van Neerven, {\em Nucl. Phys.} {\bf B297}
  (1988) 429; E: ibid., {\bf B304} (1988) 921.

\bibitem{Montagna:1997jv}
G.~Montagna, O.~Nicrosini, and F.~Piccinini, {\em Phys. Lett.} {\bf B406}
  (1997) 243--248.

\bibitem{Beenakker:1989km}
W.~Beenakker, F.~Berends, and W.~van Neerven, ``Applications of renormalization
  group methods to radiative corrections'', in {\em Proc. of the Int. Workshop
  on Radiative Corrections for $e^+e^-$ Collisions, Schlo\ss{} Ringberg,
  Tegernsee, Germany, April 1989} (J.~H. K{\"u}hn, ed.), p.~3, Springer,
  Berlin, 1989.

\bibitem{Vanc-Gruenewald:1998}
M.~W. Gr{\"u}newald, ``Combined Analysis of Precision Electroweak Results'',
  talk at ICHEP 98, July 1998, Vancouver, Canada, Berlin Humboldt-Univ.
  preprint HUB-EP-98/67 (1998).

\bibitem{home-LEPEWWG}
{LEPEWWG -- LEP electroweak working group}, {\tt http://www.cern.ch/LEPEWWG/}.

\bibitem{Caso:1998aa}
C.~Caso {\em et~al.}, {\em Eur. Phys. J.} {\bf C3} (1998) 1.

\bibitem{Kallen:1955ks}
G.~K{\"a}ll{\'e}n and A.Sabry, {\em K. Dan. Vidensk. Selsk. Mat.-Fys. Medd.}
  {\bf 17} (1955) 29.

\bibitem{Steinhauser:1998rq}
M.~Steinhauser, {\em Phys. Lett.} {\bf B429} (1998) 158--161.

\bibitem{Sirlin:1980nh}
A.~Sirlin, {\em Phys. Rev.} {\bf D22} (1980) 971--981.

\bibitem{Ross:1975fq}
D.~A. Ross and M.~Veltman, {\em Nucl. Phys.} {\bf B95} (1975) 135.

\bibitem{Veltman:1977kh}
M.~Veltman, {\em Nucl. Phys.} {\bf B123} (1977) 89.

\bibitem{Djouadi:1988di}
A.~Djouadi, {\em Nuovo Cim.} {\bf 100A} (1988) 357.

\bibitem{Halzen:1991je}
F.~Halzen and B.~A. Kniehl, {\em Nucl. Phys.} {\bf B353} (1991) 567--590.

\bibitem{Avdeev:1994db}
L.~Avdeev, J.~Fleischer, S.~Mikhailov, and O.~Tarasov, {\em Phys. Lett.} {\bf
  B336} (1994) 560--566, E: ibid., {\bf B349} (1995) 597.

\bibitem{Chetyrkin:1995ix}
K.~G. Chetyrkin, J.~H. {K\"uhn}, and M.~Steinhauser, {\em Phys. Lett.} {\bf
  B351} (1995) 331--338.

\bibitem{Fanchiotti:1991kc}
S.~Fanchiotti and A.~Sirlin, ``Higher order contributions to {$\Delta r$}'',
  New York Univ. preprint NYU-TH-91-02-04 (1991).

\bibitem{Consoli:1989pc}
M.~Consoli, W.~Hollik, and F.~Jegerlehner, ``Electroweak radiative corrections
  for {$Z$} physics'', in {\em Proc. of Workshop on Z Physics at LEP, Geneva,
  Switzerland, Feb 20-21 and May 8-9, 1989, {\rm report CERN 89--08 (1989)}}
  (G.~Altarelli, R.~Kleiss, and C.~Verzegnassi, eds.), vol.~1, p.~7.

\bibitem{Barbieri:1992nz}
R.~Barbieri, M.~Beccaria, P.~Ciafaloni, G.~Curci, and A.~Vicere, {\em Phys.
  Lett.} {\bf B288} (1992) 95--98.

\bibitem{Fleischer:1993ub}
J.~Fleischer, O.~V. Tarasov, and F.~Jegerlehner, {\em Phys. Lett.} {\bf B319}
  (1993) 249--256.

\bibitem{Degrassi:1998oo}
G.~Degrassi and P.~Gambino, private communication.

\bibitem{vanderBij:1984bw}
J.~van~der Bij and M.~Veltman, {\em Nucl. Phys.} {\bf B231} (1984) 205.

\bibitem{vanderBij:1984aj}
J.~J. van~der Bij, {\em Nucl. Phys.} {\bf B248} (1984) 141.

\bibitem{Kniehl:1995yr}
B.~A. Kniehl, ``Estimation of higher-order {QCD} effects on electroweak
  parameters'', in {\em Reports of the Working Group on Precision Calculations
  for the {$Z$} Resonance, {\rm report CERN 95--03 (1995)}} (D.~Bardin,
  W.~Hollik, and G.~Passarino, eds.), pp.~299--312.

\bibitem{Sirlin:1995yr}
A.~Sirlin, ``On the {QCD} corrections to $\delta \rho$'', in {\em Reports of
  the Working Group on Precision Calculations for the {$Z$} Resonance, {\rm
  report CERN 95--03 (1995)}} (D.~Bardin, W.~Hollik, and G.~Passarino, eds.),
  pp.~285--298.

\bibitem{Chetyrkin:1994js3}
K.~Chetyrkin, J.~K{\"u}hn, and A.~Kwiatkowski, ``{{QCD}} corrections to the
  $e^+ e^-$ cross-section and the {Z} boson decay rate'', in {\em Reports of
  the Working Group on Precision Calculations for the {$Z$} Resonance, {\rm
  report CERN 95--03 (1995)}} (D.~Bardin, W.~Hollik, and G.~Passarino, eds.),
  pp.~175--263.

\bibitem{Czarnecki:1996ei}
A.~Czarnecki and J.~H. K{\"u}hn, {\em Phys. Rev. Lett.} {\bf 77} (1996)
  3955--3958.

\bibitem{Harlander:1998zb}
R.~Harlander, T.~Seidensticker, and M.~Steinhauser, {\em Phys. Lett.} {\bf
  B426} (1998) 125--132.

\bibitem{Bardin:1986fi}
D.~Bardin, S.~Riemann, and T.~Riemann, {\em Z. Phys.} {\bf C32} (1986) 121.

\bibitem{Mann:1984}
G.~Mann and T.~Riemann, {\em Annalen Phys.} {\bf 40} (1984) 334.

\bibitem{Beenakker:1988pv}
W.~Beenakker and W.~Hollik, {\em Z. Phys.} {\bf C40} (1988) 141.

\bibitem{Bernabeu:1988me}
J.~Bernab{\'e}u, A.~Pich, and A.~Santamaria, {\em Phys. Lett.} {\bf 200B}
  (1988) 569.

\bibitem{Bernabeu:1991ws}
J.~Bernab{\'e}u, A.~Pich, and A.~Santamaria, {\em Nucl. Phys.} {\bf B363}
  (1991) 326--344.

\bibitem{Buchalla:1993zm}
G.~Buchalla and A.~J. Buras, {\em Nucl. Phys.} {\bf B398} (1993) 285--300.

\bibitem{Fleischer:1992fq}
J.~Fleischer, O.~V. Tarasov, F.~Jegerlehner, and P.~Raczka, {\em Phys. Lett.}
  {\bf B293} (1992) 437--444.

\bibitem{Degrassi:1993ij}
G.~Degrassi, {\em Nucl. Phys.} {\bf B407} (1993) 271--289.

\bibitem{Chetyrkin:1993jp}
K.~G. Chetyrkin, A.~Kwiatkowski, and M.~Steinhauser, {\em Mod. Phys. Lett.}
  {\bf A8} (1993) 2785--2792.

\bibitem{Riemann:1997tj}
T.~Riemann, ``The {$Z$} boson resonance parameters'', in {\em Irreversibility
  and Causality, Lecture Notes in Physics, vol. 504} (A.~Bohm {\em et~al.},
  eds.), pp.~157--177, Springer, Berlin, 1998, {\tt hep-ph/9709208}.

\bibitem{Bardin:1997xq}
D.~Bardin {\em et~al.}, ``Electroweak working group report'', in {\em Reports
  of the Working Group on Precision Calculations for the {$Z$} Resonance, {\rm
  report CERN 95--03 (1995)}} (D.~Bardin, W.~Hollik, and G.~Passarino, eds.),
  pp.~7--162, {\tt hep-ph/9709229}.

\bibitem{LKreptoLEP98}
L.~Kalinovskaya, {``Finite top mass effects in $e^+e^- \to {\bar f} f$ at LEP
  energies''}, talk at LEPEWWG meeting, 30 Oct 1998, obtainable from \\
   {\tt
http://www.ifh.de/$\sim$riemann/Zfitter/%
LK$\_$LEPEWWG$\_$30$\_$10$\_$98/}.


\bibitem{Boudjema:1996qg}
E.~Accomando {\em et~al.}, ``Standard model processes'', in {\em Physics at
  {LEP2}, {\rm report CERN 96--01 (1996)}} (G.~Altarelli, T.~Sj{\"o}strand, and
  F.~Zwirner, eds.), pp.~207--248, 1996, {\tt hep-ph/9601224}.

\bibitem{Djouadi:1987gn}
A.~Djouadi and C.~Verzegnassi, {\em Phys. Lett.} {\bf 195B} (1987) 265.

\bibitem{Arbuzov:1992pr}
A.~B. Arbuzov, D.~Bardin, and A.~Leike, {\em Mod. Phys. Lett.} {\bf A7} (1992)
  {2029--2038, E: ibid. {{\bf A9} (1994) 1515}}.

\bibitem{Kataev:1992dg}
A.~L. Kataev, {\em Phys. Lett.} {\bf B287} (1992) 209--212.

\bibitem{Chetyrkin:1979bj}
K.~G. Chetyrkin, A.~L. Kataev, and F.~V. Tkachev, {\em Phys. Lett.} {\bf 85B}
  (1979) 277.

\bibitem{Dine:1979qh}
M.~Dine and J.~Sapirstein, {\em Phys. Rev. Lett.} {\bf 43} (1979) 668.

\bibitem{Celmaster:1980xr}
W.~Celmaster and R.~J. Gonsalves, {\em Phys. Rev. Lett.} {\bf 44} (1980) 560.

\bibitem{Gorishnii:1991hw}
S.~G. Gorishny, A.~L. Kataev, and S.~A. Larin, {\em Phys. Lett.} {\bf B273}
  (1991) 141--144.

\bibitem{SRiemann:1991}
S.~Riemann, ``A comparison of programs used in {L3} for the analysis of
  {B}habha scattering'', Zeuthen preprint PHE 91--04 (1991),
  scanned image in KEK library, entry 9105486.

\bibitem{Beenakker:1991??}
W.~Beenakker, ``Theoretical developments in large angle {B}habha scattering'',
  in {\em Proc. of the Joint Int. Symposium on Lepton and Photon Interactions
  and Europhysics Conference on High Energy Physics, Geneva, Switzerland, 25
  July - 1 August 1991} (S.~Hegarty {\em et~al.}, eds.), vol.~1, p.~28, World
  Scientific, Singapore, 1992.

\bibitem{Bardin:1990uu}
D.~Bardin and T.~Riemann, Fortran package {\tt BHASHA} (1990), unpublished.

\bibitem{Passarino:1979jh}
G.~Passarino and M.~Veltman, {\em Nucl. Phys.} {\bf B160} (1979) 151.

\bibitem{Bardin:1984B}
D.~Bardin and V.~Dokuchaeva, {\em Nucl. Phys.} {\bf B246} (1984) 221.

\bibitem{Bardin:1986B}
D.~Bardin and V.~Dokuchaeva, ``On the radiative corrections to the neutrino
  deep inelastic scattering'', Dubna preprint JINR-E2-86-260 (1986).

\bibitem{Akhundov:1996}
A.~Akhundov, D.~Bardin, L.~Kalinovskaya, and T.~Riemann, {\em Fortsch. Phys.}
  {\bf 44} (1996) 373--482.

\bibitem{Arbuzov:1995id}
A.~Arbuzov, D.~Bardin, J.~Bl{\"u}mlein, L.~Kalinovskaya, and T.~Riemann, {\em
  Comput. Phys. Commun.} {\bf 94} (1996) 128.

\bibitem{Jadach:1994yv}
S.~Jadach, B.~F.~L. Ward, and Z.~Was, {\em Comput. Phys. Commun.} {\bf 79}
  (1994) 503--522.

\bibitem{Field:1996}
J.~H. Field and T.~Riemann, {\em Comput. Phys. Commun.} {\bf 94} (1996) 53--87.

\bibitem{Jegerlehner:1991dq}
F.~Jegerlehner, ``Renormalizing the standard model'', in {\em Testing the
  Standard Model: Proceedings of the Theoretical Advanced Study Institute in
  Elementary Particle Physics (TASI), Boulder, CO, Jun 3-29, 1990} (M.~Cvetic
  and P.~Langacker, eds.), pp.~476--590, World Scientific, Teaneck, N.J., 1991.

\bibitem{Jegerlehner:1991ed}
F.~Jegerlehner, {\em Prog. Part. Nucl. Phys.} {\bf 27} (1991) 1--76.

\bibitem{vanderBij:1987hy}
J.~J. van~der Bij and F.~Hoogeveen, {\em Nucl. Phys.} {\bf B283} (1987) 477.

\bibitem{Barbieri:1993ra}
R.~Barbieri, P.~Ciafaloni, and A.~Strumia, {\em Phys. Lett.} {\bf B317} (1993)
  381.

\bibitem{Barbieri:1999bbo}
R.~Barbieri, private communication.

\bibitem{Berman:1958}
S.~M. Berman, {\em Phys. Rev.} {\bf 112} (1958) 267.

\bibitem{Kinoshita:1959}
T.~Kinoshita and A.~Sirlin, {\em Phys. Rev.} {\bf 113} (1959) 1652.

\bibitem{Kallen:1968}
G.~K{\"a}llen, {\em Springer Tracts in Modern Physics} {\bf 46} (1968) 67--132.

\bibitem{vanRitbergen:1998yd}
T.~van Ritbergen and R.~G. Stuart, {\em Phys. Rev. Lett.} {\bf 82} (1999) 488.

\bibitem{vanRitbergen:1998hn}
T.~van Ritbergen and R.~G. Stuart, {\em Phys. Lett.} {\bf B437} (1998) 201.

\bibitem{Christova:1998tc}
P.~Christova, M.~Jack, S.~Riemann, and T.~Riemann, ``Predictions for
  fermion-pair production at LEP'', preprint DESY 98-184 (1998), to appear in
  the proceedings of RADCOR98, Sep 8-12, 1998, Barcelona, Spain,
  \href{http://xxx.lanl.gov/abs/hep-ph/9812412}{{\tt hep-ph/9812412}}.

\end{thebibliography}\endgroup

}
\end{document}